\documentclass[brazil,12pt,a4paper,openany]{book}
\usepackage{anysize}
\usepackage[utf8]{inputenc}
\usepackage[T1]{fontenc}
\usepackage[ruled,chapter]{algorithm}
\usepackage{makeidx}
\usepackage{subfig}
\usepackage{amsmath}
\usepackage{amssymb}
\usepackage{amsthm}
\usepackage{mathrsfs}
\usepackage{graphicx}
\usepackage{fancyhdr}
\usepackage{array}
\usepackage{geometry}
\usepackage{simplewick}
\usepackage{latexsym}
\usepackage[all]{xy}
\usepackage{eufrak}
\usepackage{euscript}
\usepackage{enumerate}
\usepackage{dsfont}
\usepackage{slashed}
\usepackage{booktabs}
\usepackage{xcolor}
\usepackage{titlesec}
\usepackage{enumitem}
\usepackage{acronym}
\usepackage{mathtools}
\usepackage{braket}

\makeindex

\marginsize{25mm}{25mm}{25mm}{25mm}

\definecolor{myblack}{rgb}{0.334,0.334,0.334}

\usepackage[citecolor=true,plainpages=false,pdftex,linkcolor=myblack,citecolor=blue,colorlinks=true,breaklinks=true,urlcolor=magenta]{hyperref}  

\usepackage[style=phys,biblabel=brackets,backref,hyperref = true]{biblatex}
\DefineBibliographyStrings{english}{%
  backrefpage = {cited on pp:},
  backrefpages = {cited on pps:},
}

\begin{document}

\setlength{\abovecaptionskip}{0.0cm}
\setlength{\belowcaptionskip}{0.0cm}
\setlength{\baselineskip}{24pt}

\pagestyle{fancy}
\lhead{}
\chead{}
\rhead{\thepage}
\lfoot{}
\cfoot{}
\rfoot{}

\fancypagestyle{plain}
{
	\fancyhf{}
	\lhead{}
	\chead{}
	\rhead{\thepage}
	\lfoot{}
	\cfoot{}
	\rfoot{}
}

\renewcommand{\headrulewidth}{0pt}
\newcommand\vc[1]{\vec{#1}}
\newcommand\lr[1]{\left({#1}\right)}
\newcommand\mean[1]{\left\langle{#1}\right\rangle}
\newcommand\refp[1]{(\ref{#1})}
\newcommand\be{\begin{equation}}
\newcommand\ee{\end{equation}}
\newcommand{\reffig}[1]%
{\hypersetup{linkcolor=myblack}
\ref{#1}%
\hypersetup{linkcolor=myblack}}
\newcommand\refcap[1]
{\hypersetup{linkcolor=teal}%
\ref{#1}%
\hypersetup{linkcolor=teal}}
\newcommand\refsec[1]
{\hypersetup{linkcolor=teal}%
\ref{#1}%
\hypersetup{linkcolor=teal}}
\newcommand\reftab[1]
{\hypersetup{linkcolor=teal}%
\ref{#1}%
\hypersetup{linkcolor=teal}}

\newcommand{\bea}{\begin{eqnarray}} 
\newcommand{\eea}{\end{eqnarray}}
\newcommand{\bfr}{{\bf{r}}}
\newcommand{\bfk}{{\bf{k}}}
\newcommand{\tg}{\tilde\Gamma}
\newcommand{\xidr}{\xi_{\md_{R}}}
\newcommand{\vare}{\varepsilon}
\newcommand{\md}{\mathcal{D}}
\newcommand{\mc}{\mathcal{C}}
\newcommand\norm[2]{\left\Vert {#1},{#2}\right\Vert}
\newcommand\conv[2]{\lr{{#1}\star {#2}}}

\colorlet{linkequation}{red}

\newcommand*{\SavedEqref}{}
\let\SavedEqref\eqref
\renewcommand*{\eqref}[1]{%
  \begingroup
    \hypersetup{
      linkcolor=linkequation,
      linkbordercolor=linkequation,
    }%
    \SavedEqref{#1}%
  \endgroup
}


\frontmatter 

\thispagestyle{empty}

\begin{figure}[h]
	\includegraphics[scale=0.8]{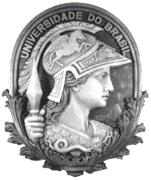}
\end{figure}

\vspace{15pt}

\begin{center}

\textbf{UNIVERSIDADE FEDERAL DO RIO DE JANEIRO}

\textbf{INSTITUTO DE FÍSICA}

\vspace{30pt}

{\Large \bf Circulation Statistics in Homogeneous and Isotropic Turbulence}

\vspace{25pt}

{\large \bf Victor de Jesus Valadão}

\vspace{35pt}

\begin{flushright}
\parbox{10.3cm}{Tese de Doutorado apresentada ao Programa de Pós-Graduação em Física do Instituto de Física da Universidade Federal do Rio de Janeiro - UFRJ, como parte dos requisitos necessários à obtenção do título de Doutor em Ciências (Física).

\vspace{18pt}

{\large \bf Orientador: Luca Moriconi}}
\end{flushright}

\vspace{90pt}

\textbf{Rio de Janeiro}

\textbf{Setembro de 2023}

\end{center}
\newpage Folha de aprovação
\newpage Ficha catalográfica


\newpage

\noindent

\vspace*{20pt}
\begin{center}
{\LARGE\bf Resumo}\\
\vspace{15pt}
{\Large\bf Estatística da Circulação em Turbulência Homogênea e Isotrópica}\\
\vspace{6pt}
{\bf Victor de Jesus Valadão}\\
\vspace{12pt}
{\bf Orientador: Luca Roberto Augusto Moriconi}\\
\vspace{20pt}
\parbox{14cm}{Resumo da Tese de Doutorado apresentada ao Programa de Pós-Graduação em Física do Instituto de Física da Universidade Federal do Rio de Janeiro - UFRJ, como parte dos requisitos necessários à obtenção do título de Doutor em Ciências (Física).}
\end{center}
\vspace*{35pt}

A turbulência é um estado de movimento caótico e imprevisível de fluidos que ocorre naturalmente e desafia previsões simples, caracterizado pelo aparecimento de estruturas vorticosas intensas (turbilhões) e flutuações que abrangem uma ampla faixa de escalas. No entanto, uma fundamentação teórica abrangente que conecte de maneira sucinta as equações da dinâmica dos fluidos às complexas propriedades estatísticas da turbulência continua a ser um desafio formidável. Esta tese tem como objetivo explorar o problema da turbulência homogênea e isotrópica à luz de uma variável historicamente esquecida: a circulação da velocidade.

Os capítulos iniciais exploram o contexto histórico da equação de Navier-Stokes, traçando seu desenvolvimento e enfatizando sua importância ao exigir uma formulação estatística para lidar com fenômenos turbulentos. O texto revisa a teoria estatística inicial da turbulência apresentada por Kolmogorov em 1941, observando tanto suas conquistas quanto suas limitações devido à omissão dos efeitos de flutuação. A emergência de estruturas alternativas, como modelos de cascata multiplicativa e várias abordagens multifractais são abordadas. Ênfase especial é dada à abordagem do Caos Multiplicativo Gaussiano (GMC).

A seção seguinte revela os resultados derivados de uma análise estatística abrangente da circulação de velocidade em simulações numéricas. Essa análise vai desde suas primeiras investigações, na metade da década de 90, até as tendências atuais nesse campo de pesquisa. Para validar os diversos aspectos fenomenológicos das estatísticas de circulação em contornos planares, uma análise numérica de um banco de dados turbulento de acesso público é realizada.

O desenvolvimento do Modelo de Gás de Vórtices (VGM) introduz uma novo arcabouço estatístico para descrever as flutuações de circulação. Este modelo é baseado em atributos estatísticos de dois constituintes fundamentais. O primeiro é um campo GMC que governa o comportamento intermitente, e o segundo é um campo Gaussiano Livre responsável pela polarização parcial dos vórtices no gás.

O modelo é revisitado em uma linguagem mais sofisticada, para incluir efeitos de vexclusão de volume entre vórtices. Essas modificações foram posteriormente validadas por meio de simulações numéricas das equações de Navier-Stokes no regime turbulento. Tal abordagem revisada harmoniza com as características multifractais exibidas pelas estatísticas de circulação, oferecendo uma explicação convincente para o fenômeno de linearização dos momentos estatísticos desta variável, observado nas simulações numéricas mais recentes.

No final, é realizada uma abordagem teórica de campo, conhecida como método funcional Martin-Siggia-Rose-Janssen-de Dominicis (MSRJD), no contexto da função de densidade de probabilidade de circulação. Essa abordagem explora os eventos extremos de circulação, frequentemente referidos como Instantons, por meio de duas metodologias distintas: a primeira investiga as soluções lineares e, por meio de um argumento de grupo de renormalização, discute uma simetria de rescala temporal. Em segundo lugar, uma estratégia numérica é implementada para lidar com as equações não-lineares de instanton na aproximação axial. Essa abordagem resulta na obtenção da topologia dos campos de velocidade responsáveis por eventos extremos de circulação.

\vspace{15pt}

\textbf{Palavras-chave:} Turbulência, Circulação, Gás de Vórtices, Caos Multiplicativo Gaussiano, Quebra de Multifractalidade, Martin-Siggia-Rose-Janssen-de Dominicis, Eventos Extremos, Instantons.


\newpage

\noindent

\vspace*{20pt}
\begin{center}
{\LARGE\bf Abstract}\\
\vspace{15pt}
{\Large\bf Circulation Statistics in Homogeneous and Isotropic Turbulence}\\
\vspace{6pt}
{\bf Victor de Jesus Valadão}\\
\vspace{12pt}
{\bf Orientador: Luca Roberto Augusto Moriconi}\\
\vspace{20pt}
\parbox{14cm}{\emph{Abstract} da Tese de Doutorado apresentada ao Programa de Pós-Graduação em Física do Instituto de Física da Universidade Federal do Rio de Janeiro - UFRJ, como parte dos requisitos necessários à obtenção do título de Doutor em Ciências (Física).} 
\end{center}
\vspace*{35pt}

Turbulence is a natural state of chaotic and unpredictable fluid motion that defies simple prediction, characterized by the formation of intense vortex structures (eddies) and fluctuations across a wide range of scales. However, a comprehensive theoretical framework that intricately links the governing fluid dynamics equations to the complex statistical properties of turbulence remains a formidable challenge. This thesis aims to explore the problem of homogeneous and isotropic turbulent flows through the light of a historically forgotten variable: The velocity circulation.

The opening chapters explore the historical context of the Navier-Stokes equation, tracing its development and emphasizing its significance in requiring a statistical formulation when dealing with turbulent phenomena. The text reviews the initial statistical theory of turbulence presented by Kolmogorov in 1941, noting both its achievements and its drawbacks due to the omission of fluctuation effects. The emergence of alternative frameworks, such as multiplicative cascade models, and various multifractal approaches. Special emphasis is given to the Gaussian Multiplicative Chaos (GMC) approach.

The following section unveils the outcomes derived from a comprehensive statistical analysis of velocity circulation in numerical simulations. This examination spans from its initial investigation in the mid-1990s to the current trends in this field of research. To validate the various phenomenological aspects of circulation statistics on planar shapes, a numerical analysis of a publicly accessible turbulent database is carried out.

The development of the Vortex Gas Model (VGM) introduces a novel statistical framework for describing the characteristics of velocity circulation. In this model, the underlying foundations rely on the statistical attributes of two fundamental constituents. The first is a GMC field that governs intermittent behavior and the second constituent is a Gaussian Free field responsible for the partial polarization of the vortices in the gas.

The model is revisited in a more sophisticated language, where volume exclusion among vortices is addressed. These additions were subsequently validated through numerical simulations of turbulent Navier-Stokes equations. This revised approach harmonizes with the multifractal characteristics exhibited by circulation statistics, offering a compelling elucidation for the phenomenon of linearization of the statistical circulation moments, observed in recent numerical simulation.

In the end, a field theoretical approach, known as Martin-Siggia-Rose-Janssen-de Dominicis (MSRJD) functional method is carried out in the context of circulation probability density function. This approach delves into the realm of extreme circulation events, often referred to as Instantons, through two distinct methodologies: The First investigates the linear solutions and, by a renormalization group argument a time-rescaling symmetry is discussed. Secondly, a numerical strategy is implemented to tackle the nonlinear instanton equations in the axisymmetric approximation. This approach addresses the typical topology exhibited by the velocity field associated with extreme circulation events.

\vspace{15pt}



\newpage

\noindent

\vspace*{20pt}

\begin{center}

{\LARGE\bf Agradecimentos}

\end{center}

\vspace*{40pt}

Existem tantas pessoas à quem quero agradecer que não consigo imaginar melhor lugar para se começar. Dito isto, agradeço à minha noiva Suellen Freitas pelo apoio emocional, companheirismo e tudo que fez por mim nesses anos que perduraram meu doutorado. À você, Suellen, obrigado! 

Agradeço também aos meus pais Cristina e Francisco, meus irmãos Leandro e Renato e minhas cunhadas Daniele e Gessica pelo apoio financeiro, uma triste necessidade real daqueles que vêem de baixa renda e optam pela carreira científica. Agradeço também pela confiança na minha capacidade ao longo destes anos de estudo. Este parágrafo de agradecimento também se aplica a minha sogra Sueli, uma segunda mãe que a vida me deu.

À minha sobrinha Ana Luiza e minha sobrinha/afilhada Giovanna, apesar de não entenderem oque faço espero que, ainda asssim, eu consiga servir de inspiração para vocês nessa infância.
 
Agradeço, ao Prof. Luca Moriconi por não somente me orientar, mas me guiar nessa jornada da científica que pode ser ardua em muitos momentos. Juntamente à ele agradeço ao Prof. Rodrigo Pereira por me ensinar e me apoiar em todos os momentos desse processo. À Gabriel Apolinário, agradeço as colaborações frutiferas e o grande acolhimento e paciência nos meus primeiros meses de estudo, à quem pedi muita ajuda.

Agradeço também aos companheiros de grupo aos quais compartilhei muitas experiencias durante nossos journal clubs semanais: Rodrigo Arouca, Elvis Soares, Leonardo Grigório, Mirlene Oliveira, Hugo Tavares e Juliana Lemos. Para Maiara Neumann e Bruno Magacho, deixo um agradecimento especial pois compartilhamos muitas experiencia sendo companheiros de escritório.

Aos inumeros amigos que fiz na graduação da UERJ, obrigado. Especialmente à Pedro Paraguassú, Rui Aquino, Vinicius Greenhalgh, Juan Leite, Luiz Felipe, Marcelle Lopes, Mylla Coffaro e, especialmente, Lucas Falcão e Pedro Bessa.

Agradeço à todos os meus amigos de longa data do ETEFV por, apesar de tomarmos caminhos bem diversos, continuarmos nossa amizade depois de tantos anos. Em especial, neste grupo, gostaria de agradecer à Leonardo Machado e Matheus Martins, dois irmãos que a vida me deu.

Aos funcionários Igor Silva, Khrisna Teixeira e Gustavo Rubini da secretaria da pós graduação, agradeço todo apoio em questões técnicas e administrativas e muitas outras soluções que não seriam possíveis sem a ajuda de vocês.

Agradeço também ao meu felino Landina, que apesar de sempre atrapalhar minhas reuniões online, sempre está comigo.

Por fim, agradeço à CAPES pelo suporte financeiro, sem o qual este doutoramento
não teria sido possível.


\newpage
\phantomsection
\addcontentsline{toc}{chapter}{Contents}
\tableofcontents

\newpage
\phantomsection
\addcontentsline{toc}{chapter}{List of Figures}
\listoffigures

\newpage
\phantomsection
\addcontentsline{toc}{chapter}{List of Tables}
\listoftables

\newpage
\phantomsection
\begin{chapter}{List of Abbreviations}
\setlength{\baselineskip}{16pt}
\begin{acronym}[MSRJD] 
    \acro{cPDF}{Circulation Probability Density Function}
    \acro{DNS}{Direct Numerical Simulation}
    \acro{ESS}{Extended Self-Similatiry}
    \acro{FFT}{Fast Fourier Transform}
    \acro{GFF}{Gaussian Free Field}
    \acro{GMC}{Gaussian Multiplicative Chaos}
    \acro{iNSE}{Incompressible Navier-Stokes Equations}
    \acro{JHTDB}{Johns Hopkins Turbulence Database}
    \acro{K41}{Kolmogorov's Theory of 1941}
    \acro{LHS}{left-hand side}
    \acro{MCM}{Multiplicative Cascade Model}
    \acro{mOK62}{modified OK62 model}
    \acro{MSRJD}{Martin-Siggia-Rose-Jansen-de Dominicis}
    \acro{OK62}{Obukhov-Kolmogorov's Theory of 1962}
    \acro{PDF}{Probability Density Function}
    \acro{RHS}{right-hand side}
    \acro{SSC}{Swirling Strength Criterion}
    \acro{VGM}{Vortex Gas Model}

\end{acronym}

\end{chapter}
\counterwithout{footnote}{chapter}

\mainmatter
\begin{chapter}{Introduction}\label{cap0}

\hspace{5 mm} 

Turbulence is a complex and fascinating fluid dynamics phenomenon, ubiquitously observed in nature. It refers to the chaotic, irregular, and often unpredictable motion of a fluid (liquid or gas) characterized by rapid changes in velocity, pressure, and other flow properties. Turbulence can be observed in various natural and man-made systems, such as water flows in rivers and oceans \cite{lueck2002oceanic}, atmospheric circulation \cite{wyngaard1992atmospheric}, combustion processes \cite{kerstein2002turbulence}, and even the mixing of ingredients in a cup of coffee.

The matter of defining turbulence is much easier by simply listing some key aspects expected in a three-dimensional fluid flow that displays turbulent behavior:
\begin{enumerate}[label=\textnormal{(P.\arabic*)}]
    \item Irregular Motion: Turbulent flows are highly irregular and unpredictable. Unlike laminar (smooth and ordered) flows, where fluid particles move along well-defined paths, turbulent flows involve intricate interactions between countless vortices, eddies, and swirls of different scales.    
    \item Enhanced Diffusion: Turbulence increases the effective diffusion of particles or properties within a fluid. This is because the rapid and random motion of fluid particles ensures that they come into contact with each other more frequently than in laminar flows.       
    \item Reynolds Number: The transition from laminar to turbulent flow is characterized by a dimensionless quantity called the Reynolds number. It represents the ratio of inertial forces to viscous forces in a fluid flow. When the Reynolds number exceeds a critical value, turbulence becomes dominant.
    \item Energy Cascades: Turbulence involves a transfer of energy across different scales. Larger vortices and eddies break down into smaller ones through a process known as energy cascade. This cascade continues until the kinetic energy is dissipated into heat due to the viscosity of the fluid.\label{pcasc}
    \item Structure Formation: Thin and elongated vortex filaments often form spontaneously due to the nonlinear interactions between different scales of motion. In this sense, turbulence can be thought of as a tangle of interacting vortices. \label{pstruc}
\end{enumerate}

The study of turbulence was historically introduced by the dynamics of a very standard variable, the velocity field. Instead, this thesis aims to offer a complementary view of turbulence through the statistics of velocity circulation. However, given the restricted number of scientists (in particular physicists) who are familiar with turbulence theory, we include in Chapters~\refcap{cap1} and~\refcap{cap2} an extensive introductory study of fluid dynamics and turbulence, respectively.

We encourage readers with prior background in turbulence to start the reading from Section~\refsec{sec2.4.4} where we introduce a mathematically formal approach to the well-known extension made by Obukhov and Kolmogorov in 1962 to the first theory of turbulence introduced by Kolmogorov in 1941.

Studying velocity circulation in turbulence has been a challenging and complex endeavor that took a significant amount of time to become feasible. The lack of theoretical tools and techniques in the early stages of turbulence studies triggered the scientific community to apply traditional fluid dynamics methods, such as velocity and/or vorticity field analysis \cite{acheson1990book}. From the experimental point of view, there is no known way to directly measure the circulation of arbitrary contours in fluid flows. In this sense, circulation is a secondary variable in a turbulence experiment. Hardware improvement in the past 30 years has not only benefited the study of circulation through numerical simulations of turbulent flows \cite{iyer2019circulation} but has also enhanced the acquisition and processing of experimental data with more accuracy and detail \cite{zhu2023circulation}. However, even with these advancements, turbulence simulations remain computationally intensive.

The historical development of the study of circulation statistics, its advantages, and drawbacks, as well as its most important results, are presented in Chapter~\refcap{cap3} through numerical processing of an open-access database.

The original contribution of this thesis to turbulence research begins in Chapter~\refcap{cap4}. Through the study of circulation statistics, we present for the first time, a unified perspective connecting multiplicative energy cascades \ref{pcasc} and the distribution of vortical structures \ref{pstruc} within turbulent flows. A model referred to as the \ac{VGM} \cite{apolinario2020vortex}, is a purely analytical tool that correctly accounts for low-order circulation statistics. The model is introduced in detail and the derivation of statistical quantities such as variance, kurtosis (or flatness), higher-order moment approximations, and probability density functions are discussed in mathematical detail. Moreover, refinements of the model, which can only be realized numerically, inspired the search for self-excluding behavior of the vortex structures at small scale \cite{moriconi2022circulation}, motivating us to reinterpret some of the results put forward in Chapter~\refcap{cap3}.

Finally, Chapter~\refcap{cap5} introduces a field-theoretic approach to the computation of extreme circulation events, where we put forward a linear solution for the extreme events equations. Such a solution is phenomenological enhanced by the utilization of old ideas of eddy viscosity and then, compared to numerical data. The second half of Chapter~\refcap{cap5} is devoted to the interpretation of numerical simulations of the nonlinear solution leading to extreme circulation events. In the end, Chapter~\refcap{cap6} summarizes the results of the thesis, providing discussions and directions for future research.

\end{chapter}
\begin{chapter}{Key Aspects of Fluid Dynamics}\label{cap1}

\hspace{5 mm} 

Almost 200 years before Isaac Newton's formulation of classical mechanics, Leonardo da Vinci was already fascinated by the dynamics of fluid motion. Based on detailed visual observations and graphic reproductions, da Vinci developed an extraordinary physical intuition about the movement of fluids, as quoted in his notebooks \cite{da2008notebooks,mossa2021recent}:
\begin{figure}[H]
\center
\includegraphics[width=\textwidth]{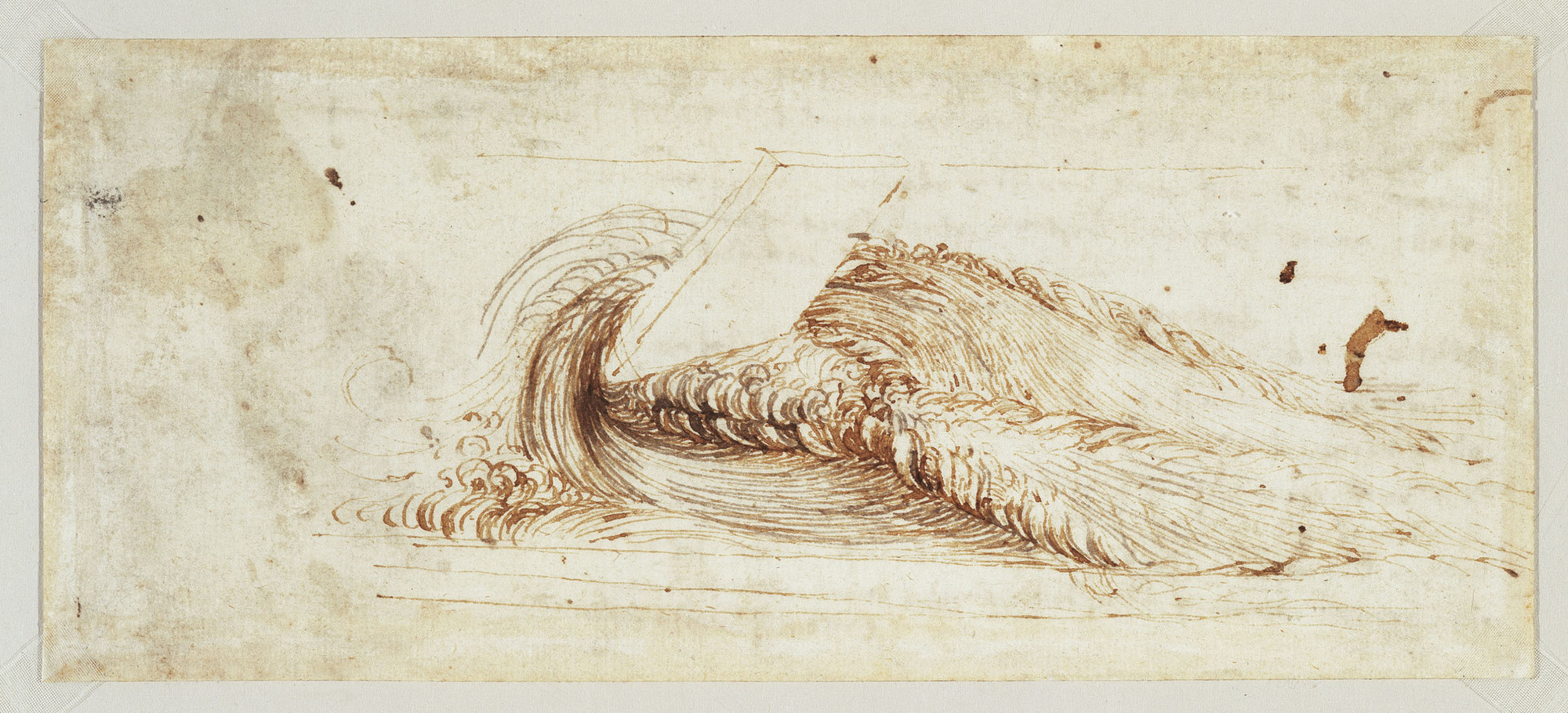}
\caption[Da Vinci's studies on fluid motion.]{Da Vinci's drawing on the fluid motion (The Royal Collection, RL 12659v), available online at \url{https://www.rct.uk/collection/912659/recto-studies-of-flowing-water-with-notes-verso-studies-of-flowing-water}.}
\label{fig1.1}
\end{figure}

``\textit{Observe the motion of the surface of the water, how it
resembles that of hair, which has two movements—one
depends on the weight of the hair, the other on the direction
of the curls; thus the water forms whirling eddies, one part
following the impetus of the chief current, and the other
following the incidental motion and return flow.}'' 

Figure~\reffig{fig1.1} is one of the several fascinating originals of da Vinci's drawings of chaotic fluid motion. It depicts the prevalence of vortical swirling structures, as well as the view of motion as a superposition of a main flow and random fluctuations. These ideas form the building rock of all modern interpretations of turbulent fluid flow phenomena. The very creation of the term ``turbulence'' can be traced back to da Vinci (``\textit{turbolenza}'' derived from ``\textit{turbare}'', meaning ``perturb'' or ``became agitated'' in Italian), although its use in scientific literature is attributed to William Thomson a.k.a. Lord Kelvin \cite{lambd1932hydro,thomson1887xlv}.

After Newton's formulation of the laws of mechanics in 1687 \cite{newton1687}, the conceptualization of several ideas changed the way we understand motion. Concepts such as space, velocity, momentum, mass, energy, and inertial frames emerged and successfully described the motion of particles and rigid bodies. Thus, the application of fluid flow was a very natural step with several technological applications at the time such as naval advancements, canal and dam constructions, and ancient challenges like the desire to fly or at least comprehend the mechanisms of flight. The theoretical challenge was also intriguing: If Newtonian mechanics elegantly and precisely described the motion of rigid bodies, how could it be applied to fluid flow? How could it account for the diverse range of behaviors and phenomena that arise under different conditions of fluid flow? Is it possible to comprise in a single equation the transition from the typical smoothly flowing stream (laminar flow) to the apparent disordered motion of a waterfall (turbulent flow)? 

In the words of the Nobel laureate Richard Feynman, written in his famous Lectures on Physics series \cite{feynman1964feynman}:

``\textit{Often, people in some unjustified fear of physics say you can't write an equation for life. Well, perhaps we can. As a matter of fact, we very possibly already have the equation to a sufficient approximation when we write the equation of quantum mechanics: $\hat{H}\psi=i\hbar\dot{\psi}$}.''

In the context of fluid dynamics and turbulence, the ``equation for life'' referred to by Feynman is believed to be the \ac{iNSE}. It is one of the few problems in history which resisted the analysis of several noble names in science like I. Newton (1663---1727), D. Bernoulli (1700---1782), L. Euler (1707---1783), J.-L. Lagrange (1736---1813), C.-L. Navier (1785---1836), Sir G. Stokes (1819---1903), Lord Kelvin (1824---1907) and many others.

The \ac{iNSE} and its energy-conservative version (Euler equations) have captured the attention of mathematicians due to their simple form. Within this context, even the fundamental question about the existence and smoothness of solutions for these equations continues to evades a mathematically formal solution. This enduring challenge motivated the Clay Mathematics Institute to attach a substantial US\$ 1 million prize to its resolution as a Millenium Prize Problem. The exact statement of the Millenium Problem can be found in a short paper by Fefferman \cite{fefferman2000existence}. In short, the question addressed is: Given smooth initial conditions, do the solutions maintain their regular and analytical characteristics for a finite duration or do they succumb to singularities?

As the \ac{iNSE} plays an important role in the dynamics of fluids, let us carry out a modern still elementary derivation of these equations, fixing notations and definitions.

\section{Incompressible Navier-Stokes Equations}\label{sec1.1}

$_{}$

All matter in the universe is constituted by particles with well-defined properties like mass, charge, etc. Disregarding particle-wave duality, a sense of intrinsic discreteness of nature is naturally introduced by the particle theory. However, the development of fluid dynamics happened much before the atomistic theory of matter ($20^{th}$ century). Instead, fluids (liquid and gases) are treated as continuous media since any small volume element in the fluid is always supposed so large that it still contains a huge number of molecules \cite{landau2013fluid}. For example, atmospheric air at ambient temperature and pressure has, in a small cube of side $10^{-3}$ cm, on the order of $3\times10^8$ molecules. This number is larger for liquids such as water since its density is about $10^3$ times air's density. In this sense, the fluid is macroscopically described by quantities that vary with continuous position $\vc x$ and time $t$ indices, such as velocity $\vc v(\vc x,t)$ and pressure $p(\vc x,t)$. The latter is called the Eulerian point of view in fluid dynamics, in which the coordinates refer to a fixed reference frame. There is also the Lagrangian frame, in which individual parcels of fluid are tracked, and the position index refers to small fluid elements.

The fluid dynamics equation derives directly from two basic concepts in Newtonian mechanics: mass conservation and the momentum equation (second Newton's law). The former can be simply stated as follows\footnote{Repeated indices are summed following Einstein's notation unless explicitly stated. Moreover, no distinction will be made among covariant $x_i$ and contravariant $x^i$ indices, since we are working with Euclidian spaces and cartesian coordinates such that the metric $g_{ij}=g^{ij}=\delta_{ij}$.},
\be\label{1.1}
\partial_t\rho+\partial_i\big(\rho v_i\big)=0\ ,\
\ee
where $\rho=\rho(\vc x,t)$ is the local fluid density and $\partial_i \equiv \partial/\partial x_i$ (similarly for temporal derivatives). The momentum equation reads,
\be\label{1.2}
\frac{d}{dt}\big(\rho v_i\big)=\partial_j\sigma_{ji}[v]\ ,\
\ee
where $\sigma_{ij}$ is the Cauchy stress tensor which describes the
interaction among neighboring fluid elements. In fact, the nature of the stress tensor relies upon the molecular interactions that compose the fluid. Different fluids can be classified regarding the dependency of the stress tensor on the velocity field. There exist fluids with nonintuitive behavior when subjected to shear and stress, some examples are cornstarch in water, silicone oils, synovial fluids, etc. The latter belong to a class called non-Newtonian fluids \cite{dealy1971extensional,wu2020recent}. Conversely, Newtonian fluids like water, air, liquid honey, blood plasma, and many others are characterized by linear responses of the Cauchy stress tensor $\sigma_{ij}$. For Newtonian fluids, the most general stress tensor assumes the form \cite{landau2013fluid},
\be\label{1.3}
\sigma_{ij}[v]=\mu\lr{\partial_i v_j+\partial_jv_i-\frac{2}{3}\delta_{ij}\partial_kv_k}+(\lambda \partial_kv_k-p)\delta_{ij}\ ,\
\ee
where $p=p(\vc x,t)$ is the pressure field and $\mu$ and $\lambda$ are the coefficients of viscosity defined in such a way that both are positive. The $\lambda$ coefficient is sometimes referred to as the second viscosity. At this point is important to assume we are dealing with barotropic fluids. A barotropic fluid is a fluid whose density is a function of pressure only $\rho=\rho(P)$. For most liquids, this assumption can be applied if the typical velocity of the fluid is not comparable to the speed of sound. Eq.~\eqref{1.1} when combined with the barotropic assumption, leads us to the condition of divergence-free velocity field $\partial_iv_i=0$. In the latter case, Eq.~\eqref{1.3} is independent of the second viscosity $\lambda$ in such a way that,
\be\label{1.4}
\partial_tv_i+v_j\partial_j v_i-\nu\partial^2v_i+\partial_iP=0\ ,\
\ee
\be\label{1.5}
\partial_iv_i=0\ ,\
\ee
where $\nu=\mu/\rho$ is the kinematic viscosity, $P(\vc x,t)=p(\vc x,t)/\rho$ is the reduced pressure (usually referred to as pressure only) and we have also used the identity for the total time derivative $\dot f(\vc x,t)=\partial_tf+v_j\partial_jf$. Examples of kinematic viscosities at $20^oC$ in S.I units are $1\times10^{-6},1.5\times10^{-5},2.2\times10^{-6},6.8\times10^{-4},1.2\times10^{-7}$ for water, air, alcohol, glycerine, and mercury, respectively. External forcing can be applied by adding a function $f_i(\vc x,t)$ on the \ac{RHS} of Eq.~\eqref{1.4}. The above equations plus appropriate initial and boundary conditions are the so-called \ac{iNSE}. Note that this set of equations models a very ideal case of the huge variety of fluid motion since we assumed the barotropic approximation for simple Newtonian fluids. However, even this very simple set of equations with a unique parameter $\nu$ is the subject of mathematical ill-posedness leading to the Millenium Prize problem \cite{fefferman2000existence}.

At first glance, Eq.~\eqref{1.4} may not seem like a closed system since it involves two different fields. However, even though the pressure is traditionally represented by the function $P(\vc x, t)$, it is not independent of the velocity field. Indeed, taking the divergence of Eq.~\eqref{1.4} and using Eq.~\eqref{1.5}, we find
\be\label{1.6}
P=-\partial^{-2}\left[\partial_i\Big(\partial_j\big(v_iv_j\big)-f_i\Big)\right]\ ,\
\ee
where we introduced the inverse of the Laplacian operator meant to satisfy the following identity $\partial^{2}(\partial^{-2}f)=f$. This shows that pressure is given in terms of velocity through a Poisson equation, analogous to the electric potential of a distribution of charges. The solution in an infinite domain can be obtained through this electromagnetic analogy, with the known solution of Gauss's law for electric potential,
\be\label{1.7}
P(\vc x,t) =\int d^3y \frac{1}{4\pi|\vc x -\vc y|}\frac{\partial}{\partial y_j}\lr{\frac{\partial}{\partial y_i}\Big(v_i(\vc y,t)v_j(\vc y,t)\Big)-f_j(\vc y,t)}\ ,\
\ee
demonstrating that we indeed have a closed system describing the evolution of the velocity field. At this point, it is interesting to note that the solution Eq.~\eqref{1.4} is non-local in space, a consequence of the classical nature of the compressibility constraint: a local change in the velocity field generates immediate effects throughout the whole space domain, instantaneously.

A different way of writing the \ac{iNSE} is by inserting Eq.~\eqref{1.6} into Eq.~\eqref{1.4} to get
\be\label{1.8}
\partial_tv_i+\mathcal{P}_{ik}\big(v_j\partial_j v_k\big)-\nu\partial^2v_i=\mathcal{P}_{ik}f_k\ ,\
\ee
where $\mathcal{P}_{ik}=\delta_{ik}-\partial^{-2}\partial_i\partial_k$ is usually referred to as Helmholtz-Leray projector operator. This operator extracts only the incompressible part of the fields since any $C^1$ three-dimensional vector field can be decomposed into ``scalar'' and ``vector'' potentials i.e. as the Helmholtz decomposition: $f_i=\partial_i\phi+\epsilon_{ijk}\partial_jA_k$, where $\epsilon_{ijk}$ is the full antisymmetric Levi-Civita symbol. It is then clear that $\mathcal{P}_{ik}(f_k)=\epsilon_{ijk}\partial_jA_k$ completely removes the compressible part of the vector $f_i$.

Eq.~\eqref{1.8} has the advantage of depending only on the velocity field, in opposition to the system of Eqs.~\eqref{1.4} and \eqref{1.5}. The cost of it is the explicit nonlinear and nonlocal formulation of \ac{iNSE} since the explicit formulation of the projection operator involves the inverse Laplacian, meaning that Eq.~\eqref{1.8} is an integral-differential equation. A rather different way of eliminating the Pressure field from the dynamic equation is to take the curl of Eq.~\eqref{1.4}, in vector notation,
\be\label{1.9}
\partial_t\vc \omega+\vc \nabla\times\big(\vc \omega \times \vc v \big)-\nu\partial^2\vc \omega=0\ ,\
\ee
\be\label{1.10}
\vc \omega=\nabla\times\vc v \ ,\
\ee
where the vector identity $v_j\partial_jv_i=\partial_i(v_jv_j/2)-\epsilon_{ijk}v_j\omega_k$ was used in the above derivation. The quantity $\vc\omega$ is the so-called vorticity, a central concept in turbulence that measures the rate of ration in a fluid element. Note that the substitution of Eq.~\eqref{1.10} in Eq.~\eqref{1.9} to have a closed $\omega$ equation, also implies in the computation of an inverse Laplacian as $v_i=\partial^{-2}(\epsilon_{ijk}\partial_j\omega_k)$ for $\partial_iv_i=0$. However, the vorticity equations are mostly used for numerical simulations of fluid dynamic equations.

\section{Symmetries and Global Conservation Laws}\label{sec1.2}

$_{}$

Noether's theorem, was formulated by mathematician and physicist Emmy Noether in 1918 \cite{noether1971invariant}. It establishes a profound link between the symmetries of physical systems and local conservation laws. Symmetries are transformations like spatial translations or rotations that keep the system's properties invariant. The theorem states that for every continuous symmetry of a physical system, there exists a corresponding conserved quantity. Conservation laws, such as energy, momentum, and angular momentum conservation, arise from these symmetries.

Mathematically, the theorem is often applied in variational mechanics, usually in Lagrangian formulations. If a system's Lagrangian remains unchanged under a continuous transformation, there exists a conserved current associated with that transformation. Integrating this current over space or spacetime gives rise to a conserved quantity. However, a variational principle leading to the hydrodynamic equations was only formulated in the ideal fluid case ($\nu=0$) \cite{arnold1966geometrie}. There are attempts to define generalized variational principles leading to the \ac{iNSE} \cite{gomes2005variational}, however, the introduction of stochasticity is necessary for the problem.

In the case of \ac{iNSE}, the symmetries must be directly accounted for by the equations of motion (Eqs.~\eqref{1.4} and~\eqref{1.5}), and the local conservation equation cannot be directly obtained by the symmetry generators in this case. Some of the known symmetries of \ac{iNSE} which are relevant for the study of turbulence are the following:
\begin{enumerate}[label=\textnormal{(P2.2.\arabic*)}]
    \item Space translation (homogeneity): \\ $t' = t \ , 
    \ \ \ \ \ x_i'=x_i+\delta x_i \ , 
    \ \ \ \ \ v_i'(x_i',-t')=v_i(x_i, t) \ ,
    \ \ \ \ \ \text{ where } \delta \vc x \in \mathbb{R}^3$ \ .\label{p121}
    \item Time translation (stationarity): \\ $t' = t+\delta t, 
    \ \ \ \ \ x_i' = x_i, 
    \ \ \ \ \ v_i'(x_i',t')=v_i(x_i, t) \ ,
    \ \ \ \ \ \text{ where } \delta t \in \mathbb{R}$ \ .\label{p122}
    \item Rotations (isotropy): \\ $t' = t \ , 
    \ \ \ \ \ x_i' = \Lambda_{ij} x_j \ , 
    \ \ \ \ \ v_i'(x_i',t')=\Lambda_{ij}v_j(x_i, t) \ , 
    \ \ \ \ \ \text{ where } \Lambda \in SO(3)$ \ .\label{p123}
    \item Galilean transformation: \\ $t' = t \ , 
    \ \ \ \ \ x_i' = x_i - U_it \ , 
    \ \ \ \ \ v_i'(x_i',t')=v_i(x_i, t)-U_i \ ,
    \ \ \ \ \ \text{ where } \vc U \in \mathbb{R}^3$ \ .\label{p124}
    \item Space reversal (parity): \\ $t'=t \ , 
    \ \ \ \ \ x_i'=-x_i \ , 
    \ \ \ \ \ v_i'(x_i',t')=-v_i(x_i, t)$ \ .\label{p125}
    \item Time reversal (only for $\nu=0$): \\ $t'=-t \ , 
    \ \ \ \ \ x_i'=x_i \ , 
    \ \ \ \ \ v_i'(x_i',t')=-v_i(x_i, t)$ \ .\label{p126}
    \item Scale invariance (only for $\nu=0$): \\ $t' = \lambda^{1-h} t \ , 
    \ \ \ \ \ x_i'=\lambda x_i \ , 
    \ \ \ \ \ v_i'(x_i',t')=\lambda^h v_i(x_i, t) \ ,
    \ \ \ \ \ \text{ where } \lambda \in \mathbb{R}_+,\ h\in \mathbb{R}$ \ .\label{p127}
\end{enumerate}
Pressure field transformations are directly derived from velocity and coordinate transformation by the use of Eq.~\eqref{1.7}.

The symmetries of points~\ref{p121} and \ref{p122} are trivially satisfied since $\partial_t=\partial_{t'}$ and $\partial_i=\partial_i'$. Space reversal is also trivial since $\partial_i=-\partial_i'$ and Eq.~\eqref{1.4} depend only on odd combinations of $v_i$ and $\partial_i$, generating a global minus sign in the whole equation.

Time reversal symmetry (point~\ref{p126}) is only satisfied if $\nu=0$ since the viscous term $\nu\partial'^2v'=-\nu\partial^2v$ while all the other terms are unchanged. The violation of time reversal symmetry is a hallmark of parabolic equations,  the so-called ``backward parabolic equations'' are not well-posed as an initial value problem. Instead, the solution often grows unboundedly at a finite time or it even fails to exist.

The more involved symmetries listed above are \ref{p123}, \ref{p124} and, \ref{p127}. Starting from the rotation symmetry, we note that if $ x'_i=\Lambda_{ij}x_j$ then $\partial'_i=\Lambda_{ij}^{-1}\partial_j$. Moreover, as $\Lambda\in SO(3)$ its inverse exists and is the same as its transpose matrix: $\Lambda^{-1}_{ij}=\Lambda_{ji}$, meaning that any contracted term satisfy $f'_ig'_i=f_ig_i$. Thus, applying the transformation \ref{p123} to Eq.~\eqref{1.4}, one gets
\be
\Lambda_{ij}\Big(\partial_tv_j+v_k\partial_kv_j+\partial_jP-\nu\partial^2v_j\Big)=0 \ ,\
\ee
then the \ac{iNSE} remains the same. For the Galilean invariance \ref{p124}, we write the equations in the transformed frame as
\be\label{1.12}
\partial_{t'}v'_i(\vc x\ ',t')\Big|_{x'_i=x_i-U_it'}+\lr{v'_j\partial'_jv'_i+\partial'_iP'-\nu\partial'_j\partial'_jv'_i}\Big|_{t'}=0 \ ,\
\ee
where vertical bars explicitly show that time/space partial derivatives are taken with constant $x'_i$/$t'$. Then, $\partial_i'=\partial_i$ and $\partial_jU_i=0$ since it is a constant vector and using Eq.~\eqref{1.12} we get,
\be\label{1.13}
\lr{\partial_{t'}v'_i(\vc x\ ',t')\Big|_{x'_i=x_i-U_it'}-U_j\partial_jv_i}+v_j\partial_jv_i=-\partial_iP+\nu\partial^2v_i \ .\
\ee
Fixing $x'_i$ for a infinitesimal time translation $t'\rightarrow t'+\delta t$ means that the unprimed frame must be transformed as $x_i\rightarrow x_i+U_i\delta t$ since $x'_i(t'+\delta t)=x'_i(t')$. The latter remark means that the time derivative in the primed variables must be calculated in moving unprimed variables, with a velocity $U_i$, producing the exact term $U_j\partial_jv_i$ needed for the symmetry to be satisfied.

The scaling symmetry can be accounted for by noting that $\partial_{t'}=\lambda^{h-1}\partial_t$ and $\partial_i'=\lambda^{-1}\partial_i$ and Eq.~\eqref{1.4} to get, 
\be
\lambda^{2h-1}\Big(\partial_tv_j+v_k\partial_kv_j+\partial_jP\Big)-\lambda^{h-2}\nu\partial^2v_j=0 \ ,\
\ee
or equivalently,
\be
\partial_tv_i+v_j\partial_jv_i+\partial_iP-\lambda^{-h-1}\nu\partial^2v_j=0 \ .\
\ee
It is clear that the inviscid ($\nu=0$) equation is scale-invariant for infinitely many $h$ exponents. A very interesting case happens if $h=-1$ when the viscous \ac{iNSE} is scale invariant. Moreover, this exponent has a particular property of leaving one important dimensional parameter related to the \ac{iNSE} invariant, however, we will introduce it later in Section~\refsec{sec1.3}. 

The symmetries pointed out in \ref{p121}---\ref{p127} are not the only word in the analysis of the conservation laws for fluid dynamics. In the words of one of the most important actors in turbulence U. Frisch \cite{frisch1995turbulence}:

``\textit{Other more local conservation laws, such as the conservation of \underline{circulation} \cite{lambd1932hydro} may be even more important but have found surprisingly few applications to turbulence (so far).}''

However, as already mentioned, the lack of a variational formulation of the \ac{iNSE} hinders a direct derivation of such currents. Instead, we will discuss only global conservation laws. Firstly, consider the total momentum per unit of volume carried by the fluid
\be\label{1.16}
P_i(t)=\rho\int d^3x\ v_i(\vc x ,t)\ .\
\ee 
Integration is often carried in $L_0$-periodic boxes, which eases the calculation, then always taking the large box limit $L_0\rightarrow\infty$, such that all the solutions decay quickly enough to ensure $L^2$-norm convergence. In any case, this is equivalent to supposing the fields are bounded in and $\mathbb{R}^3$\footnote{Of course, the one who is able to prove it will win the Millennium prize, certainly a Fields Medal and, possibly a Nobel Prize, for now on, let us just assume it.}. Now take a time derivative of Eq.~\eqref{1.16} to get,
\be\label{1.17}
\frac{dP_i}{dt}=\rho\int d^3x \lr{\partial_tv_i+v_j\partial_jv_i}=\rho\int d^3x \partial_jF_{ji}=\rho\oint dS_j F_{ji}\rightarrow0\ ,\
\ee 
where $F_{ji}=\partial_jv_i-p\delta_{ij}$ by the use of Eq.~\eqref{1.4} and the rightmost equality is found using Gauss' theorem. The convergence to zero is meant by the proposition of a bound velocity field at the borders. For the conservation of energy (per unit of mass) we do the same for $|\vc v|^2/2$ to get
\be\label{1.18}
\frac{dE}{dt}=\int d^3x\ v_i\lr{\partial_tv_i+v_j\partial_jv_i}\ ,\
\ee 
where, by the use of Eq.~\eqref{1.4}, integration by parts, and the incompressibility constraint, Eq.~\eqref{1.18} reads
\be\label{1.19}
\frac{dE}{dt}=\nu\int d^3x\ v_i\partial^2v_i\ .\
\ee 
By defining the enstrophy $\Omega$ as
\be\label{1.20}
\Omega\equiv\frac{1}{2}\int d^3x \ |\vc\omega|^2=-\frac{1}{2}\int d^3x\ v_i\partial^2v_i\ ,\
\ee 
through successively tensor contraction, integration by parts, and Gauss' theorem. In general, $\Omega\neq0$ and energy is only conserved if $\nu=0$, for inviscid ideal fluids. Another important conserved quantity is the helicity $h$, defined as
\be\label{1.21}
h\equiv\frac{1}{2}\int d^3x \ \vc v\cdot\vc \omega\ .\
\ee
Taking a time derivative of Eq.~\eqref{1.21} and applying Eqs.~\eqref{1.4},~\eqref{1.5},~\eqref{1.9} and the same strategy of integration by parts and surface integrals, one is able to show that,
\be\label{1.22}
\frac{dh}{dt}=\nu\int d^3x \ \vc{\omega}\cdot\lr{\vc\nabla\times\vc\omega}\equiv 2\nu h_{\omega}\ ,\
\ee
where $h_\omega$ is called vortical helicity. Finally, we define the last important global quantity,
\be\label{1.23}
P_\omega\equiv\frac{1}{2}\int d^3x \ |\vc\nabla\times\vc{\omega}|^2=-\frac{1}{2}\int d^3x\ \omega_i\partial^2\omega_i\ .\
\ee
The set of definitions and relation of Eqs.~\eqref{1.19}---\eqref{1.23} can be put in a very simple and symmetric form (in three dimensions):
\be\label{1.24}
\dot{E}=-2\nu\Omega\ ,\
\ee
\be\label{1.25}
\dot{h}=-2\nu h_{\omega}\ ,\
\ee
and
\be\label{1.26}
\dot\Omega=-2\nu P_\omega+\int d^3x\  \omega_i\omega_j\partial_jv_i\ .\
\ee
Note that, with the exception of enstrophy, the energy and helicity are inviscid invariants. However, nonzero values of helicity are associated with the breakdown of parity symmetry \ref{p125} but, in any case, its initial value is conserved for inviscid flow.

In the special case of Eq.~\eqref{1.26}, there is an enstrophy production term that arises from Eq.~\eqref{1.9} by equating 
\be\label{1.27}
\vc \nabla\times\big(\vc \omega \times \vc v \big)=\big(\vc v\cdot \vc \nabla\big)\vc\omega-\big(\vc \omega\cdot \vc \nabla\big)\vc v \ ,\
\ee
where the first term on the \ac{RHS} is related to vortex advection, while the second is known as vortex stretching term. In a very interesting numerical study \cite{bos2021three}, Bos works out Eq.~\eqref{1.9} without the vortex stretching term. The resulting solutions display several similarities with two-dimensional turbulence, where Eq.~\eqref{1.26} has no enstrophy production term due to the pseudo-scalar nature of vorticity. Moreover, the distribution of energy among scales is completely different from the usual 3D numerical solution, due to the absence of vortex stretching. However, there are still some differences between 3D \ac{iNSE} without vortex stretching and 2D \ac{iNSE} \cite{bos2021three,boffetta2012two}.

In some sense, it is correct to say that up to some degree, enstrophy production has the ability to phenomenologically split 2D and 3D turbulence. For example, several simplified turbulence models known as ``shell models'' differentiate 2D and 3D turbulence by imposing conservation of quantities as energy and enstrophy, or energy and helicity, respectively \cite{biferale2003shell}. Moreover, the combination of vortex stretching and conservation of circulation (to be shown in Chapter~\refcap{cap3}) is responsible for the appearance of intense vortical structures, already anticipated by da Vinci's drawings (see Fig.~\reffig{fig1.1}) and are of fundamental importance throughout this work, especially in Chapter~\refcap{cap4}.

\section{Reynolds Similarity}\label{sec1.3}

$_{}$

In a classic work from 1851,  Sir G. Stokes \cite{stokes1851effect} proposed a simple way to compare the effects of advection and dissipation. Stokes' work is considered the precursor to one of the most important concepts in fluid dynamics, the Reynolds number (Re). More precisely it associates the advection term with turbulence production and discusses the role of scales in a fluid dynamics problem.

However, this new paradigm in fluid mechanics was primarily established through the significant experimental progress of the $19^{th}$ century, especially the work of the British physicist Osborne Reynolds (1842-1912) in 1883 \cite{reynolds1883xxix}. Studying various flows in pipes, he discovered the phenomenon of transition to turbulence and demonstrated that, for a given geometry, this transition is controlled by a single dimensionless parameter, now named the Reynolds number. 

Figure~\reffig{1.2} shows the original sketches of the experiments published by Reynolds. The experimental setup is the following: a constant flow rate is pumped in a long and thin pipe and, at the same rate but in smaller volume a jet is carried by the flow. By controlling the flow rate, the geometry of the pipe, and the liquid which is flowing, Reynolds found that the relevant dimensionless parameter that controls the different flow regimes was $Re = U_0 D/\nu$, where $U_0$ is the injection velocity, $D$ is the pipe diameter and $\nu$ is the kinematic viscosity of the fluid. The emergence of vortical structures was depicted by observing the flow through light flashes generated by electric sparks. These structures appear abruptly by varying $U_0$ and only above a critical value, beyond which the transition to turbulence occurred. He also noticed that the associated critical value of Re was highly sensitive to perturbations present at the pipe entrance or in the fluid injection mechanism, demonstrating the crucial role of instabilities in turbulence creation.
\begin{figure}[t]
\center
\includegraphics[width=\textwidth]{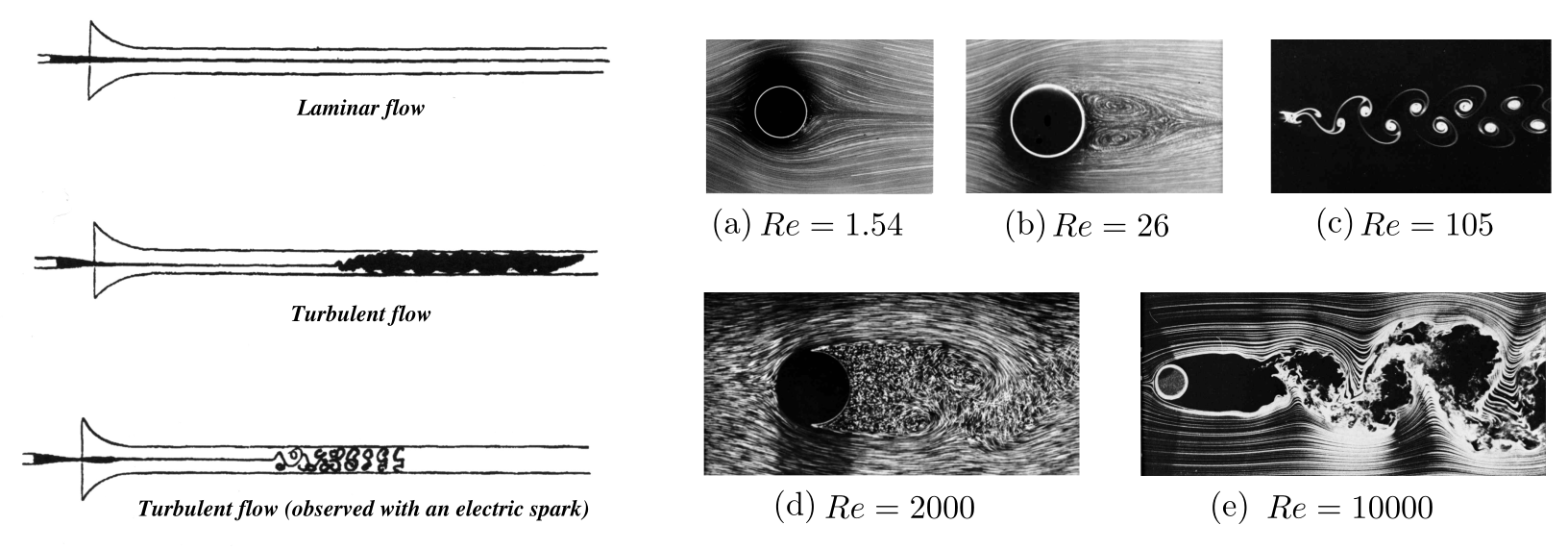}
\caption[Reynolds' sketches on his experiments and flow past a cylinder.]{Left panel shows original sketches of Reynolds' experiments (edited from \cite{reynolds1883xxix}). The right panel shows the visualization of a fluid flow passing a cylinder at different Reynolds numbers (edited from \cite{van1982album}).}
\label{fig1.2}
\end{figure}

In a more general sense, we can think of a fluid with viscosity $\nu$ where energy is pumped at a typical scale $\ell$ controlled by external force or perturbation mechanisms. This force generates velocity differences of a typical magnitude $v_0$ between the point separated by the scale $\ell$. These thus become the scales of the problem, and according to Reynolds' proposition, flow behavior should be governed by the dimensionless parameter
\be\label{1.3.11}
Re = \frac{v_0\ell}{\nu}\ .\ 
\ee
The right panel of Fig.~\reffig{fig1.2} shows different regimes of the flow passing a cylinder. The Reynolds number in this geometry is calculated based on the diameter of the cylinder. Exactly as in Reynolds' experiments, the velocity field near the cylinder starts smooth for very low flow rates (a), and, as it increases, vortices start to appear in the back of the cylinder (b). As the flow rate increases these vortices detach from the cylinder to form a vortex sheet --- also known as von Kármán's sheet --- with regular spacing (c). At (d) chaotic mixing starts to play a crucial role and, in (e) the flow past the cylinder is completely chaotic and mixing, i.e. turbulent.

To support this scaling behavior directly from \ac{iNSE}, we perform a change of variables at Eq.~\eqref{1.4} to dimensionless quantities: $\vc x\ ' = \vc x/\ell$ and $t' = t/\tau$, defining a typical time scale $\tau \equiv\ell/v_0$. In this case, derivatives transform as $\partial_i=\ell^{-1}\partial'_i$ and $\partial_t=\tau^{-1}\partial'_t$, while the transformation of physical quantities can be obtained through dimensional analysis. The dimensions of the fields, after setting $\rho = 1$, are represented by
\be\label{1.3.12}
v_i(\vc x,t) = \frac{\ell^2}{\tau^2}\bar{v}_i(\vc x\ ',t')\ ,\ 
\ee
\be\label{1.3.13}
P(\vc x,t) = \frac{\ell^2}{\tau^2}\bar{P}(\vc x\ ',t')\ ,\ 
\ee
and
\be\label{1.3.14}
f_i(\vc x,t) = \frac{\ell}{\tau^2}\bar{f}_i(\vc x\ ',t')\ ,\ 
\ee
where the bar denotes dimensionless functions, which, however, have the same functional form as their dimensional counterparts. Substituting Eqs.~\eqref{1.3.12}---\eqref{1.3.14} into Eq.~\eqref{1.4} and renaming the variables $(\vc x\ ', t')$ back to $(\vc x, t)$, we obtain the dimensionless equations:
\be\label{1.3.15}
\partial_t\bar{v}_i+\bar{v}_j\partial_j \bar{v}_i-\frac{1}{Re}\partial^2\bar{v}_i+\partial_i\bar{P}=\bar{f}_i\ .\
\ee
This equation demonstrates that the Reynolds number is indeed the sole parameter that distinguishes the possible flow outcomes from the solution. It also connects the observations of Stokes and Reynolds: when $Re \ll 1$, the dissipative term dominates over the effects of the advective term, and the solution tends to be smoother, characterizing the laminar regime. When Re$\gg 1$, the opposite is true, and the advective term prevails, causing the injected energy to perturb the entire flow due to its action, defining the turbulent regime.

Moreover, one may note that under a scaling transformation $U'=\lambda^h U$ and $L'=\lambda L$ such that
\be
\frac{U'L'}{\nu}=\lambda^{h+1}\frac{UL}{\nu} \ ,\
\ee  
where, for $h=-1$, the parameter that defines the regimes of fluid equations is scale invariant. This means that when analyzing the flow in a pipe and constructing two experimental sets where the only difference is, for example, the second set having twice the diameter and half the inflow velocity, one obtains exactly the same dynamics. This fact is widely referred to as Reynolds similarity.

From now on, we will refer to turbulence as the chaotic movement of fluids characterized by large Reynolds numbers. In pipe flows, like those observed in Reynolds' experiment, a transition in the behavior of the fluid motion is observed at a Reynolds number of approximately 2500. If Re is greater than this critical value, a sufficiently large perturbation can propagate throughout the pipe, leading to the eventual establishment of a turbulent flow. It is noteworthy that, through meticulous control of perturbations and fluid injection mechanisms, Reynolds was able to sustain a laminar pipe flow up to Re$\sim$13000 \cite{reynolds1883xxix}. In this sense, the transition to turbulence in pipe flows forms a subcritical transition where the flow is stable, at any Re, up to a finite size perturbation generating the necessary instabilities that drive the transition to turbulence \cite{eckhardt2007turbulence}. In daily situations, Re can often reach values of $10^6$ for a flow around a car moving at $80 km/h$ relative to the fluid, for instance. Wind tunnel experiments also reach impressive Reynolds numbers of order $10^7$ \cite{neuhaus2020generation} and even $10^9$ in atmospheric flows \cite{wyngaard1992atmospheric}, underscoring the importance of turbulence research in different situations. 

Identifying with the naked eye by observing an ink droplet in a tank or simply looking at a waterfall, is a considerably simpler task. This approach was even employed by da Vinci in the XV century and several of the early precursors of turbulence research. However, the question of how to define turbulence from a concise and precise scientific point of view is considerably difficult. The definition of a Reynolds number is quite nonuniversal (depending on the geometry) and reveals that the challenges in its study can be as fundamental as defining the subject itself. We will then attempt to establish a more precise definition of what turbulence is by relating it to the statistical restoration of the \ac{iNSE}'s symmetries discussed in Section~\refsec{1.2}. However, this definition cannot encompass all the richness of the turbulence phenomena and will be associated with a specific type: fully developed homogeneous and isotropic turbulence. The rigorous and definitive definition of turbulence remains an open question.

\end{chapter}
\begin{chapter}{Statistical Theory of Turbulence}\label{cap2}

\hspace{5 mm} 

The modern understanding of the problem of turbulence is based on experiments and theories developed in the $20^{th}$ century. Besides Reynolds, other important scientists in the development of the theory of turbulence range from L. Prandtl (1875-1953), L.-F. Richardson (1881-1953), T. von Kármán (1881-1963), and G. I. Taylor (1886-1975), who made the first discoveries on the statistical characterization of the laminar-turbulent transition, the formation of vortical structures, energy cascade, and others. Their contributions served as inspirations to the ones who came next such as J. von Neumann (1903---1957), A. Kolmogorov (1903---1987), L. Onsager (1903---1976), L. Landau (1908---1968), R.P. Feynman (1918---1988), U. Frisch (1940---), A.M. Polyakov (1945---), A.A. Migdal (1945---), and even G. Parisi (1948---) the 2021 Nobel laureate, among many others. 

In particular, A. Kolmogorov built the first quantitative theory to describe the turbulence phenomena in 1941 \cite{kolmogorov1941local}, which is nowadays widely referred to as \ac{K41}. This phenomenological approach has inherited in its core formulation the assumption that the symmetries presented in \ref{p121}---\ref{p127}, which are broken by viscous and forcing mechanisms when generating turbulence, are restored by internal dynamical mechanisms. This phenomenological approach defines the so-called fully developed turbulence, assumed to be homogeneous, isotropic, stationary, and scale invariant (it is also referred to as homogeneous and isotropic turbulence only). Despite its idealized nature, the results of fully developed turbulence are assumed to play a crucial role in general turbulent systems. Pipe flow, for example, can be regarded as homogeneous and isotropic turbulence if the measurements are made sufficiently far from the boundaries.

Generally, the restoration of the broken symmetries takes place in a statistical sense, and all the \ac{K41} story starts with the observed necessity of expressing turbulence in terms of statistical quantities instead of searching for a solution of \ac{iNSE}.

\section{The Need for a Statistical Description}

$_{}$

The challenge of comprehending turbulence using deterministic methods constitutes a segment of the mathematical advancement in the area until the early 20th century. The notion of deterministic evolution essentially alludes to the effort of predicting the future state of a system based on its present state, usually by solving underlying equations of motion. This concept has served as the underlying framework of differential equations theory since its conception. Nonetheless, the field of dynamical systems, pioneered by H. Poincaré (1854---1912) \cite{poincare1893methodes}, hinted at the possibility that this approach might be inadequate or might not produce the correct outcomes. The understanding of the principle of bifurcation: an abrupt change in the behavior of the solutions caused by continuous shifts in the parameters of the equations, revealed that even in simple deterministic systems, slight differences in the initial conditions could lead to entirely diverging states within a short time span. Subsequent investigations by Lorenz, Smale, and Feigenbaum \cite{lorenz1963deterministic,smale1967differentiable,feigenbaum1978quantitative} laid down the foundation of chaos theory, illustrating how dynamic systems can manifest erratic and apparently random behavior even when governed by deterministic equations of motion.

The scientific paradigm of turbulence research was dramatically reshaped by the formulation of chaos theory in dynamical systems. A novel framework has emerged, fundamentally changing our understanding of predictability. The behavior of solutions is so unpredictable and disorderly that they almost mirror randomness. Even if we determine the precise evolution of the system from an initial state $A$ to a final state $B$, the use of this knowledge would be restricted, as precise control over the state $A$ is restricted in real-world experiments such that any finite disturbance $\delta A$ generates a completely different outcome $B'$. Consequently, the central question turns out to be the identification of the aspects of a fluid's motion that are, in some sense, predictable.

Directing our attention to examples, we show in Fig.~\reffig{fig2.1}, two one-second velocity signals (relative to the mean) obtained using a hot-wire probe in a wind tunnel experiment \cite{frisch1995turbulence}. This signal represents the ``streamwise'' velocity component, aligned with the mean flow direction. The measurement of flow velocity, or a single component at a specific point over time is known as velocimetry. Diverse experimental techniques of velocimetry exist, although we will not delve into their specifics in this context. For more details on this subject, we refer to \cite{adrian1991particle,tavoularis2005measurement,gad2013advances}. 

Several noteworthy observations emerge upon inspection of Fig.~\reffig{fig2.1}:
\begin{enumerate}[label=\textnormal{(P3.1.\arabic*)}]
    \item The signal exhibits pronounced disorder, displaying structures across a wide range of time scales: Structures are perceptible to the naked eye, with temporal scales spanning approximately $1000$, $100$, $10$ sample units, and even potentially smaller intervals.\label{p211}
    \item The signal's detailed behavior appears inherently unpredictable: While the overall resemblance between the two signals remains consistent, every small detail has undergone alteration, and these alterations couldn't have been foreseen by looking at the previous illustration.\label{p212}
    \item Despite the apparent randomness, certain attributes of the signal exhibit a degree of reproducibility by the characterization of an empirical \ac{PDF}.\label{p213}
\end{enumerate}
\begin{figure}[t]
\center
\includegraphics[width=\textwidth]{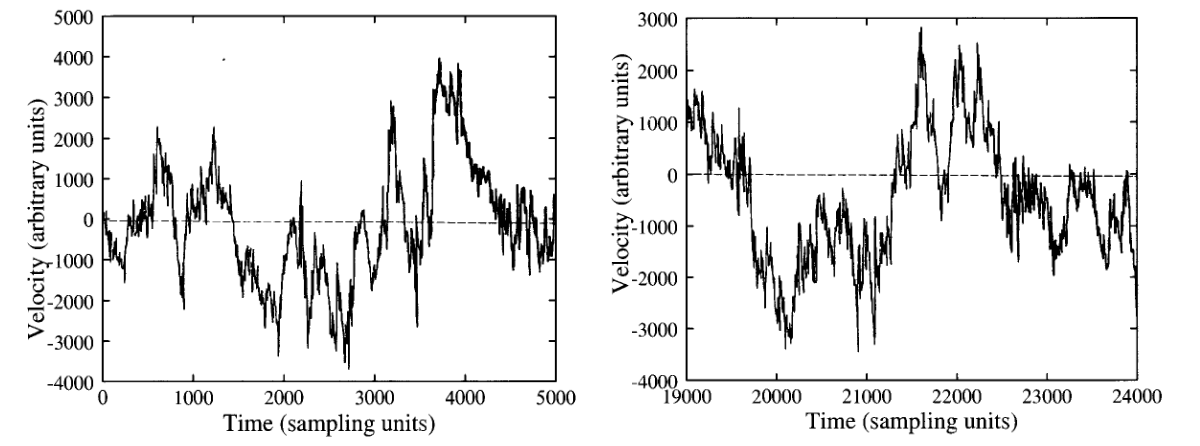}
\caption[Wind tunnel turbulent velocity signal.]{One second of a signal recorded by a hot-wire (sampled at 5 kHz) in a wind tunnel. Left and right panels show samples with different starting times but the same length. Reproduced from \cite{frisch1995turbulence}.}
\label{fig2.1}
\end{figure}
Regarding point~\ref{p213}, Fig.~\reffig{fig2.2} shows the histogram of velocity measurements. Evidently, the two histograms are indistinguishable.

We can summarize points~\ref{p211}---\ref{p213} by stating: Although the intricate attributes of the signal appear to hinder predictability, its statistical characteristics remain consistently reproducible. These observations have led theoretical scientists, in particular G.I. Taylor, to seek a probabilistic description of turbulence \cite{taylor1935statistical,taylor1938spectrum} even if the underlying dynamical equation (\ac{iNSE}) is deterministic.

A modern branch of the manifestation of randomness in a fully deterministic system is the so-called ``spontaneous stochasticity'' \cite{lorenz1969predictability}. Unlike the phenomena of chaos, spontaneous stochasticity is not related to the specific issue of indetermination of initial state errors. Instead, it is believed to be related to singular limits of the dynamical equations of motion in some specific parameter regimes \cite{eyink2020renormalization,mailybaev2023spontaneous}. In the case of \ac{iNSE}, it is related to the singular limit of $\nu\rightarrow0$ i.e., the turbulent limit Re$\rightarrow \infty$. Although several simplified models of turbulence seem to show randomness spontaneously \cite{mailybaev2016spontaneous,mailybaev2023spontaneous,thalabard2020butterfly}, a rigorous proof that the Navier-Stokes (and its inviscid counterpart) system show this property is still lacking.
\begin{figure}[t]
\center
\includegraphics[width=\textwidth]{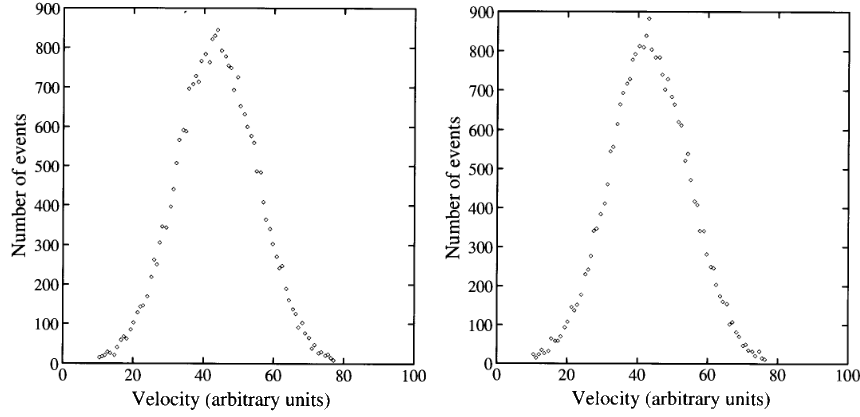}
\caption[Velocity signals histograms.]{Histograms of streamwise velocity in the same experiment that generates Fig.~\reffig{fig2.1}. The velocity was sampled $5000$ times over a time span of 150 seconds. Left and right panels are histograms delayed by a few minutes. Reproduced from \cite{frisch1995turbulence}.}
\label{fig2.2}
\end{figure}

Defining a suitable average forms the fundamental building blocks of any statistical theory and, therefore, it needs to be well-defined. From now on we will assume stationary turbulence to form an ergodic system. This means that calculating the average of any observable $O[v_i(\vc x, t)]$ (a functional of the velocity field) by the ensemble average
\be\label{2.1}
\mean{O[\vc v(\vc x ,t)]}=\lim_{N\rightarrow\infty}\sum_{n=1}^N O\big[\vc v^{(n)}(\vc x ,t)\big] \ ,\
\ee
gives exactly the same result as calculating
\be\label{2.2}
\lim_{T\rightarrow\infty}\frac{1}{T}\lim_{V\rightarrow\infty}\frac{1}{V}\int_0^{T}ds\int_Vd^3y\ O\big[\vc v(\vc x +\vc y,t+s)\big] \ ,\
\ee
where $V$ means the volume where the fluid is defined. The ergodic hypothesis means that the velocity configuration space has no prohibited regions, in other words, if we wait long enough and probe scales enough we will occasionally see, all the possible velocity field configurations needed to calculate the average.

The meaning of homogeneous and isotropic, as stated earlier, is that the symmetries of the \ac{iNSE} are restored in its statistics. In practical terms, spatial homogeneity and stationarity means that $\mean{O[\vc v(\vc x,t)]}$ does not depend on the spatial coordinates nor the time, such that,
\be\label{2.3}
\partial_t\mean{O[\vc v(\vc x ,t)]}=\partial_i\mean{O[\vc v(\vc x ,t)]}=0 \ .\
\ee
The combination of homogeneity and isotropy means
\be\label{2.4}
\mean{O_1[\vc v(\vc x_1 ,t_1)]O_2[\vc v(\vc x_2 ,t_2)]}=f(|\vc x_1-\vc x_2|,t_1-t_2) \ ,\
\ee
where the tensorial structure of the observables $O_1$ and $O_2$ must also preserve isotropy. For a simple example consider
\be\label{2.5}
A_{ij}(\vc x,t,\vc \rho)\equiv\mean{v_i(\vc x+\vc \rho,t)v_j(\vc x,t)} \ .\
\ee
The only tensorial structure compatible with isotropy, homogeneity, and incompressibility is
\be\label{2.6}
A_{ij}(\rho)=\mathcal{P}_{ij} F(\rho) \ ,\
\ee
where, $\mathcal{P}_{ij}$ is the projector operator introduced in Eq~\eqref{1.8}, $\rho=|\vc \rho|$ and $F$ is a generic function. A term $\epsilon_{ijk}\partial_kF_2(\rho)$ is also possible, but it violates parity symmetry, i.e., it has a non-zero value of helicity \cite{borue1997spectra}.

The quantity defined in Eq.~\eqref{2.5} is particularly relevant for statistical systems since it is related to the so-called isotropic energy spectrum or simply energy spectrum through,
\be\label{2.7}
E(k)\equiv 4\pi k^2\int \frac{d^3\rho}{(2\pi)^3}e^{i\vc k \cdot \vc \rho}A_{ii}(\rho)=\frac{1}{\pi}\int_0^\infty d\rho \ k \rho\ \sin(k\rho)A_{ii}(\rho) \ .\
\ee
Eq.~\eqref{2.7} comprises all the information about the energy distribution among scales and has huge experimental importance. Since isotropy means $A_{ii}=3A_{11}$ and $\vc \rho=\rho\hat{x}$ one can, without loss of generality, measure the energy spectrum by measuring a single component of the velocity in one direction, similar to the ones of Fig.~\reffig{fig2.1}. \clearpage

\section{Two Laws of Turbulence}

$_{}$

Besides discussing some fundamental concepts of fluid dynamics that provide insights into the behavior of turbulent motion, the chaotic and unpredictable nature of turbulence makes it difficult to establish simple laws that govern all aspects of such flows. However, there is a broad experimental concept regarding the energy dissipation of three-dimensional turbulent flows and, in the case of homogeneous and isotropic turbulence, a statistical law can be directly derived from the \ac{iNSE}. In order to do this, let us introduce the concept of the energy budget equation.

\subsection{Zeroth law}

$\ \ \ \ $The conservation law put forward in Section~\refsec{sec1.2} refers to global conservation laws. In the language of the ergodic hypothesis, the same laws are valid by substituting the volume integration with an ensemble average. In this sense, it is clear that
\be\label{2.8}
\frac{d}{dt}\lr{\frac{\mean{|\vc v|^2}}{2}}=-2\nu\Omega+E_{in}=0 \ ,\
\ee
where $\Omega=\mean{|\vc \omega|^2}/2$ is the enstrophy and $E_{in}=\mean{f_iv_i}$ is the energy input provided by some large scale forcing necessary to maintain the steady state. Eq.~\eqref{2.8} relates the total energy input to be completely balanced by the viscous dissipation $\vare\equiv2\nu\Omega$. It is clear that this balance equation is not sufficient to study the energy transfer between different scales. Define the low-pass filter
\be\label{2.9}
f_\Lambda^{<}(\vc x,t)\equiv\int_{|\vc k|\leq\Lambda} d^3k \hat{f}(\vc k,t)e^{i\vc k\cdot \vc x} \ ,\
\ee
and the high-pass filter 
\be\label{2.10}
f_\Lambda^{>}(\vc x,t)\equiv\int_{|\vc k|\geq\Lambda} d^3k \hat{f}(\vc k,t)e^{i\vc k\cdot \vc x} \ ,\
\ee
such that
\be\label{2.11}
f(\vc x,t)\equiv\int d^3k \hat{f}(\vc k,t)e^{i\vc k\cdot \vc x}=f_\Lambda^{<}(\vc x,t)+f_\Lambda^{>}(\vc x,t) \ .\
\ee
This kind of filtering procedure splits the spectral contribution above and below a length scale $\ell\sim\Lambda^{-1}$. Applying a low-pass filtering procedure to the \ac{iNSE} (Eq.~\eqref{1.4} plus forcing) we get,
\be\label{2.12}
\partial_tv^{<}_{i,\Lambda}+\Big[v_j\partial_j v_i\Big]^<_{\Lambda}-\nu\partial^2v_{i,\Lambda}^<+\partial_iP_\Lambda^<=f_{i,\Lambda}^{<}\ ,\
\ee
\be\label{2.13}
\partial_iv^{<}_i=0\ .\
\ee
Taking the scalar product of Eq.~\eqref{2.12} with $v_{i,\Lambda}^<$ to get,
\be\label{2.14}
\partial_t E_\Lambda+\Pi_\Lambda=-2\nu\Omega_\Lambda+F_\Lambda\ ,\
\ee
where $2E_\Lambda=\mean{|\vc v^<_\Lambda|^2}$, $2\Omega_\Lambda=\mean{|\vc \omega^<_\Lambda|^2}$, $F_\Lambda=\langle\vc v^<_\Lambda\cdot \vc f^<_\Lambda\rangle$ and,
\be\label{2.15}
\Pi_\Lambda=\Big\langle v^{<}_{i,\Lambda}v^{<}_{j,\Lambda}\partial_jv^{>}_{i,\Lambda}\Big\rangle+
\Big\langle v^{<}_{i,\Lambda}v^{>}_{j,\Lambda}\partial_jv^{>}_{i,\Lambda}\Big\rangle\ \
\ee
is the energy flux from large scales to small scales or, conversely, from small wavelengths to large wavelengths. Note for $\Lambda\rightarrow \infty$ that the low pass filter converges to the full fields while the high pass filters go to zero and, consequently, Eq.~\eqref{2.14} converges to Eq.~\eqref{2.8}. As stated in Eq.~\eqref{2.8}, the energy pumping mechanism usually has compact support at large scales, meaning that there exists a scale $L$ associated with $\Lambda_L\sim L^{-1}\ll1$ where almost all the energy is already pumped into the system. Mathematically
\be\label{2.16}
F_{\Lambda_L}=E_{in}\ .\
\ee
On the other hand, the enstrophy term is related to the energy because it satisfies the inequality $\Omega_\Lambda\leq\Lambda^2E_\Lambda$ since $2\Omega_\Lambda=\mean{|\vc \omega^<_\Lambda|^2}=-\mean{v_i\partial^2v_i}$. As the energy is positive definite and bounded, the enstrophy contribution to the energy budget equation $\Omega_\Lambda$ can be regarded as an increasing function of $\Lambda$. This suggests the existence of a scale $\eta_K$ associated with a wavelength $\eta_K\sim\Lambda_\eta^{-1}$ such that,
\be\label{2.17}
\frac{2\nu\Omega_{\Lambda_\eta}}{E_{in}}\approx0\ .\
\ee
This means that up to the scale $\Lambda_\eta$, viscous dissipation is negligible compared to the system's total energy. Supposing there exists a sufficiently large separation between these two scales $\Lambda_{\eta}\gg\Lambda_L$, allowing for the simultaneous use of Eqs.~\eqref{2.16},~\eqref{2.17}, and the stationarity assumption ($\partial_tE_\Lambda=0$), Eq.~\eqref{2.14} reads
\be\label{2.18}
\Pi_\Lambda\approx E_{in}\ ,\text{ for         } \Lambda_\eta\gg\Lambda\gg\Lambda_L\ \ \text{or, equivalently,} \ (\eta_K\ll\ell\ll L)\ .\
\ee
The definition of $\Pi_\Lambda$ as the energy flux from larger to smaller scales together with Eq.~\eqref{2.18} is the first hint of a broad phenomenological concept in 3D fully developed turbulence, the energy cascade. ``\textit{Big whorls have little whorls that feed on their velocity, and little whorls have smaller whorls and so on to viscosity — in
the molecular sense}''. This was the famous poem written by Richardson in 1922 \cite{richardson1922weather} when he first proposed the mechanism of the turbulent cascade.
\begin{figure}[t]
\center
\includegraphics[width=\textwidth]{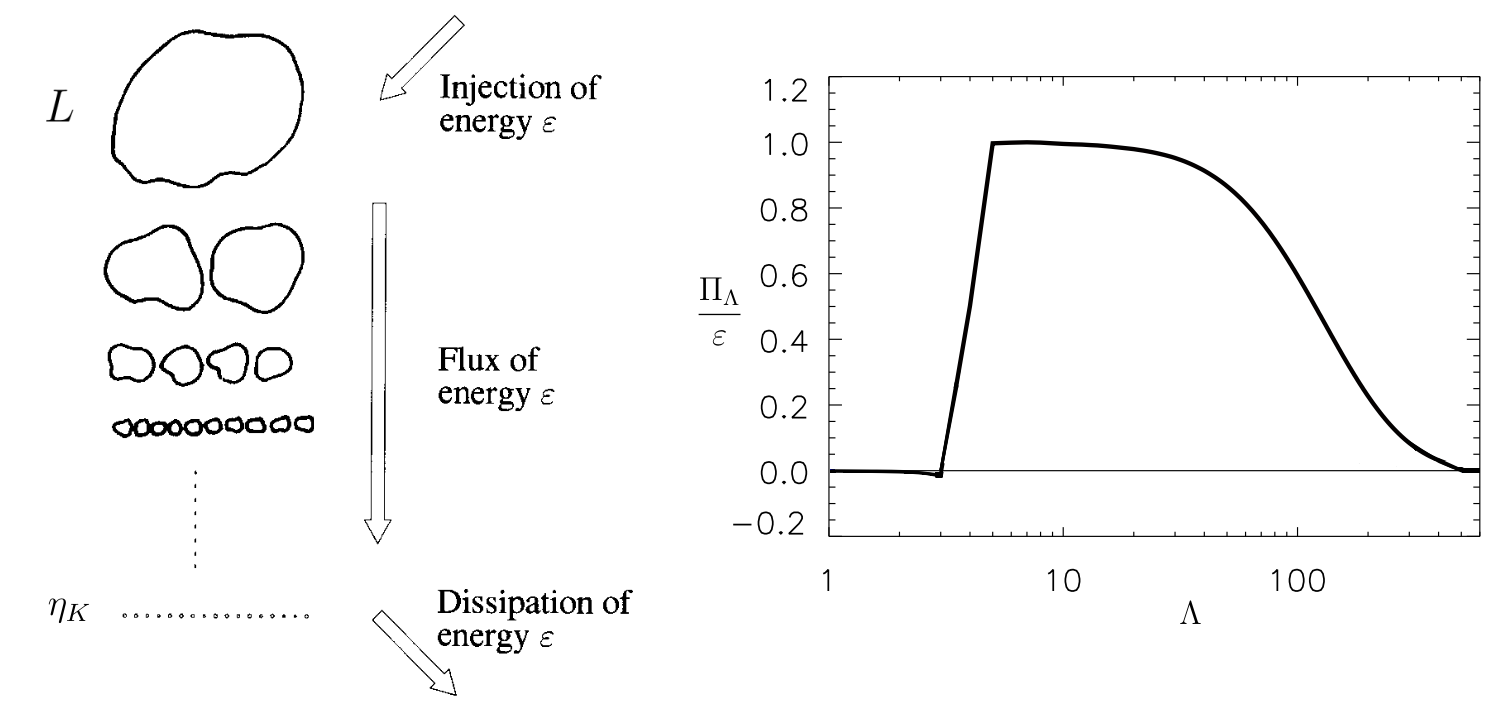}
\caption[Richardson cascade and energy flux $\Pi_\Lambda$.]{Left panel sketches the phenomenological picture of the Richardson cascade. The right panel shows the energy flux (Eq.~\eqref{2.15}) obtained by \ac{DNS} of homogeneous and isotropic turbulence \cite{alexakis2017helically}. Figures reproduced from \cite{frisch1995turbulence} and \cite{alexakis2018cascades} with minor modifications.}
\label{fig2.3}
\end{figure}

The left panel of Fig.~\reffig{fig2.3} is the standard phenomenological picture, where the energy is injected at a scale $L$, referred to as the integral length, typically enhances the motion of large-scale structures (or simply eddies). Larger eddies pump energy to smaller ones by a nonlinear and conservative transfer of energy through a wide range of scales (inertial range). Finally, the energy is suddenly dissipated into heat at the smallest dynamical scale $\eta_K$, known as Kolmogorov length scale. Moreover, the energy is dissipated at the same rate as the energy pumping $\vare=E_{in}$ since the cascade process is conservative.

The zeroth law of turbulence is related to the Richardson cascade by taking the limit $\Lambda\rightarrow\infty$ in Eq.~\eqref{2.14}. In this picture, the cascade flux ceases and the injected energy is in complete balance with dissipation. In fact, this relation was already anticipated at Eq.~\eqref{2.8} or even in Eq.~\eqref{1.24} for the global energy balance. However, this law is not related to the small-scale limit but, instead, the inviscid limit stated as follows:

\textit{``In a sequence of fully developed turbulence experiments with all parameters fixed except the viscosity $\nu$, sequentially probing smaller viscosity ($Re\rightarrow\infty$) the energy dissipation rate (per unit of mass) converges to a finite positive value.''}

The latter statement means, mathematically, that
\be\label{2.19}
\lim_{\nu\rightarrow0}\vare(\nu)=\vare_0 \ ,\
\ee
which, by the energy cascade phenomenology, 
\be\label{2.20}
2\nu\Omega=\vare_0=E_{in} \ .\
\ee
Eq.~\eqref{2.20} means that the enstrophy grows as $\Omega\sim\nu^{-1}$ and, for infinite Reynolds number, it diverges and statistics of velocity gradients (or vorticity) can be ill-behaved, developing extremely high values.

The zeroth law of turbulence, first stated by G.I. Taylor in 1935 \cite{taylor1935statistical}, is probably the first appearance of the concept of anomalies in physics. An anomaly occurs when the mechanism responsible to break the symmetry of the system is continuously restored, but the system is still breaking the symmetry. In our case, the viscous dissipation $\nu\partial^2v_i$ breaks time-reversal symmetry \ref{p126} and, even in the limit of vanishing viscosity $\nu\rightarrow0$ when the symmetry is restored, the solution of the dynamical equations is still dissipative. In this sense, the zeroth law of turbulence is usually referred to as a dissipative anomaly.

A deeper understanding of the dissipative anomaly in terms of rigorous mathematical analysis is the subject of a discussion originally stated by Onsager in 1949 \cite{onsager1949statistical}. This discussion is concerned with the language of weak solutions of \ac{iNSE} and Euler's equations. For general differential equations, weak solutions are usually an intermediate step towards the proof of regularity in the strong solutions, but this path has not been completed for the viscous and inviscid equations of fluid dynamics in three dimensions.

Onsager then noticed that, under certain conditions, the weak solutions (a function, usually nondifferentiable, which solves the equation when convolved with a compact smooth function) of the Euler equations ($\nu=0$ thus a conservative system) may display rough and irregular behavior, thus, not conserving energy. More precisely, consider the weak solution to the Euler equations $v^{(w)}_i(\vc x,t)$ such that
\be\label{2.21}
|\vc{v}^{(w)}(\vc x,t)-\vc{v}^{(w)}(\vc y,t)|\leq C|\vc x-\vc y|^h \ ,\ 
\ee
where $C$ is a constant. This condition is called H{\"o}lder continuity with an exponent $h > 0$, which measures the roughness of the field $\vc v^{(w)}$. If $h \geq 1$ the field is differentiable, while for $h < 1$, it is continuous but not a differentiable function. An example of a rough field satisfying H{\"o}lder continuity is Brownian motion, where $h=1/2$.

The so-called Onsager’s conjecture regarding the weak solutions of the Euler equations is stated as follows:

\begin{enumerate}[label=\textnormal{(Ons.\arabic*)}]
    \item If $h>1/3$, then the solution conserves energy.\label{ons1}
    \item If $h \leq1/3$, there exist weak solutions that do not conserve energy.\label{ons2}
\end{enumerate}

The \ref{ons1} statement was proved in 1994 \cite{constantin1994onsager}. The interval $h<1/3$ was proved only recently by Isett in 2018 for general dimensions $d \geq 3$. Still, the proof for the case $h=1/3$ remains open. The connection between the value $h=1/3$ and its relevance in turbulence will be clear in Section~\refsec{sec2.3} where the \ac{K41} is discussed. However, Duchon and Robert \cite{duchon2000inertial} showed that the term responsible for dissipation in the weak solutions of the Euler equation is similar to the inertial transport term $\Pi_\Lambda$ in the \ac{iNSE}. For this reason, the dissipative solutions of Euler's equations is often referred to as inertial dissipation, in the words of Onsager \cite{onsager1949statistical}:

\textit{``It is of some interest to note that, in principle, turbulent dissipation as described could take place just as readily without the final assistance by viscosity. In the absence of viscosity, the standard proof of the conservation of energy does not apply because the velocity field does not remain differentiable!''}

For this reason, it is a widely accepted hypothesis that weak solutions of the \ac{iNSE} in the singular limit of $\nu\rightarrow0$ display the same inertial dissipation mechanism as the weak Euler's solutions \cite{eyink2006onsager}.
\begin{figure}[t]
\center
\includegraphics[width=\textwidth]{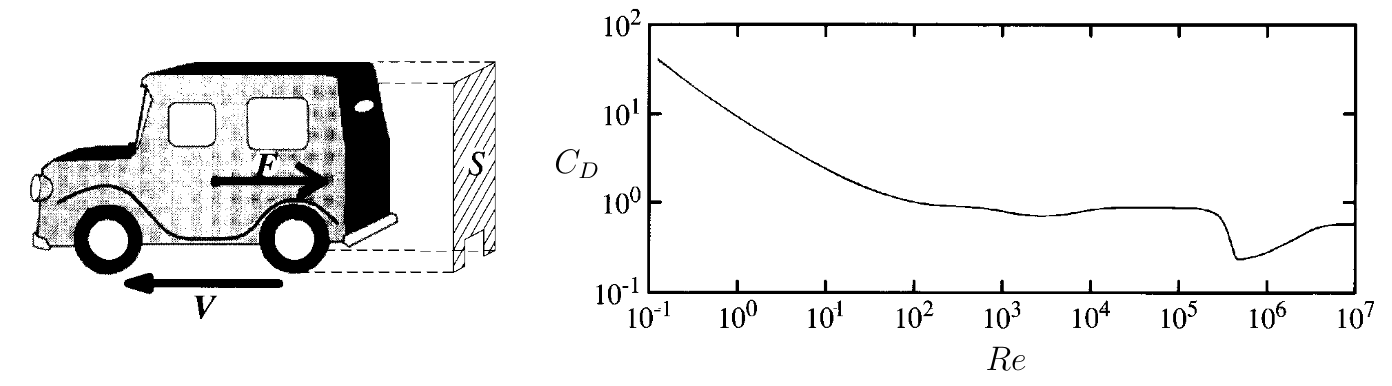}
\caption[Drag coefficient and zeroth law.]{Left panel sketches the physical picture of a car subject to turbulent drag. The right panel shows the dependency on the Reynolds number of $C_D$. Both figures were reproduced from \cite{frisch1995turbulence}.}
\label{fig2.4}
\end{figure}

A very interesting daily life consequence of the zeroth law is related to the momentum transfer from the flow to immerse objects. Fig.~\reffig{fig2.4} (left panel) shows the typical set up we are interested in, where a body with cross-section $S$ is moving relative to the fluid with density $\rho$ with velocity $V$. A drag force is applied to the body due to the momentum transfer from the fluid to the body $p$. In the latter setting the momentum transfer is the product of the relative velocity $V$ with the mass of fluid displaced by the movement of the car within a time interval $T$. Thus, $p=m V=\rho S V^2 T$, consequently
\be\label{2.22}
F=\frac{dp}{dT}=\frac{1}{2}C_D\rho S V^2\ ,\
\ee
where $C_D/2<1$ means that only a fraction of the momentum is transferred. The parameter $C_D$ is the so-called drag coefficient and, regarding Reynolds similarity, it is correct to suppose that, for objects with the same shape, $C_D$ is a function of Re only. The right panel of Fig.~\reffig{fig2.4} shows the Reynolds dependency of the drag coefficient for a cylinder. In the latter case, the Reynolds number is based on the typical length $L_S\sim\sqrt{S}$. If one estimates, for example, the energy dissipated per unit of mass per unit of time by the moving object, we have
\be\label{2.23}
\vare=\frac{FU}{\rho L^3}=\frac{1}{2}C_D\frac{U^3}{L}\ .\
\ee

The statement of $C_D$ to be Reynolds independent (at least piecewise) as shown in Fig.~\reffig{fig2.4} is the same as stating $\vare$ is independent of the Reynolds number and, consequently, of the viscosity. So the definition of a well-behaved drag coefficient relies upon the dissipation anomaly of turbulent flows. Moreover, Eq.~\eqref{2.23} can be restated as $U^3\sim \vare L$, which is the first hint on one of the most important laws of fully developed turbulence, the four-fifths law, now denoted only by 4/5's.

\subsection{Four-fifths law}

$\ \ \ \ $In the year 1941, Kolmogorov noted that an exact relation could be derived for turbulence directly from the \ac{iNSE} \cite{kolmogorov1941dissipation}. The 4/5-law can be stated as follows:

\textit{``In the limit of infinite Reynolds number, the third order (longitudinal) structure function of homogeneous isotropic turbulence, evaluated for increments $\ell$ small compared to the integral scale $L$, is given in terms of the mean energy dissipation per unit mass $\vare_0$ (assumed to remain finite and nonvanishing) by}''
\be\label{2.24}
\mean{\lr{\delta_\ell v_{\parallel}}^3}=-\frac{4}{5}\vare_0 \ell \ .\
\ee

The relation of Eq.~\eqref{2.24} assumes only homogeneity, isotropy, and the Zeroth law of turbulence dissipation. Without any further consideration, Kolmogorov derived one of the most celebrated exact relations in the whole field of turbulence. Due to its importance, we will provide a derivation similar to the one made by Kolmogorov, but in a more concise mathematical way. Firstly, we note in Kolmogorov's statement, a missing definition of what he called a ``third order (longitudinal) structure function''. To precisely define it, consider
\be\label{2.25}
S_{p,i}(\vc x,\vc \ell,t)\equiv\mean{\Big(v_i(\vc x +\vc \ell,t)-v_i(\vc x,t)\Big)^p} \ ,\
\ee
this mathematical definition does not encompass all the possible tensorial contractions of $p$ velocity fields, instead, it is already a simplified definition of structure-function which is convenient for homogeneous statistics since $S_{p,i}(\vc x,\vc \ell,t)=S_{p,i}(\vc \ell)$ only. Isotropy, in this setting, means two possible cases: the one where $\vc v \parallel \vc \ell$ and $\vc v \perp \vc \ell$. Any other case is a linear combination of these two. Then, the longitudinal structure function is defined as the one where the increment in the velocity is parallel to its direction,
\be\label{2.26}
S^L_{p}(\ell)\equiv S_{p,i}(\ell \hat{x}_i)=\mean{\Big(v_i(\vc x + \ell \hat{x_i},t)-v_i(\vc x,t)\Big)^p} \ ,\
\ee
where no sum over the $i$ index is intended. Historically, velocity increments are usually denoted by $\delta_\ell v_{(.)}=v_i(\vc x + \ell \hat{x_j},t)-v_i(\vc x,t)$, where $(.)$ is $\parallel$ or $\perp$ if $i=j$ or $i\neq j$. Now, Eq.~\eqref{2.24} is completely defined in terms of Eqs.~\eqref{2.20} and~\eqref{2.26}.

On the road to deriving the 4/5-law, three intermediate steps are required. First, the one known as the von-Kármán-Howarth-Monin relation \cite{de1938statistical,monin1959theory}:
\be\label{2.27}
\vare(\ell)\equiv-\partial_t\lr{\frac{\mean{v_i(\vc x ,t)v_i(\vc x +\vc \ell,t)}}{2}}\Bigg|_{NL}=-\frac{1}{4}\partial_{\ell_j}\mean{|\delta_\ell \vc v|^2\delta_\ell v_j}\ ,\
\ee
where the symbol $\partial_t(.)|_{NL}$ means the nonlinear contribution to the time rate of change and $\partial_{\ell_j}=\partial/\partial\ell_j$. The non-vanishing nature of the time derivative is related to the first derivation by von-Kármán and Howarth who did it in the context of decaying turbulence \cite{de1938statistical}. However, the extension made by Monin \cite{monin1959theory} in the case of steady and locally isotropic turbulence, leads to the same result as the decaying case. In this sense, let us continue with the derivation of Eq.~\eqref{2.27} for unsteady flows. Distributing the time derivative in Eq.~\eqref{2.27}, using \ac{iNSE} and collecting only the nonlinear terms we left with,
\be\label{2.28}
\vare(\ell)=\frac{\mean{v_iv'_j\partial'_jv'_i}+\mean{v'_iv_j\partial_jv_i}}{2}\ ,\
\ee
where $\partial'_i$ means the derivative with respect to $x'_i=x_i+\ell_i$ and $v'$ means the spatial argument of the velocity field is $x'$.
On the other hand, 
\begin{align}\label{2.29}
\mean{|\delta_\ell \vc v|^2\delta_\ell v_j}=\mean{\lr{v'_i-v_i}\lr{v'_i-v_i}\lr{v'_j-v_j}}
&=\mean{v_i'v_iv_j}-\mean{v_i'v_iv_j'}+\\ \nonumber
&-\mean{v_i'v_i'v_j}+\mean{v_i'v_i'v_j'}+\\ \nonumber
&-\mean{v_iv_iv_j}+\mean{v_iv_iv_j'}+\\ \nonumber
&+\mean{v_iv_i'v_j}-\mean{v_iv_i'v_j'} \ . \
\end{align}
It is clear by homogeneity that $\mean{v_iv_iv_j}=\mean{v_i'v_i'v_j'}$, hence these terms cancel. Also, the terms in the first and last lines are equal, then Eq.~\eqref{2.29} reads
\be\label{2.30}
\mean{|\delta_\ell \vc v|^2\delta\ell v_j}=
2\mean{v_i'v_iv_j}
-2\mean{v_iv_i'v_j'}
+\mean{v_iv_iv_j'}
-\mean{v_i'v_i'v_j} \ . \
\ee
Following the rightmost term in Eq.~\eqref{2.27}, an $\ell$-divergence operator must be applied to Eq.~\eqref{2.30}. Homogeneity again guarantees $\mean{g(\vc x)\partial_{\ell_j}f(\vc x+\vc \ell)}=\mean{f(\vc x)\partial_{\ell_j}g(\vc x-\vc \ell)}$ so that, using the incompressibility constraint in Eq.~\eqref{2.30}, we are led to 
\be\label{2.31}
-\frac{1}{4}\partial_{\ell_j}\mean{|\delta_\ell \vc v|^2\delta\ell v_j}=
\frac{\partial_{\ell_j}\mean{v_iv'_jv'_i}-\partial_{\ell_j}\mean{v_i'v_jv_i}}{2}\ . \
\ee
By noting that $\partial'_i=\partial_{\ell_i}=-\partial_i$ due to the chain rule, one can use Eqs.~\eqref{2.31} and \eqref{2.28} to show Eq.~\eqref{2.27}.

The second relation connects the definition of $\vare(\ell)$ to $\Pi_\Lambda$ by applying a threshold $\Lambda$ in the Fourier transform of $\vare(\ell)$,
\be\label{2.32}
\Pi_\Lambda=-\partial_t\vare_\Lambda|_{NL}=\int_{|\vc k|<\Lambda}\frac{d^3k}{(2\pi)^3}\int d^3\ell \ e^{i\vc k\cdot\vc\ell}\vare(\ell)\ , \
\ee
integrating $d^3k$ in spherical coordinates, one is able to show that,
\be\label{2.33}
\Pi_\Lambda=\frac{1}{2\pi^2}\int d^3\ell \frac{\vare(\ell)}{\ell^3}\Big(\sin(\Lambda \ell)-\Lambda \ell \cos(\Lambda \ell)\Big)\ . \
\ee
The rightmost term in Eq.~\eqref{2.33} can be integrated by parts to get
\be\label{2.34}
\Pi_\Lambda=\frac{1}{2\pi^2}\int d^3\ell \frac{\sin(\Lambda\ell)}{\ell}\lr{\frac{\vare(\ell)}{\ell^2}+\frac{1}{\ell}\frac{d\vare(\ell)}{d\ell}}=
\frac{1}{2\pi^2}\int d^3\ell \frac{\sin(\Lambda\ell)}{\ell}\vc\nabla_\ell\cdot\lr{\frac{\vc \ell}{\ell^2}\vare(\ell)}\ , \
\ee
where, by the use of the von-Kármán-Howarth-Monin relation:
\be\label{2.35}
\Pi_\Lambda=-\frac{1}{8\pi^2}\int d^3\ell \frac{\sin(\Lambda\ell)}{\ell}\partial_{\ell_i}\lr{\frac{\ell_i}{\ell^2}\partial_{\ell_j}\mean{|\delta_\ell \vc v|^2\delta_\ell v_j}}\ . \
\ee
Eq.~\eqref{2.35} is halfway on the derivation of the 4/5-law, now, the last ingredient is how to relate the mean value $\mean{|\delta_\ell \vc v|^2\delta\ell v_j}$ to the longitudinal structure function defined by Eq.~\eqref{2.26} for $p=3$. 

Note, however, that no assumption of isotropy was made up to this point. The remaining relation to be derived is the only one that requires this assumption. Isotropy means that the most general tensor structure leading to third-order correlations must be constructed in terms of $\delta_{ij}$ and the components of $\hat{\ell}=\vc\ell/\ell$ such that,
\be\label{2.36}
b_{ij,k}(\ell)\equiv\mean{v_iv_jv'_k}=C(\ell)\delta_{ij}\hat\ell_k+D(\ell)\lr{\delta_{jk}\hat\ell_i+\delta_{ik}\hat\ell_j} +F(\ell)\hat\ell_i\hat\ell_j\hat\ell_k \ , \
\ee
where $C$, $D$, and $F$ are unknown functions. Moreover, the incompressibility constraint $\partial'_kb_{ij,k}=0$ relates all those functions, in such a way that it can be expressed in terms of a single function $C$, for instance,
\be\label{2.37}
b_{ij,k}(\ell)=C\delta_{ij}\hat\ell_k+\lr{C+\frac{\ell C'}{2}}\lr{\delta_{jk}\hat\ell_i+\delta_{ik}\hat\ell_j} +\lr{\ell C'-C}\hat\ell_i\hat\ell_j\hat\ell_k \ , \
\ee
with $C'=dC/d\ell$. Within this definition, we can relate the tensor formed by increments $B_{ijk}=\mean{\delta_\ell v_i\delta_\ell v_j\delta_\ell v_k}$ to $b_{ij,k}$. After a long string of calculations, one gets,
\be\label{2.38}
B_{ijk}=6\lr{\ell C'-C}\hat\ell_i\hat\ell_j\hat\ell_k-
2\lr{\ell C'+C}\lr{\delta_{jk}\hat\ell_i+\delta_{ik}\hat\ell_j+\delta_{ij}\hat\ell_k}\ . \
\ee
The third-order longitudinal structure function is then given by,
\be\label{2.39}
S^L_3(\ell)=\mean{\lr{\delta_\ell v_\parallel}^3}=B_{ijk}\hat\ell_i\hat\ell_j\hat\ell_k=-12 C\ , \
\ee
while,
\be\label{2.40}
\mean{|\delta_\ell \vc v|^2\delta_\ell v_j}=B_{iik}\hat\ell_k\hat\ell_j=-4\lr{\frac{dC}{d\ell}+4\frac{C}{\ell}}\frac{\ell_j}{\ell}\ .\
\ee
Eq.~\eqref{2.40} forms the third and last relation needed for the presenting derivation of the 4/5-law. As the latter two (Eqs.~\eqref{2.27} and~\eqref{2.34}) combine to form Eq.~\eqref{2.35}, the definitive equation we will work is derived by using Eq.~\eqref{2.40} in \eqref{2.35} and integrating the angles in spherical coordinates (as we already assumed isotropy). The final resulting expression is given by
\be\label{2.41}
\Pi_\Lambda=-\int_0^\infty dx \frac{\sin(x)}{x}f\lr{\frac{x}{\Lambda}}\ ,\
\ee
where
\be\label{2.42}
f(x)=\lr{1+x\partial_x}\lr{3+x\partial_x}\lr{5+x\partial_x}\lr{\frac{S_3^L(x)}{6\pi x}}\ .\
\ee
Using again the fact that the forcing mechanism generating turbulence is supported at large scales, then $\Pi_{\Lambda\gg\Lambda_L}\approx\vare_0$ is only satisfied in the case where $f(x)$ is constant. This means that the third-order longitudinal structure function is related to the solutions of the following differential equation,
\be\label{2.43}
\lr{1+\frac{d}{dy}}\lr{3+\frac{d}{dy}}\lr{5+\frac{d}{dy}}s(y)=-12\vare_0\ ,\
\ee
derived from Eq.~\eqref{2.41} by the use of $\Pi_\Lambda=\vare_0$, $y\equiv\ln(x)$ and $s(y)\equiv\ell^{-1}(y)S_3^L(\ell(y))$. The general solution of Eq.~\eqref{2.43} has a homogeneous and inhomogeneous part. The former solves the same Eq.~\eqref{2.43} with $\vare_0=0$, a well-known case study of ordinary differential equations. However, the homogeneous solutions have the form $S^L_3(\ell)\propto \ell^{a}$ with $a=0$, $-2$ and $-4$. None of the latter solutions are physical since the limiting behavior of $S_p(\ell\rightarrow0)=0$ (by definition) is not fulfilled. Instead, the only allowed solution is $s(y)=s$ a constant, and, by the use of Eq.~\eqref{2.43}, we have
\be\label{2.44}
15s=-12\vare_0\Longrightarrow S^L_3(\ell)=-\frac{4}{5}\vare_0\ell\ .\
\ee
\begin{figure}[t]
\center
\includegraphics[width=\textwidth]{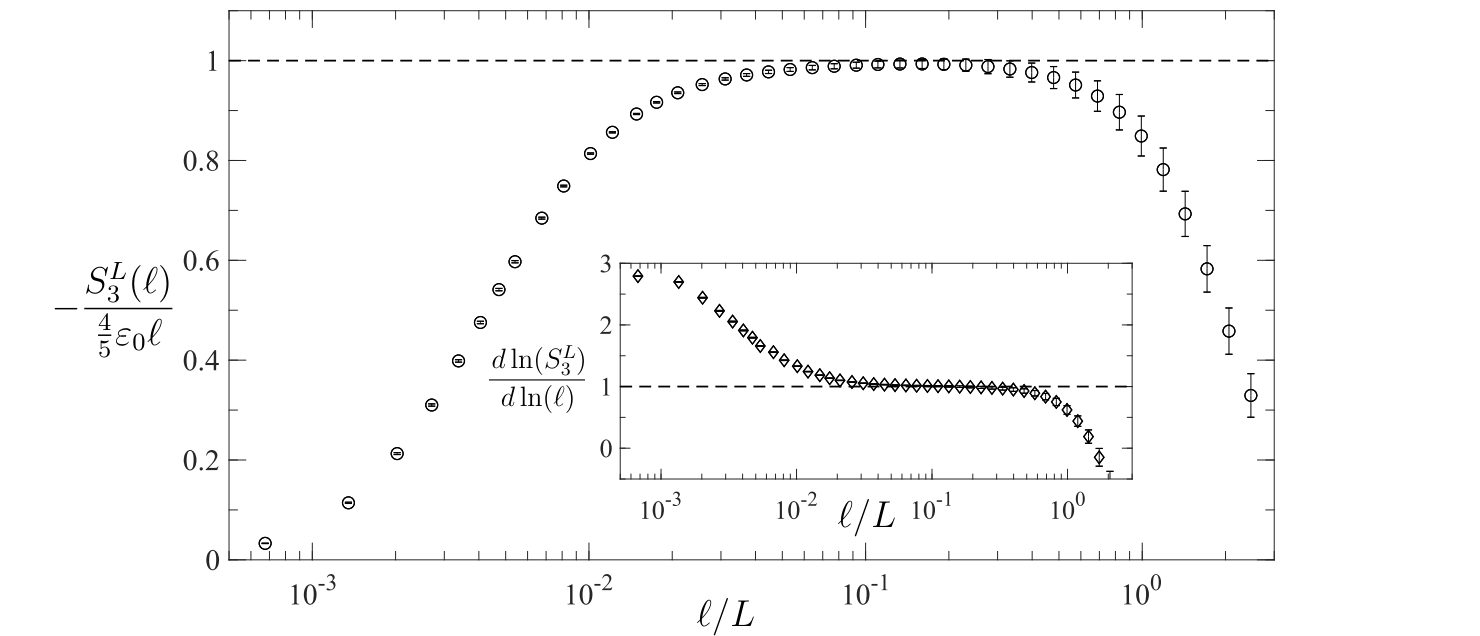}
\caption[4/5 law in simulations.]{The figure shows the validity of Eq.~\eqref{2.24} in a \ac{DNS} of homogenous and isotropic turbulence put forward by \cite{iyer2020scaling} (from where the figure was reproduced). Inset shows the logarithmic derivative of $S_3^L$ which measures the power law exponent it displays.}
\label{fig2.5}
\end{figure}

Figure~\reffig{fig2.5} shows the accordance of the 4/5 law in a \ac{DNS}s of \ac{iNSE} in the turbulent regime. The 4/5 law not only defines several quantities on turbulence, for example, the inertial range $\eta_K\ll\ell\ll L$ can be defined as the range where the 4/5 law is valid, but it also serves as a constraint that should be satisfied by any simplified turbulence model. Unless the system explicitly violates at least one of the hypotheses leading to the 4/5 law.

\section{K41 Phenomenology}\label{sec2.3}

$_{}$

The set of ideas and relations put forward by Kolmogorov in his seminal work of 1941 \cite{kolmogorov1941local} is what is nowadays known as \ac{K41}. Besides receiving the name of a ``theory'', the ideas of Kolmogorov are of phenomenological nature but still, with an impressive physical intuition, they form the solid rock of the modern understanding of the phenomena of turbulence. Two basic constituents form the pillars of the \ac{K41} idea (in a modern view). The first is the restoration of all the symmetries of the \ac{iNSE} discussed in Section~\refsec{sec1.2}, except time reversal and scale invariance which are anomalously broken. The second is the 4/5 law.

From the statistical restoration of scaling symmetry, we expect that,
\be\label{2.45}
\delta_{\lambda \ell}v_i=\lambda^h\delta_\ell v_i \ ,\
\ee
where the symbol of equality is meant to be in the distributional sense in this case (commonly referred to as equality in law). However, one can directly use this scaling relation in the 4/5 law to get, 
\be\label{2.46}
\mean{\lr{\delta_{\lambda\ell} v_\parallel}^3}=\lambda^{3h}\mean{\lr{\delta_\ell v_\parallel}^3}=-\frac{4}{5}\vare_0\lambda\ell \ ,\
\ee
which is only satisfied for the value $h=1/3$. This means that if one defines the scaling exponent of the $p^{th}$-order longitudinal structure function to be $S^L_p(\ell)=\mean{\lr{\delta_\ell v_\parallel}^p}\sim\ell^{\alpha_p}$, then scale invariance connects,
\be\label{2.47}
S^L_p(\lambda\ell)=\lambda^{p/3}S^L_p(\ell)\Longrightarrow\lr{\lambda\ell}^{\alpha_p}=\lambda^{p/3}\ell^{\alpha_p} \ ,\
\ee
then necessarily $\alpha_p=p/3$. As the combination $\vare_0\ell$ has the same dimensions of $v_i^3$, then, without loss of generality, in the inertial range we have
\be\label{2.48}
S^L_p(\ell)=C_p\vare^{p/3}\ell^{p/3}\ ,\
\ee
where $C_p$ are nondimensional quantities not necessarily universal, except from $C_3=-4/5$. Note that, recalling Onsager's conjecture, the statement of Eq.~\eqref{2.48} naturally induces the regularity of the velocity field to be associated with $h=1/3$ in Eq.~\eqref{2.21}. This means the upper maximum value of dissipative weak solutions for the Euler equation where the conjecture is not exactly proved.

Eq.~\eqref{2.48} forms the core of \ac{K41} phenomenology. The original derivation done by Kolmogorov does not involve the 4/5 law since it was not proved at the time. Instead, he conjectured the so-called \textit{first and second similarity hypotesis} which, in practical terms, states that in the limit of vanishing viscosity (or infinite Reynolds number), the only quantities responsible by the statistics in the inertial range are $\vare_0$ and the length scale itself $\ell$. Thus, by pure dimensional analysis, Kolmogorov was led to Eq.~\eqref{2.48} but with the  assumption that the coefficients $C_p$ are universal.

The hallmark of the \ac{K41} was then the result for the energy spectrum. By the use of Eqs.~\eqref{2.7},~\eqref{2.48} and the isotropic assumption, it is not difficult to show that if $S_2(\ell)\propto\ell^\alpha$ then $E(k)\propto k^{-\alpha-1}$ for $0<\alpha<2$ in three dimensions. According to \ac{K41},
\be\label{2.49}
E(k)=C_K\vare^{2/3}\ell^{-5/3}\ ,\
\ee
where $C_K$ is known as Kolmogorov's constant and is connected to $C_2$. The relation of Eq.~\eqref{2.49} it is valid only in the inertial range, where no detail on the dissipation mechanism nor the energy pumping is relevant. Furthermore, it is often referred to as Kolmogorov's 5/3 law, however, it is not a law in the same sense of the 4/5's but a simple consequence of the latter and the supposed scale invariance. Later on, we will see that Eq.~\eqref{2.49} is subject to corrections due to intense non-Gaussian fluctuation related to the breakdown of the statistical scale invariance.

The definitive experimental verification of the 5/3 law was provided more than 20 years after its first proposal \cite{grant1962turbulence}, while a feasible estimate for $C_K\approx1.6$ even later \cite{sreenivasan1995scaling}. An interesting use of Eq.~\eqref{2.49} to get information about scales in turbulence is the derivation of the Kolmogorov length scale $\eta_K$. Recalling Eq.~\eqref{1.20} in the Fourier space with a filter $\Lambda$, we get
\be\label{2.50}
\Omega_{\Lambda}\propto\int_0^{\Lambda}k^2 E(k)\sim \vare_0^{2/3}\Lambda^{4/3}\ .\
\ee
Using Eq.~\eqref{2.14} and supposing again the existence of a length scale $\eta_K\sim \Lambda_\eta^{-1}$ such that the flux $\Pi_\Lambda$ suddenly ceases, i.e., the cascade process ends, we get
\be\label{2.51}
\nu\Omega_{\Lambda_\eta}\vare^{2/3}\sim\vare\Longrightarrow\Lambda_\eta^{-1}\sim\eta_K\equiv\lr{\frac{\nu^3}{\vare_0}}^{1/4}\ .\
\ee
The physical interpretation of the Kolmogorov scale $\eta_K$ is the scale where the cascade process ceases and the field can be regarded as smooth again in the differentiable sense.

There is also a third length scale in homogeneous and isotropic turbulence, the Taylor micro-scale $\lambda$, defined as follows,
\be\label{2.52}
\lambda\equiv\frac{U}{\sqrt{\mean{\lr{\partial_1v_1}^2}}}\ ,\
\ee
where $U$ is the same velocity scale defined for the calculation of the Reynolds number. The role of the Taylor micro-scale $\lambda$ on the overall phenomenology of turbulence is not as evident as Kolmogorov's scale $\eta_K$. However, one is able to show that,
\be\label{2.53}
\frac{\mean{v_i(\vc x +\rho\hat{x}_i,t)v_i(\vc x,t)}}{\mean{(v_i(\vc x, t))^2}}=1+\frac{\rho^2}{2\lambda^2}+\mathcal{O}(\rho^3)\ ,\
\ee
where the indices $i$ are not summed in this case. Eq.~\eqref{2.53} concludes that the Taylor micro-scale is the scale that controls the decaying of the longitudinal velocity correlation functions. In this sense, one may think of $\lambda$ as the scale where if $\rho\ll\lambda$ velocity fluctuations are not affected by the dissipation mechanism. 

The importance of defining the scales in turbulence goes beyond their physical interpretation and provides a way to properly define turbulence in terms of statistically universal properties, at least in the unbounded homogeneous and isotropic case. For example, the definition of the velocity scale $U$ that composes the Reynolds number is naturally the incoming velocity on pipe flows, but it has no clear definition for unbounded turbulence since $\mean{v_i}=0$. To properly derive the relation between scales in turbulence, let us provide a clear mathematical definition of all the relevant quantities in terms of mean values of velocity fields or their derivatives. To start with, we redefine the Reynolds number 
\be\label{2.54}
Re=\frac{UL}{\nu}\ ,\
\ee
where $U^2\equiv2E/3$ is the typical velocity, $E\equiv\mean{|\vc v|^2}/2$ the mean energy, $L\equiv U^3/\vare_0$ is the integral length, $\vare_0\equiv-\nu\mean{v_i\partial^2v_i}$ is the mean dissipation rate, and $\nu$ is the kinematic viscosity. The Taylor-based Reynolds number ($R_\lambda$) is defined as
\be\label{2.55}
R_\lambda=\frac{U\lambda}{\nu}\ ,\
\ee
where $\lambda$ is the Taylor micro-scale defined in Eq.~\eqref{2.52}. Isotropy, homogeneity and incompressibility relates $\vare_0$, $\lambda$ and the enstrophy ($\Omega\equiv\mean{|\vc \omega|^2}/2$) since all these quantities depends the tensor $\mean{\partial_iv_j\partial_kv_l}$. In this way, we have
\be\label{2.56}
\Omega=\frac{\vare_0}{2\nu}=\frac{5E}{\lambda^2}\ .\
\ee
Working with Eq.~\eqref{2.54},~\eqref{2.55},~\eqref{2.56} and all the definition made above, we have
\be\label{2.57}
R_\lambda^2=15 Re\ .\
\ee
Finally the relations among separation of the scales $\eta_K^4\equiv\nu^3/\vare_0$, $\lambda$, and $L$ as functions of the Reynolds number are
\be\label{2.58}
\frac{L}{\lambda}=\frac{R_\lambda}{15}\ ,\
\frac{L}{\eta_K}=\lr{\frac{R_\lambda}{\sqrt{15}}}^{3/2}\ , \text{ and } 
\frac{\lambda}{\eta_K}=\lr{\sqrt{15} R_\lambda}^{1/2}\ .\
\ee
The set of relations in Eq.~\eqref{2.58} corroborates the hypothesis of scale separation put forward on the derivation of the turbulent cascade and the 4/5 law. Note that the scales are ordered $\eta_K<\lambda<L$, such that in the turbulent regime ($R_\lambda \gg 1$), a wide inertial range is formed.

\subsection{Laudau's Criticism}

$\ \ \ \ $Shortly after the release of Kolmogorov's work in 1941 \cite{kolmogorov1941local}, L. D. Landau criticizes the universality of the $C_p$ coefficients found by Kolmogorov. Instead, Landau argues that these constants should depend on the details of the forcing mechanism of turbulent production and the small-scale fluctuations. A ``recent'' reformulation of Laudau's arguments was developed by Kraichnan \cite{kraichnan1974kolmogorov} and can be put forward as follows: Consider a super ensemble consisting of $N>1$ different experiments with different values of $\vare_0^{(i)}>0$, with $i=1,2\cdots, N$. Suppose that the constants $C_p$ are universal, by the use of Eq.~\eqref{2.48} we have
\be\label{2.59}
S^{(i)}_p(\ell)=C_p\lr{\vare_0^{(i)}}^{p/3}\ell^{p/3}\ ,\
\ee
where we omitted the superscript $S^L$. From now on, when we refer only to the structure-function we mean the longitudinal one, except when explicitly indicated. Assuming one can define a super-ensemble average, such that
\be\label{2.60}
S^{super}_p(\ell)\equiv\frac{1}{N}\sum_{i=1}^N S^{(i)}_p(\ell)\ \text{ and }\
\vare_0^{super}=\frac{1}{N}
\sum_{i=1}^N\vare_0^{(i)}\ ,\
\ee
are the super-averaged structure function and dissipation range, respectively.
On the one hand, we expect the super ensemble to be suitable for the application of scale invariance and 4/5 law, so that we get
\be\label{2.61}
S^{super}_p(\ell)=C_p \Big(\vare_0^{super}\Big)^{p/3}\ell^{p/3}\ .\
\ee
On the other hand, the use of Eqs.~\eqref{2.59} and~\eqref{2.60} leads to
\be\label{2.62}
S^{super}_p(\ell)=C_p\frac{1}{N}\sum_{i=1}^N\lr{\vare_0^{(i)}}^{p/3}\ell^{p/3}\ ,\
\ee
where Eqs.~\eqref{2.61} and \eqref{2.62} are simultaneously true only if  
\be\label{2.63}
\lr{\sum_{i=1}^N\lr{\vare_0^{(i)}}}^{p/3}=\sum_{i=1}^N\lr{\vare_0^{(i)}}^{p/3} \ .\
\ee
As a result, the only case when Eq.~\eqref{2.63} is satisfied is $p=3$.

The original super-ensemble argument of Kraichnan can be rephrased by, instead of considering different experiments, considering different space partitions of the same experiment. One may for example divide the fluid volume into $N$ partitions of volume $L^3$ with $L$ the usual integral length to minimize correlations between different partitions. In this setting, it is clear that different $\vare_0^{(i)}$ are related to different local averages of the dissipation field, which is reasonable since enstrophy can locally be produced by the vortex stretching mechanism. 

The original objections of Landau were made at a meeting in Kazan in 1942 and in a footnote in the first edition of his textbook in fluid mechanics, published in 1944 \cite{frisch1995turbulence}. For this reason, most of the details were only accessible through conversations shared at Kazan's conference. Furthermore, their precise meaning is the subject of discussion even in the present dates \cite{frisch1995turbulence}. In a translation of the proceeding of the Kazan's meeting, there is a brief resume of a discussion between Landau and Kolmogorov \cite{spalding1991kolmogorov}:
\begin{figure}[t]
\center
\includegraphics[width=\textwidth]{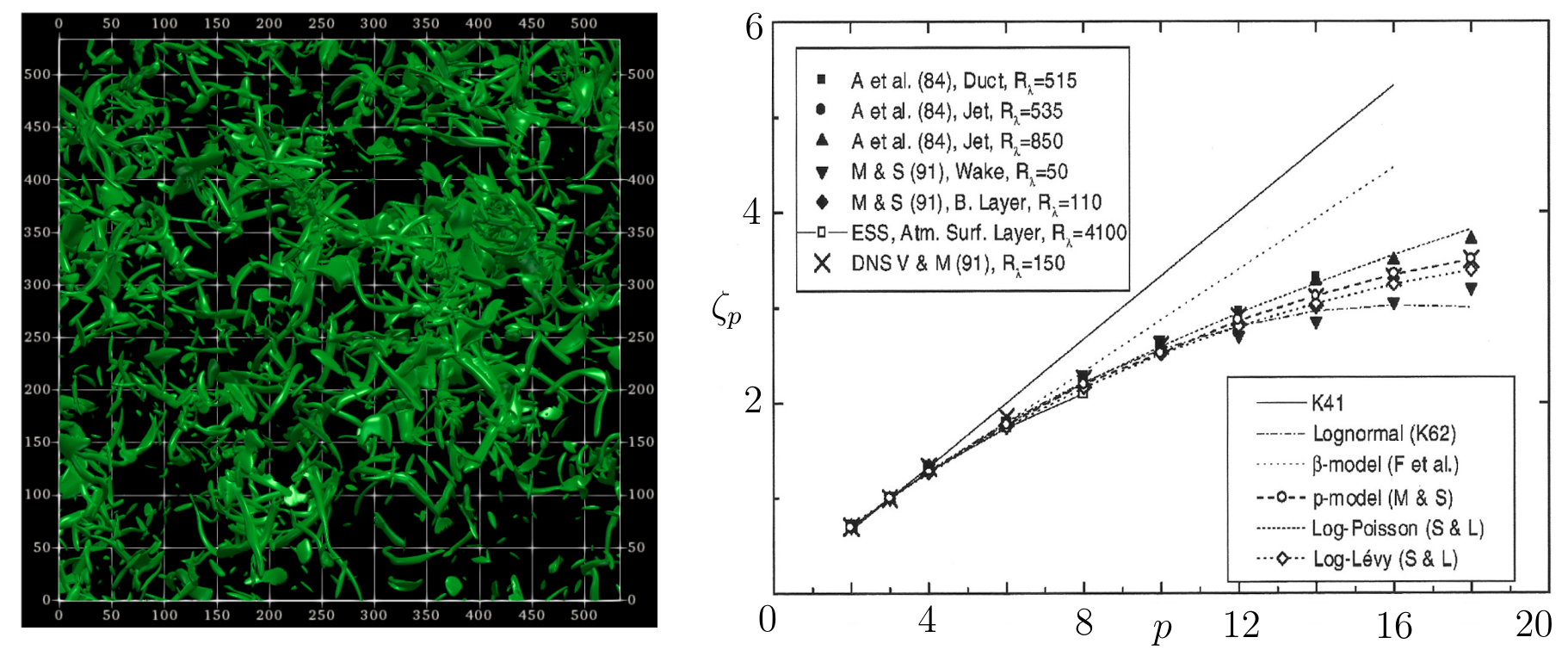}
\caption[Structural view of turbulence and fluctuating energy cascade.]{Left panel: Vorticity isosurfaces containing only $1\%$ of the most intense events for $R_\lambda=399$ (figure reproduced from \cite{ghira2022characteristics}). Right panel: Collection of several models and experiments on the scaling exponent of the longitudinal structure function $S_p(\ell)\sim\ell^{\zeta_p}$ reproduced from \cite{anselmet2001turbulent}.}
\label{fig2.6}
\end{figure}

\textit{``L. Landau remarked that A. N. Kolmogorov was the first to provide correct understanding of the local structure of a turbulent flow. As to the equation of turbulent motion, it should be constantly born in mind, in Landau's opinion, that in a turbulent field the presence of rotation in the velocity was confined to a limited region; qualitatively correct equations should lead to just such a distribution of eddies.''}

To the author's knowledge, this is the first signalization of the importance of intense localized vortex structures in turbulent flows. The concept of structural turbulence is nowadays widespread throughout the community and, is defined as the understanding of the flow as a superposition of vortex filaments with extremely high vorticity. 

Figure~\reffig{fig2.6} (left panel) shows the vortices organized as thin filaments with a core size of the order of $\eta_K$ and length of a few dozens of $\eta_K$ \cite{ghira2022characteristics}. The existence of such structures is the main precursor of the mechanism of breakdown of scale invariance that leads the structure functions to a scaling incompatible with \ac{K41} phenomenology $S_p(\ell)\sim\ell^{\zeta_p}$, with $\zeta_p\neq p/3$. The right panel of Fig.~\reffig{fig2.6} shows the comparison among different experiments and models (to be introduced in the next section) of the longitudinal structure function's scaling exponents \cite{anselmet2001turbulent}. It is worth noting that the expected universal result $\zeta_3=1$ is satisfied and, in fact, the \ac{K41} prediction holds well up to $p\leq4$. For higher orders, as they measure events that lie in the tail of the \ac{PDF}s, the nature of the non-Gaussian fluctuation starts to get shape and is reflected in the nonlinearity of $\zeta_p$. The latter phenomenon is usually referred to as ``intermittent fluctuations'' or simply intermittency.

For finite Reynolds numbers, the separation between Kolmogorov and integral scales can be, sometimes, not large enough to provide a good power law fitting of $S_p(\ell)$, for example. The left panel of Fig.~\reffig{fig2.7} shows the typical scale dependency of the third-order structure function. Note that, for the lowest Reynolds number ($R_\lambda=225$), the inertial range defined by $S_3\sim\ell$, is very short. However, something unexpected happens when $S_4(\ell)$ is plotted against $S_3(\ell)$. The right panel of Fig.~\reffig{fig2.7} shows that the details of small and large scales are exactly balanced for structure functions of different orders vis $S_4(\ell)\propto S_3^{\zeta_{(4|3)}}$. In fact, this phenomenon happens for every structure-function order in such a way that $S_p(\ell)\propto S_q^{\zeta_{(p|q)}}$. The latter phenomenon named \ac{ESS}, was empirically discovered by Benzi et al. in 1995 \cite{benzi1995scaling}. To determine the scaling exponent of the $p^{th}$-order structure functions, one must only determine $\zeta_{(p|3)}$ since $S_3\sim\ell$ from the 4/5 law. Furthermore, determining $\zeta_{(p|q)}$ means more accuracy, as it does not depend on a definition of a fitting range.
\begin{figure}[t]
\center
\includegraphics[width=\textwidth]{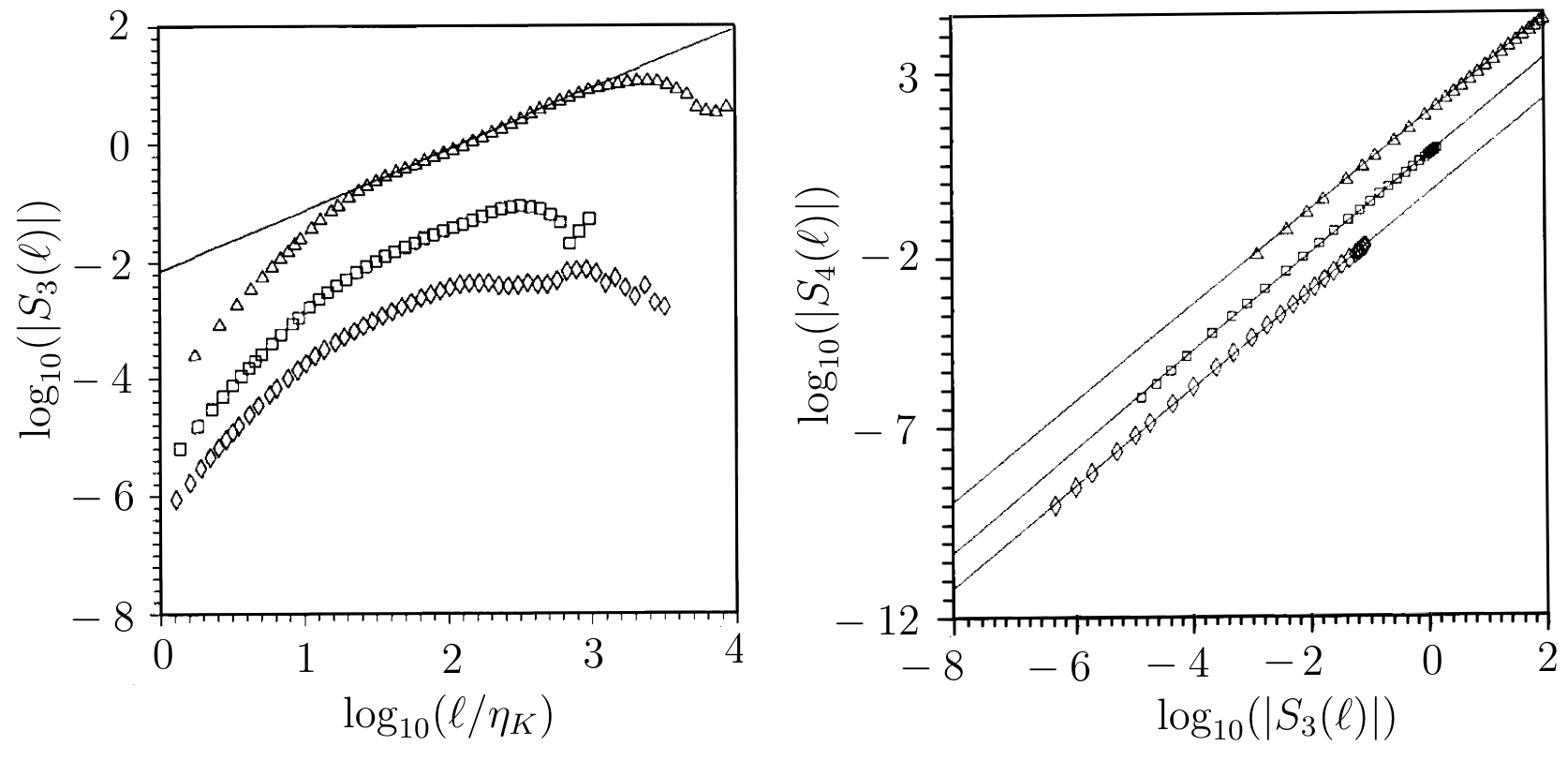}
\caption[Extended self-similar approach.]{The Left panel shows the third-order structure function as functions of the scale in log-log plot for three wind tunnel experiments with $R_\lambda=225$ (diamond), $342$ (square) and $800$ (triangle), respectively. Right panel shows $S_4(\ell)$ plotted against $S_3(\ell)$ for the same experiments. Both figures were reproduced from \cite{benzi1995scaling}.}
\label{fig2.7}
\end{figure}\clearpage

\section{Intermittency models}

$_{}$

The breakdown of the scale invariance through the formation of intense vorticity events was historically explored by the measurement of the scaling exponents $\zeta_p$. The natural way to introduce nonlinearities on $\zeta_p$ is to include fluctuations in the dissipation field. The latter can be done by imposing the cascade process to be random and scale-dependent. In order to do this, in the sense of Landau's objections, define a cubic cell of side $\ell$ centered at the point $\vc x$, then the local average energy dissipation (also denoted as coarse-grained dissipation rate) reads
\be\label{2.64}
\vare_\ell(\vc x)\equiv\frac{\nu}{2\ell^3}\int_{Cell}d^3y\Big(\partial_iv_j+\partial_jv_i\Big)^2 \ .\
\ee
Note that, due to homogeneity and Eq.~\eqref{2.20} we have, $\mean{\vare_\ell}=\vare_0$. The latter average is meant to be understood in the sense of super average as discussed in Eq.~\eqref{2.60}. However, as we are assuming ergodicity and Eq.~\eqref{2.64} defines $\vare_\ell$ locally, one may substitute it for the standard space average. 

More than 20 years after \ac{K41} and Landau's objections, Kolmogorov proposed a refinement for his first and second hypotheses which leads to the \ac{K41} behavior \cite{kolmogorov1962refinement}. The so-called \textit{refined similarity hypotheses} establishes that local fluctuation of velocity increments behaves as follows
\be\label{2.65}
\delta_\ell v=\lr{\vare_\ell \ell}^{1/3}\ ,\
\ee
meant to hold in the statistical sense. It is then clear the inspiration for this relation since the 4/5 law is satisfied $\mean{\lr{\delta_\ell v}^3}=S_3(\ell)\sim\vare_0\ell$. For higher-order structure functions, we have
\be\label{2.66}
S_p(\ell)=\mean{(\vare_\ell \ell)^{p/3}}\sim \ell^{\zeta_p}\ ,\
\ee
where
\be\label{2.67}
\zeta_p=\frac{p}{3}+\tau^{(3)}_{p}\ ,\
\ee
and
\be\label{2.68}
\mean{\lr{\vare^{1/3}_\ell}^p}\sim\ell^{\tau^{(3)}_p} \ .\
\ee
At this point, it is important to highlight that not all models for the local average energy dissipation satisfy $\mean{\lr{\sqrt[3]{\vare_\ell}}^p}\sim\ell^{\tau_{p/3}}$ if $\mean{\lr{\vare_\ell^p}}\sim\ell^{\tau_p}$, a commonly used relation in turbulence models. Assuming such a relation is the same as assuming Eq.~\eqref{2.63} holds for any $p$ besides three. However, as we will see in a moment, some of the most important intermittency models fulfill this property.

\subsection{Cascade Processes}

$\ \ \ \ $The general idea of phenomenological models of intermittency is to consider the energy cascade as a fragmentation process that occurs successively through the length scales. Fig.~\reffig{fig2.8} illustrates the process of fragmentation from the scale $\ell_0=L$ to $\ell_1= \ell_0/a$ then $\ell_2 = \ell_0/a^2$ and so on, where $a$ is a positive arbitrary parameter ($a=2$ in Fig.~\reffig{fig2.8}) indicating the step of the cascade process.
\begin{figure}[t]
\center
\includegraphics[width=\textwidth]{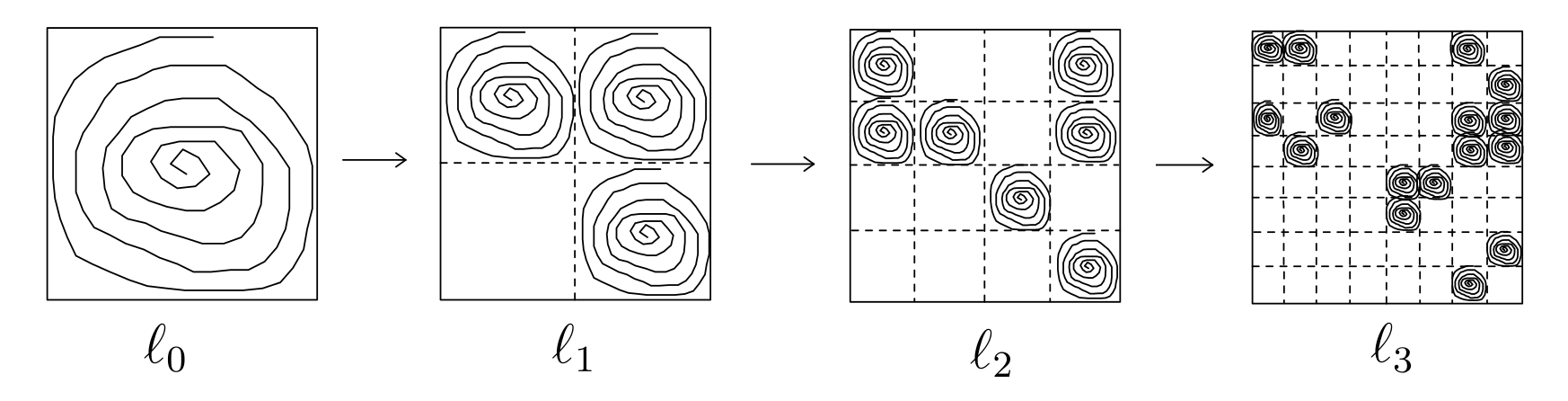}
\caption[Phenomenological picture of a random cascade process.]{Sketch of the phenomenological picture of a random cascade process, where the statistical law behind the fragmentation of vortices is the main matter of this kind of model. This figure was reproduced from L. Moriconi's lecture notes on turbulence at \url{https://www.if.ufrj.br/~moriconi/TeorEstTurb.pdf}.}
\label{fig2.8}
\end{figure}

The generation of eddies defined at the scale $\ell_n$ is associated with fluctuating local rates of energy dissipation $\vare_n$. The fragmentation of these eddies, metaphorically ``mother'' eddies, gives rise to various eddies, the ``sibling'' eddies, defined at the length scale $n+1$ associated with local rates of energy dissipation $\vare_{n+1}$. Let $W\geq0$ now be the proportionality factor between the dissipation rates of a specific sibling eddy and the dissipation rate of the parent eddy that originated it.

Thus, considering any eddy at the scale $n$, we can write its dissipation rate, by accounting for all ancestral generations until recursively reaching the first eddy defined at $L$. This means,
\be\label{2.69}
\vare_n=\vare_{n-1}W_{n}=\vare_{n-2}W_{n-1}W_{n}=\cdots=\vare_0W_1W_2\cdots W_n \ .\
\ee
Considering now $W$'s to be random independent variables that conserves the energy cascade ($\mean{W}=1$). The $p^{th}$-order moment of the cascade, supposing $W$'s to be identically distributed, we have
\be\label{2.70}
\mean{\vare^p_n}=\vare_{0}^q\mean{W^p}^n\sim \ell^{\tau_p}\ ,\
\ee
where
\be\label{2.71}
\tau_p=-\log_a\lr{\mean{W^p}}\ .\
\ee
For the derivation of Eq.~\eqref{2.70} we used the relation $\ell_n=\ell_0a^{-n}$. It is clear that in this kind of process, due to the statistical independence of the several $W$'s, a cascade of $\sqrt[3]{W}$'s has the scaling exponents of Eq.~\eqref{2.71} with $p\rightarrow p/3$ or, equivalently, $\tau^{(3)}_p=\tau^{(1)}_{p/3}$.

The model introduced by \cite{obukhov1962some,kolmogorov1962refinement}, nowadays known as \ac{OK62}, means that $W$'s are log-normally distributed, or $W=a^{-x}$ where $x$ is a Gaussian random variable with mean $\mean{x}=\bar x$ and variance $\mean{(x-\bar x)^2}=\sigma^2$. The energy conservation constraint relates $\bar x$ to $\sigma$ as
\be\label{2.72}
\bar x=\sigma^2\frac{\ln(a)}{2}\ .\
\ee
Now, by a redefinition $\mu=2\bar x$ and the use of Eqs.~\eqref{2.67} and~\eqref{2.72} we get
\be\label{2.73a}
\tau^{(1)}_p=\frac{\mu}{2}p(1-p)\ ,\
\ee
and, consequently
\be\label{2.73}
\zeta_p=\frac{p}{3}+\frac{\mu}{18}p(3-p)\ .\
\ee
The unique parameter $\mu$ is named as the intermittency parameter. The latter can be directly measured in real flows since
\be\label{2.730}
\frac{\mean{\vare_{n+1}^2}}{\mean{\vare_{n}^2}}=\mean{W^2}=a^\mu\ .\
\ee
The first measurement of the intermittency parameter in the `70s showed values as low as $0.18$ and as high as $0.7$ \cite{sreenivasan1977temperature}. With improved method of measuring the flow properties, the measure of the `80s was $\mu=0.2\pm0.2$ \cite{antonia1982statistics} and $\mu=0.25\pm0.05$ in the `90s \cite{sreenivasan1993update}. A recent measure that takes into account several different experimental setups converged to the value of $\mu=0.17\pm0.01$ \cite{tang2020scaling}.

Another interesting cascade model was introduced by She and Lévêque \cite{she1994universal} in 1994. The advantage of their approach is the absence of adjustable parameters as $\mu$ in the log-normal model. The precise derivation of She-Lévêque's scaling exponents involves postulates regarding the scaling behavior of the following variable
\be\label{2.74a}
\vare_\ell^{(\infty)}\equiv\lim_{p\rightarrow\infty}\frac{\mean{\vare^{p+1}_\ell}}{\mean{\vare^{p}_\ell}} \sim \ell^{-2/3}\ .\
\ee
This detailed derivation is not of substantial relevance in this thesis, since our focus will be on a mathematically formal derivation of the log-normal cascade. However, an interesting comment to be made at this point is that She-Lévêque's model is a particular case of a log-Poisson cascade, which means that $W=a^{\bar{x}-x}$ where $x\in \mathbb{|}^+$ and distributed as 
\be\label{2.74}
\rho(x)=\frac{c^x}{x!}e^{-c}\ .\
\ee
The postulates of \cite{she1994universal} and the conservative condition ($\mean{W}=1$) leads to the following scaling exponents
\be\label{2.75}
\tau^{(1)}_p=-\frac{2p}{3}+\lr{3-D_\infty}\left[1-\lr{\frac{7/3-D_\infty}{3-D_\infty}}^p\right]\ ,\
\ee
where $D_\infty=1$, and, consequently
\be\label{2.78}
\zeta_p=\frac{p}{9}+\lr{3-D_\infty}\left[1-\lr{\frac{7/3-D_\infty}{3-D_\infty}}^{p/3}\right]\ .\
\ee
The reason why we did not set this value of $D_\infty$ at Eq.~\eqref{2.75} will be clear in Section~\refsec{sec2.4.2}, where we will reinterpret the results of She and Lévêque in the light of the multifractal approach. Both cascade models put forward in this section, are shown in the right panel of Fig.~\reffig{fig2.6}. One can argue that both scaling exponents seem to agree with experimental and numerical data up to $p\approx10$ for $\mu=0.2$.

A direct implication of the intermittent fluctuations regards the correction to the energy spectrum of Eq.~\eqref{2.49}. Following the general expression of Eq.~\eqref{2.67}, we get,
\be\label{2.79}
E(k)\sim k^{-(5/3+\tau^{(3)}_2)}\ .\
\ee
In the log-normal model, the correction to the spectrum is $\mu/9\sim\mathcal{O}(10^{-2})$ and She-Lévêque's $14/9-2(2/3)^{2/3}\sim\mathcal{O}(10^{-2})$. In this sense, the spectrum is almost insensitive to the intermittent correction.

\subsection{General Multifractal Ideas}\label{sec2.4.2}

$\ \ \ \ $Now let us introduce a general set of ideas developed by B. Mandelbrot in the late 60s \cite{mandelbrot1967long,mandelbrot1975stochastic,mandelbrot1982fractal} and first applied by Frisch and Parisi in the context of turbulence in 1985 \cite{frisch1985turbulence}. As already discussed, Onsager's conjecture states that singularities will play an important role in the zero viscosity limit. Spatial sets with different singularities are supposedly distributed all over the flow volume.
\begin{figure}[t]
\center
\includegraphics[width=\textwidth]{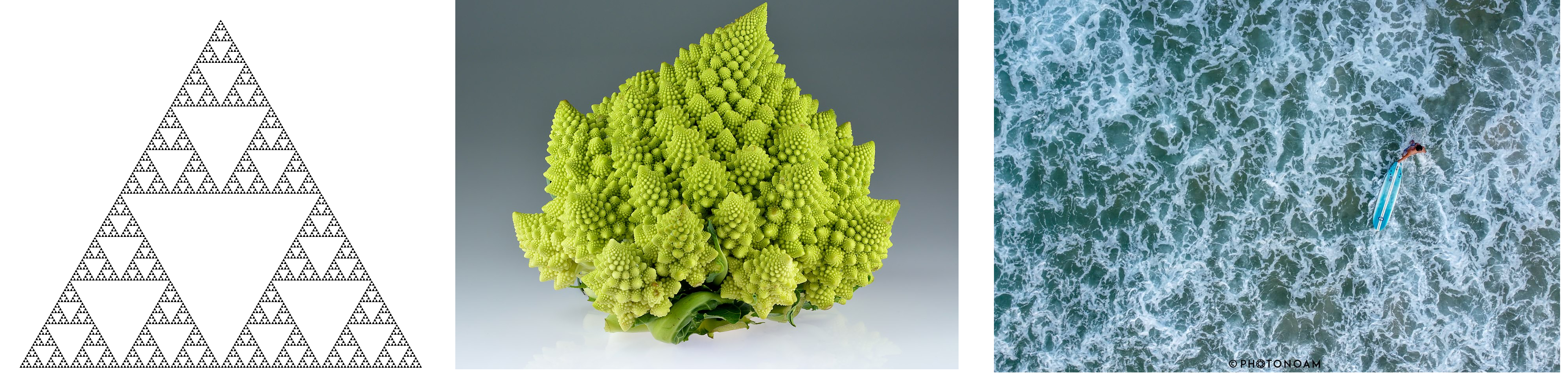}
\caption[Examples of fractals.]{Sierpinski triangle (left), Romanesco broccoli (middle), and turbulent water are examples of fractals. The Sierpinski triangle is an ideal fractal while the others are formed in nature, hence, the scale invariance is broken at either large or small enough scales.}
\label{fig2.9}
\end{figure}

Fractals are complex sets of points in space that occupy non integer fraction of it. Fig.\reffig{fig2.9} shows images of what is typically called a fractal. Defining a fractal is a very difficult subject, instead, let us list some of the expected properties of an idealized fractal:
\begin{enumerate}[label=\textnormal{(P3.4.\arabic*)}]
    \item Are scale invariant at all scales. Scale invariance can show up in the geometrical sense when the images are formed by infinitely many copies of itself, or in the statistical sense.\label{p221}
    \item They have fine and detailed structures at arbitrarily small scales.\label{p222}
    \item They have usually non integer scaling in the spatial measure sense. For example, in the Sierpienski triangle, if one tries to measure its area the result is zero naturally. However, measuring its length gives infinity since its pattern repeats infinitely. The only spatial measure that converges to a finite value is in a fractional dimensional, $D=\log_2(3)\approx1.585$.\label{p223}
    \item Local and global irregularity or roughness is typical.\label{p224}
\end{enumerate}

Examples of fractals in nature are, besides the ones illustrated in Fig.~\reffig{fig2.9}, snowflakes, coasts, rivers, lightning, etc. However, fractals in nature are often regularized by continuously zooming them (applying scaling invariance), and at some point one finds the smallest structures forming it. For example, the broccoli in Fig.~\reffig{fig2.9} is formed by fiber and cells that do not display the same irregularity as its leaves.

To precisely define the quantities in fractal geometry, let $\vc x_0$ be a point in the turbulent domain where,
\be\label{2.81}
\lim_{\vc x\rightarrow \vc x_0}\lr{\frac{|\vc v(\vc x)-\vc v(\vc x_0)|}{|\vc x-\vc x_0|^\alpha}}<\infty\ .\
\ee
The velocity field is said to be singular with H{\"o}lder exponent $h\equiv\sup(\alpha)$. Frisch and Parisi \cite{frisch1985turbulence} proposed that the set of points $\vc x_0$ with the same H{\"o}lder exponent $h$ forms a fractal with dimension $D(h)$. This means that if one subdivides the turbulence domain into small cubes of side $\ell$, the probability of finding a H{\"o}lder exponent in the interval $[h,h+dh]$ is given by
\be\label{2.82}
P_\ell(h)=f(h)\lr{\frac{\ell}{\ell_0}}^{3-D(h)}dh\ ,\
\ee
where $f(h)$ is a function of $h$ only, and $\ell_0$ is some reference scale, usually the integral length where the scale invariance is broken. the combination $d-D(h)\geq0$ is referred to as fractal co-dimension, in our three-dimensional turbulence, $d=3$. Furthermore, $f(h)$ can be regarded as a piecewise constant function, as it only indicates the positions where $h$ is inside the desired interval. Then, we write velocity fluctuation in the cube of side $\ell$ using Eq.~\eqref{2.81} as
\be\label{2.83}
\mean{\lr{\delta_\ell v}^p}\sim\int dh P_\ell(h)\ell^{hp}\sim \int dh\ell^{hq+3-D(h)}\ .\
\ee
For $\ell\ll\ell_0$ we can use the saddle point approximation (for details see Appendix~\reffig{appspa}) to get, 
\be\label{2.84}
\mean{\lr{\delta_\ell v}^p}\sim\ell^{\zeta_p}\ ,\
\ee
with
\be\label{infh}
\zeta_p=\inf_{h}\big[ph+3-D(h)\big]\ .\
\ee

\ac{K41} turbulence in terms of multifractal language, is regarded as a space-filling monofractal ($D(1/3)=3$) (a single singularity happens $h=1/3$). For She-Lévêques's model, one may note that $\zeta_p\rightarrow\infty\approx p/9+(3-D_\infty)$, then $\tau^{(1)}_p=-2p/3+3-D_\infty$ giving a very interesting phenomenological interpretation to the parameter $D_\infty=1$ and the postulate of Eq.~\eqref{2.74a} as follows: In the turbulent limit, the structures contributing to the extreme events have fractal dimension $D_\infty=1$ filaments with H{\"o}lder exponent $h=1/9$ and, consequently, highly dissipative structures.

An interesting model to build is the bifractal model, with direct applications in one-dimensional compressible turbulence (Burgers' turbulence). The idea is that the flow is composed of only two singularities $h_1$ and $h_2$ with fractal dimensions $D_1$ and $D_2$, respectively. Then, using Eq.~\eqref{2.83}, two contributions arise
\be\label{2.86}
{\delta_\ell v}^p\sim 
\alpha_1\lr{\frac{\ell}{\ell_0}}^{ph_1}\lr{\frac{\ell}{\ell_0}}^{3-D_1}+
\alpha_2\lr{\frac{\ell}{\ell_0}}^{ph_2}\lr{\frac{\ell}{\ell_0}}^{3-D_2}
\ .\
\ee
Thus, for $\ell\ll\ell_0$ the smaller exponent dominates
\be\label{2.87}
\zeta_p=\inf\left[ph_1+3-D_1,ph_2+3-D_2\right] \ . \
\ee
The main implication of the bifractal model is then the existence of a ``critical'' moment order $p_c=(D_1-D_2)/(h_1-h_2)$ where the relevant singularity abruptly changes. Note that Eq.~\eqref{infh} is the generalization of Eq.~\eqref{2.87} for continuously varying $D(h)$. Consequently, a nonlinear (concave) scaling exponent as commonly observed in turbulence experiment statistics, means that higher-order statistics are dominated by more and more discontinuous solutions. However, a structural measure of these singularities is a rather difficult task.

All the intermittency models put forward here rely upon phenomenological observation of numerical and/or experimental data. A consistent \ac{iNSE} formulation of the multifractal behavior of dissipation fluctuations in turbulence is still an ultimate goal in this field of research. However, refinements in the opposite direction (from intermittency models to the \ac{iNSE}) are always welcome since new insights can arise from modeling reformulations.

\section{Gaussian Multiplicative Chaos}\label{sec2.4.4}

$_{}$

This Section is devoted to a brief review of an extension of the log-normal modeling, the \ac{GMC}. Such formulation leads to scaling exponents similar to Eq.~\eqref{2.73a} but developed within a mathematically formal background. For a recent review on this topic, we refer to \cite{robert2010gaussian,rhodes2014gaussian}. Application of the \ac{GMC} approach goes far beyond turbulence. It dives into applied problems such as volatile modeling in stock market \cite{duchon2012forecasting} up to very abstract phenomena such as quantum gravity \cite{kupiainen2019local} and many others.

The mathematical tool used in this construction is an analog of Feynman's path integration formalism for quantum mechanics and/or quantum field theory \cite{srednicki2007quantum,zee2010quantum,itzykson2012quantum}. To be precise, the basic tool we will use is known as statistical field theory \cite{itzykson1991statistical,zinn2021quantum} which is mathematically more formal than the quantum mechanics' path integral. 

The formulation of a formal statistical measure for intermittency was first tried by Mandelbrot itself after his first ideas on fractals \cite{mandelbrot1972possible}. However, Kahane \cite{kahane1985chaos} was the one who developed this mathematical formulation and gave the name \ac{GMC}. In general, the formulation of the \ac{GMC} can be done in arbitrary space dimensions, however, we have two important reasons to formulate it in $d=2$. The first is completely practical, we will be interested in applying this approach for planar cuts of turbulent fields. The second will be commented on after the results.

The use of \ac{GMC} approach goes far beyond turbulence modeling, as already commented. In order to apply this mathematical to the problem of energy dissipation intermittency, we define the following quantity:
\be\label{2.88}
\psi^{(n)}_\ell(x)\equiv\frac{1}{A(D_\ell)}\int_\mathcal{D_\ell}d^2y\ \psi^{(n)}(y) \ ,\
\ee
where $\mathcal{D}_\ell$ is the spatial domain $|x-y|\leq \ell$, and we omitted vector notation $\vc x\rightarrow x$ for simplicity. The definition of the quantity inside the integral is then
\be\label{2.89}
\psi^{(n)}(x)\equiv\left(\frac{\vare(x)}{\mean{\vare(x)}}\right)^{1/n} \ ,\
\ee
regarded as the $n^{th}$-root of the local dissipation rate field. Note that we are seeking for formulation that allows the dissipation field to fluctuate locally, not in the same sense as the cascade process that is defined only at coarse scales. 

To do so, we define the local dissipation rate to be
\be
\vare(x)=\vare_0\frac{e^{\gamma\phi(x)}}{\mean{e^{\gamma\phi(x)}}} \ .\
\ee
In this scenario, $\vare_0=\mean{\vare(x)}$ is the mean dissipation and $\gamma$ is a parameter, later expressed in terms of the intermittency parameter $\mu$. The main actor in this formulation is the field $\phi(x)$, regarded as a \ac{GFF} with statistical average of any functional of $\phi$, $F[\phi]$ is given by
\be
\mean{F[\phi]}=\int D[\phi]F[\phi]e^{-S[\phi]} \ ,\
\ee
where,
\be\label{2.92}
S[\phi]=-\frac{1}{2}\int_{\mathcal{D}_L} d^2x\phi(x)\partial^2\phi(x) \ .\
\ee
For simplicity, we assume $\phi(x)$ to be periodic in the infrared regularized domain $\mathcal{D_L}$. Moreover, we assume a lattice regularization $\eta_K$ to uniquely define the Laplacian Green's function. The integration over $D[\phi]$ is meant to be a path/functional integration along the lines of the statistical field theory. It is a bookcase exercise to show that the Green's function of the 2D Laplacian is a logarithmic function. Within our definition of regularization, we have,
\be\label{2.93}
\mean{\phi(x)\phi(y)}=-\frac{1}{4\pi}\ln\left(\frac{|x-y|^2+\eta_K^2}{L^2}\right) \ .\
\ee
At this point, it can be important to stress that the choices of regularization, are equivalent to introduce $\int d^2xM^2\phi(x)^2$ in Eq.~\eqref{2.92} with $M\sim\eta_K/L\sim R_\lambda^{-3/2}$ and performing an integral in the whole space. In this case, it is evident that the mean-field approach: $\phi(x)\rightarrow\mean{\phi}$ converges to the statistical distribution of a simple random variable with no local spatial correlation in the standard interpretation of $\mean{\phi(x)\phi(y)}=0$. In this sense, the latter converges to the standard \ac{OK62} approach with long-range correlations among scales $\mean{\phi_\ell\phi_{\ell'}}\neq0$. 

Statistical moments of $\vare(x)$ are scale independent due to homogeneity and are related to the computation of $\mean{e^{p\gamma\phi(x)}}$, which is related to the characteristic function of $\phi(x)$. Combining the standard result for Gaussian characteristic functions, Eq.~\eqref{2.93}, and Eq.~\eqref{2.89} one can show that,
\be\label{2.94}
\psi^{(n)}(x)=\left(\frac{\eta_K}{L}\right)^{\gamma^2/4\pi n}e^{\gamma\phi(x)/n} \ .\
\ee
Now applying Eq.~\eqref{2.94} to Eq.~\eqref{2.88}, one gets
\be
\mean{\lr{\psi^{(n)}_\ell(x)}^p}\sim\ell^{-2p}
\left(\frac{\eta_K}{L}\right)^{p\gamma^2/4\pi n}
\int_{\mathcal{D}^p_\ell}\left[\prod_{i=1}^p d^2x_i \right]
\mean{e^{\gamma/n\sum_i^p\phi(x_i)}} \ .\
\ee
The computation of the characteristic function of a sum o field $\phi(x_i)$ can be straightforwardly computed by the use of the saddle point method (which is exact for Gaussian variables Appendix~\reffig{appspa}) resulting at
\be\label{2.96}
\ln\lr{\mean{e^{\gamma/n\sum_i^p\phi(x_i)}}}=-\frac{\gamma^2}{8\pi n^2}\sum_i^p\sum_j^p\ln\left(\frac{|x_i-x_j|^2+\eta_K^2}{L^2}\right) \ . \
\ee
The double summation in Eq.~\eqref{2.96} can be split into two parts, diagonal ($i=j$) and nondiagonal part. It is then clear that the diagonal contribution is linear in $p$. The nondiagonal part will serve as a source to the $p$-integrations in the domain $\md_\ell$.  Noting that interchanging $i$ and $j$ we have twice the contribution, we write
\be
\sum_i^p\sum_j^p\ln\left(\frac{|x_i-x_j|^2+\eta_K^2}{L^2}\right)=
2\sum_{i>j}^p\ln\left(\frac{|x_i-x_j|^2+\eta_K^2}{L^2}\right)
+2p\ln(\eta_K/L) \ .\
\ee
Their contribution to the space integrals can be estimated recursively. Assume $R\lambda\gg 1$ then, formally $\eta_K/L\rightarrow0$ then, for $p=2$
\be
I_2=\int_{\mathcal{D}_\ell}d^2x\int_{\mathcal{D}_\ell}d^2y \ e^{-\alpha\ln|x-y|}\sim\ell^{4-\alpha} \ ,\
\ee
with $\alpha=\gamma^2/(2\pi n^2)$. For $p=3$ we have,
\be
I_3=\int_{\mathcal{D}_\ell}d^2x\int_{\mathcal{D}_\ell}d^2y\int_{\mathcal{D}_\ell}d^2z
\ e^{-\alpha(\ln|x-y|+\ln|y-z|+\ln|z-x|)}\sim\ell^{6-3\alpha} \ .\
\ee
Note that the matter of determining the scaling is the same as counting how many different combinations among $x_i$ and $x_j$ with $i\neq j$ are possible. Thus, this is exactly the combination of $p$ positions 2 by 2, the result $p(p-1)/2$ comes naturally. The general expression is then given by,
\be
I_p=C_p\ell^{2p-\alpha p(p-1)/2} \ ,\
\ee
where $C_p$'s are related to the specific geometry of $\md$ and can also depend on the regularization $\eta_K/L$, i.e., the Reynolds number. The scaling exponents of the local average energy dissipation rate are given by
\be\label{momentsgmc}
\mean{(\psi^{(m,n)}_\ell(x))^p}=C^{(m,n)}_p
\left(\frac{\ell}{L}\right)^{\tau^{(n)}_p} \ ,\
\ee
where,
\be\label{2.104}
\tau^{(n)}_p=\frac{\gamma^2}{4\pi n^2}p(1-p) \ .\
\ee
Note that the determination of the scaling exponents is directly related to the fact that the correlations $\mean{\phi(x_i)\phi(x_j)}\sim\ln(|x_i-x_j)$. This is a very particular property of the 2D Laplacian Green's function. For $d$ dimensions, the \ac{GMC} is not defined in terms of standard differential operators, instead, fractional operators are developed to devise the property of being log correlated. That is the second reason why we choose to restrict this review on the \ac{GMC} approach to $d=2$.

A first phenomenological constraint to Eq.~\eqref{2.104} occurs when we are modeling turbulence. With the common knowledge of \ac{OK62} we have, for $n=1$, the intermittency parameter measures the second moment of coarse-grained dissipation, thus,
\be
\tau^{(1)}_2=\frac{\gamma^2}{2\pi}=\mu\Longrightarrow\gamma=\sqrt{2\pi\mu} \ .\
\ee
For turbulence modeling, the \ac{GMC} theory for coarse-grained normalized cascade gives the following result:
\be\label{GMC-scaling}
\tau^{(n)}_p=\frac{\mu}{2n^2}p(1-p) \ .\
\ee
However, this result is not exactly the same as found for the standard \ac{OK62} cascade (from now on we will refer to it as \ac{MCM}). The symmetry $\tau^{(n)}_p=\tau^{(1)}_{p/n}$ gives, for the \ac{MCM},
\be\label{MCM-scaling}
\zeta^{(n)}_p=\frac{\mu}{2n^2}p(n-p).
\ee

The difference in Eq.~\eqref{GMC-scaling} and Eq.~\eqref{MCM-scaling} has profound measurable implications on the modeling of velocity circulation to be studied in Chapter~\refcap{cap4}. Before diving into the main results of this thesis, let us contextualize the history and the recent developments regarding the statistics of velocity circulation in homogeneous and isotropic turbulence.
\end{chapter}

\begin{chapter}{Turbulent Circulation Statistics}
\label{cap3}

\hspace{5 mm} 

The existence of a unifying approach that accounts for the whole phenomenology of homogeneous and isotropic turbulence is still a debate in the turbulence community \cite{frisch1995turbulence}. Despite the 4/5 law, there are very few strong results in turbulence that are directly derived from the dynamics of \ac{iNSE}. Additionally, the efforts to understand small pieces of information about the richness of the turbulent dynamics have been mostly focused on local Galilean invariant observables such as velocity differences, velocity gradients, vorticity, strain rate, and so on \cite{pope2000turbulent}. 

Velocity circulation was a historically overlooked variable from the birth of Kolmogorov's phenomenology until the early 1990s. Even in the present dates, circulation statistics are still much less studied when compared to other traditional observables up to the present date. To point out several reasons to explore this variable, let us define the velocity circulation in a closed oriented contour as follows,
\be\label{3.1}
\Gamma[\mc]=\oint_{\mc}\vc v(\vc x,t)\cdot d\vc x \ ,\
\ee
where $\mc$ is a closed contour. In addition to the symmetries of the \ac{iNSE}, velocity circulation has several important properties such as,
\begin{enumerate}[label=\textnormal{(P4.\arabic*)}]
    \item Galilean invariance: Closed contours ensure invariance under the transformation $\vc v\rightarrow \vc v+\vc U t$ for any constant vector $\vc U$.\label{p311}
    \item Transversality: It deals only with the incompressible component (transverse modes) of the velocity field meaning that the circulation is invariant under gauge-like transformations $v_i(x)\rightarrow v_i(x)+\partial_i\alpha(x)$ for any smooth enough function $\alpha(x)$.\label{p312}
    \item Reparametrization invariance: It does not matter how your parametrization covers the contour $\mc$ the circulation is still the same.\label{p313}
    \item Multidescription: Stokes' and Gauss' theorems ensure different ways of computing circulation. It is particularly interesting to deal with numerical computations of circulation.\label{p315}
    \item It is an inviscid invariant: Kelvin-Helmholtz's theorem (usually referred to only as Kelvins') ensure the conservation of circulation in the case of $\nu=0$.\label{p316}
\end{enumerate}
Points \ref{p311}---\ref{p316} summarizes some of the advantages of the use of the circulation variable. Two of the above, \ref{p315} and \ref{p316}, points will be further stressed. Firstly, applying the Stokes' theorem, circulation can be calculated as follows,
\be\label{3.2}
\Gamma[\mc]=\oint_{\mc}\vc v(\vc x,t)\cdot d\vc x=\int_\md \vc \omega (\vc x,t)\cdot d\vc S(\vc x) \ ,\
\ee
where $\vc \omega=\vc \nabla\times\vc v$, $\md\in \mathbb{R}^3$ is a domain enclosed by the curve $\mc$, in short notation $\mc=\partial \md$, and $\vc S(\vc x):\md\rightarrow \mathbb{R}^2$ is a parametrization of the domain $\md$. Since the vorticity itself is incompressible, $\partial_i\omega_i=0$, no matter which surfaces you choose to calculate circulation, as long as the enclosed curve is the same, the circulation $\Gamma[\mc]$ will be the same. This can be easily seen from Gauss' theorem,
\be\label{3.3}
\int_V \vc \nabla \cdot \vc \omega (\vc x,t)dV =\int_{\md_1} \vc \omega (\vc x,t)\cdot d\vc S(\vc x) +\int_{\md_2} \vc \omega (\vc x,t)\cdot d\vc S(\vc x)=0\ ,\
\ee
where $\md_1+\md_2=\partial V$ is a closed outward-oriented surface, with borders equal to $\mc$, such that both surface integrals on Eq.~\eqref{3.3} have the value $\Gamma[\mc]$ with opposite signs. There are infinitely many surfaces where circulation can be calculated and this paves the way for efficient numerical designs to compute circulation. The comment to be done at this point is, as long as the contour is fixed, there exists a very special class of surfaces called ``minimal surfaces'' \cite{manfredo2003superfícies}. Those surfaces solve a minimization problem being the one with the least possible area for a given contour, it also has zero mean curvature and other very special properties.

The point~\ref{p316} is known as one of the most celebrated theorems of fluid dynamics, it was published in its current version by Lord Kelvin in the year 1869 \cite{kelvin1869vortex}, even though several of its consequences were previously stated in an axiomatic fashion by Helmholtz in 1858 \cite{helmholtz1858integrale}. The general statement is the following,

``\textit{In a barotropic, ideal fluid with conservative body forces, the circulation around a closed curve (which encloses the same fluid elements) moving with the fluid remains constant with time.}''

This theorem states that the circulation is an inviscid invariant and, as the other symmetries discussed in Chapter~\refcap{cap2} it can play an important role in the statistical properties of turbulence. Moreover, circulation is a completely local invariant, as one can shrink any continuous closed curve to a point, which means that the scaling properties of circulation can be parametrized by a contour base length scale such as the perimeter or square root of the area. Furthermore, it generates a subspace of the loop space by the transformation $t\rightarrow t'$, $\vc v(\mc,t)\rightarrow\vc v'(\mc',t')$, where $\mc'$ is obtained by the advection of the initial contour by the velocity field during the time interval $t'-t$.

In order to prove Kelvin's theorem, let us start taking the time derivative of Eq.~\eqref{3.1} to get
\be\label{3.4}
\frac{d\Gamma[\mc]}{dt}=\lim_{dt\rightarrow0}\frac{1}{dt}\lr{\oint_{\mc(t+dt)}\vc v(\vc x(t+dt),t+dt)\cdot d\vc x-
\oint_{\mc(t)}\vc v(\vc x,t)\cdot d\vc x} \ .\
\ee
Indeed, the advection of the loop can be explicitly calculated: $\mathcal{C}(t+dt) = \mathcal{C}(t) + \mathbf{v}(\mathcal{C}(t),t)dt$. Thus, the new contour shape is given by $\mathbf{x}(t+dt) = \mathbf{x}(t) + \mathbf{v}(t)dt$, and by expanding in powers of $dt$, we have $v_i(\mathbf{x}(t+dt),t+dt) = (\partial_tv_i(\mathbf{x},t) + v_j(\mathbf{x},t)\partial_jv_i(\mathbf{x},t))dt + \mathcal{O}(dt^2)$. We are left with
\be\label{3.5}
\frac{d\Gamma[\mc]}{dt}=\lim_{dt\rightarrow0}\lr{\oint_{\mc}\Big(\partial_tv_i+v_j\partial_jv_i\Big)dx_i+\oint_{\mc}v_idv_i+\mathcal{O}(dt)} \ .\
\ee
Since contour integrations of gradients of smooth functions are zero within closed domains, the rightmost term in Equation (3.5) does not contribute. Explicitly using \ac{iNSE}, we obtain,
\be\label{3.6}
\frac{d\Gamma[\mc]}{dt}=\nu\oint_{\mc} \partial^2v_idx_i+\oint_{\mc} f_idx_i\ .\
\ee
In the particular case of inviscid fluids under conservative forcing ($f_i=\partial_i U$), circulation is a conserved quantity.

The significance of Kelvin's theorem in the context of turbulence has not been extensively explored from a theoretical perspective, especially when compared to other conserved quantities such as energy dissipation in 3D turbulence and enstrophy in 2D turbulence. However, numerical studies of \cite{chen2006kelvin} suggest that homogeneous and isotropic turbulence do not locally conserve circulation due to the effect of viscous diffusion. Nonetheless, a weaker form of Kelvin's theorem is still valid. 

The general idea is that in a turbulent system, the advection of the contour $\mc$ will, for any finite time, form a random fractal curve (but still piecewise continuous at small enough scales) such that the statistical properties of this random curve have a martingale property \cite{eyink2006turbulent}. In other words, circulation's expectation value remains constant through the process of advection, but it can still have huge fluctuations. Although very interesting, this way line of research deals with several difficulties in both numerical and theoretical points of view. Numerically, a very careful analysis of the lagrangian tracking of the contour shows up to be a challenging task. From the theoretical point of view, there is much effort to be done to understand which kind of fractal curves are formed depending on the initial smooth contour. Nevertheless, connections have been made with the phenomenon of spontaneous stochasticity \cite{bernard1998slow,chaves2003lagrangian,thalabard2020butterfly}. 

The goal of this chapter is to expose all the current results widespread throughout the literature concerning circulation statistics in homogeneous and isotropic turbulence. The initial intention was to arrange the exposition chronologically, supported by results published at the time of their discovery. However, we have decided to incorporate our processed numerical data to validate our database. Consequently, readers should be aware that our current dataset offers much higher resolution compared to the datasets used in the original results from Section~\refsec{sec3.2}. These original results were computed in the late 1990s, a period when the most advanced hardware clusters were comparable to today's personal computers.

\section{Dataset Description}\label{sec3.1}

$_{}$

To investigate the statistical properties of velocity circulation we took advantage of the public availability of the \ac{JHTDB} platform \cite{perlman2007data,li2008public} where the data can be accessed and manipulated without being downloaded. We have analyzed four \ac{DNS} datasets of homogeneous and isotropic turbulent flows. All the simulations were developed in $2\pi$-periodic cubic lattices. A pseudospectral approach was applied to solve incompressible forced \ac{iNSE}s. In this approach, random forcing is typically applied in a narrow range of wavenumbers with support at large scales\footnote{For more information about the simulation method see {\url{http://turbulence.pha.jhu.edu/}}.}. 

Table~\reftab{tab3.1} shows relevant parameters for the basic homogeneous and isotropic turbulence datasets. The parameter's descriptions are the following, $N$ is the linear number of collocation points, $N_t$ is the number of accessible snapshots in the database, and $R_\lambda$ is the Reynolds-Taylor number. $L$, $\eta_K$, and $dx=2\pi/N$ are, respectively, the integral length scale, Kolmogorov length scale, and mesh parameter, $U_{rms}=(2E_{tot}/3)^{1/2}$ is the root-mean-squared velocity, and finally, $\vare_0=\nu\mean{(\partial_iv_j)^2}$ is the energy dissipation rate.
\begin{table}[t]
\centering
\caption[Johns Hopkins Database parameters.]{Parameters table of the Johns Hopkins turbulence datasets. Datasets are named in this way because they refer to isotopic simulation on $1024$, $4096$, and $8192$ collocation points. Iso-$F8196$ refers to a fine-grained resolved simulation.}\label{tab3.1}
\begin{tabular}{|c|c|c|c|c|c|c|c|c|c|} \toprule
    {Name} & {$N$} & {$N_t$} & {$R_\lambda$} & {$L/\eta_K$} & {$dx/\eta_K$} & {$U_{rms}$} & {$\vare_0$} & {$\nu\times 10^{5}$}\\ \midrule
    Iso-1024  & 1024  & 5028  & 418  & 487  & 2.191 & 0.686 & 0.103  & 1.8500\\
    Iso-4096  & 4096  & 1     & 610  & 1005 & 1.108 & 2.460 & 1.414  & 1.7320\\
    Iso-F8196 & 8196  & 1     & 613  & 939  & 0.556 & 1.577 & 1.424  & 1.7320\\
    Iso-8196  & 8196  & 5     & 1280 & 2531 & 1.565 & 1.534 & 1.339  & 0.4385\\ \bottomrule
\end{tabular}
\end{table}

Different sub-datasets were generated by calculating the circulation for two planar shapes: Squares of side $r$ and circles of radius $R$. This choice establishes the notation for various contour shapes. Note that, for the same dataset, those shapes cannot be simply compared to one another since they have different typical lengths. For reasons which will be clear in the next sections, proper length scales will be defined as $\sqrt{A(\mc)}$ where $A(\mc)$ is the minimal area of the contour which, for planar contours, coincide with the standard definition of area. Note that this is in opposition to other possible length measures like perimeter $|\mc|$, for instance, which also provide a definition of scale given the contour $\mc$. 

At this point, one may note that it should be better to compute circulation for squares of size $\sqrt{\pi}r$ and circles of radius $R$ or, conversely, squares of size $r$ and circles of radius $R/\sqrt{\pi}$ so that, in both cases, the area ratio is one. Unfortunately, it presents numerical difficulties, since very few contour points will coincide with the velocity grid, slowing down numerics due to the necessity of interpolations even in the case of squares. With this in mind, we will perform the comparison between circles of radius $R$ and squares of side $2r$, where the typical length ratio is $\sqrt{\pi/4}\approx 89\%$.

For circular contour, we employed the line integral formulation of circulation, approximating the circular contour to a polygon of $N_0$ sides, each one with a length comparable to $dx$ by equating $2\pi R\approx N_0dx$. This was possible by the use of a sixth-order Lagrangian interpolation (offered by the database itself) to get the values of the velocity field at points that do not coincide with the grid points, but it generally slows down the numerical calculations.

For the calculation in square contours, we took advantage of the spectral accuracy of the simulation to calculate it in the Fourier space as it can be written as a convolution. To do this, define an indicator function $H_f(\vc x)=1$ if $\vc x \leq (r\hat{x}+r\hat{y})$ and $H_f(\vc x)=0$ otherwise. The circulation is then given by,
\be\label{3.7}
\Gamma[\md]=\int \omega_{\perp}(\vc x)H_f(\vc x)d^2x=\mathcal{F}^{-1}\lr{\int \mathcal{F}(\omega_{\perp})\mathcal{F}(H_f)d^2k}\ ,\
\ee
where $\omega_{\perp}$ is the perpendicular component of the vorticity relative to the planar domain, $\mathcal{F}$ as well as $\mathcal{F}^{-1}$ are the Fourier transform and its inverse, respectively. The rightmost equality, known as Parseval's theorem, is commonly used to calculate observables with spectral accuracy. We note, furthermore, that efficient algorithms to compute \ac{FFT} in periodic lattices can speedup the calculations\footnote{Rough benchmarks were estimated by calculating the circulation on square contours at $M^3$ equally spaced points on the lattice with $4\leq M\leq32$. For small values of $M$, computations through velocity showed to be faster. On the other hand, vorticity convolutions were $200$ faster than the velocity line integral for the largest tested value of $M$.}.
\begin{table}[t]
\centering
\caption[Circulation datasets.]{Information table of the post-processed data. Dataset names have the following rationale: ``Circ-10'' stands for circular contours calculated based on a dataset with $N=2^10$ collocation points.}\label{tab3.2}
\begin{tabular}{|c|c|c|c|c|c|} \toprule
    Name & Original database & $\mc$-shape & Ensemble size &  $\sqrt{A}/\eta_K$\\ \midrule
    Cir-10   & Iso-1024  & Circle & $9.8\times10^6$ & $3.9\text{ --- }994$ \\
    Cir-14   & Iso-4096  & Circle & $8.0\times10^6$ & $2.0\text{ --- }2017$ \\
    Cir-15   & Iso-8196  & Circle & $4.3\times10^7$ & $1.0\text{ --- }1008$ \\
    Cir-F15  & Iso-F8196 & Circle & $8.6\times10^6$ & $2.8\text{ --- }2840$ \\
    Sqr-10   & Iso-1024  & Square & $9.4\times10^6$ & $2.2\text{ --- }280$ \\
    Sqr-14   & Iso-4096  & Square & $7.1\times10^6$ & $1.1\text{ --- }1138$\\\bottomrule
\end{tabular} 
\end{table}

Table~\reftab{tab3.2} shows the sub-datasets of circulation, spanning different shapes and sizes for the contour $\mc$ as well as an indication of the ensemble size. In the case of \ac{DNS}, the data traffic of the database is quite limited and this has forced us to calculate subsamples of circulation by homogeneously distributing the contour on the full 3D domain. To diversify the ensemble, the contour's normal vector was taken relative to the three lattice orientations, and, in the case of circular contours, also diagonal orientations were used. For the datasets Cir-10 and Sqr-10, we have picked different homogeneously distributed snapshots, as their main database offers nearly five thousand snapshots. Lastly, for the Cir-15 dataset, all five snapshots were used.

Although this composes most of the data analyzed throughout this thesis, there are other observables to be analyzed in Chapters~\ref{cap4}~and~\ref{cap5}. Their particularities and implementations will be discussed when needed, but the data are always relative to the datasets in Table~\reftab{tab3.1}. I also warn again that some of the numerical results presented in the next section are much better resolved than the ones drawn in the early studies when the computational power was considerably limited when compared to the present dates.

\section{Early Phenomenology of Circulation Statistics}\label{sec3.2}

$_{}$

Velocity circulation is, in general, a good observable to look at in turbulence because, in addition to the points~(\ref{p311}$-$\ref{p316}), it is restricted to two statistical laws. The first one is the dissipation anomaly, and the second one is the central limit theorem. Dissipation anomaly, $\nu\mean{|\vc \omega|^2}\rightarrow \vare_0=$ constant as $\nu\rightarrow0$, bounds circulation statistics at small enough (and smooth enough) scales, as it is linearly connected to the vorticity through Stokes' theorem. On the other hand, for large contours, the central limit theorem ensures the Gaussianity of the \ac{cPDF}, since circulation becomes dominated by the contribution of a large number of uncorrelated thin vortex tubes.

The consequences of both constraints are well illustrated in the left panel of Fig.~\reffig{fig3.1}. Starting from the bottom, one can see nearly Gaussian \ac{cPDF}s as the typical length is comparable to the integral length. As the contour size decreases, the \ac{cPDF}s starts to develop approximate exponential decay. The slop of these exponentials usually decreases as the contour reaches smaller scales, up to a point where it saturates due to the transition to the behavior of the vorticity \ac{PDF}, addressed by the two uppermost curves on the left panel of Fig.~\reffig{3.1}.

Nevertheless, the promised statistical simplicity of \ac{cPDF}, the first who consistently explored theoretically the problem of turbulence through the circulation point of view, to the knowledge of the presenting author, was A. A. Migdal in 1993 \cite{migdal1993loop}. He introduced a completely novel approach to the problem. However, this research line did not receive much continuity during that time \cite{umeki1993probability,cao1996properties}, primarily due to the challenging numerical access to extensive variables, which hindered both direct simulations and data processing.
\begin{figure}[t]
\center
\includegraphics[width=\textwidth]{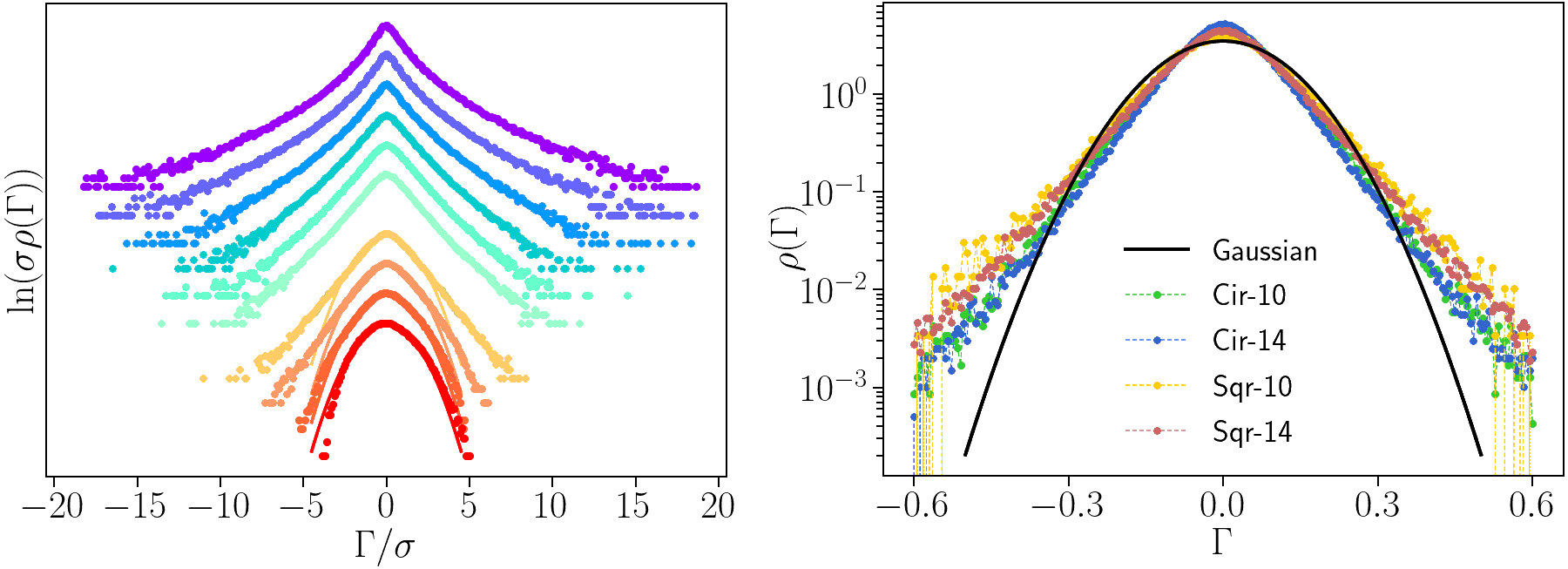}
\caption[cPDFs from the JHTDB.]{Left panel shows standardized \ac{cPDF}s from Circ-15 dataset for typical length of $\sqrt{A}/\eta_K=2.8,5.5,11,22,44,89,177,355,710,1420,\text{ and }2841$ from the top to the bottom. The vertical scale is meaningless since the data was shifted to ease visualization. One can find continuous lines denoting the Gaussian expectation for the larger contours. The right panel shows unstandardized \ac{cPDF} for selected contours of different datasets, $\sqrt{A}/\eta_K=62,63,70,71$ for the datasets Circ-10, Circ-14, Sqr-10, and Sqr-14, respectively.}
\label{fig3.1}
\end{figure}

In March 1993, Migdal published his seminal work \cite{migdal1993loop} where he applied the loop calculus developed to tackle the quark-confinement problem \cite{wilson1974confinement} to the so-called Hopf functional approach to turbulence, which associates a functional measure in the spirit of quantum field theory to random classical systems. The main object in Migdal's approach is the following observable,
\be\label{3.8}
\Phi[\mc]\equiv \mean{\exp\lr{\frac{i}{\nu}\oint_{\mc} v_idx_i}}\ ,\
\ee
the loop functional, that can be recognized as the circulation's characteristic functional $Z(\lambda=\nu^{-1})$. There are two main things to note in this formulation, first, kinematic viscosity comes up to play a role similar to the one of $\hbar$ in quantum field theory. This means that the semiclassical approach $\hbar\rightarrow0$ is the same as the turbulent limit $\nu\rightarrow0$ or $Re\rightarrow \infty$. Secondly,
the loop functional is a generator of the loop space, such that Eq.~\eqref{3.8} is parametrized by the loop shape. By employing the semiclassical approach, Migdal was led to the proposition of a semiclassical action $S[\mc]$ from where $\Phi[\mc]=e^{iS[\mc]/\nu}$, which satisfies an evolution equation,
\be\label{3.9}
\frac{dS[\mc]}{dt}=\oint_\mc dx_i\int_{\mathbb{R}^3}d^3y
\frac{y_j-x_j}{4\pi|\vc x-\vc y|} 
\frac{\delta S[\mc]}{\delta \sigma_{ki}(x)} 
\frac{\delta S[\mc]}{\delta \sigma_{kj}(y)}\ ,\
\ee
and $\delta/\delta{\sigma_{ij}}$ are called ``area derivatives''\footnote{We refer to \cite{migdal1983loop} where Migdal develops definitions of the loop calculus in the turbulence context, and \cite{kogut1979introduction} for a review in the same mathematical tool but biased by quantum field theory phenomenology and notation.}. We further stress that the precise definition of these quantities will not be of concern in this thesis, we keep our focus on drawing in general lines Migdal's results.

In addition to this completely novel approach, Migdal also provided a scaling solution to Eq.~\eqref{3.9} in the form $S[\mc]=iB\vare_0^{1/3}|A_T(\mc)|^{2/3}$, where $\vare_0$ is the standard mean energy dissipation rate, $B$ is some real positive constant, and $A_T(\mc)=\inf\left[\left|\oint_\mc x_idx_j\right|\right]$ is the modulus of the tensor minimal area of the contour $\mc$. This solution is usually referred to as tensor area law, which states that in the turbulent limit, circulation statistics (or circulation extreme events) should not depend on the shape of its enclosing contour, but only on its tensor minimal area. 

The right panel of Fig.~\reffig{fig3.1} shows \ac{cPDF}s for different datasets including different contour shapes and Reynolds numbers but with similar proper lengths. They all look like each other (despite minor differences due to differences in the exact proper lengths) supporting the idea of an area law for the circulation. However, for simply connected nonself-intersecting planar loops, there is no way to distinguish between the modulus of the tensor area and the usual scalar area.

In only three months after Migdal's first publishing of his draft, M. Umeki noted that this solution leads the \ac{cPDF}s to be Lorentzian distributed \cite{umeki1993probability}. Umeki's work did not numerically observe this phenomenon, and even considering the limited numerical resolutions available at that time, he clearly differentiated the cPDF from a Lorentzian curve and suggested data-driven adjustments to Migdal's findings. Umeki's first modification was the employment of a scalar area in opposition to the tensor area in the solution, motivated by the numerical results on the scaling of circulation variance\footnote{In the original Umeki's paper, there is a footnote which states that ``\textit{In private communication, Migdal admits the modification of the tensor-area law. ($\cdots$)}''. Then, Migdal published an extended version \cite{migdal1994loop} discussing the case of scalar area and the role of minimal surfaces. Unfortunately, he does not reach any solid conclusion about scalar area law and criticizes Umeki's work.}. 

Assuming the circulation's scaling relation adheres to \ac{K41}'s behavior, it is reasonable to anticipate that $\mean{\Gamma^2_A}^{1/2}\sim A^{2/3}$. However, ambiguity arises due to different ways to define an area. Umeki has considered the ``figure eight loop'' which is a classical example of the difference between scalar and tensor area, it is composed of two squares of lengths $L_1$ and $L_2$ sharing a common vertex with $L_1$ being anticlockwise oriented and $L_2$ clockwise oriented and, by convention, $L_1>L_2$. In this scenario, the tensor area is given by $L_1^2-L_2^2=(L_1+L_2)(L_1-L_2)$ while the scalar area is simply $L_1^2+L^2_2$. If one conditions the loop to satisfy $L_1\gg L_2$ for fixed $L_2-L_1$, it is then expected that $\mean{\Gamma_A}^{1/2}\sim A^{2/3}$ for the scalar area law while $\mean{\Gamma^2_A}^{1/2}\sim A^{1/3}$ for the tensor area law, and, Umeki has observed the scalar area scaling. 

Umeki's second observation has to do with the lack of observation of a Lorentzian-shaped \ac{PDF}. He proposed that
\be\label{3.10}
\Phi[\mc]=\exp{\lr{i\frac{S[\mc]}{\nu}}}\rightarrow \Phi[\mc]\propto \exp\lr{-\frac{|S[\mc]|^2}{\nu^2}}\ .\
\ee
From these modifications, it is not hard to show that \ac{cPDF}s will always be Gaussian with a variance compatible with \ac{K41} phenomenology $\mean{\Gamma^2}\propto\vare_0^{2/3}R^{8/3}$. 

Due to computational constraints at that time, Umeki's simulation resolution and Taylor-based Reynolds number were too low, with values of $128^3$ and $R_\lambda\approx100$, respectively. This does not allow him to see much intermittency and the \ac{cPDF}s were compatible with Gaussian ones even in the inertial range, quickly transitioning to exponential distributions at very small scales.

The scenario was slightly better in the numerical investigation of Cao et al. \cite{cao1996properties}, where collocation points reach the value of $512^3$ and $R_\lambda=216$. At that time, the authors explored another aspect of the proposed area law by computing the \ac{cPDF} for rectangular loops of equal area and different aspect ratios and, also explored higher-order statistics of circulation. To summarize the first numerical investigations on circulation statistics of homogeneous and isotropic turbulence we highlight,
\begin{enumerate}[label=\textnormal{(P4.2.\arabic*)}]
    \item Circulation statistics depends on the scalar area of the enclosed contour not on the tensor area. Moreover, it is insensitive to the change in the shape of the contour.\label{p321}
    \item \ac{cPDF}s has fat exponential tails in the inertial range, rapidly transitioning to Gaussian \ac{PDF}s when the contour typical size $A^{1/2}$ approaches the integral length.\label{p322}
    \item Circulation scaling exponents defined by $\mean{|\Gamma_A|^p}\sim A^{\lambda_{|p|}/2}$ are shown to be anomalous, not in agreement with \ac{K41} expectation $\lambda_{|p|}=4p/3$. Moreover, circulation intermittency seemed to be more intense than the ones of structure functions i.e. if $\mean{|\delta_r v|^p}\sim r^{\zeta_{|p|}}$ we have, $|\lambda_{|p|}-4p/3|\geq |\zeta_{|p|}-p/3|$.\label{p323}
\end{enumerate}
The point \eqref{p323} was of crucial importance in Cao's work since none of the intermittency models based on energy dissipation fluctuation could fit the circulation scaling exponents. At that time, the authors also pointed out that the Reynolds numbers were too low to reach a considerable scaling region, this point will be discussed later in the light of high Reynolds \ac{DNS}. To finish with, the amazing work of Yoshida and Hatakeyama \cite{yoshida2000statistical} who first studied circulation statistics through the structural perspective of turbulence\footnote{Unfortunately, this work is not widespread throughout the turbulence community. Even though it has many similarities with the model to be introduced in Chapter~\refcap{cap4}, we had only taken notice of this Yoshida's work after publishing the main results of this thesis.}. By decomposing the vorticity field into strong and weak components through the imposition of a threshold $\omega_{th}$, they were able to show that as the threshold increases, the strong vorticity field starts to be concentrated at small and sparsely distributed simply connected regions. All the quantities like density, size, and circulation carried by each spot depend on the selected threshold but they presented, typically for $\omega_{th}/\mean{\omega}\sim 2\text{---}3$ the connected region's typical radius is about $\mean{R}/\eta_K\sim3\text{---}6$, this is surprisingly compatible with the typical core size of the vortex filaments in three-dimensional turbulence \cite{ghira2022characteristics}. 

Yoshida and Hatakeyama not only brought a novel approach to deal with circulation statistics but derived completely new results at the time. Within this vortex detection procedure, they were able to show that the weak component of circulation is Gaussian with scaling exponent compatible with the \ac{K41}, and, the strong component is responsible for the intermittency. Moreover, they proposed a model where the intense vorticity spots are homogeneously distributed in space generating a Compound Poisson Process \cite{zhang2014notes}. This model could reproduce some of the general properties of circulation but it has still several limitations.

This sets up the first years of discussion about circulation statistics in homogeneous and isotropic turbulence after Migdal's work. Furthermore, it is needed to say that it inspired the investigation to go beyond such an idealized system, we refeer to \cite{sreenivasan1995scaling} for an experimental investigation on circulation in turbulent wakes and \cite{benzi1997self} for the numerical investigation of circulation in shear flows. We will not delve further into these works as this thesis focuses on homogeneous and isotropic turbulence.

\section{Current Circulation Phenomenology}\label{sec3.3}

$_{}$

In recent years, advancements in computational capabilities have revitalized the discussions initiated by Migdal, leading to renewed interest in the subject. The historical rebirth of circulation statistics is due to the efforts of Iyer et al. \cite{iyer2019circulation} in 2019, who carried out an incredibly high-Reynolds simulation to study general statistical aspects of circulation. A strong focus on the validation of the area law was given in a continuation of this work \cite{iyer2021area}, which droves Migdal himself to get new insights on the calculation he developed in the early 1990s, and drawing parallels between turbulence, string theory, and critical phenomena \cite{migdal2020clebsch,migdal2021vortex,migdal2023statistical}. 

A different approach to this line of research, our central aim in this thesis, is the effort to develop unifying models to account for the statistics of circulation at different systems such as classical turbulence \cite{apolinario2020vortex,moriconi2022circulation}, quantum turbulence \cite{muller2021intermittency,polanco2021vortex,muller2022velocity,muller20232d}, and quasi-two-dimensional turbulence \cite{zhu2023circulation,muller20232d}. All those systems seem to share many similarities when analyzed through the lens of the circulation variable at inertial range scales, even though, their underlying dynamics have entirely different physical roots.

Now, let us start with the current phenomenology of circulation statistics in homogeneous and isotropic turbulence, from its historical beginning: the area law.

\subsection{Area Law}\label{sec3.3.1}

$\ \ \ \ $To start with, we reproduce Figures 3 and 4 of \cite{iyer2021area} at Fig.~\reffig{fig3.2}. Those are the same exploration made by \cite{umeki1993probability} and \cite{cao1996properties} but with higher resolution and Reynolds number ($R_\lambda=1300$ for the left panel and $R_\lambda=650$ for the right one). Two contours with different aspect ratios and the same area collapse in the left panel of Fig.~\reffig{fig3.2} to ensure the statistics are shape independent. And, the adoption of a scalar area law in opposition to the tensor area law is clearly seen by the scaling $\mean{\Gamma_A^2}^{1/2}\sim A^{2/3}$ for the figure eight contour on the right panel of Fig.~\reffig{fig3.2}.
\begin{figure}[t]
\center
\includegraphics[width=\textwidth]{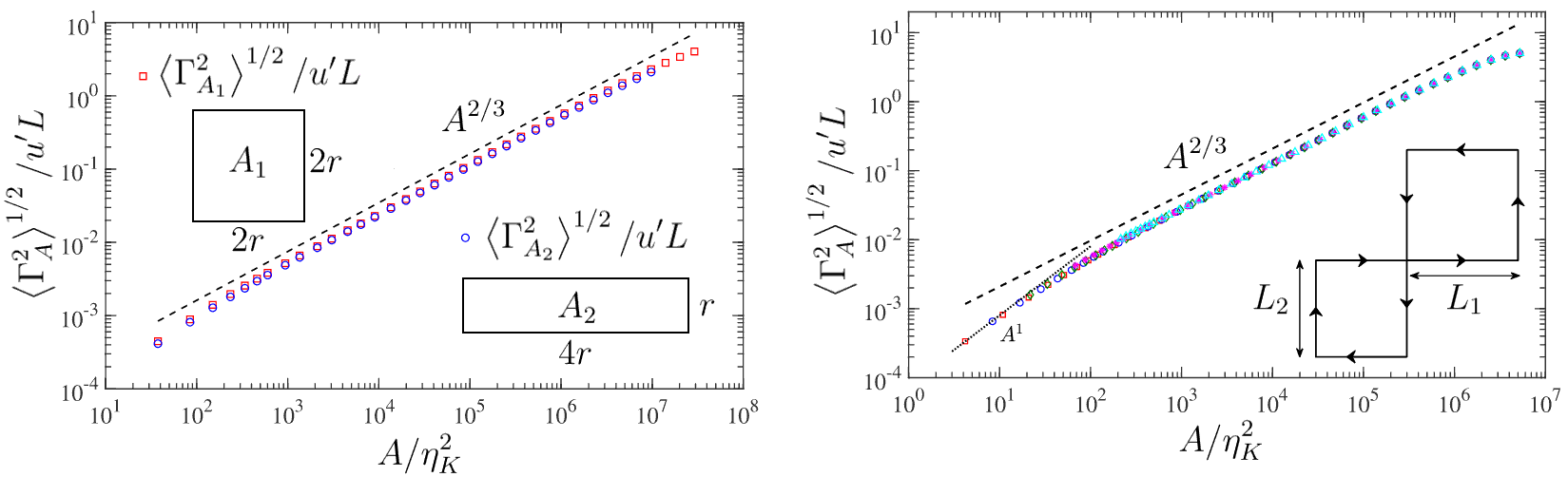}
\caption[Area law probe from Iyer et al.]{Left panel shows the standard deviation of circulation which is expected to scale as $\sim A^{2/3}$ from \ac{K41} for two contours with the same area and different aspect ratio and perimeter. The right panel shows the circulation's standard deviation for the figure eight loop in the same spirit as Umeki's work. Colors red, blue, green, pink, and cyan are corresponding to $(L_1-L_2)/\eta_K= 2, 4, 8,\text{ and } 16$, respectively. Both figures were reproduced from \cite{iyer2019circulation} with minor modifications.}
\label{fig3.2}
\end{figure}

Iyer et.al. also computed \ac{PDF}s of circulation on rectangular contours of the same area but with different aspect ratios, with both sides lying in the inertial range and with at least one side out of the inertial range. On the one hand, the resulting \ac{PDF}s which are totally in the inertial range collapse, not only at the tails but also in the bulk, meaning that Migdal's prediction about the area law is somewhat broader. On the other hand, the \ac{PDF}s out of the inertial range seems to be perimeter-dependent as their shape changes for different aspect ratios.

The tails of the collapsed \ac{PDF} was shown to be better fitted by stretched exponentials,
\be\label{3.11}
\rho(\Gamma_A)\propto e^{-c|\Gamma_A|^\upsilon} \ , \
\ee
than the usual exponential ($\upsilon=1$) found by Umeki and Cao et al., but it turns out that the parameter $\upsilon$ varies slightly depending on the aspect ratio. In a continuation of \cite{iyer2019circulation}, the authors address this question by exploring the same contour shapes, with varying aspect ratio, in a simulation with $N=8192$ linear collocation points and $R_\lambda=1300$ \cite{iyer2021area}. They found the tails of the collapsed \ac{cPDF} to be fitted within a single curve with the following shape,
\be\label{3.12}
\rho(\Gamma_A)\propto \frac{e^{-c|\Gamma_A|}}{\sqrt{|\Gamma_A|}} \ , \
\ee
where $c=21.74$ is a Reynolds-dependent parameter. The fit considers an aspect ratio ranging from $1$ to $2.25$ and a fixed area $A=32400\eta_K^2$. This basically closes the discussion about the asymptotic behavior of extreme events of circulation to be not exactly exponential but having the shape of Eq.~\eqref{3.12} with modulating square root of circulation. The statistical relevance of this finding to the modeling of circulation will be addressed in Chapter~\refcap{cap4}. 

It is thus possible to calculate the most relevant contribution for the higher-order moments for all rectangular shapes lying in the inertial range using Eq.~\eqref{3.12},
\be\label{3.13}
\mean{\Gamma^{2p}_{rec}}\approx\int_0^{\infty} d\Gamma \ \Gamma^{2p}\rho(\Gamma)\propto\frac{1}{c^{2p+1/2}}\Gamma\lr{2p+\frac{1}{2}} \ , \
\ee
where $\Gamma(n)$, the Gamma function, should not be confused with the circulations $\Gamma_R$. Using Eq.~\eqref{3.13} one is able to show that,
\be\label{3.14}
\ln\lr{\frac{\mean{\Gamma^{2p}_{rect}}^{1/2p}}{\mean{\Gamma^{2q}_{rect}}^{1/2q}}}\approx\ln\lr{\frac{p}{q}}+\frac{1}{4p}\ln\lr{4\pi p}+\frac{1}{4q}\ln\lr{4\pi q} \ , \
\ee
for sufficiently large $p$ and $q$. Note that Eq.~\eqref{3.14} is parameter-free and, by an exploration of consecutive even-order moments $(p=q+1)$, \cite{iyer2021area} were able to match exactly this expression for inertial range rectangles and $p=6$. 
\begin{figure}[t]
\includegraphics[width=\textwidth]{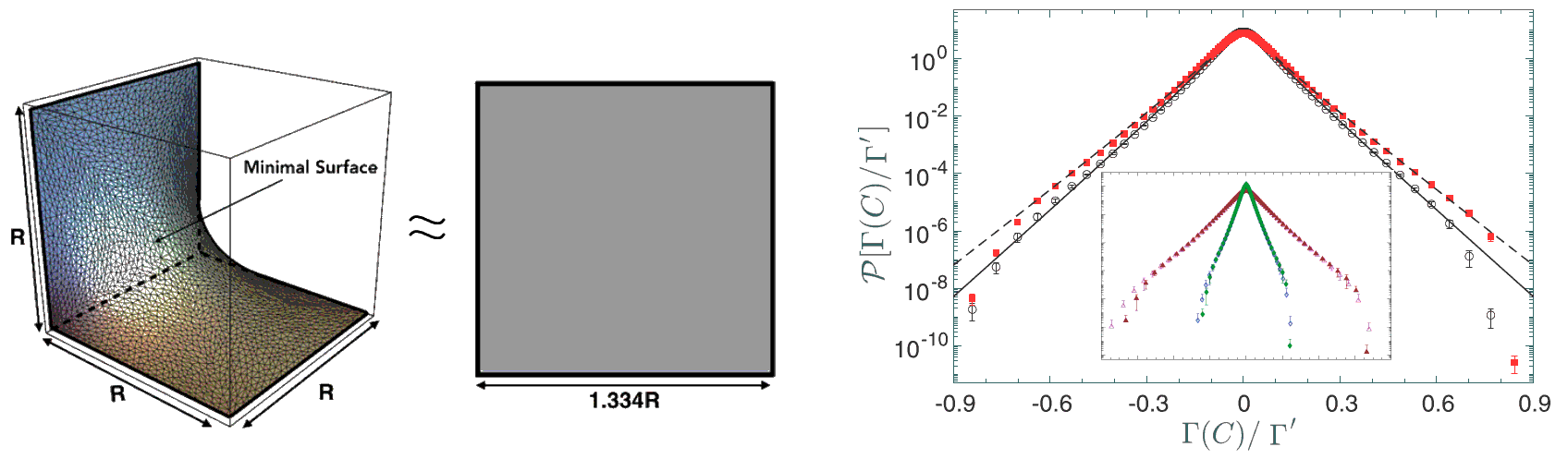}
\caption[Nonplanar area law probe from Iyer et al.]{Left panel shows the goalkeeper contour, its minimal area $A_m$, and the comparison with a square with the same area $A_m$. The right panel shows \ac{cPDF} for the planar (full red squares) and nonpĺanar (open gray circles) loop with the same area and $\Gamma'=U_{rms}L$. The inset shows the same as the main frame but with a different normalization $\Gamma'=\mean{\Gamma^p}^{1/p}$ for $p=2$ (pink/red) and $p=8$ (green/blue). Both figures were reproduced from \cite{iyer2021area} with minor modifications.}
\label{fig3.3}
\end{figure}

The first (and only) time a nonplanar loop was addressed was also only in 2021 with \cite{iyer2021area}. This is a complicated question to address numerically since, for general contours, interpolations are needed. The simple idea put forward was to consider a ``goalkeeper contour'' illustrated in the left panel Fig.~\reffig{fig3.3}. In this setting, all the pieces of straight lines that compose the contour are aligned with grid points such that, circulation is calculated through the line integral (which can also be computationally expensive). The general result found in their work is that the area law is not valid --- in its pure formulation --- for nonplanar contours, as clearly shown in the right panel of Fig.~\reffig{fig3.3}. Instead, a weaker form of area law is still valid when the \ac{cPDF} is normalized by some proper scale $\mean{\Gamma^p}^{1/p}$ (see the inset of the right panel of Fig.~\reffig{fig3.3}). It shows that, although entirely valid for planar loops, the area law is not the last word regarding circulation statistics, and, even though it is supposed to play an important role for nonplanar loops, there are missing ingredients on the Migdal's approach of the area law.

\subsection{4/5-Law, Statistical Moments, and Multifractal Aspects}\label{sec3.3.2}

$\ \ \ \ $Among all the results presented in Migdal's first investigations of circulation, he derived, in the context of the scalar area law,  that the tails of the \ac{cPDF} should depend on the combination $(\Gamma^{2\kappa_1}A^{1-2\kappa_1})^{\kappa_2}$ where $\kappa_1$ and $\kappa_2$ are unknowns. The \ac{K41} behavior is expected for $\kappa_1=3/2$ which is directly related to the scaling $\Gamma^3A^{-2}$, but it still has an infinite class of solutions depending on the exponent $\kappa_2$ which is not fixed in Migdal's prediction. One might think of this scaling relation as an instance of the 4/5 law for velocity structure function written in the circulation language. Along the same lines, one might expect some universality on the third-order exponent of circulation since it is linearly related to the velocity field. 

A properly normalized \ac{PDF} of $\Gamma^3$ is shown in the left panel of Fig.~\reffig{fig3.4} for all datasets. It is notable that all \ac{PDF}s collapse into a single shape and, its mean value, related to the third-order moment of circulation is slightly shifted off the zero value as noted by \cite{iyer2019circulation}. The inset of the left panel of Fig.~\reffig{fig3.4} shows the fitted scaling exponents $\mean{\Gamma^3_A}\sim A^{\lambda_{3}/2}$ for all datasets. Despite little deviations, they are all compatible with $\lambda_3=4$ within a $2\sigma$ confidence interval, supporting the validity of an analog of 4/5 law for circulation. Still, a general proof of this law coming from basic principles of the \ac{iNSE} is still absent.
\begin{figure}[t]
\includegraphics[width=\textwidth]{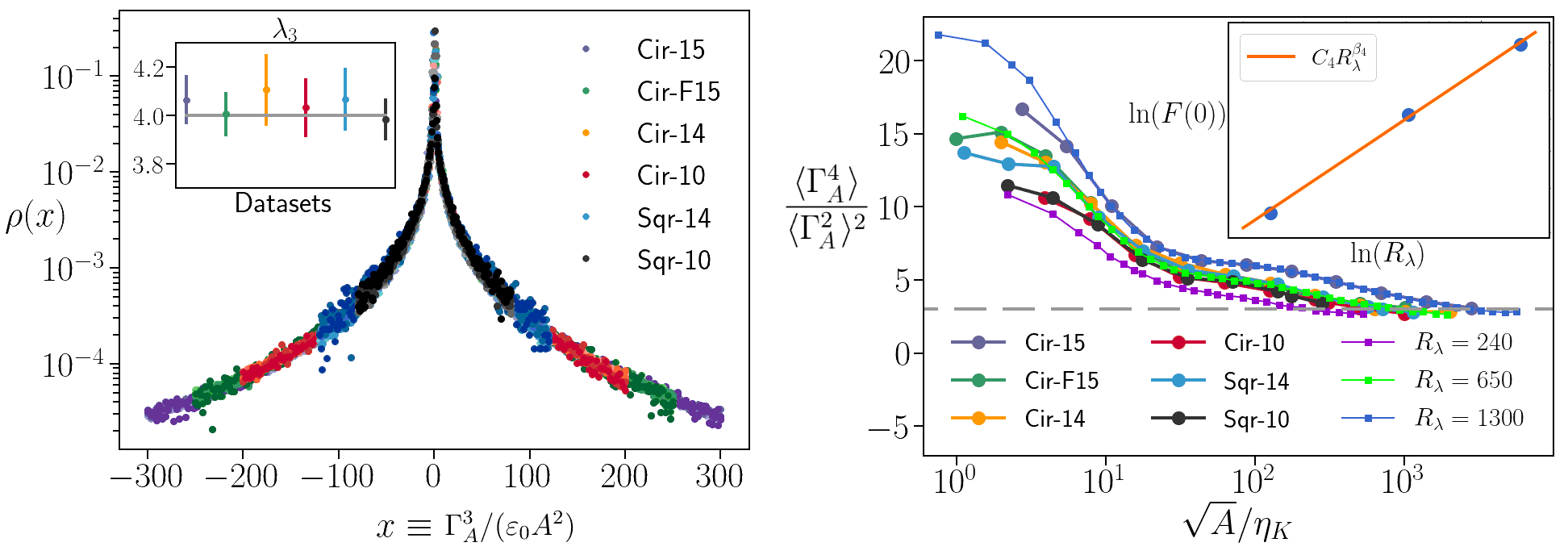}
\caption[Third order cPDF and moments.]{Left panel shows the \ac{PDF} of $\Gamma^3_A/(\vare_0 A^2)$ for all Datasets in Table~\reftab{tab3.2}. Inset shows the fitted scaling exponents of the circulation's third order moment $\lambda_3\approx4$. The right panel shows the kurtosis of circulation for different datasets in Table~\reftab{tab3.2} and \cite{iyer2019circulation}. Inset shows $F(A\rightarrow0)$ as functions of $R_\lambda$, blue dots are the data of \cite{iyer2019circulation}.}
\label{fig3.4}
\end{figure}

The first measurements of intermittency are historically related to the kurtosis, $F(A)\equiv \langle \Gamma_A^4 \rangle / \langle \Gamma_A^2 \rangle^2$ \cite{frisch1995turbulence}. The behavior of the circulation's flatness can be phenomenologically inferred by looking at Fig.~\reffig{fig3.1} again. For large contours, this quantity behaves as $F(A)\rightarrow 3$ due to the Gaussianity of the \ac{cPDF}, while in the opposite limit, it should converge to a constant related to the vorticity's kurtosis which does not depend on the area since it is pointwise defined. This is exactly what is shown in the right panel of Fig.~\reffig{fig3.4}, for all datasets in Table~\reftab{tab3.2} and the data from \cite{iyer2019circulation}. All kurtosis decays towards $3$ for a high enough contour size but in the opposite limit, a Reynolds dependency naturally appears. 

One might note that as the \ac{DNS} simulations are typically resolved up to $\mathcal{O}\lr{\eta_K}$, it turns difficult to probe the asymptotic behavior $F(A\rightarrow0)$ related to the vorticity. Notwithstanding, one can proceed to analyze the Reynolds behavior of this curve and, as the better-resolved data is the one from \cite{iyer2019circulation}, we show in the inset of the right panel of Fig.~\reffig{fig3.4} its Reynolds-dependency. A power law behavior $F(R\rightarrow0)\simeq C_4 R_\lambda^{\beta_4}$ is observed, with $C_4 \simeq 1.16$ and $\beta_4 \simeq 0.41$.
\begin{figure}[t]
\includegraphics[width=\textwidth]{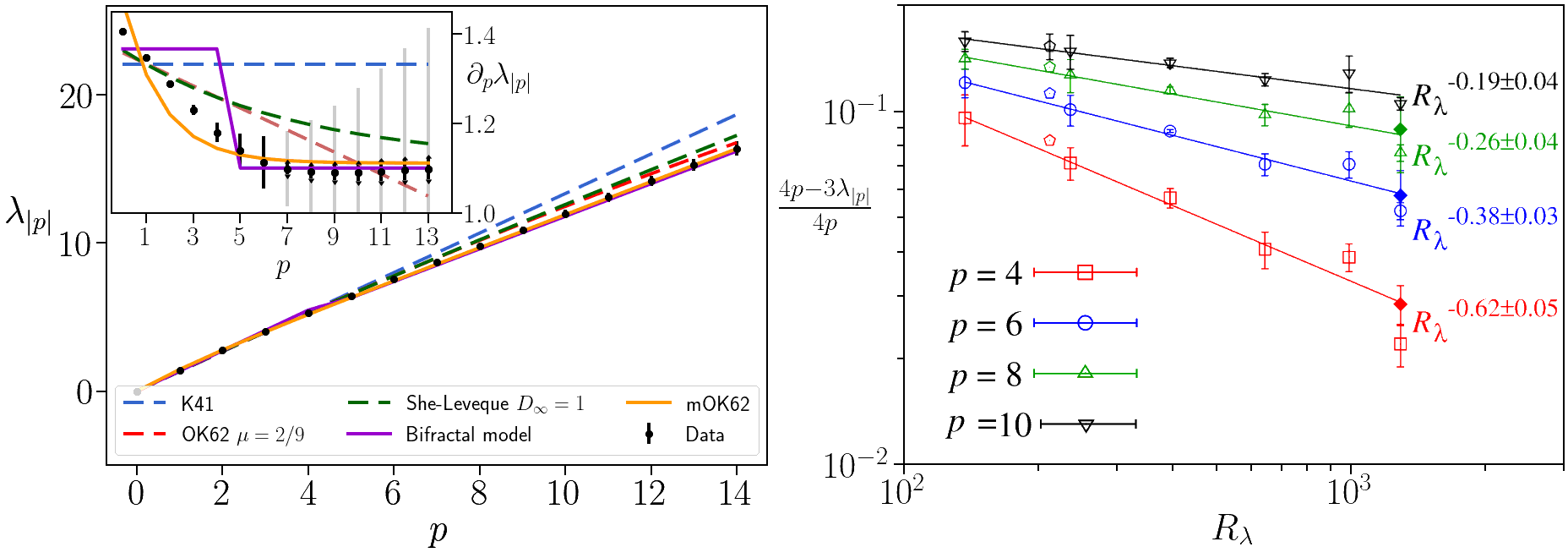}
\caption[Circulation scaling exponents and its Reynolds dependency.]{The left panel shows moments of circulation $\lambda_{|p|}$ and its comparison with \ac{K41}, two old models put forward in the 1990s (\ac{OK62} and She-Lévêque), and, current models. Inset shows the derivative of $\lambda_{|p|}$ to highlight its linearization for high enough moment order $p$. The errorbars of $\partial_p\lambda_{|p|}$ grows linearly for $|p|>6$ (gray bars) so that we plotted $\sigma_p/p$ (black bars). The right panel shows the Reynolds dependency on the relative difference to \ac{K41} (reproduced from \cite{iyer2019circulation}).}
\label{fig3.5}
\end{figure}

Another important feature of kurtosis is the tendency to the formation of a plateau in the inertial range for the highest Reynolds number. There is a clear tendency that repeats over all datasets when closed looked at, in fact, \cite{iyer2019circulation} pointed out this behavior, and by analyzing the length of the plateau they found a Reynolds dependency on it. They have also argued, based on the extrapolation of their data, that the plateau should fill the entire inertial range when $R_\lambda\approx1900$. This issue has no clear explanation in terms of the current asymptotic understanding of circulation statistics and will be further discussed in a moment.

Scaling exponents of statistical moments are of huge importance in turbulence since their exploration can highlight unnoticed phenomena to be addressed by intermittency models and other approaches. For the early phenomenology of circulation, its scaling exponents were thought to be more intermittent than the ones of structure functions, for instance. The left panel of Fig.~\reffig{fig3.5} shows scaling exponents obtained by the \ac{ESS} procedure $\lambda_{|p|}=\lambda_3\lambda_{(|p||3)}$ up to the $14^{th}$-order. We also show, by dashed lines, the predicted \ac{K41} behavior and two intermittency models based on the standard cascade models, OK62 with $\mu=2/9\approx0.22$, and She-Lévêque $D_\infty=1$, representing the comparison to the early phenomenology made by \cite{cao1996properties} and \cite{yoshida2000statistical}. From these models, it is clear that circulation deviates more than predicted in the absolute value sense. 

The belief that circulation is more intermittent than velocity difference was the common understanding of the problem until \cite{iyer2019circulation}. It turns out that, taking the relative difference from the \ac{K41} behavior, circulation is less intermittent than the velocity structure function. Rephrasing in mathematical language, if $\mean{|\delta_Rv|^p}\sim R^{\zeta_{|p|}}$ and $\mean{|\Gamma_R|^p}\sim R^{\lambda_{|p|}}$,
\be\label{3.15}
\left|\frac{p-3\zeta_{|p|}}{p}\right|\geq\left|\frac{p-3\lambda_{|p|}/4}{p}\right| \ ,\
\ee
the equality is supposed to hold for $p=3$ when both are zero. This is partially due to the completely unexpected behavior of linearization at the higher-order circulation scaling exponents reported by \cite{iyer2019circulation}. They noted that for $p>4$ the curve of scaling exponents can be well-fitted by a straight line $\lambda_{|p|}=h_\infty p+(3-D_\infty)$ while for $p<3$ it can be fitted by $\lambda_{|p|}=h^{\star}p$. This introduces the bifractal model plotted in a continuous purple line at the left panel of Fig.~\reffig{fig3.5}, with the values $h^{\star}=1.367\pm0.009$, $h_\infty=1.1$ and, $D_\infty=2.2$ (no errorbars were presented for these later ones). 

This model suggests a very simple interpretation from the multifractal perspective. At the core of the \ac{cPDF} i.e. low order moments, circulation events are dominated by space-filling, self-similar structures with H{\"o}lder exponent slightly larger than the Kolmogorov prediction. These events are supposed to be related to a smooth background vorticity field yielding to the \ac{K41} scaling in most of its occurrence and, anti-polarized vortex filaments carrying similar circulation occurring with lower probability slightly deviating the \ac{K41} behavior of the background. While extreme events of circulation $(p\geq4)$ are composed of slightly wrinkled vortex sheets since the fractal dimension $D_\infty=2.2$ is measured. These structures are also self-similar and almost differentiable structures with H{\"o}lder exponent $h_\infty=1.1$. Nevertheless, no specific study to detect such a structure was carried out up to the present date.

In the right panel of Fig.~\reffig{fig3.5}, we show an unexpected Reynolds dependency of the scaling exponents fitted by \cite{iyer2019circulation}. In the most dramatic case, the relative difference to the \ac{K41} exponent drops down from $10\%$ to $2\%$ for $p=4$, with smaller decreases for higher-order moments. From a skeptical point of view, this dependency can have two possible explanations, the first one is that it is of physical concern: The statistics of circulation have not reached the asymptotic state and this decaying behavior will persist up to a finite Reynolds number $R_\lambda^\star$ and then the ``true asymptotics'' will show up with no dependency of $\lambda_{|p|}$ on $R_\lambda$. The second explanation relies upon a numerical artifact related to physically not understood phenomena: The formation of the inertial range plateau (which grows with some power of $R_\lambda$) develops a competition among scales and, if one fits scaling exponents in a constant range, it will be gradually contaminated by the plateau scaling. None of the current works have dealt with the matter of plateau formation such that no precise conclusion of the above argumentation can be put forward.

Turning back to the bifractal model, Fig.~\reffig{fig3.6} shows $\propto\rho(\propto \Gamma_A A^{-h/2})A^{-(d-D(h))/2}$ which is supposed to collapse for self-similar scaling with well-defined H{\"o}lder exponents $h$ and fractal dimensions $D(h)$. The values of $h^{\star}$, $h_\infty$ and, $D_\infty$ are slightly different from those fitted in the original work of \cite{iyer2019circulation} for reasons which will be discussed in Section~\refsec{sec4.2}. One can see the near collapse\footnote{Even with data traffic limitations, our limited datasets show up to support the conclusions drawn by the analysis of Fig.~\reffig{fig3.6}, at least qualitatively.} of the core of the \ac{cPDF}s and their tails for different H{\"o}lder exponents and fractal dimensions, supporting the idea of the bifractal behavior of circulation. 
\begin{figure}[t]
\includegraphics[width=\textwidth]{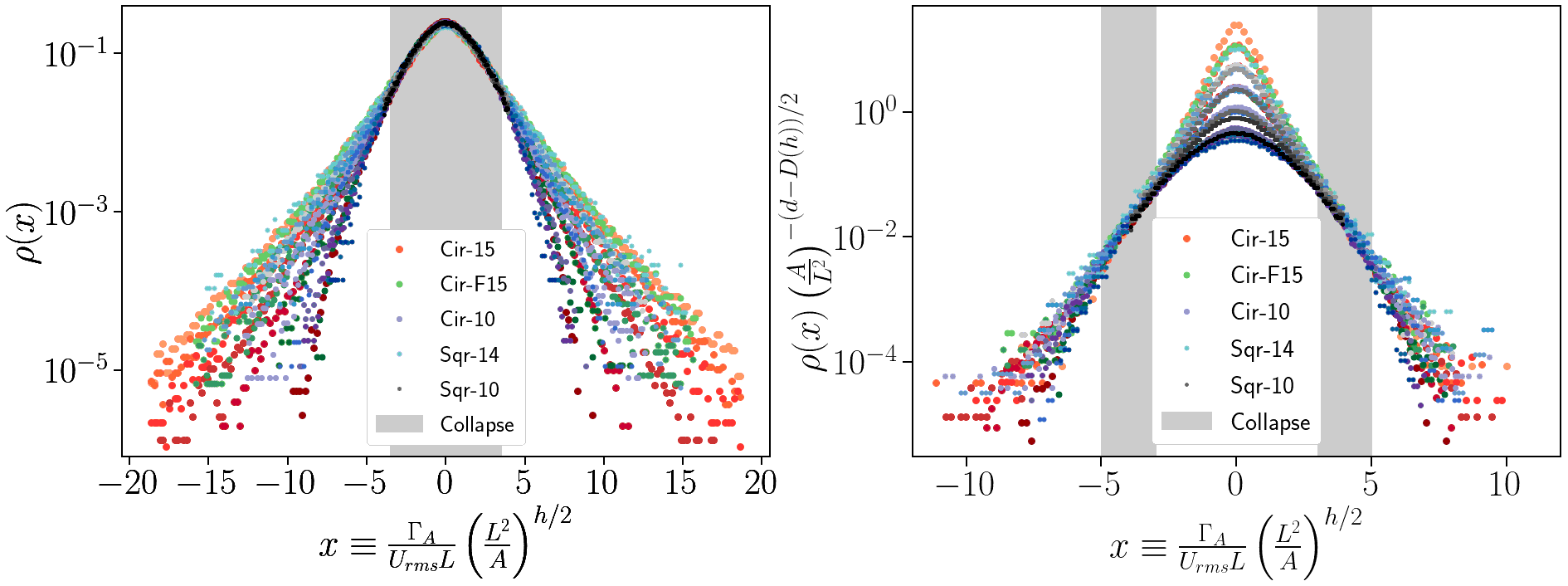}
\caption[Collapsed core/tail of cPDF.]{Left panel shows the collapse between \ac{cPDF}s of different datasets when properly rescaled according to its proper singularity spectrum, $h^\star=4/3+\mu/2\approx1.418$, and $D^\star=d$ for the left panel. The right panel uses $h_\infty\approx1.126$, and $D_\infty\approx0.866$. A proper definition of $h_0$ and the functional forms of $D$ and $h$ will be clarified in Section~\refsec{sec4.2}.}
\label{fig3.6}
\end{figure}

Even though this approach offers a very simple interpretation of the phenomena, a closer inspection of the scaling exponents around $p=4$ shows a clear discrepancy to the measured data and, moreover, $\lambda_3\neq4$ which is not consistent with the 4/5 law for circulation. This motivates the introduction of a different multifractal model \ac{mOK62} put forward in the context of quantum circulation statistics (quantum and classical circulation statistics were shown to share the same scaling exponents when probed on their respective inertial range \cite{muller2021intermittency}). Although referred to as allusive to the \ac{OK62}, this model has scaling exponents based on the She-Lévêque model with $D_\infty=2.2$ instead of $D_\infty=1$. This represents a smooth transition to a scaling equivalent to the one of \cite{iyer2019circulation} at large enough moment order, more precisely, $\lambda_{|p|}=4p/3 +\tau_{SL}(p/3)\approx 10p/9+(3-D_\infty)$ for high enough $p$. The \ac{mOK62} is shown in Fig.~\reffig{fig3.5} as a continuous yellow line, note that it visually catches much better the transition between the two linear scalings and satisfies $\lambda_3=4$ exactly. However, a closer inspection of the inset of the left panel of Fig.~\reffig{fig3.5} shows that it still does not reproduce very well the low order moments. 

That basically finishes the discussions made so far and, to resume the results of this chapter, we will provide an outline encapsulating three decades of research into turbulent circulation statistics in a few points:
\begin{enumerate}[label=\textnormal{(P4.3.\arabic*)}]
    \item Circulation statistics is bounded by the vorticity and dissipation anomaly at very small scales and related to the law of large numbers at scales comparable to the integral length. \label{p331}
    \item Extreme events of circulation in planar contours do not follow stretched exponential decay, but simple exponentials modulated by a prefactor $1/\sqrt{|\Gamma|}$ as in Eq.~\eqref{3.12}, almost independently of the aspect ratio. \label{p332}
    \item The scalar-area law does not hold widely in its pure formulation, even for planar loops. At the level of \ac{PDF}s, it holds as point~\ref{p332}. At the moments level, systematic deviations of $\lesssim5\%$ have been observed in inertial range scales.\label{p3331}
    \item Nonplanar loop area law is completely flawed in the pure sense. However, it is restored --- at the level of \ac{PDF}s --- when circulation is properly normalized by an inner variable like $\mean{(\Gamma[\mc])^p}^{1/p}$. \label{p3332}
    \item An equivalent of 4/5 law for circulation is assumed to hold, without rigorous proof. \label{p334}
    \item Circulation's kurtosis has a nontrivial shape with an asymptotic tendency to constancy at the inertial range (the plateau formation). \label{p335}
    \item Circulation's scaling exponents crossover to linear behavior at high enough moment orders. \label{p337}
    \item Points \ref{p335} and \ref{p337} are shown to be Reynolds dependent. \label{p338}
\end{enumerate}

The absence of a unifying picture, even if phenomenological, to give a reason for the above points is the main motivation for the present work. From now on, we will explicitly tackle the problem of circulation in turbulence flows by employing pioneering modeling which is still very robust and mathematically well-posed. 

\end{chapter}

\begin{chapter}{Vortex Gas Modeling of Circulation Statistics}
\label{cap4}

\hspace{5 mm} 

In the general numerical and experimental phenomenology of 3D homogeneous and isotropic turbulence, the flow is noticed to organize itself as a tangle of thin and elongated vortex filaments \cite{she1990intermittent,farge2001coherent,ishihara2009study}. Such a structural view of turbulence was the motivating scenario behind Yoshida's work \cite{yoshida2000statistical}. The circulation computed for any closed line is, by Stokes' theorem, the sum of the circulation carried out by each of the filamentary structures that crosses the circulation domain. These observations pave the way for the introduction of simple vortex modeling to reproduce the statistical behavior of velocity circulation.

The rationale behind the construction of the \ac{VGM} \cite{apolinario2020vortex} is based on a very simple phenomenological analysis of the \ac{K41} expectation for the statistical moments of circulation, $\langle \Gamma_R^n\rangle\propto \vare^{n/3}R^{4n/3}$. The second-order moment can be rewritten as follows, 
\be\label{gdecomp}
\langle \Gamma_R^2 \rangle \propto \left ( \frac{R}{\eta_K} \right )^4 \left [ \eta_K^2 \sqrt{ \frac{\vare_0}{ \nu} } \right ]^2 \left ( \frac{\eta_K}{R} \right )^{\frac{4}{3}}\ ,\
\ee  
where $\eta_K$, $\nu$, and $\vare_0$ follow the usual definitions of Chapter~\refcap{cap2}. The decomposition in Eq.~\eqref{gdecomp} together with the structural view of turbulence, suggests that the total circulation is composed of,
\begin{enumerate}[label=\textnormal{(P5.\arabic*)}]
    \item $N \propto (R / \eta_K)^2$ planar vortices of typical size on the order of $\eta_K$, \label{(P4.1)}
    \item each carrying an r.m.s. vorticity of the order of $\sqrt{\vare / \nu }$,\label{(P4.2)}
    \item with correlations that decay as $R^{-4/3}$ for scale separations $R\gg \eta_K$.\label{(P4.3)}
\end{enumerate}
Up to the present date, the definitions of the \ac{VGM} are only restricted to planar contours. This constraint turns out to be impossible to explore minimal area phenomenology since a continuous contour $\mc_R$ in 2D splits the plane into two planar regions: inside and outside the contour. We then define the inner contour domain $\md_R$, satisfying $\mc_R=\partial\md_R$, to be the only way to compute circulation via Stokes' theorem. From the latter remark, it follows that $\md_R$ always has a minimum area.

For the model definitions, suppose the contour $\mathcal{C}_R$ lays on a plane $\Psi$ that slices the whole turbulent domain. Considering now that each vortex filament intersects the plane $\Psi$ at points denoted as $\vc x_i$, the contribution of the $i^{th}$ structure to the contour integral is $\Gamma_i$. In this scenario, it is clear that,
\be\label{cfil}
\Gamma_R=\sum_{i\in \Psi}\Gamma_i(\mathcal{D}_R) \ .\
\ee
Note that the sum is applied for the whole 2D slice $\Psi$, not only to the vortices inside the domain $\md_R$. As long as the spatial shape of the vortex is not addressed, all the structures in the plane $\Psi$ will contribute to the final circulation, some of them less than the others depending on how close they are from contour $\mc_R$. Within these remarks, for each vortex contribution, we propose,
\be\label{ccindiv}
\Gamma_i(\mathcal{D}_R)=\int_{\mathcal{D}_R}d^2y\tg(\vc x_i)G(\vc y-\vc x_i) \ , \
\ee
where $\tg(\vc x)$ is a \ac{GFF}. The two-point correlation function fully the statistic of $\tg(\vc x)$ and, in accordance with point~\ref{(P4.3)} we set: 
\be\label{gtilcoorcases}
\begin{cases}
    \mean{ \tg(\vc x) \tg(\vc y)}\sim |\vc x - \vc{y}|^{-\alpha} &,\ \text{ if }|\vc x -\vc y|\gg\eta_K \ ,\ \\
    \mean{ \tg(\vc x) \tg(\vc y)}\sim 1 &,\ \text{ if }|\vc x -\vc y|\ll\eta_K\ .\
\end{cases}
\ee
Also based on point~\ref{(P4.3)}, one can impose $\alpha=4/3$. Therefore, we will not constrain it yet since \ac{K41} expectations are subjected to small intermittency correction. The fundamental importance of these corrections is the subject of Section~\refsec{sec4.4}.

From now on, we will shorten all the vector notations $\vc x\rightarrow x$ for aesthetical reasons. The field $G(x- x_i)$ comprises, not only the shape but the intensity $\sqrt{\vare(x_i)/\nu}$ of the planar structure in consonance with point~\ref{(P4.2)}. Employing the continuum limit of Eq.~\eqref{cfil}, and using Eq.~\eqref{ccindiv} we get,
\be\label{circmodel}
\Gamma_R=\int_\Psi d^2x\int_{\mathcal{D}_R}d^2y\tg(x)\sigma(x)G(y-x)\ ,\
\ee
where $\sigma(x)$ is the local surface density of vortex structures. Fig.~\reffig{fig4.1} shows the vertical spots formed by the intersection of vortex filaments with a 2D slice of 3D homogeneous and isotropic turbulent flow. The local well-defined spot observed in Fig.~\reffig{fig4.1} that represents $\sigma(x)$ was generated by applying a vortex identification method named \ac{SSC} \cite{zhou1999mechanisms,chakraborty2005relationships}. This method basically consists in calculating imaginary parts of the eigenvalues of the velocity gradient matrix $\partial_iv_j(x)$ denoted as $\lambda_{ci}(x)$ and, by applying a threshold $\lambda_T$, intense and localized structures are identified (see Appendix~\ref{appssc} for details).

Vortices are clearly not homogeneously distributed since clusters and voids are visually seen. Moreover, the structures have a huge variety of intensities, shapes, and sizes, which are supposed to be accounted for by $\sigma(x)G(y-x)$ in Eq.~\eqref{circmodel}. The vortex distribution can be a very difficult modeling task to address in general, but it must be constrained by the point~\ref{(P4.1)}, meaning that at least its first order moment $\mean{\sigma(x)}$ is independent of the contour's typical length $R$.
\begin{figure}[t]
\center
\includegraphics[width=\textwidth]{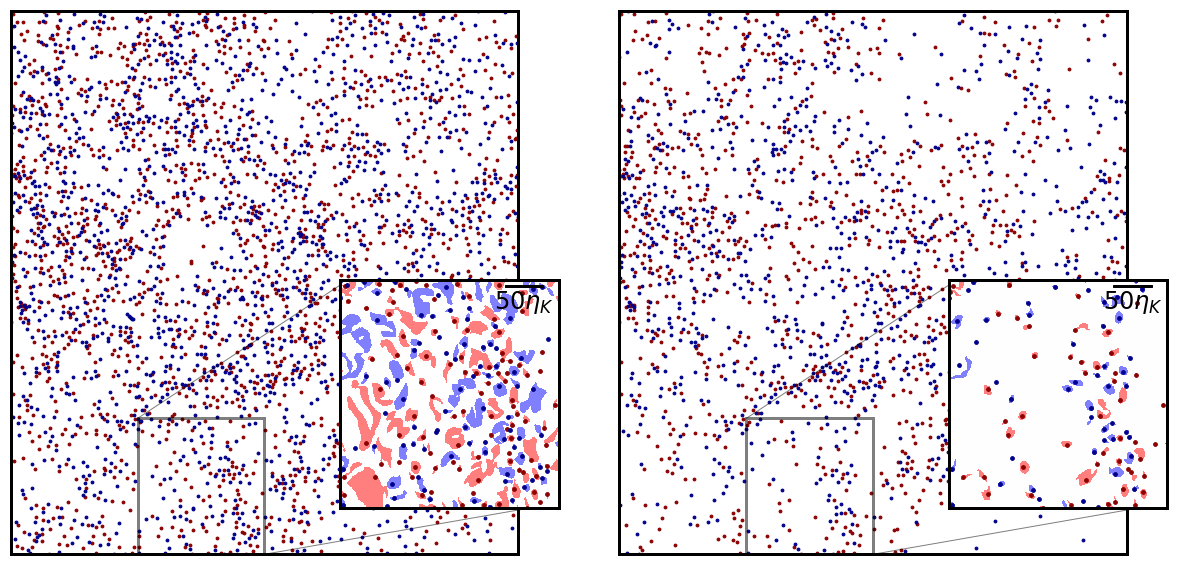}
\caption[Vortex distribution using SSC.]{Typical vortex distribution for $\lambda_{T}=0.125\sigma_{\lambda_{ci}}$ (left panel) and $\lambda_{T}=1.50\sigma_{\lambda_{ci}}$ (right panel), see Appendix~\ref{appssc} for details. Red/blue dots show the peak position of the detected structure for positive/negative circulation. Inset shows a zoom with a detailed shape of the detected region. Both images were produced using dataset Iso-1024 (see Table~\reftab{tab3.1}) with an arbitrarily chosen snapshot, 2D slice, and orientation.}
\label{fig4.1}
\end{figure}

Before discussing the issues related to the vortex distribution and its shape, let us formally construct the \ac{GFF} $\tg (x)$ with the desired correlation function by writing \cite{barnsley1988science,prakash1992structural,javerzat2020topological},
\be\label{gtil}
\tg ( x ) = \frac{\eta_K^{\alpha/2}}{\sqrt{2\pi\Gamma(\alpha)}}\int d^2 k \, \psi (k) |k|^{\frac{\alpha}{2}-1} \exp \left ( i k \cdot x - |k| \frac{\eta_K}{2} \right )\ ,\
\ee
where $\psi(k)$ is a complex \ac{GFF} such that $\mean{\psi(k)}=0$ and $\mean{\psi(k)\psi(k')^\star}=\delta^2(k-k')$. For $\alpha>0$, the correlation function can be explicitly computed in terms of special functions
\be\label{gtilcorr}
\mean{\tg (x) \tg (y)}={}_2F_1\lr{\frac{\frac{\alpha}{2},\frac{1+\alpha}{2}}{1};-\frac{|x-y|^2}{\eta_K^2}}\ , \
\ee
where ${}_2F_1$ is the hypergeometric function of $2+1$ entries. This special function is dominated by an algebraic decay $|x -y|^{-\alpha}$ when $|x -y|\gg\eta_K$, but has a regular behavior in the opposite limit since ${}_2F_1(a,b,c,0)=1$, exactly as desired for our model.

The construction of the density field $\sigma(x)$ and the shape field $G(x)$ is phenomenologically more involved than $\tg(x)$. It must attain intermittent fluctuations of intensity and the statistical distribution of vortices, incorporating the description made in points~\ref{(P4.1)} and \ref{(P4.2)} at once. For this purpose, we put forward two basic assumptions to end up with an analytically tractable system, without being too idealistic. 

The first assumption parametrizes the shape of the vortex structure by the product of a Gaussian packet $g_\eta(x)=\exp(-|\vc x|^2/2\eta^2)$, with a fluctuating intensity field $\xi(x)\sim \sqrt{\vare(x)}$, where $\eta=a\eta_K$ with $a$ an order unit parameter to be adjusted. The statistical behavior of coarse-grained dissipation becomes readily tractable using the modeling ideas put forward in Chapter~\refcap{cap2}. To achieve this, we directly replace the local dissipation rate with its coarse-grained counterpart, leading Eq.~\eqref{circmodel} to,
\be\label{circmodelcoarse}
\Gamma_R=\xidr\sqrt{\frac{\vare_0}{3\nu}}\int_\Psi d^2x\int_{\mathcal{D}_R}d^2y\tg(x)\sigma(x)g_\eta(y-x)\ ,\
\ee
where,
\be\label{xid}
\xidr=\frac{1}{A(\mathcal{D}_R)}\int_{\mathcal{D}_R}d^2x\sqrt{\frac{\vare(x)}{\vare_0}}\ ,\
\ee
and $A(\md_R)$ is the area of the domain $\md_R$.
The dimensional prefactor $d^2x\sqrt{\vare_0/3\nu}$ in Eq.~\eqref{circmodelcoarse} ensures the correct circulation units. Moreover, all the aforementioned fields are assumed to be statistically independent i.e. the mean value of any combination of functions of the fields $\mean{f_1(\xidr)f_2(\tg)f_3(\sigma)}=\mean{f_1(\xidr)}_{\xidr}\langle f_2(\tg)\rangle_{\tg}$ $\mean{f_3(\sigma)}_\sigma$.

The direct substitution of the intensity field for its coarse-grained version relies on robust mathematical reasoning, utilizing the central value theorem (or Rolle's theorem) as outlined in \cite{ballantine2002simple}. This theorem can be stated as follows, consider the integral of the product of two continuous functions $f_1(x)$ and $f_2(x)$ in a domain $V$, there exists a ``central point'' $x_0$ which satisfies,
\be\label{rolle}
\int_V f_1(x)f_2(x)d^dx=f_1(x_0)\int_Vf_2(x)d^dx=f_2(x_0)\int_Vf_1(x)d^dx \ .\
\ee
Since we are dealing with the random field and not with continuous functions, Eq.~\eqref{rolle} specifically holds for homogeneous distributed central points $x_0$ over $V$ in the asymptotic limits $R\ll\eta_K$ and $R\gg\eta_K$, as discussed in \cite{apolinario2020vortex}. Although not proved, the validity of Eq.~\eqref{rolle} in the inertial range scale seems to be very reasonable since correlations of the dissipation field are long-ranged when compared to the correlations of the \ac{GFF} $\tg(x)$ \cite{yeung2012dissipation}.

Eq.~\eqref{circmodelcoarse} is the fundamental block of the \ac{VGM}. The next sections will be devoted to the use of the dilute approximation, where the vortex distribution $\sigma(x)$ can be conducted analytically. We also discuss the modeling of intermittent fluctuation from a phenomenological adapted perspective of the \ac{GMC} theory discussed in Chapter~\refcap{cap2}.

\section{Dilute Approximation}

$_{}$

Visualizations of the vortex structures in Fig.~\reffig{fig4.1}, for instance, explicitly show the appearance of voids and clusters as already pointed out. The existence of complex long-ranged interactions among vortices is the general source of the complex and multifractal spatial behavior of such structures \cite{she1990intermittent,farge2001coherent,ishihara2009study}. However, the volume occupied by vortex tubes is usually a very small fraction of the total fluid volume, especially when the Reynolds number is very high. If the mean intervortex distance of the detected structures is much larger than the proper size of the structures, the dilute approximation becomes a reasonable assumption. Let us suppose the vortex density to be $\sigma(x)=\bar \sigma + \phi(x)$, where $\bar \sigma$ is the mean density and $\phi(x)$ is the fluctuating density, which is statistically homogeneous as a first approximation. The latter assumptions can be translated mathematically by the imposition that $\phi(x)$ obey Poissonian distribution i.e., up to the fourth-order,
\be\label{dens}
\begin{cases}
    \Big\langle \phi(x_1) \phi(x_2) \Big\rangle &= \bar \sigma \delta_{[12]},\\
    \Big\langle \phi(x_1) \phi(x_2) \phi(x_3) \Big\rangle &= \bar \sigma \delta_{[12]} \delta_{[13]},\\
    \Big\langle \phi(x_1) \phi(x_2) \phi(x_3) \phi(x_4) \Big\rangle &= \bar \sigma \delta_{[12]}\delta_{[13]}\delta_{[14]}+ \bar \sigma^2 ( \delta_{[12]} \delta_ {[34]} + \delta_{[13]} \delta_ {[24]} + \delta_{[14]} \delta_ {[23]} )\ ,\
\end{cases}
\ee
where $\delta_{[ij]} \equiv \delta^2(x_i - x_j)$ was used to ease notation.

For the statistical modeling of the $\xidr$ field, we use the approach of Section~\refsec{sec2.4.4} on its unbounded version, where the dissipation field is described by a log-normal distribution. Unbounded \ac{GMC} field formulation is fully aligned with the inertial range phenomenology of circulation discussed in Chapter~\refcap{cap3}, as the dilute approximation is confined to low-order moments, where the circulation scaling exponents $\lambda_p$ are not yet linearized.

From the modeling perspective, the \ac{GMC} approach needs to be adapted to the circulation problem. This adaptation is necessary due to the existence of well-established phenomenological results for this problem. For instance, Gaussianity is observed at very large scales, and a scale-independent kurtosis is observed at very small scales. Motivating a crossover between two scale dependencies on the problem.

Additionally, in the original paper \cite{apolinario2020vortex}, this crossover was not problematic because the cascade modeling was inspired by a \ac{MCM} of the \ac{OK62} model which is very simple to adapt. However, the difference between the mean-field and the full \ac{GMC} approach will be explored in greater detail in Section~\refsec{sec4.2} through the light of multifractal ideas. For now, one can assume the scaling exponents of $\xidr$ to be as follows,
\be\label{paramb}
\mean{\xidr^p}=\xi^p_0\lr{\frac{R_\lambda}{\sqrt{15}}}^{\mu p(3p-1)/16}\lr{\frac{b R}{\eta_K}+1}^{\mu p(1-p)/8} \ ,\
\ee
where $\xi_0$ and $b$ are modeling parameters to be fixed\footnote{In principle, we should consider a set of parameters $b_p$'s because it should be related to the geometrical factors $c_p$ in Eq.~\eqref{momentsgmc}. However, for our purposes, only $b_2$ and $b_4$ will be important, and, as we will see later, the model is well fitted for $b_2=b_4=b=2.0$.}. The explicit Reynolds dependency is directly related to the original \ac{GMC} formulation (Eq.~\eqref{momentsgmc}) by writing $L=Re^{3/4}\eta_K$ and $Re=R_\lambda^2/15$. The only difference between Eq.~\eqref{momentsgmc} and Eq.~\eqref{paramb} is the introduction of a modeling parameter $b$. At small scales, $b$ combines with $\xi_0$ to control the asymptotic behavior of circulation which is related to the vorticity scaling. At sufficiently large scales, $b$ controls the length scale where $\xidr$ has a Gaussian flatness.

The dilute gas approximation in addition to the use of circular contours of radius $R$ turns out to be analytically tractable in the asymptotic limits $R\ll\eta_K$ and $R\gg\eta_K$ \cite{apolinario2020vortex}. The main idea here is to relate the open parameters $a$, $b$, $\xi_0$, and $\bar \sigma$ to the only turbulence control parameter $R_\lambda$. The latter is done in terms of the phenomenological constraints discussed in Chapter~\refcap{cap3}.

\subsection{Small Contour Asymptotics}\label{smallcontour}

$\ \ \ \ $Statistical moments of circulation can be directly calculated using Eqs.~\eqref{circmodelcoarse}, \eqref{dens} and, \eqref{paramb}. For the variance, for instance, we write
\be\label{second}
\mean{\Gamma_R^2}=\frac{\vare_0}{3\nu}\mean{\xi^2_{\mathcal{D}_R}}
\int_{\mathcal{D}_R}d^2x_1d^2x_2\int_\Psi d^2x'_1d^2x'_2 \  g_\eta^{1,1'}g_\eta^{2,2'}
\Big\langle\sigma(x'_1)\sigma(x'_2)\Big\rangle
\mean{\tg(x'_1)\tg(x'_2)}\ , \
\ee
where $g_\eta^{i,j'}$ is a shorthand notation for $g_\eta(x_i-x'_j)$. The Eqs.~\eqref{gtilcorr}, \eqref{dens} and, \eqref{paramb} can be used to calculate circulation's variance explicitly. The rationale behind the asymptotic limit $R\ll\eta_K$ is to Taylor-expand $g_\eta^{i,j'}$ as powers of $x_i$ and explicitly integrate on the primed variable. A technical warning must be done at this point: to easily perform the plane integration using the Gaussian integration theorem, one may substitute $\mean{\tg (x) \tg (0)}\approx g_{\delta\eta_K}(x)$ for $|x|\ll\eta_K$, representing a Gaussian pack with a width of $\delta\eta_K$ and 
 $\delta^2=2/\alpha(1+\alpha)$. By considering only first-order terms in $\bar\sigma$ and the second nontrivial order in $R/\eta_K$, we get using Eq.~\eqref{second}
\be\label{varapp}
\mean{\Gamma^2_R}=\mean{\xidr^2}\frac{\vare_0}{3\nu} N_\eta
\lr{\pi R^2}^2
\lr{1-\frac{1}{4a^2}\frac{R^2}{\eta^2_K}}
+\mathcal{O}\lr{\frac{R}{\eta_K}}^{8} \ ,\
\ee
where $N_\eta=\bar\sigma \pi a^2 \eta_K^2$ is the mean number vortices in a disk of radius $\eta=a\eta_K$. The above result allows one to relate the parameter $\xi_0$ to the Reynolds number by the assumption of the dissipation anomaly. In the limit of very small contours, $R\ll\eta_K$, the circulation variance is directly related to the vorticity variance by
\be
\lim_{R\rightarrow 0}\frac{\mean{\Gamma_R^2}}{(\pi R^2)^2}=\mean{|\omega|^2}=\frac{\vare_0}{3\nu}\Longrightarrow
\xi^2_0=\frac{1}{N_\eta}\lr{\frac{R_\lambda}{\sqrt{15}}}^{-5\mu/8}\ .\
\ee

For the calculation of the fourth-order moment, one can follow a similar rationale, the only difference is that the calculations are more involved since we must deal with four-point correlation functions. Using again Eqs.~\eqref{circmodelcoarse}, \eqref{dens} and, \eqref{paramb} we get,
\begin{align}\label{fourth}
\mean{\Gamma_R^4}=&\mean{\xidr^4}
\frac{\vare_0^2}{9\nu^2}
\int_{\mathcal{D}_R}\lr{\prod^4_{i=1} d^2x_i}
\int_{\Psi}\lr{\prod^4_{j=1} d^2x'_j}
g_\eta^{1,1'}g_\eta^{2,2'}g_\eta^{3,3'}g_\eta^{4,4'}\times \nonumber \\
\times&\Big\langle\sigma(x'_1)\sigma(x'_2)\sigma(x'_3)\sigma(x'_4)\Big\rangle
\mean{\tg(x'_1)\tg(x'_2)\tg(x'_3)\tg(x'_4) }\ , \
\end{align}
using Eqs.~\eqref{dens}, it is not difficult to show that, up to the second order in the mean density,
\be\label{fourth-dens}
\Big\langle\sigma(x_1)\sigma(x_2)\sigma(x_3)\sigma(x_4)\Big\rangle=
\bar{\sigma}\delta_{12}\delta_{13}\delta_{14}+
4\bar{\sigma}^2\delta_{12}\delta_{13}+
3\bar{\sigma}^2\delta_{12}\delta_{34}\ . \
\ee
Eq.~\eqref{fourth-dens} in combination with Eq.~\eqref{fourth} reduces the four-point correlation functions to a combination of products of two-point correlations which, by the use of Gaussian factorization property, are written as follows
\be\label{wicks}
\begin{cases}
    \mean{\tg^4(x)}=3\mean{\tg^2(x)}=3\ , \\
    \mean{\tg^3(x_1)\tg(x_2)}=3\mean{\tg^2(x_1)}\mean{\tg(x_1)\tg(x_2)}\approx 3g_{\delta\eta_K}^{1,2}\ , \\
    \mean{\tg^2(x_1)\tg^2(x_2)}=\mean{\tg^2(x_1)}\mean{\tg^2(x_2)}+2\mean{\tg(x_1)\tg(x_2)}^2\approx 1+2g_{\delta\eta_K/\sqrt{2}}^{1,2}\ .\
\end{cases}
\ee
We can translate the calculation of the fourth-order circulation moment into the computation of 4 integrals,
\begin{align}
    I_1^<&=\int_{\mathcal{D}_R}\lr{\prod^4_{i=1} d^2x_i}\int_{\Psi}d^2x'_1\
    g_\eta^{1,1'}g_\eta^{2,1'}g_\eta^{3,1'}g_\eta^{4,1'}\ , \\
    I_2^<&=\int_{\mathcal{D}_R}\lr{\prod^4_{i=1} d^2x_i}\int_{\Psi}d^2x'_1d^2x'_2\
    g_\eta^{1,1'}g_\eta^{2,1'}g_\eta^{3,2'}g_\eta^{4,2'}\ , \\
    I_3^<&=\int_{\mathcal{D}_R}\lr{\prod^4_{i=1} d^2x_i}\int_{\Psi}d^2x'_1d^2x'_2\
    g_\eta^{1,1'}g_\eta^{2,1'}g_\eta^{3,2'}g_\eta^{4,2'}g_{\delta\eta_K/\sqrt{2}}^{1',2'}\ , \\
    I_4^<&=\int_{\mathcal{D}_R}\lr{\prod^4_{i=1} d^2x_i}\int_{\Psi}d^2x'_1d^2x'_2\
    g_\eta^{1,1'}g_\eta^{2,1'}g_\eta^{3,1'}g_\eta^{4,2'}g_{\delta\eta_K}^{1',2'}\ ,\
\end{align}
such that,
\be\label{19}
\langle \Gamma_R^4\rangle=
\frac{\vare_0^2}{9\nu^2}\mean{\xi^4_{\mathcal{D}_R}}3\bar\sigma\Big(
I_1^<+\bar{\sigma} \lr{I_2^<+2 I_3^<+4 I_4^<}\Big)\ .\
\ee
In order to be consistent with the truncation of Eq.~\eqref{varapp} up to $\mathcal{O}(\bar\sigma^1)$, we simply compute $I_1^<$. In this case, the fourth-order circulation moment is written as
\be\label{fourapp}
\mean{\Gamma^4_R}=\mean{\xidr^4}\lr{\frac{\vare_0}{3\nu}}^2 \frac{3 N_\eta}{2}\lr{\pi R^2}^4 \lr{1-\frac{3}{4a^2}\frac{R^2}{\eta^2_K}}
+\mathcal{O}\lr{\frac{R}{\eta_K}}^{12} \ .\
\ee
Note that, however, the substitution of the ${}_2F_1$ function to a Gaussian curve $g_{\delta\eta_K}$ does not affect the first-order density approximation. The circulation kurtosis is calculated from Eqs.~\eqref{varapp} and \eqref{fourapp} as
\be\label{kurtapp}
\frac{\langle \Gamma_R^4\rangle}{\langle \Gamma_R^2\rangle^2}=
\frac{\mean{\xi^4_{\mathcal{D}_R}}}{\mean{\xi^2_{\mathcal{D}_R}}^2}
\frac{3}{2 N_\eta}\frac{1-3R^2/4a^2\eta^2_K}{\Big(1-R^2/4a^2\eta^2_K\Big)^2}\approx
\frac{R_{\lambda}^{3\mu/2}}{15^{3\mu/4}}
\frac{3}{2 N_\eta}\lr{1-\frac{1}{4a^2}\frac{R^2}{\eta^2_K}}+\mathcal{O}\lr{\frac{R}{\eta_K}}^{4}\ .\
\ee
Two major properties of Eq.~\eqref{kurtapp} must be emphasized. Firstly, there is an asymptotic limit where the kurtosis depends solely on the Reynolds number and the mean number of vortices $N_\eta$. Secondly, the asymptotic curvature of the kurtosis in the limit $R\rightarrow0$ is controlled by the typical size of the Gaussian packets $\eta=a\eta_K$. As a consequence of those observations, it is possible to recall \ac{DNS} phenomenological result to fix $a$ and $\bar\sigma$ as functions of the Reynolds number. Using the asymptotic result discussed in Section~\refsec{3.3} ,$F(R\rightarrow0)\simeq C_4 R_\lambda^{\beta_4}$, and Eq.~\eqref{kurtapp} we find,
\be\label{4.24}
N_\eta=\frac{3}{2C_4}\frac{R_\lambda^{3\mu/2-\beta_4}}{ 15^{3\mu/4}}\approx0.91\times R_\lambda^{-0.15} \ . \
\ee
The larger the Reynolds number, the better the dilute approximation. For the smallest Reynolds number analyzed ($R_\lambda=240$), Eq.~\eqref{4.24} gives $\bar\sigma\pi\eta^2\simeq0.39$. The mean intervortex distance can be estimated by $\bar{\sigma}^{-1/2}\simeq 2.8 \eta$. When the modeling assumption to fix the shape of the vortex structure as a Gaussian curve is made, $95\%$ of the vortex intensity is concentrated in a disk of radius $2\eta$ around the vortex center point. In this sense, the overlapping area between two disks of radius $2\eta$ separated by a distance $2.8\eta$ is $\approx 19\%$ of the disk area, supporting the validity of the dilute approximation. 

The left panel of Fig.~\reffig{fig4.2} shows the comparison between Eq.\eqref{kurtapp} and the \ac{DNS} data from \cite{iyer2019circulation}. The parameter $a=3.3$ is found by the minimization of the $L^2$-norm between the model and data for the points $R\leq 2\eta_K$. Slightly different values of $a$ and $\bar\sigma$ are still visually consistent with the data presented in Fig.~\reffig{fig4.2}. However, this ambiguity is included in the error bars of the fit $F(R\rightarrow0)=C_4 R_\lambda^{\beta_4}$ which is made by taking only three data points. The large contour asymptotic is also shown in Fig.~\reffig{fig4.2} and the derivation of this result follows in the next subsection.

\subsection{Large Contour Asymptotics and Inertial Range Moments}

$\ \ \ \ $In the case of a large contour radius, $R\gg\eta_K$ one may use the very same expressions~\eqref{second} and \eqref{fourth} for the second and fourth-order circulation moments, respectively. The only difference is the limiting behavior of the Gaussian-shaped packet $g_\eta(x)\approx (2\pi\eta^2)\delta(x)$. This is possible because the relative dimension $\eta/R$ is so small that the packets can be seen as small point vortices. Combining $g_\eta(x)\approx (2\pi\eta^2)\delta(x)$ with Eq.~\eqref{second}, we get
\be
\mean{\Gamma_R^2}=\frac{\vare_0}{3\nu}\mean{\xi^2_{\mathcal{D}_R}}\lr{2\pi\eta^2}^2
\int_{\mathcal{D}_R}d^2xd^2y\Big\langle\sigma(x)\sigma(y)\Big\rangle
\mean{\tg(x)\tg(y)}\ . \
\ee
In this case, the hypergeometric form of the \ac{GFF} correlation will play an important role in the calculations from now on, as the contour radius lies in the region of the algebraic decay $\tg(x)$. Reminding the reader that $\sigma(x)=\bar\sigma+\phi(x)$, where the statistics of $\phi(x)$ is given by Eq.~\eqref{dens}, it is easy to show that
\be\label{second-large}
\mean{\Gamma_R^2}=\frac{\vare_0}{3\nu}\mean{\xi^2_{\mathcal{D}_R}}\lr{2\pi\eta^2}^2
N_R\lr{1+N_R\int_{\mathcal{D}_R}\frac{d^2x}{\lr{\pi R^2}}\frac{d^2y}{\lr{\pi R^2}}\mean{\tg(x)\tg(y)}}\ , \
\ee
where $N_R=\bar\sigma \pi R^2$ is the mean number of structures inside a disk of radius $R$. For the moment, let us keep the integral untouched and turn our attention to the fourth-order moment.
\begin{figure}[t]
\center
\includegraphics[width=\textwidth]{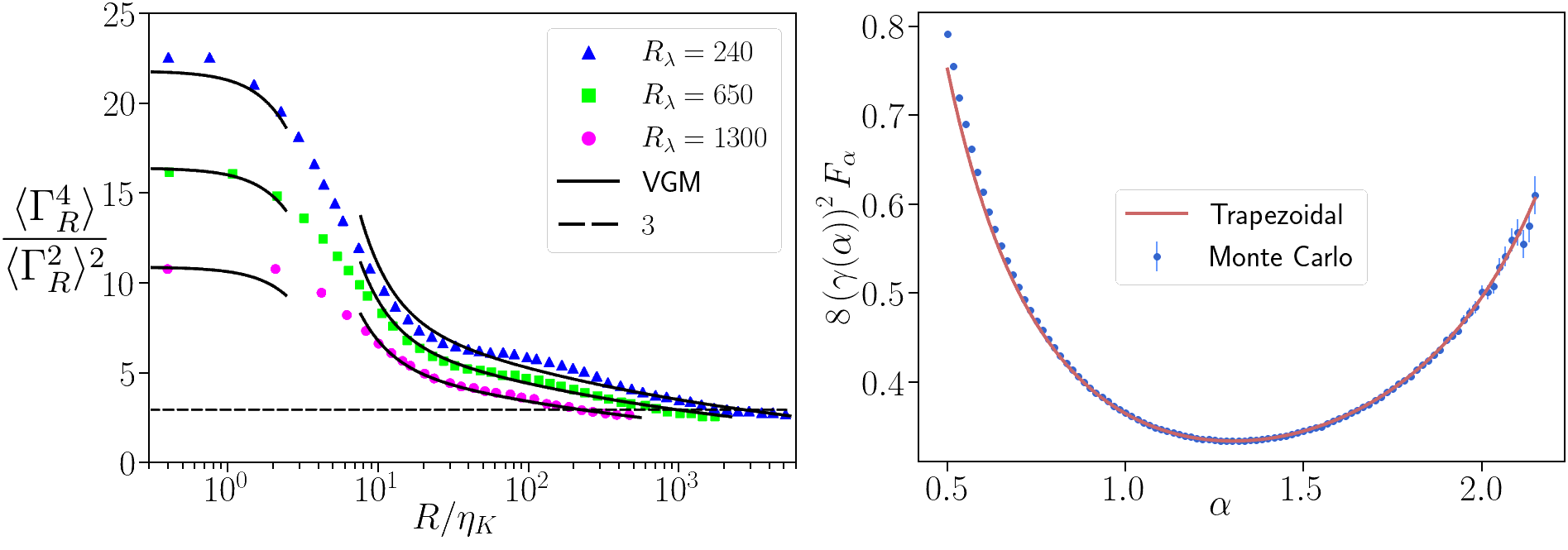}
\caption[Circulation kurtosis and numerical integration of $F_\alpha$.]{Left panel shows the comparison between the circulation kurtosis of the present model with $a=3.3$ and $b=2.0$, and \ac{DNS} data from \cite{iyer2019circulation} for different Reynolds numbers, reproduced from \cite{apolinario2020vortex} with minor modifications. The right panel shows the numerical integration of $F_\alpha$ for a set of values of $\alpha$ from two different numerical methods.}
\label{fig4.2}
\end{figure}

In addition to the expression in Eq.~\eqref{fourth-dens} we have two terms of higher order in the density which are $6\bar\sigma^3 \delta_{12}$ and $\bar\sigma^4$, respectively. In the case of large contours, higher-order terms can play a role since the contour may, in general, enclose a large number of vortex structures. The general expression for the fourth-order moment can be compressed in the following equation,
\be\label{fourth-large}
\mean{\Gamma_R^4}=\lr{\frac{\vare_0}{3\nu}}^2\mean{\xidr^4}\lr{2\pi\eta^2}^4
3N_R^2
\Bigg(1+N_R^{-1}+
2I_2^>+4I_1^>+
2N_R \Big(I_1^>+2I_3^>\Big)+
N_R^2\lr{I_1^>}^2\Bigg)
\ , \
\ee
where we define
\begin{align}
    I_1^>&=\int_{\mathcal{D}_R}\frac{d^2x}{\lr{\pi R^2}}\frac{d^2y}{\lr{\pi R^2}}
    \mean{\tg(x)\tg(y)}\ \label{I1}, \\
    I_2^>&=\int_{\mathcal{D}_R}\frac{d^2x}{\lr{\pi R^2}}\frac{d^2y}{\lr{\pi R^2}}
    \lr{\mean{\tg(x)\tg(y)}}^2\ , \\
    I_3^>&=\int_{\mathcal{D}_R}\frac{d^2x}{\lr{\pi R^2}}\frac{d^2y}{\lr{\pi R^2}}\frac{d^2z}{\lr{\pi R^2}}
    \mean{\tg(x)\tg(y)}\mean{\tg(x)\tg(z)}\label{I3}\ . \
\end{align}
One may note that using Eq.~\eqref{second-large} and Eq.~\eqref{I1} we get $\mean{\tg^2_R}\propto(1+N_RI_1^>)$. Each integral has different $R$-scaling properties that must be computed separately. The Fourier representation of the \ac{GFF} correlation function is very useful in this case,
\be\label{fouriercorr}
\mean{\tg(x)\tg(y)}=\frac{\eta_K^\alpha}{2\pi \Gamma(\alpha)}\int d^2k |k|^{\alpha-2}e^{ik \cdot(x -y)-|k|\eta_K}\ ,\
\ee
where $\Gamma(\alpha)$ is the Gamma function. Combining Eq.~\eqref{I1} and \eqref{fouriercorr}, one is able to integrate Eq.~\eqref{fouriercorr} writing the space integrals in polar coordinates, and using hypergeometric recursive relations to get,
\be
I_1^>={}_3F_2\Bigg(\frac{\frac{3}{2},\frac{1+\alpha}{2},\frac{\alpha}{2}}{2,3};-4\frac{R^2}{\eta_K^2}\Bigg)\ . \
\ee
which is particularly interesting in the limit of large contours, by the use of a Pfaff-like transformation\footnote{More about this transformation can be found at \url{https://functions.wolfram.com/HypergeometricFunctions/Hypergeometric3F2/17/02/04/0001/}.}
\be
I_1^>=
E_\alpha\lr{\frac{R}{\eta_K}}^{-\alpha}+
\mathcal{O}\lr{\frac{R}{\eta_K}}^{-1-\alpha}\ ,\
\ee
where subdominant contributions were neglected. The prefactor $E_\alpha$ can be written in terms of a combination of Gamma functions,
\be
E_\alpha=2^{2-\alpha}\frac{\Gamma(\frac{3-\alpha}{2})}
{\Gamma(\frac{4-\alpha}{2})
\Gamma(\frac{6-\alpha}{2})
\Gamma(\frac{1+\alpha}{2})}\ . \
\ee
The second term, $I_2^>$, can be calculated using the same strategy of integrating in polar coordinates, the integration over the angles is made explicit to get
\be\label{i2inter}
I_2^>=\frac{\eta_K^{2\alpha}}{\lr{\Gamma(\alpha)}^2}\int_0^R\frac{dr dr'}{\lr{\pi R^2}^2}\int d^2kd^2k'rr' \lr{|k||k'|}^{\alpha-2} J_0(|k+k'|r)J_0(|k+k'|r')e^{-\eta_K(|k|+|k'|)}\ , \
\ee
where $J_0(x)$ is the Bessel function of the first kind. The integration over $dr$ and $dr'$ are easily done in terms of Bessel functions' identities. It is now convenient to change variables to $k_\pm=(k\pm k')/R \sqrt{2}$, expressing Eq.~\eqref{i2inter} as follows,
\begin{align}\label{i2inter2}
I_2^>=\frac{2^{2-\beta}}{\pi^2 \lr{\Gamma(\alpha)}^2}&\lr{\frac{R}{\eta_K}}^{-2\alpha}
\int d^2k_+\lr{\frac{J_1(k_+)}{k_+}}^2 
\int d^2k_- \lr{|k_++k_-||k_+-k_-|}^{\alpha-2}\times \nonumber \\
&\times\exp\lr{-\frac{\eta_K}{R}\frac{(|k_++k_-|+|k_+-k_-|)}{\sqrt{2}}}\ , \
\end{align}
the above integral is difficult to solve without any approximation. Having in mind the asymptotic limit $R\gg\eta_K$, we argue that the most relevant contribution for the large-scale behavior happens when $k\approx k'$\footnote{When analyzing Eq.~\eqref{i2inter} we note that $r$ and $r'$ are bounded by the contour radius, such that the contributions for the integral when $|k-k'|\gg1/R$ are rapidly oscillating producing pointwise cancelations to $(|k ||k'|)^{\alpha-2}\exp(-\eta_K(|k|+|k'|)$.}, in this case, $|k_-|\gg|k_+|$ and the integration in Eq.~\eqref{i2inter2} simplifies to
\be
I_2^>\approx
G_\alpha\lr{\frac{R}{\eta_K}}^{-2}\ ,\
\ee
with
\be
G_\alpha=2^{4-2\alpha}\frac{\Gamma(2\alpha-2)}{\lr{\Gamma(\alpha)}^2}\ . \
\ee
Finally, the third contribution can be calculated with the very same script: Writing the spatial integrations in polar coordinates, using Bessel function integrations, and taking asymptotic limits which, in this case, is done by approximating $e^{-k\eta_K/R}\approx1$. After a long string of calculations,
\be
I_3^>\approx
F_\alpha\lr{\frac{R}{\eta_K}}^{-2\alpha}\ ,\
\ee
with,
\be\label{abovintegral}
F_\alpha=\frac{2^3}{\lr{\Gamma(\alpha)}^2}\int_0^1 dx \ x\lr{\int_0^\infty dy \ y^{\alpha-2}J_1(y)J_0(xy)}^2\ . \
\ee
Eq.~\eqref{abovintegral} cannot be expressed in terms of standard functions, therefore, we resorted to numerical methods for its evaluation. In the right panel of Fig.~\reffig{fig4.2} we show the results of the numerical integration of Eq.\eqref{abovintegral} by Monte Carlo methods and standard trapezoidal sum. In the former, we performed a change of variables $t=y/(1-y)$ to compactify the $\mathbb{R}^+$ set to the unit, while in the latter we simply truncated the interval to $[0,100]$. For the sake of completeness, the reference value for $\alpha=4/3$ is $F_{4/3}\approx0.33\times 2^{-3}/(\Gamma(4/3))^{-2}$ while $\alpha=2$ can be analytically integrated to $F_2\approx 2^{-4}/(\Gamma(4/3))^{-2}$.

With all the ingredients in hand, the circulation kurtosis can be directly calculated using Eqs.~\eqref{fourth-large} and \eqref{second-large},
\be\label{k4largeR}
\frac{\mean{\Gamma_R^4}}{\mean{\Gamma_R^2}^2} = 
3\frac{\mean{\xi^4_{\mathcal{D}_R}}}{\mean{\xi^2_{\mathcal{D}_R}}^2}
\lr{1+
\frac{(N^{-1}_K+2G_\alpha)\lr{\frac{R}{\eta_K}}^{-2}+
4E_\alpha\lr{\frac{R}{\eta_K}}^{-\alpha}+
2N_KF_\alpha\lr{\frac{R}{\eta_K}}^{-2\alpha}}{\lr{1+N_K E_\alpha\lr{\frac{R}{\eta_K}}^{2-\alpha}}^2}
}\ , \
\ee
which models the asymptotic curvature near the Gaussian value $F(R\gg \eta_K)\approx3$. A closer inspection of Eq.~\eqref{k4largeR} shows that for a large very contour radius (comparable to the integral length), circulation kurtosis is only dependent on the $\xidr$-field kurtosis, which brings the Reynolds number dependency of the model. Tracing back $\xidr$'s moments at Eq.~\eqref{paramb} we can reinforce the role of the parameter $b$ at the large scale expansion to be addressed as follows: When $b$ combines to $\xi_0$ in the large scale limit, it produces a Reynolds dependent scale $R_0(R_\lambda)$ such that, its the vicinity $R_0\pm\delta R$, the fluctuations of $\xidr$ can be regarded as Gaussians. Above $R_0$, our model does not prevent the flatness to be lesser than $3$. This is a modeling artifact since we did not impose any infrared constraint at the fluctuations of $\xidr$. The resulting asymptotics is shown in Fig.~\reffig{fig4.2} for $b=2.0$ which was fixed by the minimization of an $L^2$-norm between data and model.

In the case of general circulation moment, it is, in principle, possible to calculate all the contributions in the large-scale asymptotics, using the same strategy employed for the second and fourth-order moments. However, the phenomenology explored in Chapter~\refcap{cap3} of the linearization of the circulation scaling exponents is particularly difficult to address within the present model. This can be clarified by calculating the dominant contribution to the $p^{th}$-order moment scaling exponent in this large-scale limit. The latter can be done easily by arguing that, at inertial range scales, small-scale particularities of $\sigma(x)$ are irrelevant, and, effectively,
\be\label{modelnew}
\Gamma_R\sim\xidr\int_{{\mathcal{D}_R}}d^2x
\tg (x)\ ,\ 
\ee
such that the scaling exponent of the $p^{th}$ order moment is dominated by $\mean{|\Gamma_R|^p}\sim R^{\lambda_{|p|}}$, where:
\be \label{lambdap}
\lambda_{|p|} = (4-\alpha)\frac{p}{2} + \frac{\mu}{8} p (1-p) \ .\ 
\ee
Eq.~\eqref{lambdap} is compatible with the numerical results discussed in Chapter~\refcap{cap3} for low order moment if $\alpha=4/3-\mu/2$, however, it still missing a fundamental mechanism of linearization for $p\gtrsim6$. 

Another intriguing question arises when we analyze the validity ranges of the dilute approximation through Fig.~\reffig{fig4.2}. One may note that, in opposition to the small contour asymptotics, the large contour behavior remains valid for a range of scales that includes a significant part of the inertial range (it is valid $\approx [30\eta_K,300\eta_K]$). This supports the usage of the large-scale expansion to the derivation of Eq.~\eqref{lambdap}, however, there is no a priori reason for this to happen. A naive explanation for this observation is that we kept almost all terms for $R\gg\eta_K$ and only first-order terms in the small contour expansion. Thus, we have a clear motivation to explore higher-order density corrections in the small contour limit.

\subsection{Density Corrections and Model Validation}

$\ \ \ \ $To further investigate the validity of the \ac{VGM} at small scales, we performed a Monte Carlo simulation based on Eq.~\eqref{circmodelcoarse} using the prescriptions given by Eqs.~\eqref{gtil}, \eqref{dens} and, \eqref{paramb}\footnote{We did not employ a simulation of a \ac{GMC} process at the time of the original paper \cite{moriconi2022circulation}. Instead, we used its mean field version, which produces different scaling exponents (see Section~\refsec{sec2.4.4}). However, for low order moments, they are in good agreement with \ac{DNS} data as discussed in Section~\refsec{cap3}.}. The grid resolution used in Monte Carlo simulations was comparable to the ones of \ac{DNS} data. Square contours of area $A=r^2$ were probed since the computations match exactly the grid points. 
\begin{figure}[t]
\center
\includegraphics[width=\textwidth]{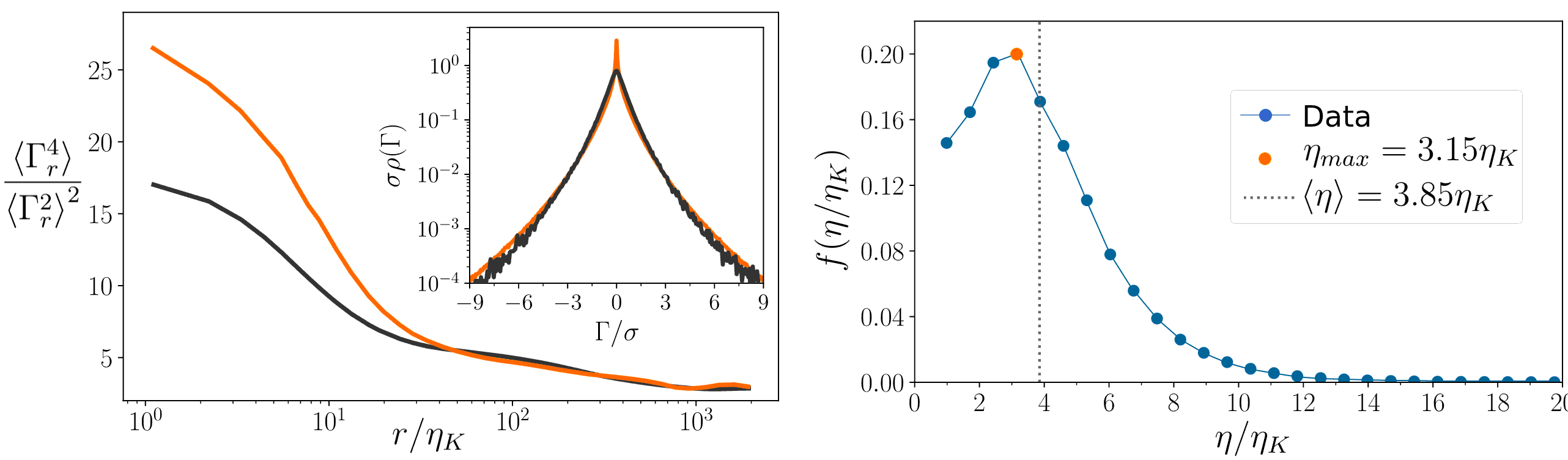}
\caption[Comparison between Monte Carlo and DNS and vortex size distribution.]{The main frame of the left panel shows the comparison between the kurtosis of Monte Carlo simulations of Eq.~\eqref{circmodelcoarse} (orange) and \ac{DNS} data (black) from dataset Sqr-14. Inset shows standardized \ac{cPDF} for $r=2.2\eta_K$. The right panel shows the distribution of vortex radius, its mean value, and its most frequent value. Both figures were reproduced from \cite{moriconi2022statistics} with minor modifications.}
\label{fig4.3}
\end{figure}

Figure~\reffig{fig4.3} (left panel) shows an unexpected mismatch between Monte Carlo simulation and \ac{DNS} data at small scales. While the model closely follows the \ac{DNS} curve for large contours (down to $r\approx 40\eta_K$), its circulation fluctuations become too intermittent for smaller contours. The resulting kurtosis is larger than the ones of \ac{DNS}, being translated into sharply peaked \ac{cPDF}s. A first attempt to cope with this problem from the modeling point of view, we calculated density corrections at small contours. Now, up to the second order in the mean density, the strategy of Section~\refsec{smallcontour} lead us to
\be\label{varapp-sec}
\mean{\Gamma^2_R}=\mean{\xidr^2}\frac{\vare_0}{3\nu} N_\eta
\lr{\pi R^2}^2
\lr{f_0-f_1\frac{R^2}{\eta^2_K}}
+\mathcal{O}\lr{\frac{R}{\eta_K}}^{8} \ ,\
\ee
where,
\be
f_0=
1+4N_\eta\lr{
\frac{\delta^2}{\delta^2+2a^2}}+\mathcal{O}\lr{N_\eta}^{2}\ .\
\ee
and,
\be
f_1=\frac{1}{4 a^2}+2N_\eta
\lr{\frac{\delta^2}{\lr{\delta^2+2a^2}^2}}+\mathcal{O}\lr{N_\eta}^{2} \ . \
\ee
The combinations of $\delta$ and $a$ arise from successive Gaussian convolutions that are only present at higher-order corrections due to the approximation $\mean{\tg (0)\tg(x)}\approx g_{\delta\eta_K})$. Note that, corrections up to the order $N_\eta^p$ are present in the $p^{th}$-order moment, at all orders of $R/\eta_K$. In this sense, it differs from the large contour expansion where density is not present in some of the coefficients so that the above expressions can be regarded as a virial-like expansion of the statistical moments of circulations, which are supposed to be small if the vortex gas is sufficiently dilute.

Similar calculations are carried out for the fourth-order moment to get, after a sequence of Gaussian convolutions,
\be\label{fourapp-sec}
\mean{\Gamma^4_R}=\mean{\xidr^4}\lr{\frac{\vare_0}{3\nu}}^2 \frac{3 N_\eta}{2}\lr{\pi R^2}^4 \lr{f_2-f_3\frac{R^2}{\eta^2_K}}
+\mathcal{O}\lr{\frac{R}{\eta_K}}^{12} \ ,\
\ee
where,
\be
f_2=1+2N_\eta\lr{
1+\frac{16\delta^2}{4a^2+3\delta^2}+\frac{2\delta^2}{2a^2+\delta^2}} +\mathcal{O}\lr{N_\eta}^{2} \ ,\
\ee
and,
\be
f_3=\frac{3}{4 a^2}+\frac{N_\eta}{a^2}\lr{
1+48\delta^2
\frac{\delta^2+2a^2}{(4a^2+3\delta^2)^2}+
4\delta^2\frac{\delta^2+3a^2}{(2a^2+\delta^2)^2}}
+\mathcal{O}\lr{N_\eta}^{2} \ .\
\ee
Then, the circulation kurtosis, up to the second order on density correction is
\be\label{kurtapp-sec}
\frac{\langle \Gamma_R^4\rangle}{\langle \Gamma_R^2\rangle^2}\approx
\frac{\mean{\xi^4_{\mathcal{D}_R}}}{\mean{\xi^2_{\mathcal{D}_R}}^2}
\frac{3f_0^{-2}f_2}{2 N_\eta}\lr{
1-\left(\frac{f_3}{f_2}-2\frac{f_1}{f_0}\right)\left(\frac{R}{\eta_K}\right)^2}
+\mathcal{O}\lr{\frac{R}{\eta_K}}^{4}\ .\
\ee
The procedure of parameter fixing is exactly the same as in Sec.~\eqref{smallcontour}. Firstly, we fix $\xi_0$ by constraining the limiting behavior of the circulation's variance using the energy dissipation anomaly to get, $\xi^{-2}_0=N_\eta f_0$. Secondly, we examine the asymptotic numerical power law $F(R)\approx C_4 R_{\lambda}^{\beta_4}$ to fix the mean density as a function of the Reynolds number using Eq.~\eqref{kurtapp-sec}. The latter leads to the following relation,
\be\label{eqseila}
\frac{f_2}{f_0^{2}N_\eta}=\frac{2C_4}{3}15^{3\mu/4} R_{\lambda}^{\beta_4-3\mu/2}\equiv F_0 \ , \
\ee
where $F_0=1.10\times R_\lambda^{0.15}$ using the values of $\beta_4$, $C_4$ and $\mu=0.17$ \cite{tang2020scaling}. As $f_0$ and $f_2$ are linearized functions of $N_\eta$ we denote then by to ease notation $f_{0,2}=1+c^{(0,2)}N_\eta$. Eq.~\eqref{eqseila} is rearranged to show that $N_\eta$ are the roots of a third-order polynomial,
\be\label{poly}
F_0(c^{(0)})^2N_\eta^3+2F_0c^{(0)}N_\eta^2+(F_0-c^{(2)})N_\eta-1=0 \ .\
\ee
Lastly, the expression $f_3/f_2-2f_1/f_0$ controls the curvature of the parabolic approximation of the flatness in the limit $R\rightarrow0$, which is highly sensitive to the value of the parameter $a$.

Simple numerical computation of the unique positive real solution of Eq.~\eqref{poly} shows a huge increase in $\bar{\sigma}$ compared to the first-order solution. In the best case, we have a multiplicative factor of $4.7$ in the mean density for $R_\lambda=1300$, being worse as the Reynolds decreases. The estimation of the mean intervortex distance is then $\bar{\sigma}^{-1/2}\approx1.57\eta$ for the lowest Reynolds number $(R_\lambda=240)$ signalizing an overlapping area ratio of about $50\%$. These results are incompatible with visualizations of the structures shown in Fig.~\reffig{fig4.1}. 

Moreover, a surprising decrease in the parameter $a$ is observed (from $a=3.3$ to $a\approx2.9$). For the sake of comparison, we show in the right panel of Fig.~\reffig{fig4.3}, the distribution of the estimation of the vortex radius $\eta=\sqrt{A/\pi}$, calculated from the dataset Iso-4096. The observed mean value of vortex radius estimation was $\mean{\eta}=3.85\eta_K$, while the distribution has a peak at $\eta_{max}=3.15\eta_K$. Although the first-order dilute approximation reasonably accounts for the statistical properties of circulation, the higher-order corrections are diverging from the observed numerical result instead of getting closer. This motivates us to revisit the fundamental modeling assumptions of the \ac{VGM}, in order to understand such deviations. 
\begin{figure}[t]
\includegraphics[width=\textwidth]{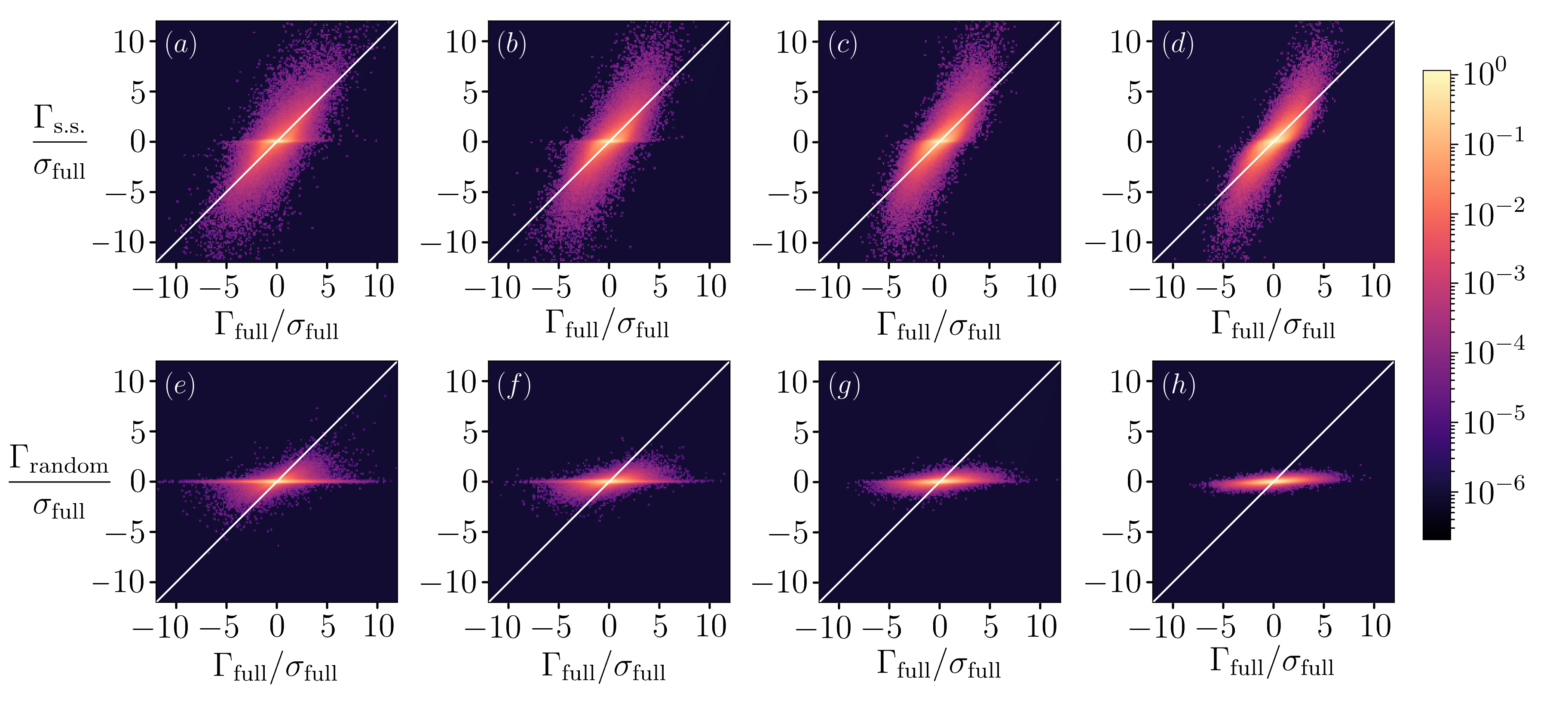}
\caption[Detected and full circulation joint density.]{Joint density plots for the circulation pairs
($\Gamma_\mathrm{full}$, $\Gamma_\mathrm{s.s.}$), panels (a-d), and ($\Gamma_\mathrm{full}$, $\Gamma_\mathrm{random}$), panels (e-h), over square contours positioned on the grid. All the circulations values are normalized by the full circulation's standard deviation $\sigma_\mathrm{full}$. Contours sides correspond, from left to right, to $r/\eta_K=10$, $19$, $37$ and $72$.}
\label{fig4.4}
\end{figure}

The first hypothesis of the \ac{VGM} is that most of the vortex structures account for circulation fluctuations. We test this idea by combining \ac{DNS} data and the model's main definitions as follows. A two-dimensional vorticity field is created, representing a \ac{DNS} slice, by adding Gaussian packets $g_\eta(x - x_i)$ centered on every position $x_i$ where a vortex was detected on the slice through the \ac{SSC} (Appendix~\ref{appssc}). Each Gaussian has the estimated vortex radius $\eta$ as its width and the maximum vorticity inside the vortex as its amplitude. We then compare the circulation around square contours when computed from this swirling strength originated vorticity field ($\Gamma_\mathrm{s.s.}$) with those computed from the full \ac{DNS} field ($\Gamma_\mathrm{full}$).

We have analyzed $192$ 2D slices of the dataset Iso-4096 (see Table~\reftab{tab3.1}) where the slices are homogeneously spaced to ensure decorrelation between the ensembles and the three axes direction are probed. The \ac{SSC} identification method with $\lambda_T=1.5\sigma_{\lambda_{ci}}$ was employed (see Appendix~\ref{appssc}) to detect $N_i$ structures for each slab, which fluctuates around $N_i\sim2.7\times 10^4$. In Figs.~\reffig{fig4.4}(a-d) we show the obtained joint density plots for contours with increasing typical lengths. Vertical axis correspond to $\Gamma_\mathrm{s.s.}$, while horizontal ones corresponds to $\Gamma_\mathrm{full}$. Data from all the slices are plotted together and the white lines $y=x$ serve as a reference of fully correlated circulations. One sees a grouping tendency around the lines, showing that circulation is well captured by the combined and independent contributions of identified vortex structures.

To verify that this strong correlation is not a product of a simple generic sampling of the vorticity field, and hence that the detected structures are indeed the main actors in play, we now build a different vorticity field to serve as a null hypothesis test. First, we pick random homogeneously distributed positions on the \ac{DNS} slices, saving the normal component of the vorticity at each one. This is done for around 27 thousand points per slice, which is roughly the mean number of detected structures on a \ac{DNS} slice. Then, Gaussian functions centered on those positions are superposed, using the measured vorticities as amplitudes and the mean detected vortex radius, $\langle \eta\rangle = 3.85\eta_K$ as widths. Joint density plots of Figs.~\reffig{fig4.4}(e-h) once again compare $\Gamma_\mathrm{full}$ with the circulation computed with this newly constructed field $\Gamma_\mathrm{random}$, for the same contours as before. The lack of correlation concludes that structures created randomly from the vorticity field do not account for the circulation fluctuations. In particular, it is seen that large fluctuations are completely suppressed.

To further quantify these correlations, we compute for each one of the cases reported in Figs.~\ref{fig4.4} the Pearson coefficient,
\be
\rho_{XY} = \frac{\langle \left(X-\langle X \rangle \right) \left(Y- \langle Y \rangle \right) \rangle}{\sigma_X\sigma_Y} \ , \
\ee
where $\sigma_X$ and $\sigma_Y$ are the standard deviations of the random samples $X$ and $Y$, respectively.
Results, summarized in Table~\reftab{tab4.1}, further point to strong correlations between the circulation observed in \ac{DNS} with that measured from the field associated with structures detected with the swirling strength criterion. Correlations are, in contrast, significantly reduced when structures are randomly sampled (Poisson distribution) across the flow.

The above observations corroborate the idea that circulation is accounted for by the thin vortex structures, especially for larger contours. Now, to tackle the incompatibility of the higher-order corrections to the dilute approximation, we will explore more involving formulations of vortex distribution.
\begin{table}[t]
\centering
\caption[Correlations between detected and full circulations]{Pearson correlation coefficients between $\Gamma_\mathrm{full}$ and both $\Gamma_\mathrm{s.s.}$ and $\Gamma_\mathrm{random}$ over contours of varying sides.}\label{tab4.1}
\begin{tabular}{|c|c|c|c|c|} \toprule
    {$(X,Y)$} & {$10\eta_K$} & {$19\eta_K$} & {$37\eta_K$} & {$72\eta_K$} \\ \midrule
    $(\Gamma_\mathrm{full},\Gamma_\mathrm{s.s.})$ & 0.70 & 0.73 & 0.76 & 0.81\\
    $(\Gamma_\mathrm{full},\Gamma_\mathrm{random})$ & 0.32 & 0.39 & 0.45 & 0.52\\ \bottomrule
\end{tabular}
\end{table}

\section{Hard Disk Model and the Multifractality Breakdown}\label{sec4.2}

$_{}$

The essential phenomenological ingredient that explains the overrated intermittency at small scales, and the catastrophic failure of higher-order corrections are exclusion effects. If there is no intersection between vortices, the first-order dilute approximation is naturally dominant in the density expansion. Volume expulsion effects can be accomplished by replacing the homogeneous distribution $\sigma (x)$ for hard disk distributions with excluding radius $\eta=a\eta_K$, such point process is also known as a hard core Gibbs point process \cite{ripley2005spatial}. The generation of a statistical ensembles of mutually excluding disks is a well-discussed topic in the literature of statistical mechanics \cite{metropolis1953equation,alder1962phase,isobe2016hard}, and the development of efficient computational algorithms has advanced in recent years \cite{isobe1999simple,bernard2009event,anderson2013massively,hollmer2020jellyfysh,li2021multithreaded,bernard2011two}. In a statistical sense, the existence of short-distance vortex repulsion should not be very surprising, since small-scale clusters of thin vortex tubes are likely to be polarized \cite{burger2012vortices,treib2012turbulence,polanco2021vortex}.

We benefit, for our statistical analyses, from the publicly available code reported in \cite{bernard2011two,li2021multithreaded} to generate a statistical set of hard disks with maximum entropy. Additionally, we covered the center position of the hard disks with the Gaussian-shaped packet $g_\eta(x)$ and prescribed the fields Eqs.~\eqref{gtil} and \eqref{paramb}. Spanning a wide range of parameters $(a,b,\bar \sigma)$, we were able to find the best ones through independent optimizations between \ac{DNS} and Monte Carlo data. The best-fitted parameters are shown in Table~\reftab{tab4.2} together with their dilute \ac{VGM} counterparts.
\begin{table}[b]
    \caption[Optimal values for the VGM and Hard disk model]{Comparison between modeling parameters of the hard disk and the homogeneous gas.}\label{tab4.2}
    \centering
    \begin{tabular}{ |c||c|c|c|c|c|  }
        \hline
        \multicolumn{6}{|c|}{Hard disk gas}\\
        \hline
        \hline
        $R_\lambda$ & $240$ & $433$ & $610$ & $650$ & $1300$\\
        \hline
        $a$ & $3.3$ & $3.3$ & $3.3$ & $3.3$ & $3.3$ \\
        $b$ & $1.6$ & $1.7$ & $1.7$ & $1.8$ & $1.8$ \\
        $\bar\sigma \pi \eta^2$ & $0.35$ & $0.31$ & $0.29$ & $0.31$ & $0.29$\\
        \hline
    \end{tabular}
    \begin{tabular}{ |c|c|c|c|c|  }
        \hline
        \multicolumn{5}{|c|}{Homogeneous gas}\\
        \hline
        \hline
        $240$ & $433$ & $610$ & $650$ & $1300$\\
        \hline
        $3.3$ & $3.3$ & $3.3$ & $3.3$ & $3.3$ \\
        $2.0$ & $2.0$ & $2.0$ & $2.0$ & $2.0$ \\
        $0.39$ & $0.36$ & $0.34$ & $0.34$ & $0.30$\\
        \hline
    \end{tabular}
\end{table}

Figure~\reffig{fig4.5} (left panel) shows excellent agreement between \ac{DNS} data and the hard disk model for a wide range of Reynolds numbers with very small modifications on the parameters. These results indicate the existence of small-scale interaction effects between structures that may be crucial for the spatial organization of vorticity filaments. We then proceed to analyze \ac{DNS} data in order to find the signatures of these small-scale repulsive interactions by defining the following quantities,
\be\label{m2}
\mathcal{M}_2 = \frac{\mean{\lr{N_r-\mean{N_r}}^2}}{\mean{N_r}}\ , \
\ee
\be
\mathcal{M}_3 =\mean{N_r}^{1/2}\frac{\mean{\lr{N_r-\mean{N_r}}^3}}{\mean{\lr{N_r-\mean{N_r}}^2}^{3/2}}\ , \
\ee
\be\label{m4}
\mathcal{M}_4 = \mean{N_r} \lr{\frac{\mean{\lr{N_r-\mean{N_r}}^4}}{\mean{\lr{N_r-\mean{N_r}}^2}^{2}}-3}\ ,\
\ee
where $N_r$ is the mean number of vortices inside a square of side $r$. The quantities on Eqs.~\eqref{m2}---\eqref{m4} are normalized in such a way that $\mathcal{M}_i=1$ for homogeneous point process (Poisson process). The right panel of Fig.~\reffig{fig4.5} shows the comparison between processed \ac{DNS} data (from the dataset Iso-4096 through the \ac{SSC} with $\lambda_{T}=0.125\sigma_{\lambda_{ci}}$) and the prescribed values of $\mathcal{M}_i$. The data agrees very well with a hard disk behavior up to $r\lesssim7\eta_K$. Large scale deviations are interpreted as follows: They only mean that vortex structures develop further correlations at larger scales, reflected in the vortex gas model through the specific definitions provided by the other fields. We refer, more particularly, to $\xidr$ which is expected to have an important role in the statistical properties of $N_r$ in the inertial range \cite{moriconi2022statistics}.
\begin{figure}[t]
\center
\includegraphics[width=\textwidth]{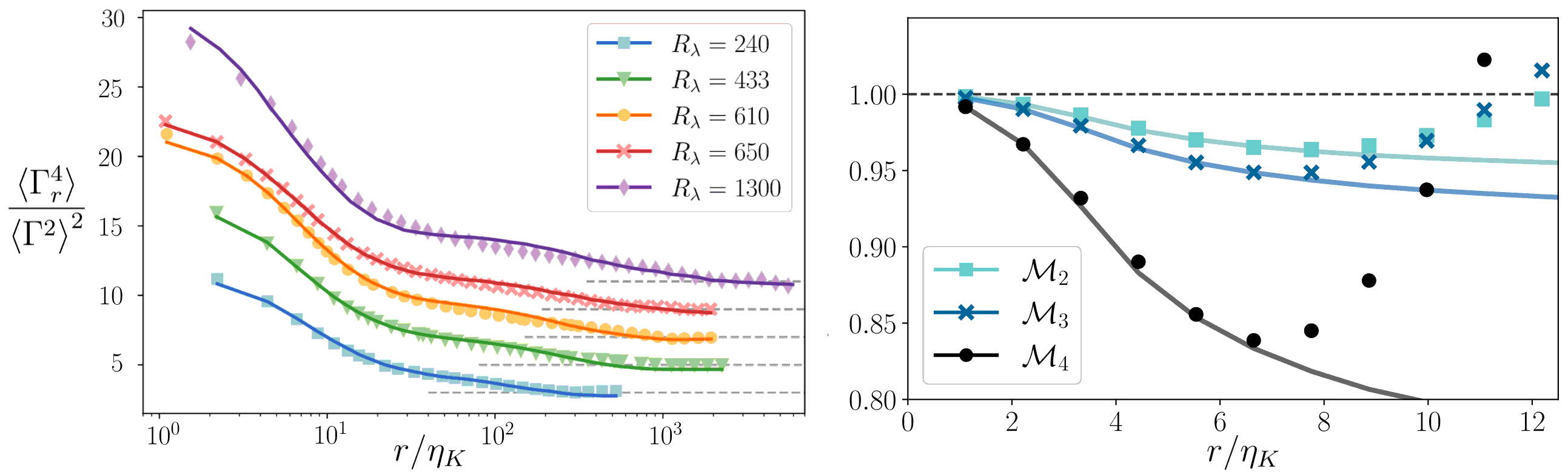}
\caption[Hard disk's kurtosis and statistics of vortex distributions.]{Left panel shows the kurtosis calculated by Monte Carlo simulation of the hard disk model (symbols) against \ac{DNS} data (lines) for several Reynolds numbers. $R_\lambda=433$ belongs to Sqr-10, $R_\lambda=610$ belongs to Sqr-14 and the other curves were reproduced from \cite{iyer2019circulation}. Curves are vertically displaced to ease visualization and, dashed grey lines represent the Gaussian value. The right panel shows normalized moments of a number of vortices in a square of size $N_r$, points are \ac{DNS} from Iso-4096 and lines are the values expected from a hard disk gas. Both figures were reproduced from \cite{moriconi2022statistics} with minor modifications.}
\label{fig4.5}
\end{figure}

The fact that circulation's scaling exponents are correctly accounted for by Eq.~\eqref{modelnew} at inertial range scales suggests, in addition to points (\ref{(P4.1)}-\ref{(P4.3)}), that the coarse-grained square root dissipation field $\xidr$ should be connected to the mean number of vortices lying in the contour $N_r$. Suppose, for example, the circulation is statistically reproduced by counting this number of structures times the circulation carried out by each one of the vortices, then, using Eq.~\eqref{modelnew} we have, 
\be\label{discu}
\mean{\Gamma_R^p}\sim \mean{N^p_R}_\sigma R^{2p} \mean{|\tg|^p}\sim\mean{\xi^p_{\mathcal{D}_R}}\mean{|\tg|^p}\sim R^{\lambda_{|p|}}\ . \
\ee
Based on Eq.~\eqref{discu}, we putatively argue that $N_R\sim R^{-2}\xidr$ in the inertial range. In this sense, the intermittent cascade fluctuations (in our case ruled by the \ac{GMC} approach) are intrinsically connected with the multifractal properties of the number of vortices inside the contour. In Fig.~\reffig{fig4.6} we corroborate this idea by carrying out the statistical moments $\mean{N^p_r}\propto r^{\zeta_p}$. The data show good agreement with the expected relationship $\tau_p=\zeta_p-2p$, supporting the above discussion which was first noted by Moriconi and Pereira \cite{moriconi2022circulation}. 

The interpretation of the \ac{GMC} field as the mean number of structures inside a loop, directly leads to an interesting interpretation of the linearization effect on circulation moments through the microscopic behavior of the hard disk structures. As the exclusion effect comes into play, it is natural to expect that there exists a number of structures $\bar N_0$ inside the domain $\md_R$, such that there is not enough room to packet $\bar N_0+1$ vortices in $\md_R$. The right panel of Fig.~\reffig{fig4.6} roughly illustrates this procedure. When comparing to nonexcluding distributions, it is always possible to reorganize the distribution to keep it pointwise homogeneous with $N+1$ point vortices i.e. there is no bound for the number of structures and its density can be arbitrarily high. 

The idea is that the maximum packing fraction produces a natural bound to the \ac{GMC} field and, in the multifractal formalism, it means that the scaling exponent of $\xi^p_{\mathcal{D}_R}$ gets linearized above certain critical moment order $p_c$. This linearization effect was already put forward by \cite{moriconi2021multifractality} through extensive Monte Carlo simulation of bounded \ac{GMC}. In that work, the application to the case of circulation statistics, the conclusion drawn from this work was addressed only for the case of \ac{MCM} scaling exponents. Although very similar, the \ac{MCM} multifractal breakdown is not fully compatible with the \ac{GMC} one since multifractal properties such as singularity spectrum and H{\"o}lder exponents are different for these two systems.
\begin{figure}[t]
\center
\includegraphics[width=\textwidth]{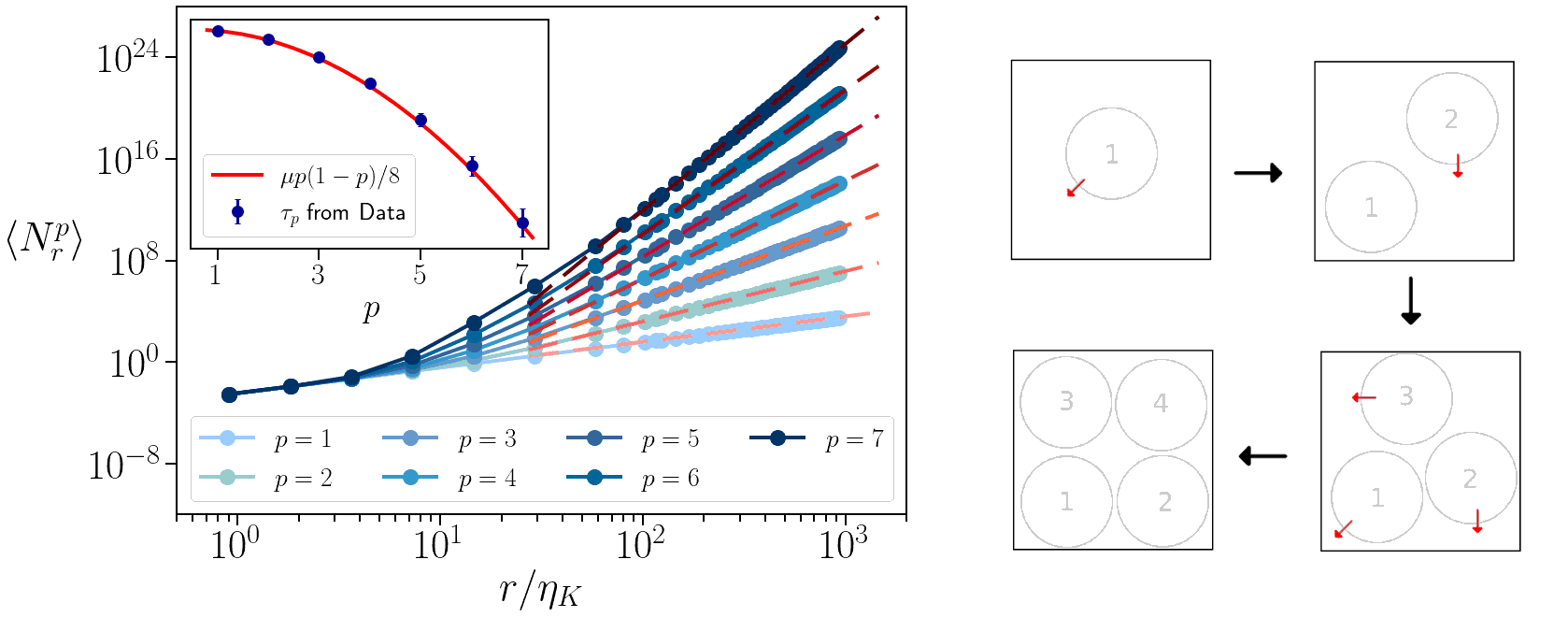}
\caption[Circle packing in a square and statistical moments of $N_r$.]{The left panel's main frame shows the statistical moments of $N_r$ as functions of the scale $r$ for Iso-4096 with $\lambda_{T}=0.125\sigma_{\lambda_{ci}}$, red dotted lines represent the \ac{GMC} power laws. Inset shows its scaling exponents where the systematic error was partially accounted for by sliding the fitting range. This figure is equivalent to the one found in \cite{moriconi2022statistics}. Moreover, the right panel illustrates the procedure of packing rigid disks, it is clear that there has a maximum packing fraction.}
\label{fig4.6}
\end{figure}

In order to fill this gap, let us put forward the idea of a bounded singularity spectrum for both systems. From the \ac{MCM} of \ac{OK62},
\be\label{MCM-codim}
D^{(MCM)}(h)=-\frac{h^2}{2\mu}+\frac{h}{2}+d-\frac{\mu}{8} \ , \
\ee
represents the correct singularity spectrum leading to the scaling exponent of $\mean{\vare^p_\ell}\sim\ell^{\tau_p}$ with $\tau_p$ given by Eq.~\eqref{MCM-scaling} through the use of Eq.~\eqref{infh}. If one restricts the parabolic behavior of Eq.~\eqref{MCM-codim} for $|h|\leq h_0$, there exists a ``critical'' moment order $p=p_c$ where the minimizer of Eq.~\eqref{MCM-codim} is always at the borders of the domain $h\in[-h_0,h_0]$. Mathematically,
\be\label{eqqc}
p_c h_0+d-D(h_0)\leq \inf_{h}\left[p_ch+d-D(h)\right]\ .\
\ee
Substituting Eq.~\eqref{MCM-codim} into Eq.~\eqref{eqqc} one gets,
\be\label{critq}
p_c=\frac{1}{2}+\frac{|h_0|}{\mu} \ , \
\ee
such that for $p>p_c$ the scaling exponents are linearized. For multiplicative cascades of $\vare^{1/n}_\ell$, the absence of spatial correlations throughout the cascade implies that the substitution $\tau_p\rightarrow \tau_{p/n}$ as $p\rightarrow p/n$ is valid. The consequence for the critical moment order of the $n^{th}$-rooted cascade is the relation
\be
p^{(n),MCM}_c=n\lr{\frac{1}{2}+\frac{|h_0|}{\mu}} \ .\ 
\ee
The general scaling exponent, taking the bound effect into account, is
\be\label{tmcm}
\tau^{(n,MCM)}_p=
\begin{cases}
   \frac{\mu}{2n^2} p(n-p) &,\ \text{ if } p<p^{(n)}_c\ , \\\
    \\
    \frac{\mu}{2n^2}\Big(p\big(n-2p^{(n)}_c\big)+\lr{p_c^{(n)}}^2\Big) &,\ \text{ if } p\geq p^{(n)}_c\ ,\
\end{cases}
\ee
where, for $n=2$ and an estimate $h_0\simeq0.5$ given by \cite{moriconi2021multifractality} based on the analysis of \cite{meneveau1987simple,chhabra1989direct}. The resulting critical moment order $p^{(2)}_c\approx6.88$ is found to be in good agreement with \ac{DNS} data.

In the case of a bounded \ac{GMC} field, the multiplicative cascade is now spatially correlated, and the substitution $p\rightarrow p/n$ dos not connect different scaling exponents $\tau_p$ and $\tau_{p/n}$. The only way to account for the \ac{GMC} scaling exponents is by considering an $n$-dependent singularity spectrum, that is, the singularity spectrum of the $n$-root cascade cannot be transformed in the $m$-root cascade through a simple transformation. The correct spectrum that reproduces the \ac{GMC} scaling exponent of Eq.~\eqref{GMC-scaling} is 
\be\label{GMC-codim}
D^{(GMC)}_n(h)=-\frac{h^2n^2}{2\mu}+\frac{h}{2}+d-\frac{\mu}{8n^2}\ .\
\ee
Exploiting the comments made above, one may note when $n=1$ the singularity spectra of \ac{MCM} and \ac{GMC} are exactly equal. The general solution for the scaling exponent of a bounded $h$-domain is then,
\begin{figure}[t]
\center
\includegraphics[width=\textwidth]{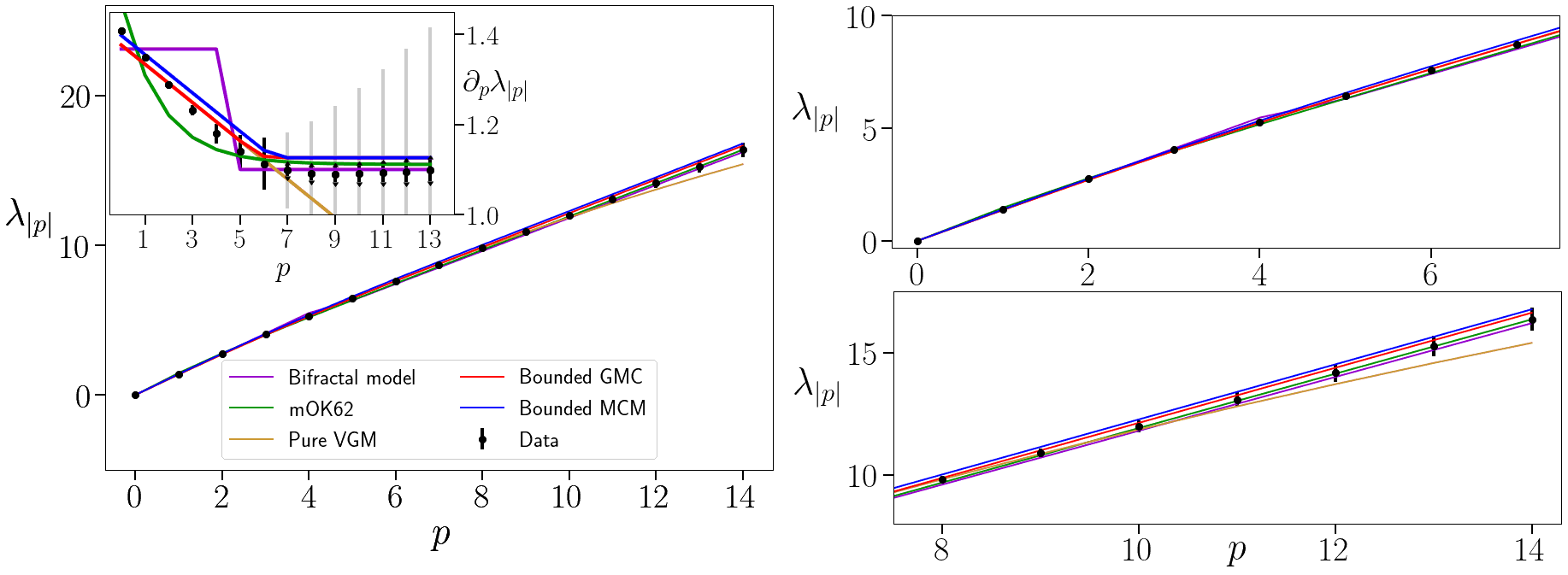}
\caption[Comparison among models and the data.]{Left panel: circulation scaling exponents $\lambda_{|p|}$ for different models, bifractal fit, and bounded \ac{MCM} refers to \cite{iyer2019circulation} and \cite{moriconi2021multifractality}, respectively. The data is the same as in Fig.~\reffig{fig3.5} and the inset follows the same reasoning, the local tangent for each model. The right panel shows the same as the left panel with zoomed-on low-order moments (upper right panel) and high-order moments (lower right panel) for better comparison among the models.}
\label{fig4.7}
\end{figure}
\be\label{tgmc}
\tau^{(n,GMC)}_p=
\begin{cases}
    \frac{\mu}{2n^2}p(1-p) &,\ \text{ if } p<p^{(n)}_c\ , \\\
    \\
    \frac{\mu}{2n^2}\Big(p(1-2p^{(n)}_c)+\lr{p_c^{(n)}}^2\Big) &,\ \text{ if } p\geq p^{(n)}_c\ ,\
\end{cases}
\ee
where,
\be\label{pmcm}
p^{(n),GMC}_c=\frac{1}{2}+n^2\frac{|h^{(n)}_0|}{\mu} \ .\ 
\ee
Despite similarities between the \ac{MCM} and \ac{GMC} multifractal formalism are clearly evident, their fundamentals are different. The former approach is completely based on a three-dimensional analysis of turbulence flows to the estimation of $h_0$. The latter is completely constructed in terms of two dimensions quantities and, is basically related to the coarse-grained number of structures inside a loop. Owing to this fact, we have no arguments to set $h_0$ as the same in \ac{MCM}, but if one can still believe this to work for $D^{(GMC)}_1(h)$. It follows from these remarks that $h^{(n)}=h_0/n$, such that Eq.~\eqref{pmcm} can be rephrased as
\be\label{critqgmc}
p^{(n),GMC}_c=\frac{1}{2}+n\frac{|h_0|}{\mu} \ . \
\ee

Fig.~\reffig{fig4.7} shows the comparison among \ac{mOK62}, bifractal model, \ac{DNS}, \ac{VGM}, the bounded \ac{MCM} put forward by \cite{moriconi2021multifractality} $p^{(2,MCM)}_c\approx6.88$ and the one discussed above with $p^{(2,GMC)}_c\approx6.38$. The \ac{GMC} multifractality breakdown seems to improve slightly the \ac{MCM} approach. However, it lacks a better estimate for the maximum H{\"o}lder exponent $h^{(n)}_0$. Moreover, both bounded \ac{GMC} and \ac{MCM} depart from the bifractal fit of \cite{iyer2019circulation} and the \ac{mOK62} of \cite{muller2021intermittency} for the highest moment orders. On the other hand, the transition between the low order scaling and the asymptotic one is much better addressed by the bounded fields, since $\partial_p\lambda_{|p|}$ is shown to be mostly linear up to $p\lesssim 6$ (Fig.\reffig{fig4.7} left panel's inset). 

Tracing back the conclusions drawn by the discussion of Fig.~\reffig{fig3.6}, we note that the values $h_\infty$ and $D_\infty$ are directly motivated by the use of Eqs.~\eqref{GMC-codim},~\eqref{tgmc}~and~\eqref{critqgmc} in the case $p>p^{(2)}_c$,
\be\label{4.67}
\lambda_{|p|}=\lr{\frac{4-\alpha}{2}}p+\tau_p^{(2,GMC)}=h_\infty p+(d-D_\infty) \ ,\
\ee
with,
\be
h_{\infty}=\frac{4}{3}+\frac{\mu}{4}-|h_0^{(2)}| \ ,\
\ee
and,
\be
D_{\infty}=d-D_2^{(GMC)}\lr{-|h_0^{(2)}|}=\frac{\mu}{32}+\frac{1}{8}+\frac{1}{8\mu} \ ,\
\ee
which justifies the values $h_\infty\approx1.126$ and $D_\infty\approx0.866$ used in Section~\refsec{sec3.3.2} for $\mu=0.17$ and $|h_0^{(2)}|=h_0/2\approx1/4$. 

The arguments in this section collectively support a straightforward yet highly effective phenomenological view of the statistical characteristics of circulation in turbulent flows. The contribution of non-overlapping, strong vortical structures to the overall phenomenon is the manifestation of an asymptotic self-similarity based on the maximum packing fraction interpretation. In order to address the discussion of Section~\refsec{sec3.3.2} for the low-order scaling exponents, let us turn our attention to the derivation of \ac{cPDF}s.

\section{Probability Density Function(al)}

$_{}$

An interesting way to start the derivation of the \ac{cPDF} associated with the \ac{VGM} is to take advantage of the field-theoretical formulation of the \ac{GMC}. In order to get a more general expression, let us redefine the fields in such a way to cope with the locality of the \ac{GMC} field,
\be\label{4.72}
\psi(x)=\sqrt{\frac{\vare_0}{3\nu}}\int_{\Psi}d^2y\xi(x)g_\eta(y-x)\sigma(x)\ .\
\ee
The above expression is directly motivated by Eqs.~\eqref{circmodel}~and~\eqref{circmodelcoarse}. In this sense, a mean-field approach to $\psi(x)$ means to replace the local \ac{GMC} field $\xi(x)$, for its uncorrelated version modeled by the \ac{MCM} approach $\xidr$. The \ac{cPDF} can be derived directly from the identity
\be
\rho(\Gamma_R) = \mean{\delta\left[\Gamma_R-\int_{\mathcal{D}_R}d^2x\tg(x)\psi(x)\right]}_{\lr{\tg,\psi}}=\int \frac{d\lambda}{2\pi}\mean{e^{i\lambda\lr{\Gamma_R-\int d^2x\tg(x)\psi(x)}}}_{\lr{\tg,\psi}} \ . \
\ee
The average over the \ac{GFF} can be directly performed to get,
\be\label{flucvar}
\rho(\Gamma_R) = \int \frac{d\lambda}{2\pi}e^{i\lambda\Gamma_R}\mean{e^{-\frac{\lambda^2}{2}\Sigma^2[\psi]}}_{\lr{\psi}}=\mean{\frac{e^{-\Gamma^2_R/2\Sigma^2[\psi]}}{\sqrt{2\pi\Sigma^2[\psi]}}}_{\lr{\psi}} \ , \
\ee
where the last equality was obtained by the Gaussian integration in $d\lambda$ and,
\be
\Sigma^2[\psi]=\int_{\mathcal{D}_R}d^2x\int_{\mathcal{D}_R}d^2y\psi(x)\Delta(x-y)\psi(y)\ . \
\ee
We denote by $\Delta(x-y)=\mean{\tg(x)\tg(y)}$, the known two-point correlation function. The above formulation bridges Umeki's modifications to Migdal's asymptotic result (Eq.~\eqref{3.10}) to the language of our model \cite{umeki1993probability}. Moreover, one may note that computing Eq.~\eqref{flucvar} is the same as a Gaussian shape with fluctuating variance. Umeki's case is then restored by excluding those fluctuations and imposing a \ac{K41}-like relation for the variance $\Sigma^2\propto A^{4/3}$. 

Once we included intermittency, $\Sigma^2$ fluctuations are then modeled by the underlying statistical measure density related to the $\psi$ field. Along the lines of the \ac{GMC} approach, the remaining mean value to be taken can be accounted for by the path integral \cite{zinn2021quantum,itzykson1991statistical}
\be
\rho(\Gamma_R)=\int D[\phi]e^{-S[\phi]} \ ,\ 
\ee
where, $S[\phi]$ reads
\be\label{act-full}
S[\phi]=\frac{1}{2}\int_{\mathcal{D}_L}d^2x\Big(\partial_i\phi\partial_i\phi +V(\phi)\Big)+\frac{\Gamma^2_R}{2}\Big(\Sigma^2\left[\psi_0e^{\gamma\phi}\right]\Big)^{-1}+\frac{1}{2}\ln\Big(2\pi\Sigma^2\left[\psi_0e^{\gamma\phi}\right]\Big)  \ , \
\ee
where again $\gamma^2=2\pi\mu$ in order to model turbulent fluctuation and $\psi_0$ is a constant related to $\vare_0$. The ``potential'' $V(\phi)$ defines the bound to the fluctuation of the \ac{GMC} field \cite{moriconi2021multifractality,moriconi2022circulation},
\be\label{poten}
V(\phi)=
\begin{cases}
    0 &,\ \text{ if } \phi<\phi_c\ , \\
    V_0\rightarrow\infty &,\ \text{ if } \phi\geq \phi_c\ .\
\end{cases}
\ee
The value of $\phi_c\approx2.7$ is not far from the estimation for $p_c\approx6.88$ in the \ac{MCM} model put forward by \cite{moriconi2021multifractality,moriconi2022statistics} but a reasonable estimate for the value $p_c\approx6.38$ is still in order.

\subsection{Mean Field Approach to the cPDF's Core}

$\ \ \ \ $There is no analytical method to integrate Eq.~\eqref{act-full}, and not even power expansions are proven to converge. However, one can try to tackle this problem numerically, but it is still hard work to do since there are a lot of parameters and boundary conditions which are not fixed. Instead, let us take the mean-field approach in the asymptotic region $R\gg\eta$. In this limit, $g_\eta(x)\approx 2\pi\eta^2\delta^2(x)$, and the measure density converges to the lognormal measure density $f_R(x)$ of a random variable $x\sim\mathcal{N}\lr{\bar{x}_R,\bar{x}_R}$ with $\bar{x}_R\sim \ln(R^{-\mu/4})$ \cite{apolinario2020vortex,crow1987lognormal}, such that
\be
\rho(\Gamma_R) = \int_0^{\infty}\frac{dx}{x} f_R(x)\mean{\frac{e^{-\Gamma^2_R/2x^2\Omega}}{\sqrt{2\pi\Omega}}}_{\lr{\Omega}} \ , \
\ee
where, 
\be
\Omega=\frac{\vare_0}{3\nu}\int_{\mathcal{D}_R}d^2xd^2y\ \sigma(x)\Delta(x -y)\sigma(y),
\ee
and
\be\label{fexpl}
f_R(x)=\frac{1}{x\sqrt{2\pi\bar x_R}}\exp\lr{-\lr{\frac{(\ln(x/x_0)-\bar x_R)^2}{2\bar x_R}}} \ ,\
\ee
where $x_0$ is an unimportant constant related to $\xi_0$. Careful calculation of the relative error of $\Omega$ through Eqs.~\eqref{I1}---\eqref{I3} shows that $(\Omega-\mean{\Omega})^2/\mean{\Omega}^2\propto (R/\eta_K)^{\alpha-2}$, which quickly converges to zero in the large contour limit. This allows us to treat $\Omega$ as deterministic in the large scale limit $\Omega\approx \mean{\Omega}=N_K^2E_\alpha(R/\eta_K)^{4-\alpha}$. Within this approximation, we get
\be\label{cpdffin}
\rho(\Gamma_R) = \frac{1}{\sqrt{2\pi\mean{\Omega}}}\int_0^{\infty}\frac{dx}{x} f_R(x)\exp{\lr{-\frac{\Gamma^2_R}{2x^2\mean\Omega}}} \ . \
\ee
Using Eqs.~\eqref{cpdffin} and \eqref{fexpl}, one can change the integration variables $x=x'\Gamma_R/\sqrt{2\mean{\Omega}}$ to get
\be\label{cpdffin2}
\rho(\Gamma_R) = \frac{1}{\Gamma_R\sqrt{2\pi^2\bar x_R}}\int_0^{\infty}\frac{dx}{x^2}\exp{\lr{-\frac{1}{x^2}-\frac{1}{2\bar x_R}
\lr{
\ln\lr{
\frac{x\Gamma_R e^{\bar x_R}}{x_0\sqrt{2\mean\Omega}}}}^2
}} \ . \
\ee
As a consequence, $\rho(\Gamma_R)$ can be regarded as a function of $\Gamma_R e^{\bar{x}_R}\mean{\Omega}^{-1/2}\sim\Gamma_R R^{-(4-\alpha)/2-\mu/4}$ only.
\begin{figure}[t]
\center
\includegraphics[width=\textwidth]{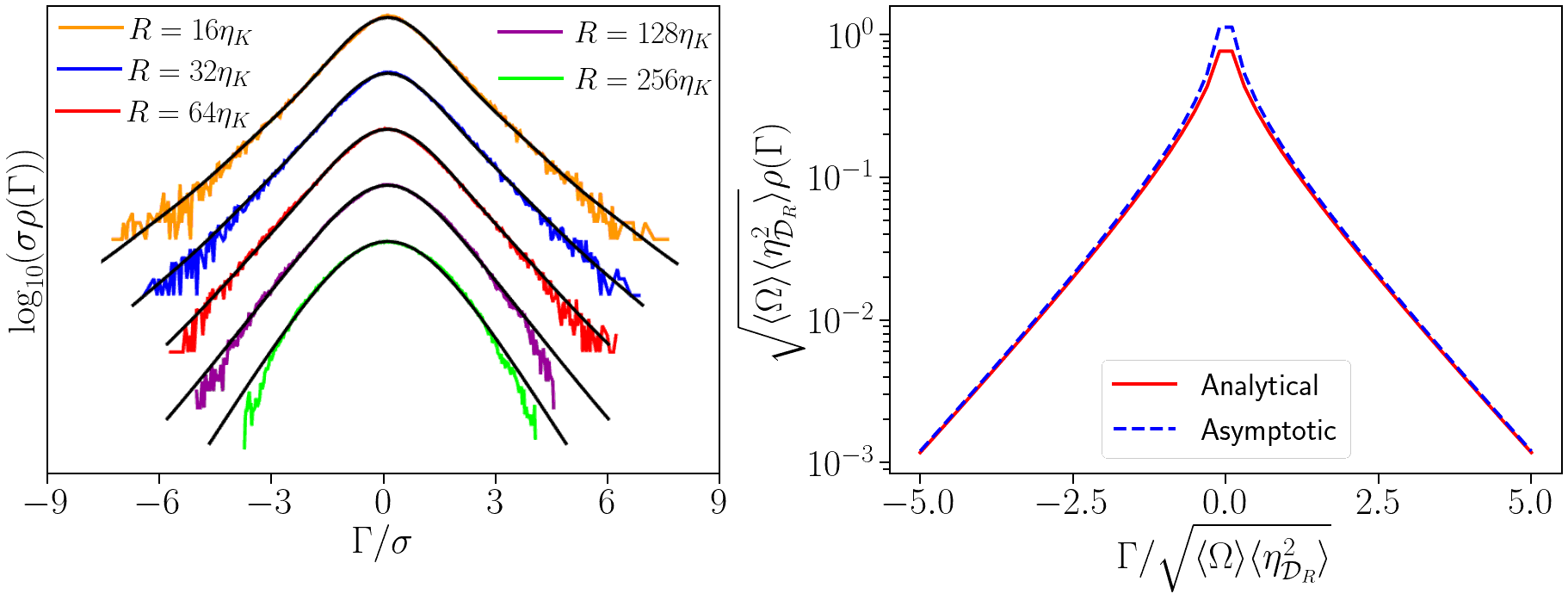}
\caption[Comparison of cPDFs from data and the model.]{Comparison between \ac{cPDF}s obtained by the dataset Cir-1024 (colored lines) and by numerical integration of Eq.~\eqref{cpdffin} (black lines) is shown in the left panel, curves are vertically displaced to ease visualization. The right panel shows the comparison between normalized \ac{PDF}s of the product of Gaussians. The figure on the left can be found in \cite{apolinario2020vortex}.}
\label{fig4.8}
\end{figure}

The above result explains the observed collapse for the core of the \ac{cPDF} discussed in Section~\refsec{sec3.3.2} at Fig.~\reffig{fig3.6} for the rescaled variable $\Gamma_R R^{-h^\star}$. The value $h^\star=4/3+\mu/2\approx1.418$ seems to be a little closer than $h\approx 1.376$ found by \cite{apolinario2020vortex} to the reference value $h\approx1.4$ reported by \cite{iyer2019circulation} for the extremely high Reynolds \ac{DNS} data. The mismatch between \cite{apolinario2020vortex} and the presenting value of $h$ occurs because of the correction to $\alpha$ due to the 4/5 law. In consonance with the presented approach, the collapse cannot be observed in the tails of the \ac{cPDF} since the fluctuations of the intermittent field are bounded for extreme events. Instead, a collapsing exponent $h_\infty$ for the tail was related to the asymptotic limiting H{\"o}lder exponent $\partial_p\lambda_{|p|}$ for $p\rightarrow \infty$ in Section~\refsec{sec4.2}.

The associated fractal dimension $D^\star$ cannot be simply ruled out without an explicit analytical integration of Eq.~\eqref{cpdffin2} and, up to this point, the space-filling nature of the core circulation fluctuations $D^\star=d$ is only empirical. Numerical integration of Eq.~\eqref{cpdffin} is shown in the left panel of Fig.~\reffig{fig4.8}, it is in surprising accordance with \ac{DNS} regarding the analytical simplicity of the mean-field approach employed here.

\subsection{Mean Field Approach to the cPDF's Far Tail}

$\ \ \ \ $For the opposite asymptotic limit, when circulation achieves extremely high values compared to its standard deviation, it is very likely that the \ac{GMC} field is under bounded circumstances. In other words, the extreme events of circulation are supposed to be composed of a partially polarized tangle of vortices and not by a single ultra-extreme vortical structure. Moriconi and Pereira \cite{moriconi2022statistics} explored this limiting behavior of the bounded \ac{GMC} field performing Monte Carlo simulations. Their results indicate the bounded \ac{GMC} field to be modeled as random perturbations around its, mean i.e., $\xidr=\mean{\xidr}+\eta_{\mathcal{D}_R}$, such that $\eta_{\mathcal{D}_R}$ has Gaussian centered \ac{PDF}. Thus, the circulation is given by,
\be\label{decomp}
\Gamma_R\sim\mean{\xidr}\int d^2x \tg(x)+\eta_{\mathcal{D}_R}\int d^2x \tg(x) \ . \
\ee
The asymptotics of Eq.~\eqref{decomp} are dominated by the behavior of the rightmost term, since the first is clearly Gaussian. The probability distribution of the product of independent Gaussian variables is a well-established case study in statistics \cite{springer1966distribution} and the resulting probability distribution can be expressed in terms of Bessel functions.
Using Eq.~\eqref{cpdffin}, in the case that $f_R(x)$ converges to a Gaussian function, we end up with,
\be
\rho(\Gamma_R) \sim \int dx e^{-s(x)} \ , \
\ee
with
\be
s(x)=\frac{x^2}{2\mean{\eta_{\mathcal{D}_R}^2}} +\frac{\Gamma}{2x^2\mean{\Omega}}+\frac{1}{2}\ln\lr{2\pi x^2\mean{\Omega}}\ , \
\ee
where $\mean{\eta_{\mathcal{D}_R}^2}$ is an unknown variable that completely characterizes the distribution of $\eta_{\mathcal{D}_R}$. The resulting asymptotic probability distribution found through the saddle point method (see Appendix.~\ref{appspa}) is
\be
\rho\lr{x\equiv\frac{\Gamma_R}{\sqrt{\mean{\Omega}\mean{\eta_{\mathcal{D}_R}^2}}}\gg\frac{1}{2}} \sim \frac{1}{(2\pi)^{1/2}\lr{\mean{\Omega}\mean{\eta_{\mathcal{D}_R}^2}}^{1/2}}\frac{e^{-|x|}}{\sqrt{|x|}} \ . \
\ee
The right panel of Fig.~\reffig{fig4.8} shows the comparison between analytical result and saddle point approximation. The $p^{th}$-order moment can be calculated with the help of Eq.~\eqref{3.13} in this asymptotic limit. The final result gives
\be\label{fixeta}
\mean{|\Gamma_R|^p}\sim \Big(\mean{\Omega}\mean{\eta_{\mathcal{D}_R}^2}\Big)^{p/2}\ ,\
\ee
in the case that $p\rightarrow\infty$. The scaling dependence $\sim R^{\lambda_{|p|}}$ of the left hand side of Eq.~\eqref{fixeta} is given in terms of Eq.~\eqref{4.67} while $\mean{\Omega}\sim R^{4-\alpha}$. The latter can be used to fix the scale dependence of $\eta_{\mathcal{D}_R}$'s variance in the limit $p\rightarrow\infty$,
\be
\mean{\eta_{\mathcal{D}_R}^2}\sim R^{\mu(1-2p^{(2,GMC)}_c)/4}\sim R^{|h_0|}\ ,\
\ee
which is the maximum H{\"o}lder exponent in the bounded \ac{GMC} approach as expected.

Our rather simple mean-field approach to the \ac{cPDF} shows up to be consistent with the multifractal breakdown and the several pieces of information along the circulation statistics literature \cite{apolinario2020vortex,moriconi2021multifractality,moriconi2022circulation,moriconi2022statistics}. However, the necessity of an efficient numerical or analytical computation of Eq.~\eqref{act-full} can reveal details about the modeling parameters like $\phi_c$ and $h_0$, along with their relations to the spatial vortex distribution and other phenomenological quantities introduced by the hard disk approach.

The modeling of the circulation statistics here presented, provides a use of standard statistical tools for the problem of classical turbulence in a promising way. Almost all the phenomenological pieces of information that build the fundamental rock of the gas model led to unpredicted new phenomena that are being observed in numerical simulations. The only remaining piece is the physical meaning of the \ac{GFF} $\tg(x)$. This is the main issue for the last, but not least, section of this chapter.

\section{Vortex Polarization}\label{sec4.4}

$_{}$

In the last few years, the study of circulation statistics in classical homogeneous and isotropic turbulent flows motivated the exploration of this variable in other fluid dynamical systems. A particularly interesting system is a quantum fluid, where the vortex structures have quantized circulation. The collective behavior of quantized vortices, when analyzed through the circulation variable shows up to have a very similar statistical behavior as classical turbulence (within inertial range scales \cite{muller2021intermittency}). A very simple, but still robust model of biased random walk was introduced by \cite{polanco2021vortex} to understand the role of self-similar and intermittent behaviors of circulation statistics. Introducing the one-dimensional biased random walk model, one can write the circulation as follows,
\be
\Gamma_n=\sum_{i}^n s_i \ ,\ 
\ee
where $s_i=\pm1$ represents the quantized circulation of the $i^{th}$ vortex in a chain of $n$ vortices. The first vortex $s_1$ can be regarded as a boundary condition to be chosen with equal probability. The bias is introduced by letting $s_{i+1}$ to be correlated to the whole past chain of vortices $\Gamma_{i}$, then, the probability to find $s_{i+1}$ positive is given by $p_{i+1}=\lr{1+\beta\Gamma_i/i}/2$, where $\beta\in [0,1]$ is an adjustable parameter.

The resulting process is a Markov chain for the $\Gamma_n$ variable which only depends on $n$ and $\Gamma_{n-1}$. Owing to a master's equation formulation of this memoryless process, \cite{polanco2021vortex} showed that the general scaling exponents are given by,
\be\label{polscal}
\mean{|\Gamma_n|^p}\propto n^{\beta p} \ ,\ 
\ee
for large enough $n$, and $\beta\in(1/2,1)$. One may note that $\beta=1/2$ represents a random walk, since $\langle\Gamma_n^2\rangle\propto n$ while $\beta=1$ is a full polarized chain as $\langle\Gamma_n^2\rangle\propto n^2$. In fact, the general scaling exponent is given by $p\inf[1/2,f'(0)]$ where $f(x)$ is any continuous odd function, defined by $p_{i+1}=(1+f(\Gamma_i/i))/2$. The choice $f(x)=\beta x$ is a mere simple case \cite{polanco2021vortex}.

The great turning point achieved within this formulation is the procedure of ``cluster summation'' \cite{polanco2021vortex}. With this procedure, the circulation is computed by simply summing the contribution of $n$ neighboring detected vortices in a \ac{DNS} of quantum turbulence. Since $n\sim R^2$ in a first approximation, Polanco et al. were able to find a perfect self-similar scaling relation with $\beta=2/3$ which correctly reproduces \ac{K41} behavior of the circulation statistics, while the intermittent behavior comes basically from the heterogeneous vortex distribution \cite{polanco2021vortex}. It resembles the approach made by \cite{yoshida2000statistical} to the statistics of classical turbulence when splitting the vorticity into strong and weak components. Nevertheless, cluster summation seems to be a more suitable procedure for quantum turbulence since it does not need a threshold. The advantage of the cluster summation is that it does not fix the contour shape, nor its area, such that the maximal packing fraction can be circumvented by changing the contour shape/area every time a vortex is added to the summation.

In the case of classical turbulence, such a summation procedure is far more involving than the quantum case, as the circulation is not quantized. It can assume any value on the real line and, there can be strong correlations among the amplitude, signs, and spatial distribution of vortices. Let us assume the polarization of vortices is roughly accounted for by the \ac{GFF} $\tg(x)$, and that the scaling exponents $\lambda_{|p|}$ and $\beta$ are related through Eqs.~\eqref{polscal} and the linear part of \eqref{lambdap}. For $p=2$, we get 
\be\label{2momclus}
4\beta=\lambda^{\text{linear}}_2\rightarrow \beta=\frac{2}{3}+\frac{\mu}{8}\approx 0.688\ ,\ 
\ee
where the factor $4$ in the leftmost side of Eq.~\eqref{2momclus} is achieved by using the relation $n\sim R^2$. Note that the former relation is valid, in principle, for the full correlation of the $\tg(x)$ field (including its amplitudes). 

Suppose we have a 2D slice of a 3D turbulent flow, with $N$ vortices at the positions $\vec{x}_i$, each of which carrying a circulation $\Gamma_i$, we define the cluster summation as follows \cite{polanco2021vortex},
\be\label{clusum}
\Gamma^{(f)}_i(n)=\sum_{j=1}^{n} f\left(\Gamma^{(i)}_j\right)\ ,\ 
\ee
where $n\leq N$ is the size of the cluster spanning an area with a typical size of $r\sim n^{1/2}$. Also, $\Gamma^{(i)}_j$ is the circulation $\Gamma_j$ of the $j^{th}$ vortex, when sorted by distance from the $i^{th}$ vortex, $r_{ij}=|\vec{x}_j-\vec{x}_i|$. The function $f:\mathbb{R}\rightarrow\mathbb{R}$ defines two different types of cluster summation which can be interesting for us. We refer to ``binarized summation'' as the procedure of summing only the signs of the vortices ($f(x)=\text{sign}(x)$). This method basically measures the degree of polarization of the vortex cluster in the same sense as the quantum vortices. In opposition to this method, we define it as ``continuum summation'' when it takes the full value of circulation ($f(x)=x$) carried by each vortex. Note that, this choice in spite of $f(x)=|x|$, does not totally exclude the polarization effects. The scaling exponent of these two different series must be connected in some sense.

We have analyzed $192$ 2D slices of the dataset Iso-4096 (see Table~\reftab{tab3.1}) where the slices were homogeneously spaced to ensure decorrelation and the three axes direction were used. The \ac{SSC} identification method with $\lambda_T=0.125\sigma_{\lambda_{ci}}$ was employed (see Appendix~\ref{appssc}) to detect $N_i$ structures for each slab, a number which fluctuates around $N_i\sim5\times 10^4$ vortices. An ensemble is generated by starting the summation at all possible vortices, thus, the average at each slice is defined as
\be
\mean{\Gamma_n}=\frac{1}{N}\sum_{i=1}^N \Gamma^{(f)}_i(n)\ ,\
\ee
where the superscript $f$ was omitted for simplicity and the indications about the summation method will be explicit. Meanwhile, an additional average can be taken by considering all the $192$ cuts. For memory-saving reasons, we restricted the cluster sizes up to $10^4$ which spans an area with linear size of the order of the integral length scale.
\begin{figure}[t]
\center
\includegraphics[width=\textwidth]{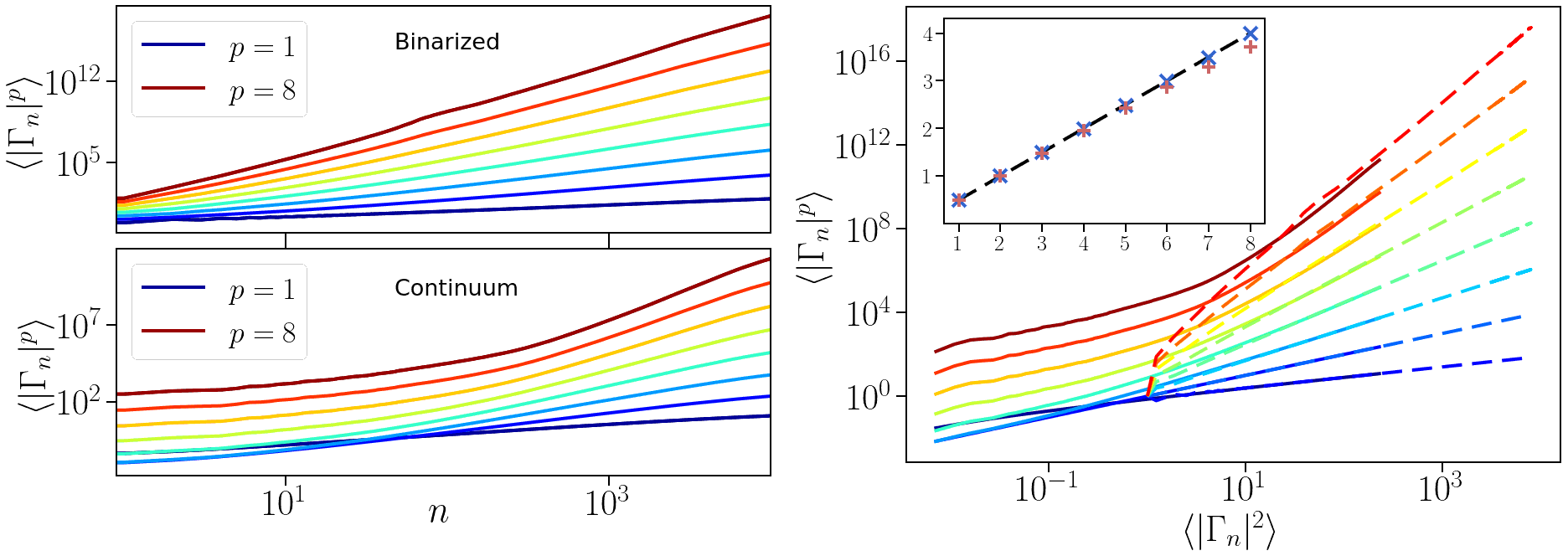}
\caption[Moments of cluster summation.]{Statistical moments of $|\Gamma_n|$ as functions of the cluster size $n$ in the binarized version (upper left) and the continuum process (lower left). The right panel shows the \ac{ESS} approach for the binarized method (dashed lines) and continuum method (full lines). Inset is showing the \ac{ESS} exponent $\zeta(p|2)$ for binarized (blue $\times$), continuum (red $+$), and Gaussian (dashed lines) statistics $\zeta(p|2)=p/2$.}
\label{fig4.9}
\end{figure}

Figure~\reffig{fig4.9} shows the moments of the cluster summation as functions of the cluster size for the binarized and continuum summation methods. The binarized processes show very clear scaling in the whole cluster size range studied. The resulting scaling exponents are given by Eq.~\eqref{polscal} with $\beta=0.56\pm0.02$. The latter means that the tangle of point vortices in classical turbulence is less polarized than in quantum turbulence since Polanco et al. found $\beta=2/3$ \cite{polanco2021vortex}. In this sense, for classical turbulence, not only the polarization of the vortices, but their amplitude correlations account for the \ac{K41} scaling. The latter remarks mean that the full scaling of the continuum process can be more fundamental than the sole polarization of the vortices.

In striking contrast to its binarized version, the continuum cluster summation presents two scaling regions. The relevant region for nontrivial scaling is the one for larger cluster sizes $n\gtrsim10^3$ which will be referred to as ``far-n scaling range''. Moreover, the size of this range seems to be dependent on the moment order such that, the higher the moment order the shorter it is. The determination of the scaling exponent in the far-n scaling range is rather difficult because of the huge contamination of the small-scale scaling. Regardless of this fact, we can try to improve the scaling range by noting that, at very small contours, the scaling is trivially related to the regularization of $\tg(x\rightarrow0)$. Indeed, the introduction of regularization was the leading mechanism used by \cite{apolinario2020vortex} for the saturation of the kurtosis for $R\sim \eta_K$.

To stress this issue using the language of cluster summation, we note that a scaling $\mean{|\Gamma_n|^p}\sim n$ is trivially achieved by splitting the correlation function into diagonal and off-diagonal contributions. Consider the cluster summation's variance, for instance,
\be
\mean{\Gamma_n^2}=\sum_{i,j}^n\mean{\tg^2 (x_i))}+2\sum_{i}^n\sum_{j>i}^n\mean{\tg(x_i)\tg(x_j)} \ ,\
\ee
for,
\be
\Gamma_n=\sum_i^{n}\tg(x_i) \ . \
\ee
Assuming now the correlation function of the \ac{GFF} is regularized as
\be\label{tmcmnew}
\mean{\tg(x_i)\tg(x_j)}=
\begin{cases}
   \Gamma_0^2 &,\ \text{ if } i=j\ , \\\
    \\
    \Gamma_0^2|x_i- x_j|^{-\alpha}&,\ \text{ if } i\neq j\ .\
\end{cases}
\ee
The cluster summation's variance is shown to behave as
\be\label{combined}
\mean{\Gamma_n^2}=\Gamma_0^2\lr{n+\frac{4}{2-\alpha}\lr{\pi\bar\sigma}^{\alpha/2}n^{(4-\alpha)/2}} \ .\
\ee
Eq.~\eqref{2momclus} relates $(4-\alpha)/2=2\beta$ as expected. In order to understand the far-n range contamination, one defines
\be\label{n0}
\bar n_2\equiv\left(\frac{4}{2-\alpha}(\pi\bar\sigma)^{\alpha/2}\right)^{2/(\alpha-2)}\ ,\
\ee
such that
\be\label{gamavarclu}
\mean{\Gamma^2_n}=\Gamma_0^2 \ n\lr{1+\lr{\frac{n}{\bar n_2}}^{(2\beta-1)}}\rightarrow
 n^{2\beta}\text{ for }n\gg \bar n_2 \ .\
\ee
From the estimation of $\bar\sigma$ given by our simulations, we have $\bar \sigma \approx 5\times 10^4/(4096^2)\approx3\times10^{-3}$ (lattice units), and $\bar n_2\approx28$.
At first glance, one can think of $n\sim 10^4$ as a good asymptotic value. However, we will show below that $\bar n$ grows very fast with the moment order. Invoking the Gaussianity of $\tg(x)$ and the factorization property, one can write that
\be\label{smomsp}
\mean{\Gamma_n^{2p}}\sim n^p\lr{1+\lr{\frac{n}{\bar n_2}}^{(2\beta-1)}}^{p} \ .\
\ee
Restricting $\beta\geq1/2$, the binomial expansion can be applied to Eq.~\eqref{smomsp}, the dominant contribution for the limit $n\rightarrow \infty$ will be $\mean{\Gamma_n^{2p}}\sim n^{2\beta p}$. Moreover, the larger crossover scale (the scale where the asymptotics can be applied) is $\bar n_{2p}=\bar{n}_2^p$. For example, for two consecutive even-moment orders, we have $\bar n_{2p+2}\approx750\bar n_{2p}$ which is almost three orders of magnitude greater than the previous scale.

The progressive shrinking of the fitting range is highlighted by the right panel of Fig.~\reffig{fig4.9}, where we show the \ac{ESS} approach to the cluster summation scaling. The moment ratio seems to converge to the same as the binarized summation method visually seen. Moreover, the binarized moment ratio obeys the self-similar relation where $\zeta(p|2)=p/2$ is expected from Gaussian variables. These results advocate in favor of the Gaussianity --- at sufficiently large scales --- of the cluster summation. The remaining piece of information that is missing for the full characterization of this Gaussian is the variance scaling exponent.

Figure~\reffig{fig4.10} shows the scaling of the variance of the circulation among clusters (left panel) and the \ac{PDF} of clusters of size $n$ (right panel). There is a numerical coincidence between Eq.~\eqref{gamavarclu} and the power law $n^{5/4}$ for the numerical values of parameters $\bar\sigma$, $\alpha$, and, $\Gamma^2_0$ (shown in the lower inset of Fig.~\reffig{fig4.10}). The resulting scaling exponent fitted from the data is compatible with the ones of Eq.~\eqref{2momclus} with a numerical value of $\beta=0.686\pm0.018$ for the scaling exponent of $\mean{\tg(x_i)\tg(x_j)}$, and $\beta=0.615\pm0.011$ for the scaling exponent of $\mean{\Gamma^2_n}$. The former can be regarded as the ``true'' correlation function of the cluster summation while the latter is the fitted scaling exponent spoiled by the regularization. 

For a large enough cluster size, the scaling exponent of $\mean{n^p}$ is compatible with a monofractal scaling, expected by the Gaussianity of $\tg(x)$. The H{\"o}lder exponent of this monofractal is compatible with the prediction of the \ac{VGM} with correction due to the 4/5 law to the circulation. We highlight that the intermittent behavior of this field is partially due to non-Gaussian fluctuations of the intensity of each vortex structure at very small scales.
\begin{figure}[t]
\center
\includegraphics[width=\textwidth]{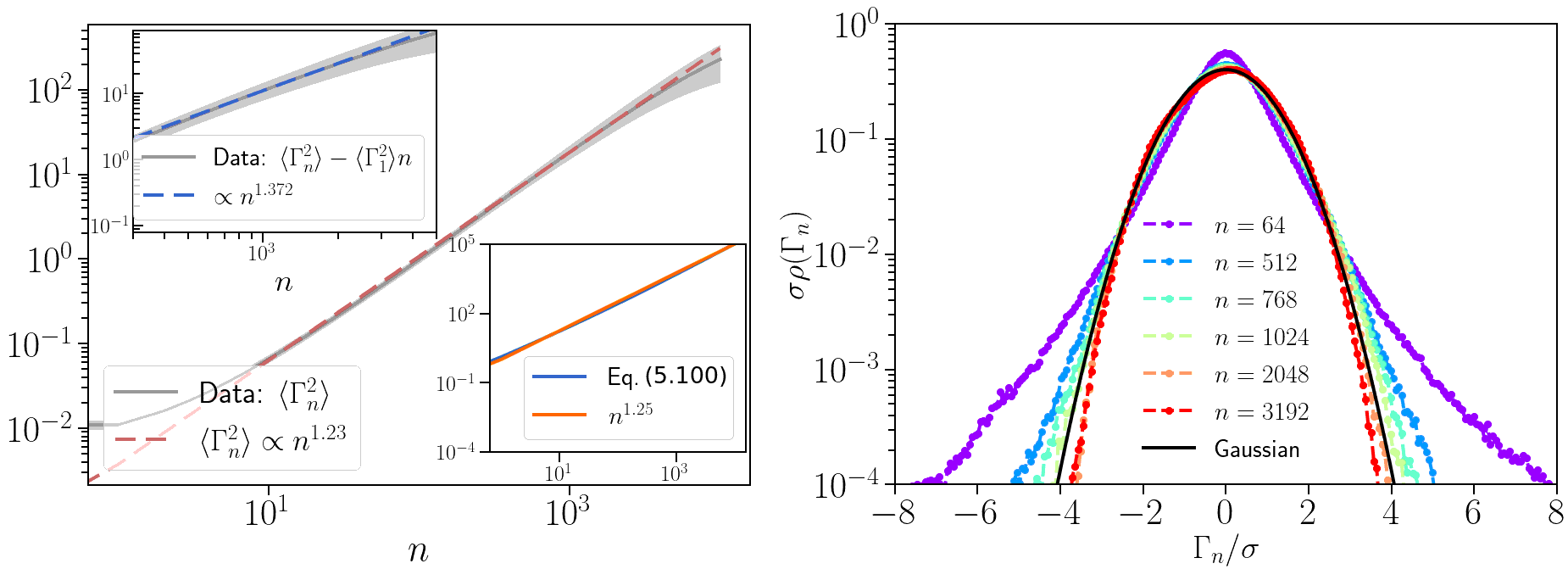}
\caption[Cluster summation variance and PDF.]{Left panel shows the circulation variance with (main frame), and without (upper inset) the regularization component and their respective power law fit. The lower inset shows the numerical coincidence between Eq.~\eqref{combined} and a power law $n^{5/4}$. The right panel shows the \ac{PDF} of the circulation calculated through the cluster summation.}
\label{fig4.10}
\end{figure}

The fundamentals of the \ac{VGM} revealed important structural components of the almost singular vortex structures that compose a turbulent flow. The current findings signalize fundamental differences between quantum and classical vortex tangles. For instance, the degree of vortex polarization in classical viscous fluids is found to be clearly incompatible with superfluid vortices: With $\mean{\Gamma^2_n}\sim n^{1.120\pm0.022}$, in comparison to the quantum scaling of $\sim n^{4/3}$.

Notably, we found decaying exponents to be compatible with the correction predicted by the \ac{GMC} approach introduced in the \ac{VGM} of $\langle \tg(x)\tg(y)\rangle\sim n^{2\beta}$ with $\beta=0.686\pm0.018$ in striking agreement with the theoretical value $\beta=2/3+\mu/8\approx0.688$. For the models based on the standard machinery of random cascades as $\vare^{1/3}$ cascade or \ac{mOK62}, for instance, the correction $\mu/8$ is not present, since the scaling exponents of the coarse-grained dissipation exactly vanish for $p=3$ due to the $\tau_{p}\longleftrightarrow\tau_{p/3}$ symmetry. Moreover, the bifractality of the turbulent circulation statistics is put into question as our findings fit all (and sometimes better) the circulation phenomenology by the use of the multifractality breakdown mechanism to the \ac{VGM}. The predicted  $\lambda_p\approx(4-\alpha)p/2=1.376p$ for $p\rightarrow0$ is in complete consonance with the data reported by \cite{iyer2019circulation}.

A natural way to proceed is to understand how intermittency and extreme events are interconnected through the direct perspective of the dynamical equations of fluid motion. From this point onward, we dedicate this thesis to the study of extreme circulation events, considering a functional approach to the statistics of noise-induced classical dynamics.

\end{chapter}

\begin{chapter}{Circulation Instantons}\label{cap5}

\hspace{5 mm} 

Instantons, in the context of stochastic processes\footnote{The reader who is more familiar with quantum field theory can give a rather different notion to the word Instanton. I ask for those, to stick to the definition given above as a simple nomenclature problem of different subject areas that got touched in the past.}, are the set of specific space-time realizations of some stochastic dynamics that generate observable values deviating significantly from the typical events of these dynamics. In general, such events are called ``extremes'' because they are very unlikely, lying in the tail of the \ac{PDF}s. The term ``extreme'' is also mathematically suitable since they are found to be the extrema (minima) of some generating functional.

The collection of techniques used to study extreme events is referred to as the theory of large deviations \cite{dembo2009large}. On the road to applying this technique to explore intermittent fluctuations of the velocity circulation in turbulent flows, we will first introduce a functional formalism that becomes very convenient for our approach.

\section{MSRJD Formalism}

$_{}$

The \ac{MSRJD} formalism was formulated in the mid-1970s by \cite{martin1973statistical,janssen1976lagrangean,dedominicis1976technics}. This formalism provides a representation of the mean value of any observable which satisfies an additive stochastic process as a functional integral measure. This mathematical tool bears several similarities to the standard Feynman path integration as the statistical weight on calculating averages is the exponential of an ``action''. However, it doesn't carry the same physical interpretation as the latter. This is because its associated ``Lagrangian density'' is not the familiar physical quantity --- kinetic energy minus potential energy---. Instead, it is directly linked to the underlying stochastic differential equations. We proceed on the derivation of this functional measure adopting a simplified and modern version of the \ac{MSRJD} action introduced by E. V. Ivashkevich \cite{ivashkevich1997symmetries}. Consider the following stochastic partial differential equation,
\be\label{5.1}
\partial_t v=b[v]+f \ ,\ 
\ee
where ${v(\vec{x},t):(\mathbb{R}^d\otimes[-T,0])\rightarrow \mathbb{R}^d}$, $b[v]$ is a functional of the $v(\vec{x},t)$ field, usually a time-local, nonlinear mapping from the space of $n$-vector functions to itself. Finally, $f=f(\vec{x},t)$ is a \ac{GFF} with zero-mean, used to model the energy pumping at large scales. The two-point correlation function of the forcing term is given by
\be\label{5.2}
\mean{f_i(\vec{x},t)f_j(\vec{y},s)}=\delta_{ij}\chi(|\vec{x}-\vec{y}|)\delta(t-s)\ .\
\ee
In general, the exact form of the spatial correlation $\chi(|\vec{x}-\vec{y}|)$ is not relevant since its correlation length $L$, is usually much larger than the length scales we are interested in. A precise definition of the correlation length depends on the specific application. In general, it refers to a length scale beyond which the correlation becomes small ($\chi(|\vc x| > L) \rightarrow 0$). In our particular case, we will define it in terms of the shape of the correlation function itself $L^2 = \chi(0) / \chi''(0)$ \cite{grafke2015instanton}. Finally to fully characterize the important properties of the correlation function, one can note that the ``energy'' injected into the system due to the stochastic forcing $f(\vec{x},t)$ is
\be\label{5.3}
\mean{v_i(\vec{x},t)f_i(\vec{x},t)}=\int_{\mathbb{R^d}} d^dy\  \chi(|\vec{x}-\vec{y}|)\frac{\delta v_i(\vec{x},t)}{\delta v_i(\vec{y},t)}=\chi(0)d   \ ,\
\ee
due to the Gaussianity of $f$. In the case of the \ac{iNSE}, Eq.~\eqref{5.3} is related to the energy balance required for statistical homogeneity with fixes $\chi(0)d=-\vare_0$. Once defined, the correlation function in Eq.~\eqref{5.2} completely characterizes the probability measure of the stochastic field $f(\vec{x},t)$. Thus, the expected values of any functional $O[v]$ can be calculated through the following functional measure,
\be\label{5.4}
\mean{O[v[f]]}=\int_{f(\vc{x},-T)=0} \frac{D[f]}{\mathcal{N}} \ O[v[f]]\exp\lr{-\frac{1}{2}\int_{-T}^0dt \norm{f}{\conv{\chi^{-1}}{f}}}\ ,\
\ee
where,
\be\label{5.5}
\mathcal{N}=\int D[f]\exp\lr{-\frac{1}{2}\int_{-T}^0dt \norm{f}{\conv{\chi^{-1}}{f}}}\
\ee
is an unimportant normalization constant (to be omitted hereafter).
\be\label{5.6}
\norm{a}{b}(t)=\int_{\mathbb{R}^d}d^dx \ a_i(\vec{x},t)b_i(\vec{x},t) \ ,\
\ee
is an $L^2$-based inner product, such that the $L^2$-norm is defined by $|a|_{L^2}^2=\norm{a}{a}$. We also define a spatial convolution in short notation by,
\be\label{5.7}
\conv{a}{b}(\vc x,t)=\int_{\mathbb{R}^d}d^dy\ a_i(\vec{x}-\vec{y},t)b_i(\vec{y},t) \ .\
\ee
From now on, the omission of integration intervals is made by simplicity. It is, in general, more convenient to use the direct correlation function than its inverse, thus we introduce a conjugate field by inserting a functional identity into Eq.~\eqref{5.4}:
\be\label{5.8}
\mean{O[v[f]]}=\int D[f]D[p] \ O[v[f]]\exp\lr{
-\int dt \lr{\frac{1}{2}\norm{p}{\conv{\chi}{p}}-i\norm{p}{f}}}\ .\
\ee
Note by integrating over $D[p]$ in the above equation, we get exactly Eq.~\eqref{5.4}. Now, by the use of Eq.~\eqref{5.1}, one can write realization of the stochastic process $f(\vc x,t)$ in terms of our desired field $v(\vc x,t)$, at the cost of a functional Jacobian
\be\label{5.9}
J[v]=\det\left[\frac{\delta f(\vc x ,t)}{\delta v(\vc y ,s)}\right]=\det\left[\delta_{ij}\delta^d(\vc x- \vc y)\partial_t\delta(t-s)-\frac{\delta b_i[v](\vc x,t)}{\delta v_j(\vc y,s)}\right]\ .\
\ee
The above operator determinant can be formally calculated in terms of $b[v]$ by writing it as follows, 
\be\label{5.10}
J[v]=\det\left[\frac{d}{dt}\lr{\delta_{ij}\delta^d(\vc x- \vc y)\delta(t-s)-\Theta(t-s)\frac{\delta b_i[v](\vc y,s)}{\delta v_j(\vc x,s)}}\right]\ .\
\ee
The extraction of a time delta function from the rightmost term in Eq.~\eqref{5.10} is due to the supposed locality of the functional $b[v]$. Now, one may note that the problem reduces to the computation of $\det(1+\hat{K})$ which, by the use of the identity $\ln{\det(\hat{A})}=\text{Tr}\ln(\hat{A})$, one can power expand the logarithm to show
\be\label{5.11}
\det(1+\hat{K})=\exp\lr{\sum_{n=1}^{\infty}\frac{(-1)^{n-1}}{n}\text{Tr}[K^n]}\ .\
\ee
We switched the computation of a functional determinant into all traces of powers of $\hat K$ or, in terms of $b[v]$,
\be\label{5.12}
\text{Tr}\left[\lr{\Theta\frac{\delta b[v]}{\delta v}}^n\right]=\int \lr{\prod_{m=1}^n d^dx_mdt_m} \Theta(t_1-t_2) \cdots
\Theta(t_{n-1}-t_{n}) \Theta(t_n-t_1)
\text{Tr}\left[\lr{\frac{\delta b[v]}{\delta v}}^n\right]\ .\
\ee
At first, it seems difficult to realize what is the advantage of this change, but as $\hat{K}$ has an explicit Heaviside function, it can be shown that the only nonzero contribution is $n=1$. Then the functional Jacobian will be related to the trace of an ill-defined quantity $\Theta(0)$, which is usually fixed by the choice of prescription of the stochastic process \cite{arenas2010functional}. Thus, we finally arrive at
\be\label{5.13}
J[v]=\det\left[\frac{d}{dt}\right]\exp\lr{-\Theta(0)\text{Tr}\left[\frac{\delta b[v]}{\delta v}\right]}\ .\
\ee
In the specific case of \ac{iNSE}, we have $b_i[v]=\nu\partial^2v_i-\lr{\delta_{ij}-\partial_i\partial_j\partial^{-2}}\lr{v_k\partial_kv_j}$ and the functional Jacobian results at
\be\label{5.14}
\text{Tr}\left[\frac{\delta b_i[v]}{\delta v_l}\right]=\int d^dx dt\Bigg(\nu \delta_{il}\partial^2\delta^d(0)-\lr{\delta_{ij}-\partial_i\partial_j\partial^{-2}}\lr{\delta_{jl}v_k\partial_k\delta^d(0)-v_j\partial_l\delta^d(0)}\Bigg)\ .\
\ee
The second term at Eq.~\eqref{5.14}  is linear in the velocity field, thus, by invoking homogeneity it must be zero. Since the remaining calculation of the Jacobian does not involve the dynamical field $v_i(x,t)$, it can be completely absorbed into the definition of the normalization constant at Eq.~\eqref{5.5}. 

From now on, we assume that the Jacobian does not produce any dynamic contribution to the path integration, as in the case of \ac{iNSE}. One can explicitly use Eqs.\eqref{5.1}, \eqref{5.8} and, \eqref{5.13} to switch the integration to the $v$ field. Moreover, an interesting way to use this approach is by calculating probability densities. By constraining the observation of a $v$-dependent observable $F[\vc v(\vc x,t)]$, to be developed by a noise-induced effect at $T \rightarrow -\infty$, one can measure an arbitrary value $F[\vc v(\vc x,0)]=a$ at the time $t=0$. Mathematically, it translates into the computation of,
\be\label{5.15}
\mean{\delta\big(F[v]-a\big)}=\rho(a)=\int D[v]D[p]\int_{-\infty}^{\infty} \frac{d\lambda}{2\pi} 
e^{
-S[v,p,\lambda]}\ ,\
\ee
where the \ac{MSRJD} action is defined by
\be\label{5.16}
S[v,p,\lambda]=\int dt \lr{\frac{1}{2}\norm{p}{\conv{\chi}{p}}-i\norm{p}{\partial_tv-b[v]}-i\delta(t)\lambda\lr{a-F[v]}} \ . \
\ee
The variable $\lambda$ can be regarded as a conjugate variable in the sense of the Legendre transforms, $\lambda(a)=dS(a)/da$. On the one hand, one may note that this statistical measure is Gaussian in the $p$-variable, meaning that it is integrable. On the other hand, it is usually highly nonlinear in the $v$-variables for most of the applications since $b[v]$ and/or $F[v]$ are usually nonlinear and/or singular functionals of $\vc v(\vc x,t)$. The extreme events are dominated by the saddle point solution of Eq.~\eqref{5.16} with $\delta S=0$ (see Appendix~\ref{appspa}), explicitly:
\be\label{5.17}
\frac{\delta S}{\delta p}=0 \Longrightarrow \partial_tv-b[v]=-i\conv{\chi}{p} \ , \
\ee
\be\label{5.18}
\frac{\delta S}{\delta v}=0 \Longrightarrow \partial_tp+p\frac{\delta b[v]}{\delta v}=-\lambda\delta(t)\frac{\delta F[v]}{\delta v(\vc x,t)} \ . \
\ee
Eqs.~\eqref{5.17} and \eqref{5.18} are the so-called ``instanton's equations'', a set of coupled nonlinear deterministic partial differential equations with vanishing conditions at $-T$. One may note the \ac{RHS} of Eq.~\eqref{5.18} to be singular in the time variable. This singularity can be regarded as homogeneous dynamics (with the \ac{RHS} equals zero) plus a final condition to be imposed to the solution $p(\vc x,t\rightarrow0^-)=\lambda\delta F[v]/\delta v(\vc x,0)$ by the integration of Eq.~\eqref{5.18} in an infinitesimal time interval around $t=0$.

Other constraints can be imposed on the instanton's equations though, an obvious one is a constraint related to the minimization of $\lambda$-variable in the action, relating $F[v]=a$. Additional dynamical constraints can be added to the theory by a similar strategy as the $a$-variable, by the change $D[v]\rightarrow D[v]\delta[\text{Constraint}]$. This is completely equivalent to the insertion of linear Lagrange multipliers to the action functional at Eq.~\eqref{5.16}. In the particular case of \ac{iNSE}, incompressibility condition means the addition of $i\norm{P}{\partial_ip_i}$ and $i\norm{Q}{\partial_iv_i}$ to the \ac{MSRJD} Lagrangian density, where $Q(\vc x, t)$ and $P(\vc x,t)$ are thought of as Pressure fields.

An interesting digression about the dynamical system of Eqs.~\eqref{5.17} and~\eqref{5.18} follows in terms of a constrained Hamiltonian formulation. Shortly, the transformation $p\rightarrow ip$ and $\lambda\rightarrow i\lambda$ let the \ac{MSRJD} action to $S=\int dt \lr{\norm{p}{\dot v}-H[p,v]}$ where $p$ and $v$ work as conjugate variables and, $H=\norm{p}{\conv{\chi}{p}}/2+\norm{p}{b[v]}$ has the interpretation of an Hamiltonian. There are two zero energy solutions, the trivial one $p=0$, is the solution of the deterministic system. The nontrivial solution $p=-2\chi^{-1}b[v]$ is related to time-reversed dynamics $\dot v=-b[v]$, which is only consistent with the evolution of $p$ if the nonlinear functional derivative $\delta b[v]/\delta v$ is self-adjoint as showed by \cite{grafke2015instanton}. Self-adjointness is satisfied if the nonlinear term $b[v]$ has potential formulation $b[v]=-\delta U[v]/\delta v$, which form a very special class of problems with well-defined asymptotic limits \cite{alqahtani2021instantons} where minimum action paths are a synonym for minimum energy paths.

Turning back to the extreme event problem, the probability density induced by the instanton solution $p_0,v_0,\lambda_0$ can be written as,
\be\label{5.19}
\rho(a)\propto
\exp\lr{\frac{1}{2}\int dt \norm{p_0}{\conv{\chi}{p_0}}}\ ,\
\ee
where the action is a function of the fixed multiplier $\lambda_0$ and the parameters of the dynamical system (Reynolds number in the case of \ac{iNSE}). The multiplier $\lambda_0=\lambda_0(a)$ is supposed to be invertible in terms of the observable $a$ by the use of the constraint $F[v_0(\vec x,0)]=a$. It is then trivial to note that, if $b[v]$ is a linear functional, the system is solvable with $v_0\propto \mathcal{L} p_0$ where $\mathcal{L}$ is some linear operator, thus both $v_0$ and $p_0$ are proportional to $\lambda_0$. The linear solution generates always a $\lambda_0$-quadratic probability, therefore, it does not mean the observable $a$ to be a Gaussian as the mapping $\lambda_0(a)$ can be highly nonlinear depending on the functional form of the observable $F[v]$.

\section{Circulation Instantons for iNSE}

$_{}$

Interested in velocity circulation probabilities of three-dimensional fluid flow, we choose $b_i[v]=\nu\partial^2v_i-v_j\partial_jv_i$, $F[v]=\oint_{\mc}dx_iv_i$, $a=\Gamma$ and, $\delta b_j[v]/\delta v_i=\nu\partial^2+v_j\partial_i+\delta_{ij}v_k\partial_k$. When all those transformations are applied to Eq.~\eqref{5.16}, one gets the following action,
\begin{align}\label{5.20}
    S&=\int dt \Bigg(\frac{1}{2}\norm{\vc p}{\conv{\chi}{\vc p}}-i\norm{\vc p}{\partial_t\vc v+(\vc v\cdot\vc \nabla)\vc v-\nu\nabla^2\vc v}+\\ \nonumber
    &+i\norm{Q}{\vc\nabla\cdot \vc v}+i\norm{P}{\vc \nabla\cdot \vc p}-i\delta(t)\lambda\lr{\Gamma-\oint_\mc v_idx_i}\Bigg) \ .\
\end{align}
As turbulence phenomenology states that turbulent flows are dependent only on the Reynolds number, this must be the case in the construction of the functional measure. An interesting way of explicitly showing the Reynolds dependency of the \ac{MSRJD} functional, is by defining $g^2=\chi(0)L^4\nu^{-3}=(UL/\nu)^3=Re^3$ and rescaling all the variables to:
$
x\rightarrow x' L;
t\rightarrow t' \tau_\nu;
v_i\rightarrow v_i' \nu/L;
P\rightarrow P' \nu^2/L^2;
\Gamma\rightarrow \Gamma' \nu;
\lambda\rightarrow i\lambda' /(\nu g^2);
p_i\rightarrow ip_i'/(\nu g^2L^2);
Q\rightarrow -iQ'\nu/(g^2 L)
$. The effect of this nondimensionalization is to produce a parameter-independent action,
\begin{align}\label{5.21}
    \tilde S&=\int dt' \Bigg(-\frac{1}{2}\norm{\vc p^{\ \prime}}{\conv{\tilde \chi}{\vc p^{\ \prime}}}
    +\norm{\vc p^{\ \prime}}{\partial_t\vc v^{\ \prime}+(\vc v^{\ \prime}\cdot\vc \nabla')\vc v^{\ \prime}-\nabla'^2\vc v^{\ \prime}}+\\ \nonumber
    &+\norm{Q'}{\vc\nabla'\cdot \vc v^{\ \prime}}
    -i\norm{P'}{\vc \nabla'\cdot \vc p^{\ \prime}}
    -\delta(t')\lambda'\lr{\Gamma'-\oint_{\mc'} v_i'dx'_i}\Bigg) \ ,\
\end{align}
where $\tilde\chi$ is the same correlation function as $\chi$ but with unit correlation length $\tilde\chi(0)/\tilde\chi''(0)=1$ and unit amplitude $\chi(0)=1$. The relation between Eqs.~\eqref{5.20} and \eqref{5.21} is simple as $S=g^{-2}\tilde S$. The explicit Reynolds number dependency of the rescaled action is not the only advantage of using it. In this rescaled version, the instanton's equations become parameter-free:
\be\label{5.22}
(\partial_t-\partial^2)v_i+v_j\partial_j v_i+\partial_iP=
\conv{\tilde\chi}{p_i}\ ,\
\ee
\be\label{5.23}
(\partial_t+\partial^2)p_i+ v_j\partial_j p_i+\partial_iQ=-v_j\partial_i p_j\ ,\
\ee
\be\label{5.24}
p_i(\vc x,t\rightarrow0^{-})=-\lambda\oint_\mc dy_i\delta^3(\vc x-\vc y)\ ,\
\ee
\be\label{5.25}
\Gamma[\mc]=\oint_{\mathcal{C}}v_i(\vc x,0)dx_i\ ,\
\ee
where primed variables are implicitly understood. Eqs.~\eqref{5.22} and \eqref{5.23} are subjected to the compressibility constraint. On one hand, Eq.~\eqref{5.22} is exactly the \ac{iNSE} with a dynamical forcing given by Eq.~\eqref{5.23}. On the other hand, dynamics of the auxiliary field $p_i$ is not exactly an \ac{iNSE} nor an advection-diffusion equation due to the term on the \ac{RHS}. It is still linear in the field $p_i$ but the constraints difficult an analytical exploration of the solution. Moreover, one might note that the sign of the diffusive term is positive. This implies that Eq.~\eqref{5.23} is a focusing parabolic one, requiring it to be propagated backward in time to accurately capture its numerically diffusive properties.

The particularities of the circulation variable appear in Eq.~\eqref{5.25}, where the contour shape fixes the spatial form of the auxiliary field $p_i$ at $t=0$, regarded as an initial condition to its backward propagation. From now on, we will fix the contour shape to a circle of radius $R$ due to its analytical simplicity. Without loss of generality, one can choose the contour to lie in the plane $z=0$ and centered at $r=0$, within this definition, it is clear that $p_3(\vc x,0^{-})=0$ from Eq.~\eqref{5.24}. Now define $l(\vc x)=p_1(\vc x,0^{-})+ip_2(\vc x,0^{-})$, in the complex plane we have
\be\label{5.26}
l(\vc x)=\lambda \delta(x_3)\oint_\mc \delta(x_1-\Re(w))\delta(x_2-\Im(w))dw \ ,\
\ee
where $w=y_1+iy_2$. Using polar complex coordinates $w=Re^{i\theta}$ and changing the cartesian variables $x_i$ to polar coordinates, the final result of these transformations gives,
\be\label{5.27}
l(\vc x)=i\lambda \delta(x_3)\delta(x_\perp-R)\int_0^{2\pi}\delta\lr{\arctan\lr{\frac{x_2}{x_1}}-\theta_w}e^{i\theta_w}d\theta_w\ .\
\ee
The remaining integral is almost tautological if one interprets $\theta_x=\arctan(x_2/x_1)$ as the polar angle of the $\vc x$ vector. The result is $e^{i\theta_x}=(x_1+ix_2)/x_\perp$ with $x_{\perp}=\sqrt{x_1^2+x^2_2}$. Taking the real and imaginary part we finally arrive at, 
\be\label{5.28}
p_i(\vc x,0^-)=\lambda\epsilon_{ji3}\frac{x_j}{R}\delta(x_\perp-R)\delta(x_3) \ .\
\ee
Using Eq.~\eqref{5.28} it is not difficult to show that $\partial_ip_i(\vc x,0^-)=0$ due to symmetric-antisymmetric contractions. This result is not an obvious result because when looking at Eq.~\eqref{5.24}, the contour integral is calculated in a singular field, invalidating Stokes' theorem. 

The specific choice of a circular contour motivates the search of solutions for Eqs.~\eqref{5.22}---\eqref{5.25} which obeys the same symmetries of $\mc$ since the singular forcing field in Eq.~\eqref{5.28} is driving the whole system. However, regarding the highly nonlinear and nonlocal nature of hydrodynamic interactions, there would be no a priori reason to consider axisymmetric solutions in opposition to any other. Nevertheless, we expect the symmetric solution to play an important role on the road to extreme events.

\subsection{Linear Instanton}

$\ \ \ \ $As the system of Eqs.~\eqref{5.22}---\eqref{5.25} has no known method to be solved, we have first worked with the linearized equations in order to analytically extract some information about the statistical properties of circulation. At first glance, the linear instanton may not seem too relevant in the context of extreme events. However, under certain conditions, it can reproduce the low-order statistics of circulation. This is particularly true for the dependence of variance on the typical length of the contour $\mathcal{C}$. To work out the linear instantons, we introduce the diffusion Green's function with vanishing boundary conditions in $\mathbb{R}^3$:
\be\label{5.29}
G(\vec{x},t)=\frac{1}{(4\pi|t|)^{-\frac{3}{2}}}\exp \left (- \frac{|\vec{x}|^2}{4|t|} \right ) \ . \ 
\ee 
The solutions $\lr{v_i^{(1)}, p_i^{(1)}, \lambda_0 }$ of the linearized problem can be written as follows,
\be\label{5.30}
v^{(1)}_i(\vec{x},t)=\int_{-T}^tds\int d^3yd^3z
G(\vec{x}-\vec{y},t-s) \tilde{\chi}(|\vec{y}-\vec{z}|)p_i^{(1)}(\vec{z},s)\ ,\
\ee
\be\label{5.31}
p^{(1)}_i(\vec{x},t)=\lambda_0\oint_{\mathcal{C}}dz_i\int_{t}^0
ds\int d^3y
G(\vec{x}-\vec{y},t-s)\delta(s)\delta^3(\vec{y}-\vec{z})\ ,\
\ee
note that the linearity of the equations implies the conservation of the incompressible initial/final condition of Eq.~\eqref{5.28}. Using Eqs.~\eqref{5.19} and \eqref{5.22}---\eqref{5.25} in the linear regime (simply discarding any nonlinear contribution to the dynamics), the instanton action is calculated explicitly in terms of the circulation,
\be\label{5.32}
S^{(1)}=\frac{1}{2g^2}\int dt \norm{\vc p}{(\partial_t-\nabla^2)\vc v}=
-\frac{1}{2g^2}\int dt \norm{\vc v}{(\partial_t+ \nabla^2)\vc p}=\frac{\lambda_0}{2g^2}\oint_\mc v^{(1)}_i(\vc x ,0)dx_i\ .\
\ee
A natural choice for the correlation function $\tilde{\chi}(r)$ is a Gaussian curve with a unit standard deviation. This choice simplifies most of the calculations through the use of the Gaussian convolution theorem. The limit of infinite waiting time, $T\rightarrow \infty$, is also applied. Under these conditions, the expression for the velocity field is reduced to:
\be\label{5.33}
v^{(1)}_i(\vec{x},t)=\lambda_0\pi\sqrt{\frac{R}{r}}\epsilon_{3ji}\frac{x_j}{r}
\int_0^{\sqrt{\frac{r R}{1-2t}}} du \ e^{-\left(\frac{R^2+r^2+z^2}{2rR}\right)u^2}I_1(u^2)\ ,\
\ee
where $I_1(x)$ represents the modified Bessel function of the first kind. It is important to highlight that, due to the nondimensionalization of the equations, time is measured using viscous time scales and lengths are proportional to the integral length scale. The integral in Eq.~\eqref{5.33} cannot be computed for a general spatial coordinate using standard functions. However, it has a closed form when calculated along the curve $\mc$,
\be\label{5.34}
\vec{v}^{(1)}(\mathcal C,t)=\lambda_0\frac{\pi}{6} \frac{R^3}{(1-2t)^{3/2}}\ {}_2F_2\left(\frac{3}{2},\frac{3}{2},\frac{5}{2},3;-2\frac{R^2}{1-2t} \right)\hat{\theta}\ ,\
\ee
where ${}_2F_2$ is the hypergeometric function of $2+2$ entries. We can define time-dependent circulation as follows,
\be\label{5.35}
\Gamma^{(1)}_R(t)=\oint_{\mathcal C}\vec{v}^{(1)}(\vec{x},t)\cdot d\vec{x}=\lambda_0\frac{\pi^2}{3} \frac{R^4}{(1-2t)^{3/2}}\ {}_2F_2\left(\frac{3}{2},\frac{3}{2},\frac{5}{2},3;-2\frac{R^2}{1-2t} \right)\ .\
\ee
At this point, the reader must be warned that the circulation $\Gamma[\mc]$ fixed by the Lagrange multiplier $\lambda_0$ is only $\Gamma^{(1)}_R(0)$. In this sense, the quantity in Eq.~\ref{5.35} has only statistical relevance when calculated at $t=0$, as explicitly shown by the rightmost term in Eq.~\eqref{5.32}. However, one can interpret it as the most probable time evolution which leads to the extreme event $\Gamma^{(1)}_R(0)$, fixed by the $\lambda_0$.

The linearity of the evolution equations and the circulation functional $F[v]$ means, in this particular case, that \ac{cPDF}s are Gaussian distributed being fully characterized by its second-order moment. Using Eq.~\eqref{5.32} and restoring dimensional units for the original variable, circulation's variance is written as follows,
\be\label{5.36}
\frac{\sigma^2}{\nu^2 Re^3}=\frac{\pi^2}{3} \frac{R^4}{L^4}\ {}_2F_2\left(\frac{3}{2},\frac{3}{2},\frac{5}{2},3;-2\frac{R^2}{L^2} \right)\ .\
\ee
We have, for $R \ll L$, 
\be\label{5.37}
{}_2F_2\left(\frac{3}{2},\frac{3}{2},\frac{5}{2},3;-2\frac{R^2}{L^2} \right) = 1 - \frac{3}{5} \lr{\frac{R}{L}}^2 + \mathcal{O}\lr{\frac{R}{L}}^4 \ , \
\ee
such that, in this asymptotic limit, 
\be\label{5.38}
\tilde \sigma_R^2 \equiv\frac{3 \sigma_R^2 L^4}{(\pi R^2)^2 \nu^2 {\hbox{Re}}^3} \simeq 1 +\mathcal{O}\lr{\frac{R}{L}}^2\ . \  
\ee
Although Eq.~\eqref{5.38} is expected to hold only for very viscous flows, it may be possibly relevant even for higher Reynolds number solutions,
once scales are probed and the \ac{MSRJD} path-integration is dominated by smooth velocity field configurations. A plausible guess that generalizes Eq.~\eqref{5.38} to turbulent flows sustained by alternative forcing mechanisms, where $L$ and Re are not defined in terms of Gaussian correlation functions, reads
\be\label{5.39}
\frac{3 \sigma_R^2 L^4}{(\pi R^2)^2 \nu^2 {\hbox{Re}}^3} \simeq {\cal{O}}(1) \ . \ 
\ee
To investigate the correctness of Eq.~\eqref{5.39}, we have worked this expression with the \ac{DNS} data of Table~\reftab{tab3.2}. The results reported in the left panel Fig.~\reffig{fig5.1} fully corroborate our expectation of Eq.~\eqref{5.39} for $R \sim \eta_K$. It should be noted that small-scale evaluations of $\sigma_R$ are subject to relatively significant errors, due to the inaccuracy associated with the representation of a circular contour in a square lattice, as a closed polygonal curve.

Figure~\reffig{fig5.1} also shows that the \ac{LHS} of Eq.~\eqref{5.39} leads, incidentally, to an excellent collapse of data across the inertial range scales $30 < R / \eta_K < 300$. Furthermore, the circulation variance has, in this range, a scaling dependence with $R$ which is very well approximated by the \ac{K41} scaling exponent $\sigma_R^2\sim R^{8/3}$, as indicated by a dashed line. Note also, for fixed integral scale and fixed energy injection rate per unit mass, $\vare_0 = U^3/L=Re^3\nu^3/L^4$, when combined with Eq.~\eqref{5.39} implies in $\sigma^2_R \propto \nu^{-1} R^4$. Supposing that an analogous result should hold in the inertial range due to the observed collapse, we replace $\nu$ in this last relation by the scale-dependent eddy viscosity \cite{smith1998renormalization},
\be\label{5.40}
\nu_R = \nu_0 \left ( \frac{R}{R_0} \right )^\frac{4}{3} \ , \
\ee
where $\nu_0$ and $R_0$ are some reference viscosity and length scale parameters, to obtain the observed scaling $\sigma^2_R \propto \nu_R^{-1} R^4 \propto R^{8/3}$.

The idea of ``turbulent viscosity'' was first presented by Boussinesq \cite{boussinesq1877essai} and formalized in the language of differential equations by \cite{kraichnan1964decay}. This subject is nowadays understood through the light of the dynamic renormalization group techniques \cite{forster1977large,yakhot1986renormalization}. The basic idea is that an effective transport equation arises from successive integration of the smaller degrees of freedom. Under certain hypotheses put forward by \cite{smith1998renormalization}, this procedure leads to a diffusion equation with scale-dependent viscosity $\nu(k)\propto k^{-4/3}$, where $k=|\vec{k}|$ is the length of a wave vector within the inertial range. This is the main physical motivation for a number of numerically tractable models of turbulence, as the celebrated Smagorinsky sub-grid formulation of the Navier-Stokes equations \cite{smagorinsky1963general,scotti1993generalized}, which underlies the whole field of large eddy simulations \cite{germano1991dynamic,mason1994large,piomelli1999large,piomelli2014large}.
\begin{figure}[t]
\center
\includegraphics[width=\textwidth]{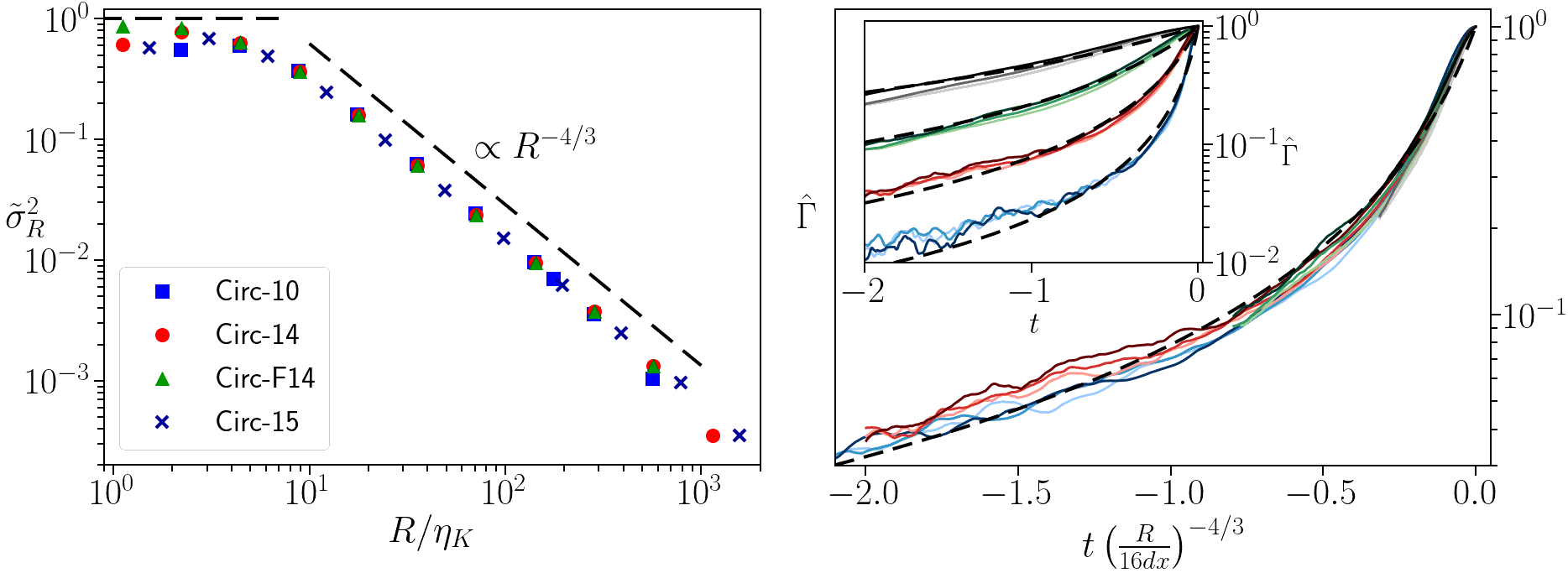}
\caption[Circulation variance scaling and filtered instanton.]{Left panel shows circulation's variance as functions of the scale $R$. The right panel shows filtered circulation events for $R/dx=8,16,32,64$ in blue, red, green, and gray shades, respectively, and the linear instanton decay as a dashed black line. Lighter to darker colors mean $\bar \Gamma_R(0)=n\sigma_R$ for $n=2,3,4$. Both figures can be found in \cite{apolinario2022eddy} with slight modification.}\label{fig5.1}
\end{figure}

The replacement of the molecular viscosity $\nu$ by the eddy viscosity $\nu_R$ in Eq.~\eqref{5.36} sounds at this point like a purely rhetorical remark. It is possible, however, that the linear instantons, when combined with eddy viscosity ideas, can in fact be used to model the time evolution of high Reynolds number circulation's instanton. The argument goes as follows. Consider the set of all flow realizations which have evolved from the remote past and ended with extreme circulation 
\be\label{5.41}
\Gamma_R \in [ \bar \Gamma_R(0) - \delta \Gamma, \bar \Gamma_R(0) + \delta \Gamma] \ ,\
\ee
at time $t=0$, where 
$\bar \Gamma_R(0) \gg \delta \Gamma > 0$ ($\delta \Gamma$ is just a measurement bin). Let now $\bar \Gamma_R(t)$ represent the time-averaged circulation taken over the conditioned ensemble of all flows satisfying Eq.~\eqref{5.41}, which we refer to as the {\it{filtered instanton}}. 
Resorting to dynamic similarity and to the fact that $\eta_K/L \ll 1$ (i.e., Reynolds number is high), we write down the dimensionless circulation ratio, units are restored from now on,
\be\label{5.42}
\hat \Gamma_R(t) \equiv \frac{\bar \Gamma_R(t)}{\bar \Gamma_R(0)} = f\left(\frac{t}{\tau_\nu}, \frac{R}{L}, \frac{\eta_K}{ L}\right) \simeq 
f\left(\frac{t}{\tau_\nu}, \frac{R}{L}, 0\right) \equiv \tilde f\left(\frac{t}{\tau_\nu}, \frac{R}{L}\right)  \ . \
\ee
where we used the identity $\eta_K/L=Re^{-3/4}$ and the limit of large Reynolds numbers. It follows from Eq.~\eqref{5.42},
\be\label{5.43}
\hat \Gamma_R\lr{\frac{\nu_R t} {\nu}} \simeq \tilde f\left(\frac{t}{\tau_{\nu_R}}, \frac{R}{L}\right) \ .\
\ee
In line with the phenomenology of turbulent viscosity, we assume that Eq.~\eqref{5.43} is scale-invariant, i.e.,
\be\ \label{5.44}
\frac{d}{dR} \tilde f\left(\frac{t}{\tau_{\nu_R}}, \frac{R}{L}\right) =  0 \ , \
\ee
note that the Eq.~\eqref{5.43} is a renormalization group flow equation \cite{smith1998renormalization}, valid in the inertial range, and has as a general solution the functional relationship:
\be\label{5.45}
\tilde f\left(\frac{t}{\tau_{\nu_R}}, \frac{R}{L}\right) \equiv h\left(\frac{t}{\tau_\nu}\right) \ . \
\ee
and from Eq.~\eqref{5.43}, we get,
\be\label{5.46}
\hat \Gamma_R(t) \simeq
\tilde f\left(\frac{t}{\tau_{\nu}}, \frac{R}{L}\right) =h\left(\frac{t'}{\tau_{\nu}}\right) \ ,\
\ee
with,
\be\label{5.47}
t' = t \frac{\nu}{\nu_R} = t \frac{\nu}{\nu_0} R^{-\frac{4}{3}} \ . \ 
\ee
Where analogously, we have for the solution of viscous instanton Eq.~\eqref{5.35}
\be\label{5.48}
\hat \Gamma^{(1)}_R(t) \equiv  \frac{\Gamma^{(1)}_R(t)}{\Gamma^{(1)}_R(0)} = g\left(\frac{t}{\tau_{\nu}}, \frac{R}{L}\right) \equiv 
\frac{1}{(1-2t/\tau_\nu)^{3/2}}
\frac{{}_2F_2\left(\frac{3}{2},\frac{3}{2},\frac{5}{2},3;-2\frac{R^2/L^2}{1-2t/\tau_\nu} \right)}{{}_2F_2\left(\frac{3}{2},\frac{3}{2},\frac{5}{2},3;-2\frac{R^2}{L^2} \right)} \ , \
\ee
consequently,
\be\label{5.49}
g\left(\frac{t}{\tau_{\nu}}, \frac{\eta_K}{L}\right)
\simeq g\left(\frac{t}{\tau_{\nu}}, 0\right) \equiv \tilde g(t/\tau_\nu) =\frac{1}{(1-2t/\tau_\nu)^{3/2}} \ , \
\ee
considering the validity of Eq.~\eqref{5.46} to dissipative scales, we conjecture that the viscous and filtered instantons are equal, leading us to
\be\label{5.50}
\tilde f(t/\tau_{\nu_R}, R/L) =  \tilde f(t/\tau_{\nu}, \eta_K/L)=h(t/\tau_\nu) \simeq   \tilde g(t/\tau_\nu)  \ , \
\ee
therefore, $\hat{\Gamma}_R(t)$ can be modeled by the viscous instanton solution $\tilde{g}(t'/\tau_\nu)$, i.e.
\be\label{5.51}
\hat \Gamma_R(t) = \hat \Gamma^{(1)}_{0}(t') =
\frac{1}{(1-2t'/\tau_\nu)^{3/2}}
\frac{{}_2F_2\left(\frac{3}{2},\frac{3}{2},\frac{5}{2},3;0;-2\frac{R^2/L^2}{1-2t'/\tau_\nu}
\right )}{{}_2F_2\left(\frac{3}{2},\frac{3}{2},\frac{5}{2},3;0 \right)} \ . \
\ee
In order to explore the validity of Eq.~\eqref{5.51}, we have implemented a numerical procedure of filtering defined by Eq.~\eqref{5.41}.

The only possible dataset for the exploration of time evolution is the Iso-1024 Dataset. We calculated the time series of circulation for circular contours centered at equally spaced grid points for a whole turnover time. Letting $dx$ be the lattice parameter ($dx\approx 2.1 \eta_K$), we have studied circulation fluctuations for four different radii, namely, $R= 8dx$, $16dx$, $32dx$, and $64dx$. For each given radius, we considered $32^3$ contours oriented normally to each Cartesian direction, for a total of $3 \times 32^3$ contours per radius.

We conventionally define the set $\Lambda_n$ of extreme events as the ensemble of circulations events whose absolute values $|\Gamma_R|$ reach some multiple $n$ of the circulation standard deviation, $\sigma_R=\sqrt{\langle \Gamma_R^2 \rangle}$, within a small tolerance window. In the language of Eq.~\eqref{5.41}, we consider the interval defined by $\bar\Gamma_R(0) = n \sigma_R$ with a $0.5\%$ tolerance, that is, {\hbox{$\delta\Gamma = 5 \times 10^{-3} \bar\Gamma_R(0)$}}.

Once an extreme circulation event belonging to a given set $\Lambda_n$ is identified, we assign it the observation time instant $t=0$ and save its earlier time evolution. Then, a time-dependent average $\bar \Gamma_R(t)$ over all saved series for each set $\Lambda_n$ is computed. Similar filtering procedures have been applied in instanton studies of Burgers turbulence \cite{grafke2013instanton}, Lagrangian turbulence models \cite{grigorio2017instantons,apolinario2019instantons}, and rogue wave formation \cite{dematteis2018rogue,dematteis2019experimental}. 

Our results for $\hat \Gamma_R(t)$ are shown in the right panel of Fig.~\reffig{fig5.1}. Taking $R_0 = 16 dx$ as an arbitrary reference length scale, we find that plots of $\hat \Gamma_R(t)$ as a function $t(R/16 dx)^{-4/3}$ collapse reasonably well for all the investigated radii and sets $\Lambda_n$, as predicted by Eq.~\eqref{5.46}. Following now Eq.~\eqref{5.51}, we carry out an $L^2$-norm minimization of
\be\label{5.52}
\sum_R \int\lr{\hat\Gamma_R\lr{\frac{\nu_R t}{\nu}} - \hat \Gamma^{(1)}_0(t)}^2 dt\ , \ 
\ee
to adjust $\nu_0 = 4.5 \times 10^{-5} \nu$ as the reference viscosity in Eq.~\eqref{5.40}.

It is surprisingly unexpected that the particular symmetry of the linear solution on the rescaling $\nu\rightarrow\nu_R$ while $t\rightarrow R^{-4/3}t$ is also a symmetry of the nonlinear solution. The fitted curve based on Eq.~\eqref{5.51}, plotted as dashed lines, seems to be better suited for the tail i.e. far times than in the vicinity of the measured circulation $t=0$. This implies that the decaying shape of these events follows roughly an algebraic decay $t^{-3/2}$, but the concave approach at $t=0$ is more connected to the nonlinearity of the equations.
\begin{figure}[t]
\center
\includegraphics[width=\textwidth]{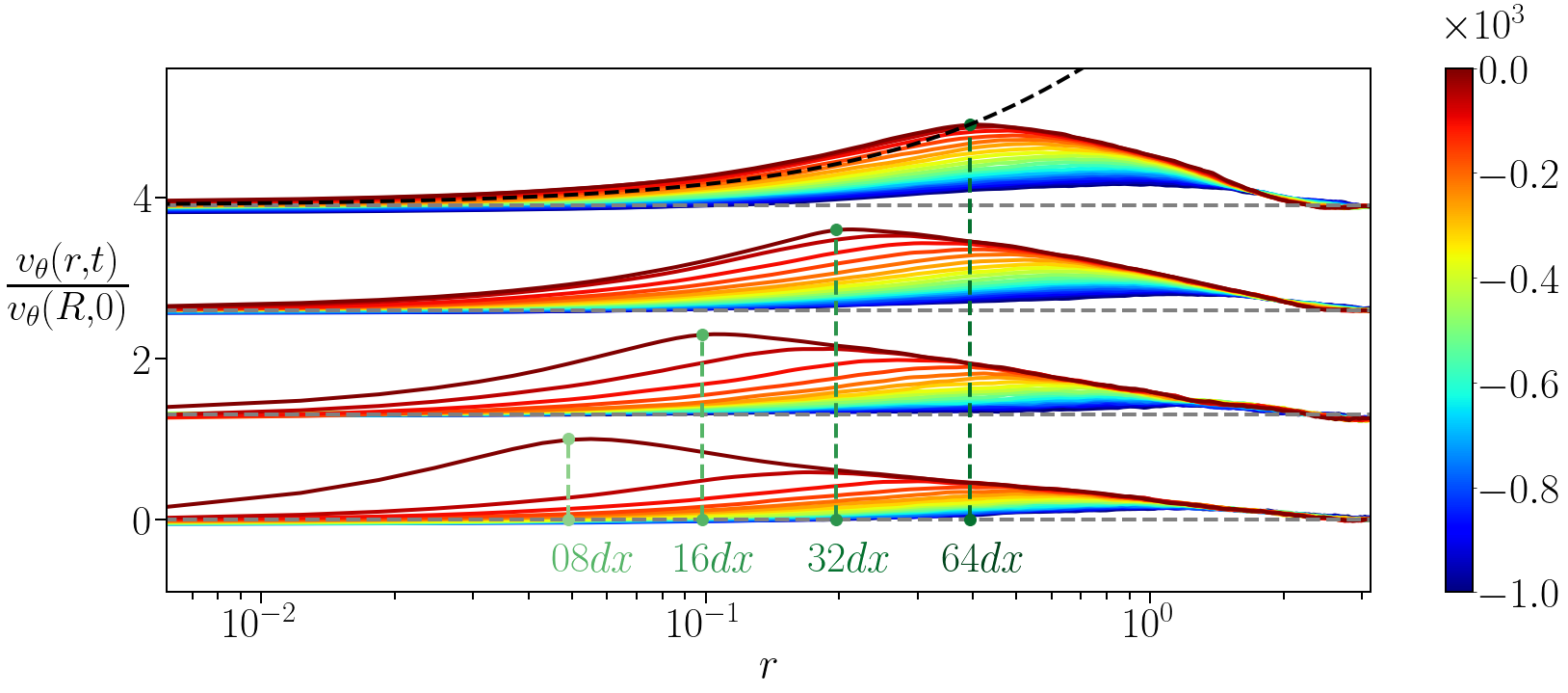}
\caption[Filtered velocity.]{Figure shows the time evolution of the filtered velocity field resulting in the filtered circulation's instanton in $\Lambda_4$ for $R/dx=8,16,32,64$, from the bottom to the top. The time scale (colors) is measured in units of simulation time steps. Horizontal dashed lines indicate zero since the images are vertically shifted. Green vertical dashed lines are $r=R$. Black dashes show the numerical integration of Eq.~\eqref{5.33} for $R=64dx$ and $t=0$. This figure is a variation of the one found in \cite{apolinario2022eddy}.}\label{fig5.2}
\end{figure}

In addition to the above discussion, one might wonder if the functional forms of the filtered velocity field related to its corresponding filtered circulation instanton could also be effectively modeled by viscous instantons. With this aim, we take the symmetry axis of each prescribed circular contour as the $z$-axis of a local Cartesian reference system and oriented it in such a way that velocity circulations are rendered positive at the event occurrence time. Then, we perform a similar time-dependent averaging procedure over all events in $\Lambda_n$, as we did for $\hat\Gamma_R(t)$, but now for velocity field configurations in cylindrical coordinates, $\vec v = v_r \hat r + v_\theta \hat \theta + v_z \hat z$.
Due to isotropy, one expects the mean velocity components to be functions of $r$, $z$, and $t$ only, i.e., $\langle v_r \rangle \equiv \bar v_r(r,z,t)$, $\langle v_\theta \rangle \equiv \bar v_\theta(r,z,t)$, and $\langle v_z \rangle \equiv \bar v_z(r,z,t)$. 

Unfortunately, the radial and axial components $\bar v_r(r,z,t)$ and $\bar v_z(r,z,t)$ turn out to develop relatively small intensities, which prevents us from extracting a clear behavior out of the noise within our relatively limited data. The mean azimuthal component $\bar v_\theta(r,0,t)$, on the other hand, was found to be well-resolved in all studied cases. Fig.~\reffig{fig5.2} shows comparisons between the numerically filtered $\bar v_\theta(r,t)$ (normalized by $\bar v_\theta(R,0)$) and the analogous normalized azimuthal velocity obtained from the viscous solution Eq.~\eqref{5.33}. It is seen that the eddy viscous modeling only provides an adequate description of $\bar v_\theta(r,t)$ for $r \leq R$, which is just the core region of the filtered instantons. Notwithstanding such a limitation, extreme circulations events are in fact well accounted for by eddy viscous modeling as shown by the intersection among the darker greenish dashed line, black dashed line, and the data for $v(R=64dx,t=0)$.

Conversely, as the filtering results indicate, $\bar v_\theta(r,0)$ gets its maximum value at $r=R$, as expected from the singularity of Eq.~\eqref{5.28}. Explicit calculation of $\partial_rv_\theta(r=R,t=0)$ can be made to show using Eq.~\eqref{5.33} and, it turns out that the unique positive root of this equation happens only if $R\approx 2.4$ which seems to be inconsistent with the observed peak of the filtered velocity (Fig.~\reffig{fig5.2}).

To further investigate the detailed structure of the circulation instantons moving beyond the viscous solutions, one must deal with the full nonlinear instanton equations Eqs.~\eqref{5.22}---\eqref{5.24}, this issue will be numerically addressed in the next subsection. 

\subsection{Nonlinear Axisymnetric Instanton}

$\ \ \ \ $A convenient numerical scheme to solve the hydrodynamic instanton's equations, similar to Eqs.~\eqref{5.22}---\eqref{5.24}, was introduced by Chernykh and Stepanov \cite{chernykh2001large} in the context of Burgers' turbulence. In our case, we adopted the original scheme of Chernykh and Stepanov sketched in the left panel of Fig.~\reffig{fig5.3} and resumed as follows. Step $1$: at the iteration step $\bar n$, a given velocity field $v^{(\bar{n}-1)}_i(\vc x,t)$ is inserted at Eq.~\eqref{5.23} to solve $p_i^{(\bar n)}(\vc x,t)$ backward in time, from $t=0$ to $t=-T$, using the numerical approach put forward in Appendix~\ref{appspectral}. The initial condition is given by Eq.~\eqref{5.28}. The final time $-T$ must be as close as possible to $p_i^{(\bar n)}\approx0$, in other words, the initial condition must get completely diffused in the space domain. Step $2$: the $p^{(\bar{n})}_i(\vc x,t)$ solution defines the forcing field of Eq.~\eqref{5.22} which is solved forward in time, also with the numerical approach of Appendix~\ref{appspectral}. Step $3$: define a prescription of convergence, in our case, we define it to be the relative difference of velocity circulation,
\be\label{5.53}
\left|\frac{\Gamma^{(\bar n)}-\Gamma_{\bar n-1}}{\Gamma_{\bar n-1}}\right|<10^{-4} \ ,\ 
\ee
where $\Gamma^{(\bar n)}=2\pi Rv^{(\bar n)}_\theta(R,z=0,t=0)$. Step $4$: return $\Gamma^{(\bar n)},v_i^{(\bar n)},p_i^{(\bar n)}$ if the convergence condition is fulfilled or goes to the next cycle $\bar n+1$, otherwise.

The right panel of Fig.~\reffig{5.3} shows three different kinds of convergence for the proposed algorithm. One may note that all the evolutions do not start at zero because we have used previous numerical simulations to start the scheme. For example, as a sequence of Lagrange multipliers $\{\lambda_i\}$, we start the simulation for $\lambda_i$, with the final numerical solution of $\lambda_{i-1}$. The latter scheme speeds up the convergence of the code, especially for high values of $\lambda$.

In Fig.~\reffig{5.3}, we denote by ``smooth'' convergence, the typical one observed for low $\lambda$, where it rapidly reaches the final value in the first steps. The velocity fields associated with this convergence share several similarities with the linear solution. Secondly, the relaxed convergence means the solution starts to behave nonlinearly such that the previous numerical solution inserted to speed up the code needs a relaxation process before properly converging. Lastly, the divergent is an example of a simulation that did not converge in the Chernykh-Stepanov sense.
\begin{figure}[t]
\center
\includegraphics[width=\textwidth]{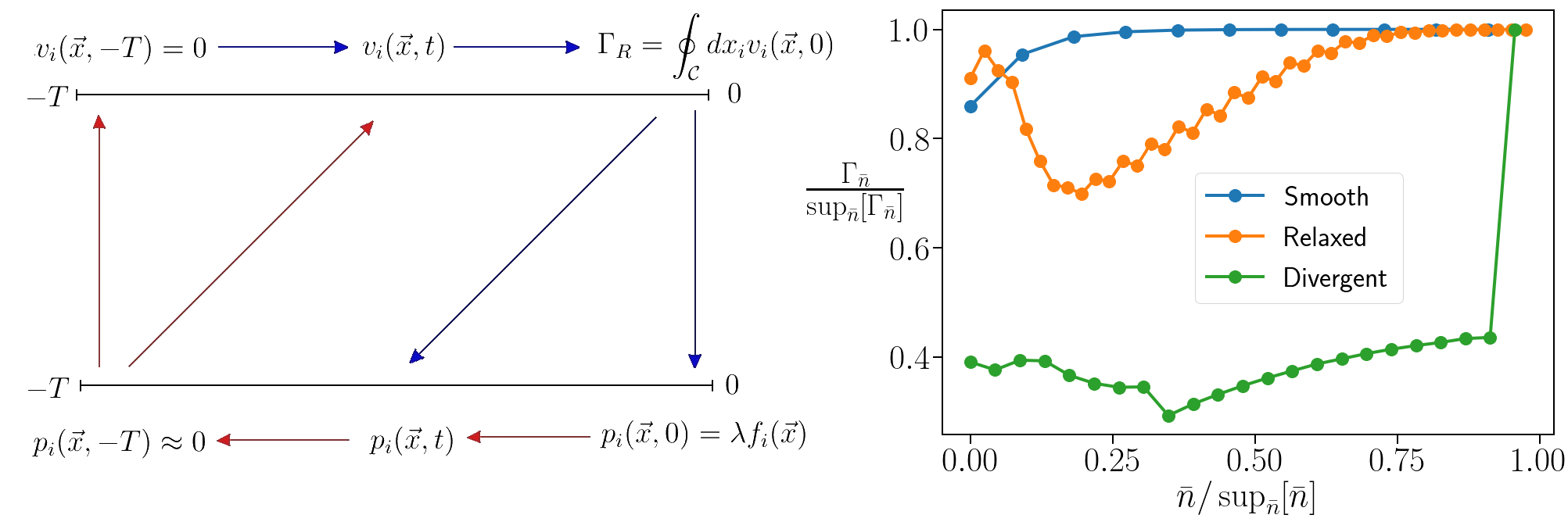}
\caption[Chernykh-Stepanov algorithm and convergence.]{Left panel shows a sketch of Chernykh-Stepanov's algorithm. The right panel shows the evolution of velocity circulation as functions of $\bar n$. As different scales were probed, we normalized all the curves by the final circulation value to ease comparison.}\label{fig5.3}
\end{figure}

One may call attention to the oscillations present in the relaxed convergence. This numerical artifact was already noted by \cite{chernykh2001large} and its nature relies on oscillations between two local minima. A simple way to cope with this problem is to substitute $v^{(\bar n)}\rightarrow \alpha v^{(\bar n)}+(1-\alpha)v^{(\bar n-1)}$, with $\alpha\in(0,1)$ before starting the Step $3$. This substitution does not affect the fixed point of the cycle, it just effectively slows down the cycle step ensuring better convergence at the cost of more cycles needed for the convergence. In fact, the simulation shown in Fig.~\reffig{fig5.3} is already using this substitution in an adaptive fashion to be explained as follows. Consider the following variable,
\be\label{5.54}
K(\bar n)=\lr{\frac{d \Gamma^{(\bar n)}}{d\bar n}}^{-1}\frac{d^2 \Gamma^{(\bar n)}}{d\bar n^2} \ ,\
\ee
if $K(\bar n)=0$ it is a saddle point (in the pure sense), then $\alpha$ is still the same. $K(\bar n)<0$ means the curvature and the tangent have opposite signs, it is an indication of convergence so that $\alpha$ is still the same again. Lastly, $K(\bar n)>0$ means an oscillation, in this case, we substitute $\alpha\rightarrow0.995\alpha$. The sequence of simulations usually starts with $\alpha=0.7$ for the first $\lambda$ in the sequence $\{\lambda_i\}$. For the simulations shown in Fig.~\reffig{5.3}, the final values were $\alpha=0.7,0.47,0.1$ in the smooth, relaxed, and divergent cases, respectively.
\begin{table}[t]
\centering
\caption[Instanton's parameter table.]{Parameters table of the instanton simulations. The reasoning for the dataset labels relies upon the contour radius. For instance, Inst-R06 refers to $R=\cos(\pi (N_r/2-6)/(N_r-1)$.}\label{tab5.1}
\begin{tabular}{|c|c|c|c|c|c|c|c|c|} \toprule
    {Name} & {$R/L$} & {$N_r$} & {$N_z$} & {$N_t$} & {$dt$} & {$\lambda_{min}$} & {$\lambda_{max}$} \\ \midrule
    Inst-R03 & $0.097$  & $256$ & $128$ & $400$ & $3.75\times10^{-3}$ & $5\times10^4$ & $1\times10^6$\\
    Inst-R06 & $0.213$  & $256$ & $128$ & $400$ & $3.75\times10^{-3}$ & $1\times10^4$ & $2\times10^6$\\
    Inst-R09 & $0.328$  & $256$ & $128$ & $400$ & $3.75\times10^{-3}$ & $1\times10^4$ & $3\times10^6$\\
    Inst-R12 & $0.443$  & $256$ & $128$ & $400$ & $3.75\times10^{-3}$ & $1\times10^4$ & $3\times10^6$\\ \bottomrule
\end{tabular}
\end{table}

The matter of instanton method is to solve the equations to find a solution $S(\lambda)$ and a map $\lambda=\lambda(\Gamma)$ such that one can express \ac{cPDF} as $\ln(\rho(\Gamma))\sim -S(\lambda(\Gamma))$. The existence/uniqueness of such a solution heavily relies upon the existence of the cumulant generating function (see Appendix~\ref{appspa}). In fact, if the action is any nonconvex function of $\Gamma$, the cumulant generating function is ill-defined and the instanton method does not converge to the correct asymptotic limit \cite{alqahtani2021instantons}. Putting forward a practical example, let us suppose one can solve the path integration by other methods and the result gives, for $\Gamma>0$,
\be\label{5.55}
S(\Gamma)\propto\begin{cases}
    \Gamma^2 &,\ \text{ if } \Gamma<\Gamma_0 \ , \\
    2\Gamma_0\Gamma &,\ \text{ if } \Gamma\geq\Gamma_0 \ .\
\end{cases}
\ee
This function is discontinuous only on the second derivative and, is nonconvex since it has infinitely many inflection points for $\Gamma>\Gamma_0$. From the definition of Lagrange multipliers, it follows,
\be\label{5.56}
\lambda=\frac{dS}{d\Gamma}\propto\begin{cases}
    2\Gamma &,\ \text{ if } \Gamma<\Gamma_0 \ , \\
    2\Gamma_0 &,\ \text{ if } \Gamma\geq\Gamma_0 \ . \
\end{cases}
\ee
The function $\lambda$ cannot be inverted in terms of $\Gamma$ since the mapping is not one-to-one after $\Gamma_0$. What happens if we naively simulate the governing instanton equations for this system for $\lambda>2\Gamma_0$? How could we know if we are working with nonconvex problems since we don't know it beforehand?

These questions are old issues on the instanton's approach an only recently were addressed by the work of Alqahtani and Grafke \cite{alqahtani2021instantons}. The authors reformulated the large deviation principle for nonlinear Legendre transforms in order to force the convexification of the action. In practical terms, to adopt this procedure is the same as replacing $\lambda_0\rightarrow \lambda_0 df(\Gamma)/d\Gamma$ at Eq.~\eqref{5.33}, where $f(\Gamma)$ is some nonlinear function defining the nonlinear Legendre transform. In most cases studied by \cite{alqahtani2021instantons}, the choice $f(\Gamma)=\ln(\Gamma)$ showed up to convexify the action, this is due to the fact that the inverse function $f^{-1}$ (exponential) is heavily convex.
\begin{figure}[t]
\center
\includegraphics[width=\textwidth]{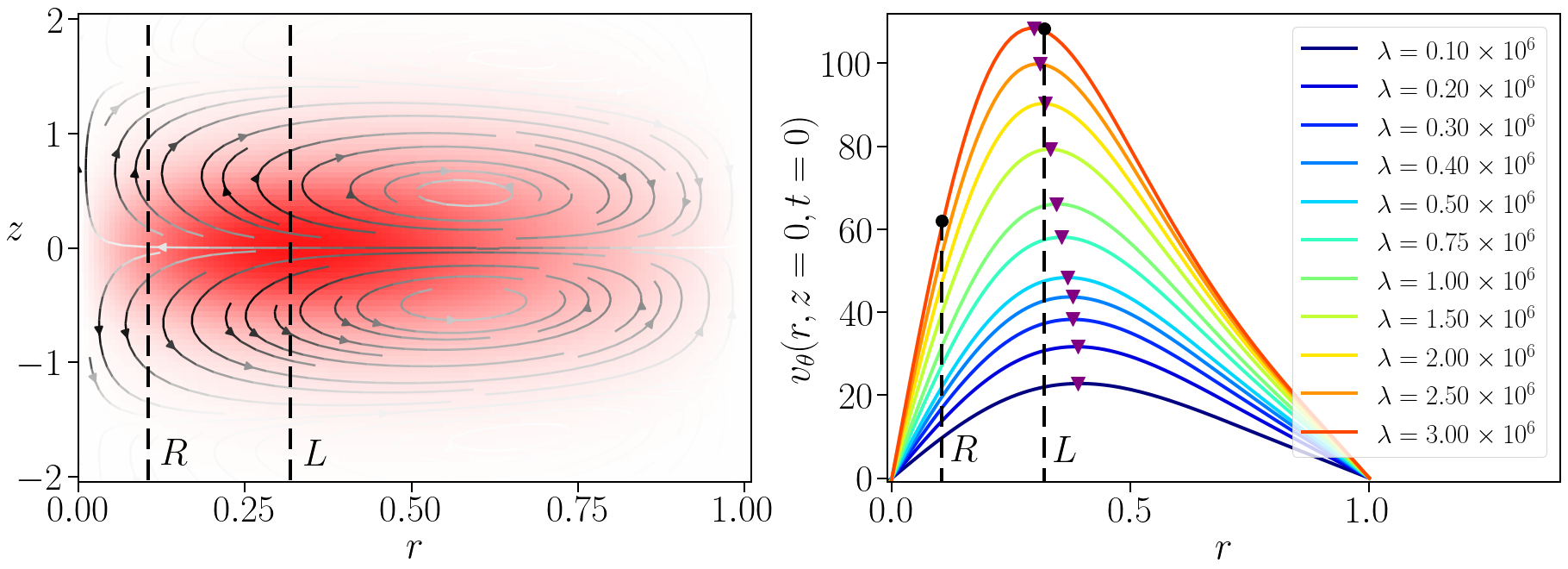}
\caption[Instantons' topology.]{Left panel shows the typical velocity field topology associated with the circulation instanton for the dataset Inst-R06 (see Table~\reftab{tab5.1}). The right panel shows the monotonic drift in the peak position towards the contour radius also for Inst-R06.}\label{fig5.4}
\end{figure}

After all the above-mentioned considerations, we have worked out solutions for the circulation instantons in cylindrical coordinates using the integration method described in Appendix~\ref{appspectral}. A broad range of $\lambda$'s was simulated for four contour radii with fixed integral length $L=\pi^{-1}$ (more can be found at Table~\reftab{tab5.1}). Dirichlet boundary conditions at $|r|=1$ are imposed and all the fields are $z$-periodic. The extremization of the minimization was carried out by the use of Chernykh-Stepanov's algorithm outlined above, enhanced at larger values of $\lambda$, by the convexification function $f(\Gamma)=\ln(|\Gamma|)$ in order to avoid possible spoiling inflection effects.

The left panel of Fig.~\reffig{fig5.4} shows the typical topology of the velocity field, which can be seen as a triple vortex structure. The main vorticity is aligned with the $z$-axis and it is responsible for the $v_\theta$ field. In addition, two ring-shaped vortex structures can be seen in the $r\otimes z$ plane as closed streamlines. Moreover, instanton's velocity field does not reproduce the alignment between the peak position of $v_\theta$ and the contour radius, showing only a slow tendency towards alignment as the field intensity $\lambda$ grows (right panel of Fig.~\reffig{fig5.4}). Rough estimates for the value of $\lambda$ which satisfies $\sup_{\vec{x}} [v_\theta]\approx v_\theta (R,z=0)$ gives $\lambda=10^{10}$ which is far beyond the computational limitations of the present methodology.

We find, thus, that the direct numerical schemes presented here were only able to address the parabolic cores of \ac{cPDF}s. It is then compatible with $S(\Gamma)\approx \Gamma^2/2\sigma^2 + \mathcal{O}(\Gamma^3)$ for $\sigma \approx 6.43 \times 10^{-5} \times (R/L)^{1.98}$, which is actually close to the result of the viscous action. The description of the \ac{cPDF}'s tails remains a numerically challenging problem to tackle. A promising strategy to circumvent the stiffness of the instanton equations is space-time reparametrizations \cite{grafke2014arclength,grafke2015instanton,grigorio2020parametric} and minimization techniques other than the fixed point Chernykh-Stepanov's method \cite{wright2006numerical}. Reparametrization techniques are shown to have several computational advantages in the case of one-dimensional equations: Burgers, Ornstein–Uhlenbeck, $\phi^4$-potential, and Lagrangian passive scalar, for instance. While augmented Lagrangian methods (which avoid the computation of the $\lambda\longleftrightarrow\Gamma$ mapping) are shown to be computationally efficient in graphical cards since memory-saving schemes were recently developed \cite{schorlepp2022spontaneous}.

The only other case when three-dimensional hydrodynamic instanton is addressed, to the author's knowledge, is related to vorticity and strain observables \cite{schorlepp2022spontaneous}. The authors adopted a similar numerical technique to solve instanton's equations with and without the axisymmetric assumption. They reported a spontaneous break in the continuous rotational symmetry along the preferential axis, for sufficiently extreme values of $\omega_n\gtrsim 85$, $\partial_nv_n\gtrsim 14$, and $\partial_nv_n\lesssim-38$. The continuous symmetries are found to be substituted for 2-fold or 4-fold symmetries depending on how extreme the event is. This observation suggests some effects of cartesian grid discretization \cite{thampi2013isotropic}, but the authors presented no study on this issue. If one believes symmetry breaking to be a genuine instanton property, its consequences for circulation can dramatically spoil numerical computations. For example, the transition between symmetric and nonsymmetric solutions might depend nontrivially on $R$ being very short for most of the contour scales.

The fact that we could only address the Gaussian cores of the \ac{cPDF}s hinders the understanding of the vortical structures which composes extreme events of circulation. The robustness of the typical topology of the velocity field presented in Fig.~\reffig{fig5.4} and its topological compatibility with vorticity instantons found by \cite{schorlepp2022spontaneous} in the axisymmetric regime, points into the direction that extreme circulation events are composed by single vortex structures in contrast with the discussion put forward in Section~\refsec{sec4.4}. These connections must be further investigated through the application of accurate, efficient, and scalable numerical methods to the instantons equation, preferentially on large GPU clusters since these computations are usually stiff.

\end{chapter}

\begin{chapter}{Conclusion}\label{cap6}

\hspace{5 mm} 

In this thesis, two distinct approaches to the statistics of velocity circulation in three-dimensional homogeneous and isotropic turbulence have been pursued. The first approach introduces a vortex gas model (VGM) that, for the first time, unifies the structural approach to turbulence with the multiplicative nature of energy cascade fluctuations. The second approach employs functional methods to uncover connections between the time evolution of extreme circulation events, eddy viscosity, and the underlying topology of velocity fields leading to extreme circulation events.

The two fluctuating fields comprising the \ac{VGM} are $\xidr$ (a \ac{GMC} field)  and $\tg_R$ (a \ac{GFF}). These fields reveal intriguing physical interpretations when examined on sufficiently large scales, as supported by processed data from \ac{DNS} at exceedingly high Reynolds numbers. The \ac{GMC} field is associated with the mean number of vortical structures inside the circulation contour $\mc$, while the Gaussian field is related to the partial polarization of vortices in the same domain. These interpretations enable a reexamination of the phenomenon of linearization of circulation's scaling exponent $\lambda_{p\rightarrow\infty}\approx h_{\infty}p+D_\infty$, where the \ac{VGM} presents a phenomenological picture of a bounded \ac{GMC} approach inspired by hard-disk-like interactions at small scales. This interpretation is further corroborated by numerical data.

The use of \ac{GMC} in place of the standard \ac{MCM} has measurable implications for vortex polarization. A correction of $\mathcal{O}(10^{-2})$ is predicted for its scaling exponent and it was accurately addressed through the cluster summation procedure. This detection is of summary importance in distinguishing circulation statistics in classical and quantum turbulence. While these two systems manifest statistical distinction in terms of vortex spatial distribution and polarization at inertial range scales, they converge when evaluating circulation statistics, which emerges roughly as a product between the number of structures ($\xidr$) and the partial polarization ($\tg_R$).

Moreover, the \ac{GMC} approach introduces a field-theoretical framework, different from the standard \ac{MSRJD} approach, to the calculation of \ac{cPDF}s. The resulting \ac{PDF} obtained from the \ac{GMC} approach is a Gaussian curve with fluctuating variance according to the specific realization of the $\xidr$ field. Conversely, the \ac{cPDF} obtained from the standard \ac{MSRJD} approach is related to the entire dynamics of the system, where its extreme events are dominated by the instantons partially addressed in Chapter~\refcap{cap5}. The connection between these two field theoretical frameworks is still a matter of further investigation.

Reflecting on the points \ref{p331}---\ref{p338} discussed at the end of Chapter~\refcap{cap3}, we have engaged with several of them with the formulation of the \ac{VGM}. Particularly \ref{p332} and \ref{p337} are addressed on their totality, and the points \ref{p335} and \ref{p338} are partially discussed. Point \ref{p332} was addressed by computing the mean-field at \ac{GMC}'s field theoretical approach to \ac{cPDF}s. Point \ref{p337} is related to the breakdown of multifractality, where we provide expressions to $h_\infty$ and $D_\infty$, such that its Reynolds dependency is a natural consequence of the dependency of the hard disk exclusion radius on $\eta=a\eta_K\sim R_\lambda^{-3/2}$ (for fixed integral length), affecting the maximum packing fraction. The phenomenological nature of the inertial range plateau in \ref{p335} is still yet unknown, but the asymptotic limits ($R\gg\eta_K$ and $R\ll\eta_K$) are accurately described by the \ac{VGM}, through analytical formulation and hard-disk simulations.

Several other points remain unanswered on the topic of turbulent circulation statistics. The first natural question that arises in the \ac{VGM} is how to generalize it for non-planar loops and how the minimal area plays its role (or not) in the model. A first try at this generalization to substitute the integration measure in $d^2x\rightarrow d^2x \sqrt{\det(\hat{g})}$ where $\hat{g}$ is the metric tensor related to the surface curvature. However, the analysis of this system, in the numerical sense, is not an easy task to do, and the comparison to \ac{DNS} data requires very precise data. In this sense, the data traffic limitations of \ac{JHTDB} are a central source of limitations. In the ideal case, an accessible database (or an own generated) where the circulation is calculated in all points of the simulation can be used for the comparison of \ac{DNS} data to Monte-Carlo simulations of non-planar \ac{VGM}.

A particularly interesting contour shape can shed light on the role of the area law of circulation. The catenoid surface can be generated through a rotation of a catenary curve. This surface is the only non-planar minimal surface of revolution. Then if the circulation depends only on the minimal area, there is a completion between the area of the catenoid surface $A_{cat}$ and the combined area of two disks of radius $R$. This is due to the fact that circulation can be calculated as the sum of the vorticity of the ``lateral'' area (the catenoid) or, equivalently the area of the upper and lower cover (two disks). As the separation of the disks $H$ grows, $A_{cat}$ grows while $2\times \pi R^2$ is $h$ independent. In this sense, one expects the existence of a $H_0$ where $A_{cat}(H_0)=2\pi R^2$, then the area of the disks dominates. This fact can be regarded as a ``phase transition'' and the exploration of this particular setting can be very interesting for the phenomenology of the area law.

The understanding of circulation in turbulent flows can be an alternative way of representing turbulence phenomenology. As circulation naturally interplays between vorticity and velocity differences, it can offer a multidescriptive interpretation of results which is not possible by the sole use of velocity structure functions. However, there is a long way to go before the statistics of circulation get spread throughout the turbulence research community and others. Our findings claim cooperation between different areas of knowledge such as numerical and optimization methods, statistical physics, complex systems, field theory approaches, and many others. Moreover, the main contribution of this thesis is the fact that it can serve as introductory material for researchers or students who want to tackle the problem of circulation statistics in turbulent flows.

\end{chapter}



\newpage
\phantomsection
\addcontentsline{toc}{chapter}{Bibliographycal References}

\printbibliography[title={Bibliographycal References}]


\appendix
\begin{chapter}{Saddle Point Approximation}\label{appspa}

$_{}$

Saddle point, Laplace's method, Steepest descent, Instanton, Stationary phase method, and many others names refer to a specific approximation method to calculate integrals. Consider integrals of the following form,
\be\label{A1}
I_\epsilon=\int_{x_i}^{x_f}\ dx f(x)e^{-S(x)/\epsilon^2}\ , \
\ee
for $f,S:\mathbb{R}\rightarrow \mathbb{R}$, analytic functions in the domain $x_i\leq x\leq x_f$ and $\epsilon>0$. The saddle point method provides an approximation for Eq.~\eqref{A1} in the limiting case of $\epsilon\rightarrow 0$ to be associated with the properties of the function $S(x)$. Let $x_0$ be the global minimum of $S(x)$ inside the integration interval, i.e. $S'(x_0)=0$ and $S''(x_0)>0$, centering the minimum $x=x_0+\epsilon y$ we get,
\be\label{A2}
I_\epsilon=\epsilon\int_{y_i}^{y_f}dy \ f(x_0+\epsilon y)e^{-\epsilon^{-2} S(x_0+\epsilon y)}\ . \
\ee
As $x_0$ is the minimum\footnote{The solution $x_0$ for the minimization problem is usually referred to as the saddle point solution. Indeed this is not a saddle point in the pure sense since $S''(x_0)\neq0$.} of $S(x)$, the exponential factor $e^{-\epsilon^{-2} S(x)}$ is maximized around $x_0$, meaning that the exponential decays quickly to zero for points $x$ that are sufficiently far from $x=x_0$. With the later remark, one can power expand the functions $f,S$ in the $y$-variable around $y=0$,
\be\label{A3}
\epsilon^{-2} S(x)\approx \epsilon^{-2} S(x_0)+\frac{S''(x_0)}{2}y^2+\frac{S'''(x_0)}{6}\epsilon y^3+\mathcal{O}\lr{\epsilon^2 y^4}\ , \
\ee
and for $f(x_0)\neq0$
\be\label{A4}
f(x)\approx f(x_0)\lr{1+
\frac{f'(x_0)}{f(x_0)}\epsilon y+
\frac{f''(x_0)}{2f(x_0)}\epsilon^2 y^2
+\mathcal{O}\lr{\epsilon^3 y^3}}\ . \
\ee
Using Eqs.~\eqref{A2},~\eqref{A4} and,~\eqref{A4}, one gets
\be\label{A5}
I_\epsilon\approx\epsilon f(x_0)e^{-\epsilon^{-2} S(x_0)}\int_{y_i}^{y_f}\ dy e^{-S''(x_0)y^2/2}
\lr{1+
\frac{f'(x_0)}{f(x_0)}\epsilon y+
\frac{f''(x_0)}{2f(x_0)}\epsilon^2 y^2+
\mathcal{O}\lr{\epsilon^3 y^3}}\ . \
\ee
Note that the third-order term in Eq.~\eqref{A3} was omitted in Eq.~\eqref{A5}. The former is linear in $\epsilon$, leading to $e^{ay^3}\approx 1+ay^3$ which does not contribute up to $\mathcal{O}(y^2)$.

It is convenient to set $x_i\rightarrow-\infty$ and $x_f\rightarrow\infty$ since we will employ this method for the calculation of probability-based integrals. In this case, it is clear that only even orders contribute to the integral since it is the integration of symmetric Gaussian fluctuations around $x_0$. In this case, we write Eq.~\eqref{A5} as follows,
\be\label{A6}
I_\epsilon=\epsilon f(x_0)e^{-\epsilon^{-2} S(x_0)}\sqrt{\frac{2\pi}{S''(x_0)}}
\lr{1+
\sum_{n=1}^\infty A_{2n}\epsilon^{2n}}\ . \
\ee
where all the coefficients $A_{2n}$ are given in terms of the derivatives of the functions $f(x)$ and $S(x)$ at the saddle point $x_0$, for example, $A_2=f''(x_0)/(2f(x_0)S''(x_0))$. 

An interesting application of this technique is related to the large deviation principle. This principle states that the probability of observing a rare event decays exponentially, and its exponential scaling is given by the minima of a corresponding rate function $I(x)$ \cite{dembo2009large}. This theory has its basins on the Gärtner-Ellis theorem \cite{ellis1984large}, which is a powerful mathematical result that establishes the existence of a large deviation principle for processes where the cumulant generating function tends towards a well-behaved limit, in other words, if
\be\label{A7}
G(\lambda)=\lim_{\epsilon\rightarrow0}\epsilon^2 \ln\lr{\mean{e^{\lambda x/\epsilon^2}}}\ ,\
\ee
exists for each $\lambda\in \mathbb{R}$ and is differentiable in $\lambda$. Then, there exists a rate function $I(x)$, such that the tails of the probability measure associated with Eq.~\eqref{A7} decays as,
\be\label{A8}
P^\epsilon \sim e^{-\epsilon^{-2}\inf_x[I(x)]}\ , \
\ee
as $\epsilon\rightarrow 0$. The saddle point approximation appears in this context to provide the exact formulation of the rate function. Consider Eq.~\eqref{A1} with $f(x)=e^{\lambda x /\epsilon}$ and $S(x)=-\epsilon^2\ln(\rho(x))$, i.e. the probability measure can be expressed in terms of an analog of ``action''. Then,
\be\label{A9}
G(\lambda)=\lim_{\epsilon\rightarrow0}\epsilon^2 \ln\lr{I_\epsilon}\ , \
\ee
using Eq.~\eqref{A6} we get,
\be\label{A10}
G(\lambda)=\lambda x_0-S(x_0)+\lim_{\epsilon\rightarrow0}\epsilon^2 \ln\lr{\epsilon\sqrt{\frac{2\pi}{S''(x_0)}}
\lr{1+\sum_{n=1}^\infty A_{2n}\epsilon^{2n}}}=\lambda x_0-S(x_0)\ , \
\ee
which is well-behaved in the $\lambda$ variable. In the definitions we made, the rate function and the ``action'' function $S(x)$ have the same meaning.

Generalization of the above discussion for more complex tensor structures and space domains is straightforward in the case of analytic functions. For example, $f, S:\mathbb{C}^N\rightarrow \mathbb{R}^M$ is analytic with $N,M\in\mathbb{Z}$ one can define a well-suited inner product such that the conclusions are the same \cite{bender1999advanced}. This is not the case when the extremum of $S$ is a degenerate saddle point, defined as $\partial_i S(z_0)=0$, $\det[\partial_i\partial_j S(z_0)]=0$, for $z_0\in \mathbb{C}^N$ --- a real saddle ---, the solution to this problem relies on a mathematical formulation called as the catastrophe theory which is out of the scope of our applications \cite{poston2014catastrophe}.

In the context of functional approaches as in Chapter~\refcap{cap5}, as the number of degrees of freedom diverges, Eq.~\eqref{A6} involves the determinant of differential operators, which are usually problematic. However, one may note that the G{\"a}rtner-Ellis Theorem is blind to such a correction since the limit $\epsilon\rightarrow0$ (far tail limit) is taken. What has been done in the literature, is to stop the expansions of Eqs.~\eqref{A3} and \eqref{A4} at first order without integrating the fluctuation around the saddle point solution. The fluctuations were integrated only in very special cases as in Burgers' equation \cite{apolinario2019onset}, for instance.

\end{chapter}

\begin{chapter}{Swirling Strength Criterion}\label{appssc}

$_{}$

Vortex identification is a classic topic in the turbulence literature, especially active in wall-bounded flows \cite{chong1990general,jeong1995identification,zhou1999mechanisms,chakraborty2005relationships,wu2006population,elsas2017vortex,chen2018contributions,zhang2018selected}. Here, we adopt a very simple and the widely diffused identification method, the Swirling Strength Criterion \ac{SSC} \cite{zhou1999mechanisms,chakraborty2005relationships}.

The \ac{SSC} consists in computing the eigenvalues of the velocity gradient tensor $A_{ij}(\vc x) = \partial_j v_i(\vc x)$, associating vortices to regions where the imaginary part of an eigenvalue is non-zero. Of course, since $A_{ij}$ is a $3\times 3$ traceless tensor, its eigenvalues fulfill the condition $\lambda_1=-\lambda_2-\lambda_3$. Moreover, since the velocity field is real, the related characteristic polynomial has real coefficients implying, for example, if $\lambda_3=\lambda_{cr}+i\lambda_{ci}$ is an eigenvalue of $A_{ij}$, then $\lambda_3^\star$ also is. Combining the latter remark with the traceless constraint, we get $\lambda_1=-2\lambda_{cr}$ and $\lambda^\star_2=\lambda_3=\lambda_{cr}+i\lambda_{ci}$, in this sense, $\lambda_{cr}$ and $\lambda_{ci}$ fully characterizes the eigenvalues of $A_{ij}$.

The eigenvalues of the velocity gradient tensor are of fundamental importance on the \ac{SSC} as already mentioned. The rationale behind the method is that the evolution of the tracer particles following frozen streamlines follows $\dot x_i = A_{ij}(\vc x_0) (x-x_0)_j$ when linearized. In this sense, complex eigenvalues are related to spiraling orbits. The criterion then defines vortex structures as domains where the $\lambda_{ci}$ field is non-zero. A threshold on $\lambda_{ci}$ is often employed to avoid noise and get smoother results. In the main text, we not only followed \cite{wu2006population,chen2018contributions}, adopting the threshold $\lambda_T=1.5\sigma_{\lambda_{ci}}$, where $\sigma_{\lambda_{ci}}$ is the standard deviation of $\lambda_{ci}$, but we also used a smaller one $\lambda_T=0.125\sigma_{\lambda_{ci}}$, which is better suited for the computations in Sections~\refsec{sec4.2} and \refsec{sec4.4}.

We are interested in 2D slabs of 3D turbulent velocity fields. In this case, the calculation of the velocity gradient determinant is simplified since
\be\label{B1}
\det(\partial_jv_i-\lambda\delta_{ij})=\det(\partial_jv_i)+\det(-\lambda\delta_{ij})+\text{Tr}(-\lambda\delta_{ij})\text{Tr}(\partial_jv_i)-\text{Tr}(-\lambda\delta_{ik}\partial_jv_k)\ .\
\ee
The above identity is only valid for $2\times2$ matrices, the resulting characteristic polynomial associated with this determinant is
\be\label{B2}
\lambda^2-\lambda\text{Tr}(\hat{A})+\det(\hat{A})=0 \ , \
\ee
whose solution is, in general,
\be\label{B3}
\lambda_{\pm}=\frac{\text{Tr}(\hat{A})\pm\sqrt{(\text{Tr}(\hat{A}))^2-4\det(\hat{A})}}{2} \ . \
\ee
Note that, in this case, $\text{Tr}(\hat{A})\neq 0$ since we are dealing only with the components projected into the plane. This means that $\text{Tr}(\hat{A})=-\partial_zv_z$, where $v_z$ is the normal component of the velocity field and the trace of the 2D velocity gradient tensor is related to the normal strain rate of the local flow. 

The definition of $\lambda_{ci}$ in addition to Eq.~\eqref{B3} can be interpreted as follows: The vortex regions detected by the \ac{SSC} are the ones where the plane vorticity (defined by the antisymmetric part of $\hat{A}$ also related to its determinant) is much higher than the normal strain rate. The most suitable condition for the existence of a $\lambda_{ci}$ is $-4\partial_xv_y\partial_yv_x>(\partial_xv_x-\partial_yv_y)$ which is usually satisfied in the latter statement. However, it is also satisfied in other special cases depending on the local strain rate. This observation is the key aspect of a generalization of the \ac{SSC} put forward by \cite{elsas2017vortex} usually referred to as $\lambda_\omega$ criterion.

\end{chapter}
\begin{chapter}{Spectral Approach for the iNSE in Cylindrical Coordinates}\label{appspectral}

$_{}$

In order to numerically solve the set of Eqs.~\eqref{5.22}---\eqref{5.25} in the axisymmetric regime we rewrite it in cylindrical coordinates,
\be\label{C1}
\begin{cases}
    \hat{\mathcal{L}}^-v_r+v_r/r^2-v_\theta^2/r+\partial_rP=(\tilde{\chi}\star p_r)\ , \\
    \hat{\mathcal{L}}^-v_\theta+v_\theta/r^2+v_\theta v_r/r=(\tilde{\chi}\star p_\theta)\ , \\
    \hat{\mathcal{L}}^-v_z+\partial_zP=(\tilde{\chi}\star p_z)\ , \\
    \partial_r(rv_r)/r+\partial_zv_z=0\ , \\
    \hat{\mathcal{L}}^+p_r-p_r/r^2-v_\theta p_\theta/r+v_r\partial_r p_r+v_\theta\partial_r p_\theta+v_z\partial_r p_z+\partial_rQ=0\ , \\
    \hat{\mathcal{L}}^+p_\theta-p_\theta/r^2+(2v_\theta p_r-v_rp_\theta)/r=0\ , \\
    \hat{\mathcal{L}}^+p_z+v_r\partial_z p_r+v_\theta\partial_z p_\theta+v_z\partial_z p_z+\partial_zQ=0\ , \\
    \partial_r(rp_r)/r+\partial_zp_z=0\ , \ 
\end{cases}
\ee
where $\hat{\mathcal{L}}^\pm\psi=\partial_t\psi+v_z\partial_z\psi+v_r\partial_r\psi\pm(\partial_z^2\psi+\partial_r(r\partial_r\psi)/r)$. We have implemented a pseudo-spectral approach where the fields are expanded in a truncated Fourier-Chebyshev series of orders $N$ and $M$, respectively. The collocation points are defined as
\be\label{C2}
z_j=\frac{2\pi j}{N} \ , \ 
\ee
and
\be\label{C3}
r_m=\cos{\left(\frac{\pi m}{M-1}\right)}\ , \ 
\ee
where $j$ and $m$ are integers in the ranges $j \in [-N/2,N/2-1]$ and $m \in [0,M-1]$. Along the longitudinal direction, usual collocation points for Fourier series were used (Eq.~\eqref{C2}), while in the radial direction, we employed Gauss-Lobatto collocation points defined at Eq.~\eqref{C3}, corresponding to the extrema of the $M^{th}$ order Chebyshev polynomials \cite{canuto2007spectral}.
This choice of a non-homogeneous grid hinders a direct comparison among simulations with different resolutions. However, it eases the implementation of boundary conditions in the radial direction \cite{peyret2013spectral}. Moreover, for even $M$ the coordinate singularity at $r=0$ is explicitly avoided on the collocation, with the drawback of requiring the radial interval to be duplicated, as $r \in [0,1]\rightarrow r\in [-1,1]$, as a requirement of the Chebyshev polynomial expansion.

All fields in Eq.~\eqref{C1} are given by a truncated polynomial series of order $N$ in the Fourier basis and $M$ in the Chebyshev basis,
\be\label{C4}
\phi(r_m,z_j)\equiv\phi_{mj}=\sum_{l=0}^{M-1}\sum_{k=-N/2}^{N/2-1}\hat{\phi}_{lk}T_l(r_m)e^{ikz_j}\ , \ 
\ee
where $T_l(x)$ is the $l^{th}$ Chebyshev polynomial. Exceptions to this discretization are the pressure fields $Q(\vec{x},t)$ and $P(\vec{x},t)$, which are of order $M-2$ in the Chebyshev expansion. This approach is known as the $\mathbb{P}_N-\mathbb{P}_{N-2}$ approximation having several advantages in numerical stability \cite{peyret2013spectral}. Radial derivatives are related to a linear transformation computed using recursion relations of the Chebyshev polynomials, namely,
\be\label{C5}
(\partial_r\phi)_{mj}=\sum_{l=0}^{M-1}D_{ml}\phi_{lj}\ , \ 
\ee
where the $M\times M$ matrix $D$ has, for all fields but the pressure-related ones, the entries
\be\label{C6}
D_{mj}=
\begin{cases}
    (-1)^{m+j}c_m/(c_j(r_m-r_j)) &\ , \  m\neq j\ ,\\
    -r_m/(2(1-r_m^2)) &\ , \  m = j, \ m,j\neq [0,(M-1)] \ ,\\
    \pm (2(N-1)^2+1)/6 &\ , \  m=j=0 \text{ or } m=j=(M-1),
\end{cases}
\ee
with $c_0=c_{M-1}=2$ and $c_i=1$ otherwise. As for the pressure fields, which are 2 degrees lower in the Chebyshev expansion, they must be interpolated from the $M-2$ Gauss-Lobatto grid to the $M$ grid where the velocity fields are defined. This can be done through a Lagrange interpolation (see Ref.~\cite{peyret2013spectral} for details),
\be\label{C7}
D^{PQ}_{mj}=
\begin{cases}
    (-1)^{j+m}(1-r_j^2)/((r_m-r_j)(1-r_m^2)) &\ , \  m\neq j, \ m,j \in [1,M-2]\ ,  \\
    3 r_m/(2(1-r_m^2)) &\ , \ m=j, \ m \in [1,M-2] \ .\
\end{cases}
\ee

By virtue of the explicit form of these matrices, $r$-derivatives are evaluated in physical space while $z$-derivatives are better performed in Fourier space. Ultimately, in order to consistently represent all fields in the mirrored radial domain $r\in [-1,1]$, the fields must satisfy the following reflection properties:
\be\label{C8}
\begin{cases}
    \phi_{mj}=\phi_{(-m)j} \text{ for scalar fields,} \\
    \phi_{z,mj}=\phi_{z,(-m)j} \text{ for the }z\text{-component of vector fields,} \\
    \phi_{\beta,mj}=-\phi_{\beta,(-m)j} \text{ for the radial and azimuthal components (}\beta \text{ is either } r\text{ or }\theta).
\end{cases}
\ee
The time discretization follows a combined Adams-Bashforth/Implicit Backward Differentiation method of the second order which consists of implicit evaluations of the linear terms and explicit evaluations of the nonlinear terms.
As an illustration, consider the equation
\be\label{C9}
\partial_t \phi=L(\phi)+N(\phi)+J\ , \
\ee
where, $L(\phi)$ and $N(\phi)$ stand for the linear and nonlinear terms of the differential equation, respectively, and $J$ accounts for either a forcing term or pressure gradient. Eq.~\eqref{C9} is discretized at regularly spaced time steps $t_n=n dt$ as
\be\label{C10}
\frac{3\phi^{(n+1)}-4\phi^{(n)}+\phi^{(n-1)}}{2dt}
=L\phi^{(n+1)}
+2N(\phi^{(n)})-N(\phi^{(n-1)})
+J^{(n+1)}\ , \ 
\ee
where $\phi^{(n)}=\phi(t_n)$. In the first time step, we set $\phi^{(-1)}=\phi^{(0)}$ and change $dt\rightarrow 3dt/2$, reducing the scheme to the usual Euler discretization.
The choice of discretizing the pressure gradient terms as $\partial_i P^{(n+1)}$ leads to a Stokes problem which is solvable by the Uzawa method \cite{uzawa1958iterative}. We note that similar discretization setups were successfully applied to the Navier-Stokes equations in cylindrical coordinates with a few different boundary conditions \cite{raspo2002spectral,lopez1998efficient,peres20123d}.

Applying the above discretization procedure to the system of Eqs.~\eqref{C1} supplemented by Dirichlet boundary conditions in the radial direction, one finds a system of coupled equations for each independent Fourier mode $k\in[N/2,N/2-1]$ for the forward and backward time integration of the fields,
\be\label{C11}
\begin{cases}
    \hat{L}_{r,k}\ket{v_{r,k}^n}=-\hat{D}^{PQ}\ket{P_k^n}+\ket{f_{r,k}^n} &\ , \ \in \Omega_s-\partial\Omega_s\ , \\
    \hat{L}_{r,k}\ket{v_{\theta,k}^n}=\ket{f_{\theta,k}^n} &\ , \ \in \Omega_s-\partial\Omega_s\ , \\
    \hat{L}_{z,k}\ket{v_{z,k}^n}=-ik\ket{P_k^n}+\ket{f_{z,k}^n} &\ , \ \in \Omega_s-\partial\Omega_s\ , \\
    \hat{R}^{-1}\hat{D}\hat{R}\ket{v_{r,k}^n}+ik\ket{v_{z,k}^n}=0 &\ , \ \in \Omega_s\ , \\
    (\ket{v_{r,k}^n},\ket{v_{\theta,k}^n},\ket{v_{z,k}^n})=0 &\ , \  \in \partial\Omega_s\ ,\
\end{cases}
\ee
and,
\be\label{C12}
\begin{cases}    \hat{L}_{r,k}\ket{p_{r,k}^n}=\hat{D}^{PQ}\ket{Q_k^n}+\ket{g_{r,k}^n} &\ , \ \in \Omega_s-\partial\Omega_s\ , \\
    \hat{L}_{r,k}\ket{p_{\theta,k}^n}=\ket{g_{\theta,k}^n} &\ , \ \in \Omega_s-\partial\Omega_s\ , \\
    \hat{L}_{z,k}\ket{p_{z,k}^n}=ik\ket{Q_k^n}+\ket{g_{z,k}^n} &\ , \ \in \Omega_s-\partial\Omega_s\ , \\
    \hat{R}^{-1}\hat{D}\hat{R}\ket{p_{r,k}^n}+ik\ket{p_{z,k}^n}=0 &\ , \ \in \Omega_s\ , \\
    (\ket{p_{r,k}^n},\ket{p_{\theta,k}^n},\ket{p_{z,k}^n})=0 &\ , \ \in \partial\Omega_s\ , \\
\end{cases}
\ee
where the ket notation represents the $M$-dimensional vector $\ket{\phi_{i,k}^n}=\{\phi_{i,jk}^n\}$ with $j\in [0,M-1]$, the domain $\Omega_s=\{j\in\mathbb{N}:j\in[0,M-1]\}$ and $\partial\Omega_s=\{0,M-1\}$. The $M\times M$ matrices $\hat{R}=\mathrm{diag}(r_0,r_1,\cdots,r_{M-1})$, $\hat{L}_{r,k}=(3/(2|dt|)+k^2)\hat{\mathbb{I}}-\hat{D}^2-\hat{R}^{-1}\hat{D}+\hat{R}^{-2}$ and $\hat{L}_{z,k}=\hat{L}_{r,k}-\hat{R}^{-2}$ can be efficiently inverted and stored in a pre-processing stage. The $M$ vectors $(\ket{f_{r,k}^n},\ket{f_{\theta,k}^n},\ket{f_{z,k}^n})$ and $(\ket{g_{r,k}^n},\ket{g_{\theta,k}^n},\ket{g_{z,k}^n})$ are, respectively, the explicit part of the discretized Eqs.~\eqref{C1}. For instance, in Eq.~\eqref{C10} one has $\ket{f_k^n}=(4\phi^{(n-1)}-\phi^{(n-2)}+2N(\phi^{(n-1)})-N(\phi^{(n-2)}))/3$.

The algebraic system defined by Eqs.~\eqref{C11} and \eqref{C12} has a unique solution for every $k\neq0$. The $\mathbb{P}_N-\mathbb{P}_{N-2}$ approximation prevents zero eigenvalues of the Uzawa operator\footnote{The Uzawa operator is obtained by solving formally the pressure field by setting the momentum equations into the incompressibility constraint, in this case, $\hat{Z}_k=\hat{R}^{-1}\hat{D}\hat{R}\hat{L}_r^{-1}\hat{D}^{PQ}-k^2\hat{L}_z^{-1}$.} for $k=0$ and avoids the requirement of prescribing boundary conditions to the pressure field. Indeed, a source of non-uniqueness related to the pressure fields is the fact that they are defined up to a constant. Thus, simple calculations show the unique consistent solution regarding the Dirichlet boundary conditions and incompressibility constraint is $\ket{v_{r,0}^{n}}=0$ for $k=0$.

The convolutions in Eq.~\eqref{C1} and the action integral (Eq.~\eqref{5.19}) can be efficiently computed using the explicit form of the Fourier transformed correlation $\tilde{\chi}$ and the inverted Chebyshev derivative matrix, 
\be\label{C13}
\mathcal{F}_z\big[(\tilde{\chi}\star p_{\beta})\big]_{jk}^{(n)}=(2\pi)^{3/2} L e^{-\frac{r_j^2}{2L^2}}
e^{-\frac{k^2L^2}{2}}\sum_{l}\left(\hat{D}^{-1}_{M/2,l}-\hat{D}^{-1}_{0,l}\right)U_{jl,\beta}
f^{(n)}_{l,\beta}\ , \
\ee
\be\label{C14}
S=(2\pi)^{7/2} L \sum_{n,k,l,j,\beta}e^{-\frac{k^2 L^2}{2}}
\left(\hat{D}^{-1}_{M/2,l}-\hat{D}^{-1}_{0,l}\right)\left(\hat{D}^{-1}_{M/2,j}-\hat{D}^{-1}_{0,j}\right)
f^{(n)}_{lk,\beta}
U_{jl,\beta}
\left(f^{(n)}_{jk,\beta}\right)^{\star}\ , \
\ee
where $\beta=(r,\theta,z)$, $f^{(n)}_{lk,\beta}=r_l \exp{(-r_l^2/2L^2)}\mathcal{F}_z[p_\beta]^{(n)}_{lk}$, $U_{jk,\beta}=I_0(r_jr_k/L^2)$ for $\beta=z$, and $U_{jk,\beta}=I_1(r_jr_k/L^2)$ for $\beta=(r,\theta)$, with $I_0,I_1$ being modified Bessel functions of the first kind. 

In order to validate the numerical method, we performed several numerical experiments of the linear instanton. The left panel of Fig.~\reffig{figC1} shows 
\be\label{C15}
E_\theta=\left|\frac{\hat{v}^{num}_\theta(r,z,t=0)-\hat{v}^{(1)}_\theta(r,z,t=0)}{\hat{v}^{(1)}_\theta(r,z,t=0)}\right|\ ,\
\ee
where the hat notation means the field is normalized by its maximum value and, $\hat{v}^{(1)}$ is obtained by the numerical integration of Eq.~\eqref{5.33} with standard trapezoidal techniques. We first note that the Dirichlet boundary condition has a relevant influence on the solution for $r \gtrsim 0.5$ (right panel of Fig.~\reffig{figC1}). This is not surprising though, since Eq.~\eqref{5.33} holds for unbounded domains, thus a slower decay is expected. Variations of $\pm 5\%$ in the peak position are observed depending on how far to the boundaries one sets $L$ and $R$. We stick to the value $L=1/\pi$ to minimize boundary effects and ease comparison with $\pi$-periodic functions.
\begin{figure}[t]
\hspace{0.0cm}
\center
\includegraphics[width=\textwidth]{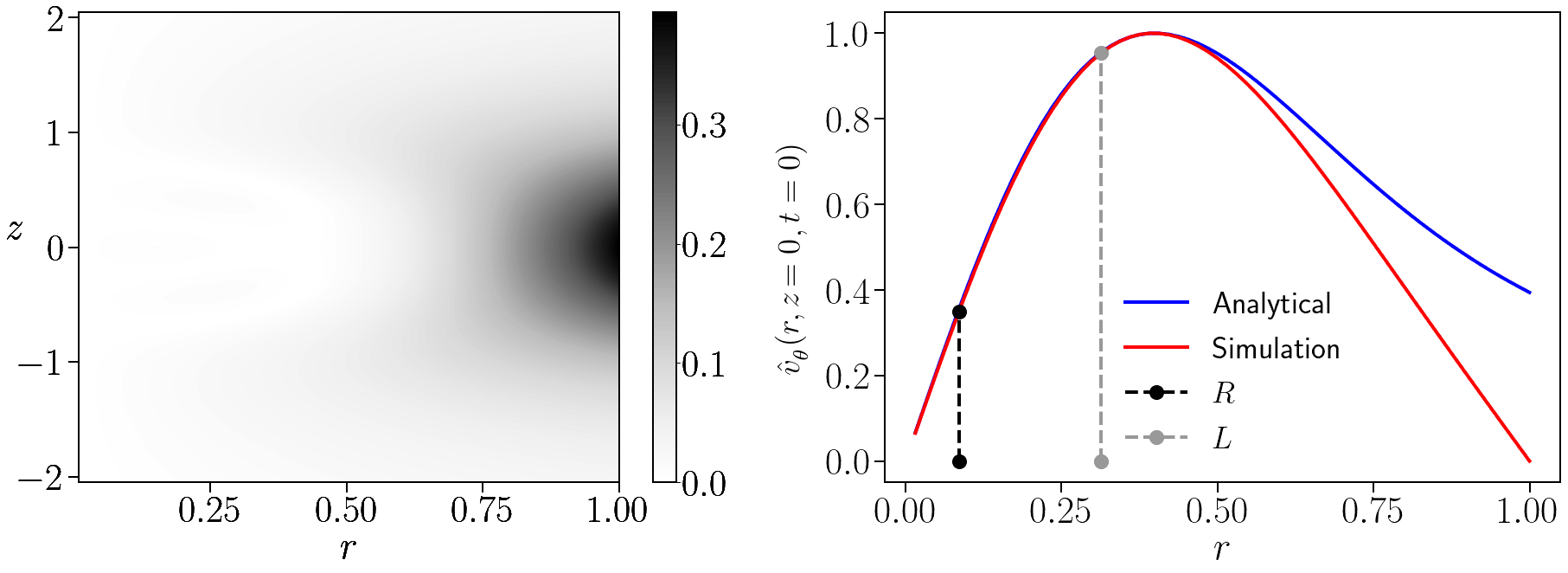}
\caption[Comparison between analytical and numerical viscous solution.]{Left panel shows Eq.~\eqref{C15} as functions of the space domain $r$ and $z$. The right panel shows normalized azimuthal velocity component for $z=0$ and $t=0$.}\label{figC1}
\end{figure}

The viscous action was calculated by means of Eq.~\eqref{C14}, and it was found to be compatible with a Gaussian shape $S=\Gamma^{\alpha_\Gamma}/(2\sigma^2)$ with $\alpha_\Gamma=1.99999993(5)$ and $\sigma=(\beta_\sigma/N) R^{1.95(2)}$ as expected. The errors were estimated by computing the action for different grid resolutions, both in time and space. Small variations of $\beta_\sigma$ are seen when $dt$ and/or $M$ are changed, but one must keep in mind that the radial collocation is not regular, and hence direct comparisons among different resolutions in $M$ are not perfect. Regarding the $dt$ dependence, finite time effects are in play, since decreasing $dt$ for a fixed number of total timesteps $N_t$ also causes the total simulation time to decrease, so the boundary conditions $p_i(\vec{x},T)=v_i(\vec{x},T)=0$ are effectively imposed on different time instants $T=-N_t dt$. In our tests, we worked with all combinations of $N_t=100$, $200$, and $400$, with $dt=0.0025$, $0.005$, and $0.01$.

The conclusion drawn from this set of numerical experiments is that both spatial and statistical properties are accurately captured by the numerical algorithm presented here, at least far from the boundaries. Furthermore, the schematic numerical discretization presented above does not cope with the particular parabolic behavior of the instanton equations, nor the numerical approach to solve the $\lambda$ constraint, those issues will be further discussed in the main text.

\end{chapter}


\end{document}


\title{SUPPLEMENTAL MATERIAL \\
~~~ \\
Vortex Gas Modeling of Turbulent Circulation Statistics}

\author{G.B. Apolinário$^{1}$, L. Moriconi$^{1}$, R.M. Pereira$^{2}$, and V.J. Valadão$^{1}$}
\affiliation{1 - Instituto de F\'\i sica, Universidade Federal do Rio de Janeiro, \\
C.P. 68528, CEP: 21945-970, Rio de Janeiro, RJ, Brazil}
\affiliation{2 - Instituto de Física, Universidade Federal Fluminense, 24210-346 Niterói, RJ, Brazil}

\begin{widetext}

\begin{appendix}

\section{SUPPLEMENTAL MATERIAL}

\section{Evaluation of the $A_n$ and $B_n$ Coefficients in Eq. (\ref{k4largeR})}

We obtain, from (\ref{p2}-\ref{fbm}), (\ref{circ4}), and taking $\Gamma_R$ in units of $\sqrt{\epsilon_0/(3 \nu)}$,
\bea
&&\langle \Gamma_R^2 \rangle = \langle \xi_R^2 \rangle \int_{\mathcal{D}} d^2 \bfr_1 \int_{\mathcal{D}} d^2 \bfr_2 \langle \tilde \omega (\bfr_1) \tilde \omega(\bfr_2)
\langle \sigma(\bfr_1) \sigma(\bfr_2) \rangle \nonumber \\
&&= \langle \xi_R^2 \rangle 
\int_{\mathcal{D}} d^2 \int_{\mathcal{D}} \bfr_1 d^2 \bfr_2 
\langle \tilde \omega (\bfr_1) \tilde \omega(\bfr_2)
(\bar \sigma \delta_{12} + \bar \sigma^2 ) \nonumber \\
&&= \langle \xi_R^2 \rangle  \left [ \bar \sigma  \pi R^2
+ \bar \sigma^2 
\int_{\mathcal{D}} d^2 \bfr_1 \int_{\mathcal{D}} d^2 \bfr_2 \langle \tilde \omega(\bfr_1) \tilde \omega(\bfr_2) \rangle \right ] \nonumber S1  
\eea
and
\bea
&&\langle \Gamma^4 \rangle = \left ( \frac{\epsilon_0}{3 \nu} \right)^2 (2 \pi \eta^2)^4 \int d^2 x_1 d^2 x_2 d^2 x_3 d^2 x_4 
\langle u(x_1) u(x_2) u(x_3) u(x_4) \rangle \times \nonumber \\
&&\times  \langle \xi(x_1) \xi(x_2) \xi(x_3) \xi(x_4) \rangle
\langle \sigma(x_1) \sigma(x_2) \sigma(x_3) \sigma(x_4) \rangle \nonumber \\
&&= \left ( \frac{\epsilon_0}{3 \nu} \right)^2 (2 \pi \eta^2)^4 \int d^2 x_1 d^2 x_2 d^2 x_3 d^2 x_4 
\langle u(x_1) u(x_2) u(x_3) u(x_4) \rangle  \langle \xi(x_1) \xi(x_2) \xi(x_3) \xi(x_4) \rangle \times \nonumber \\
&&\times 
[\bar \sigma^4 + 6 \bar \sigma^3 \delta^2_{12} + \bar \sigma \delta^2_{12} \delta^2_{13} \delta^2_{14}
+ 3 \bar \sigma^2 \delta^2_{12}\delta^2_{34} ] \nonumber \\
&&= \left ( \frac{\epsilon_0}{3 \nu} \right)^2 (2 \pi \eta^2)^4 \{ \bar \sigma \langle \xi^4 \rangle \pi R^2 + 3 \bar \sigma^2 
\langle \xi^2 \rangle^2 [ (\pi R^2)^2 + 2  (c^2/\delta^{2/3}) G R^2 ] + \nonumber \\
&& + 6 \langle \xi^2 \rangle \langle \xi \rangle^2  \bar \sigma^3 [\pi c E(4/3) R^{14/3} + 2 c^2  F R^{10/3}  ] + 
3\bar \sigma^4 \langle \xi \rangle^4 c^2 E^2(4/3) R^{16/3} \} \ . \ 
\eea

\section{Evaluation of the Kurtosis in the limit $\bar{R}\ll 1$}

Let's start from equation (3).

\be
\omega(\textbf{r})=\sqrt{\frac{\epsilon_0}{3\nu}}
\int_{\mathcal{R}^2} d^2 \textbf{r}' g_\eta(\textbf{r}-\textbf{r}') \xi(\textbf{r}')\tilde{\omega}(\textbf{r}')\sigma(\textbf{r}')
\ee

One would expect that the circulation is given by.

\be
\Gamma_R=\sqrt{\frac{\epsilon_0}{3\nu}}
\int_{\mathcal{D}} d^2 \textbf{r}\int_{\mathcal{R}^2} d^2 \textbf{r}' g_\eta(\textbf{r}-\textbf{r}') \xi(\textbf{r}')\tilde{\omega}(\textbf{r}')\sigma(\textbf{r}')
\ee

Region ${\mathcal{D}}$ is a circle of radius $R$, circulation's variance for slow varying $\xi(\textbf{r})$ can be written as follows.

\be
\langle \Gamma_R^2\rangle=
\frac{\epsilon_0}{3\nu}\langle \xi^2_R\rangle
\int_{\mathcal{D}}d^2\textbf{x}d^2\textbf{y}
\int_{\mathcal{R}^2}d^2\textbf{x}'d^2\textbf{y}'
g_\eta(\textbf{x}-\textbf{x}')g_\eta(\textbf{y}-\textbf{y}')
\langle\sigma(\textbf{x}')\sigma(\textbf{y}') \rangle
\langle\tilde{\omega}(\textbf{x}')\tilde{\omega}(\textbf{y}') \rangle
\ee

The exact form of  $\langle\tilde{\omega}(\textbf{x}')\tilde{\omega}(\textbf{y}') \rangle$ call it $I(\textbf{x}'-\textbf{y}')$ can be calculated with (7).

\be
I(\textbf{r})=2F1(4/6,7/6,1,-r^2/\eta_K^2)
\ee

The physical picture that we have to draw is the following, region ${\mathcal{D}}$ is much smaller than a circle of radius $\eta_K$, it means that the contribution of $I(\textbf{x}-\textbf{y})$ will be relevant when $\textbf{x}\approx \textbf{y}$, we can approximate the Gaussian packs too and perform the integral in a small region of radius $2\eta$, but its interesting to keep the packs and the integral in the whole plane in order to use all the stuff used in Gaussian convolutions by approximating.

\be
I(\textbf{r})\approx 1 -\frac{\textbf{r}^2}{2\Delta^2}+\mathcal{O}(\textbf{r}^4)\approx e^{-\frac{\textbf{r}^2}{2\Delta^2}}=g_\Delta(\textbf{r})
\ee

Where $\Delta^2=\alpha^2\eta_K^2$ and $\alpha^2=9/14$. Using relation (4) from the paper we get that the circulation's variance in units of $\epsilon_0\langle \xi^2_R \rangle/3\nu$ is written as follows.

\be
\langle \Gamma_R^2\rangle=
\int_{\mathcal{D}}d^2\textbf{x}d^2\textbf{y}
\int_{\mathcal{R}^2}d^2\textbf{x}'d^2\textbf{y}'
g_\eta(\textbf{x}-\textbf{x}')g_\eta(\textbf{y}-\textbf{y}')
g_\Delta(\textbf{x}'-\textbf{y}')
(\bar{\sigma}^2+\bar{\sigma}\delta^2(\textbf{x}'-\textbf{y}'))
\ee

Two contribution of $\bar{\sigma}$ are seen in the variance, the lowest order is the trivial one, second order give us corrections to the diluted gas model.

\be
\langle \Gamma_R^2\rangle=\bar{\sigma}^2
\int_{\mathcal{D}}d^2\textbf{x}d^2\textbf{y}
\int_{\mathcal{R}^2}d^2\textbf{x}'g_\eta(\textbf{x}-\textbf{x}')
\int_{\mathcal{R}^2}d^2\textbf{y}'
g_\eta(\textbf{y}-\textbf{y}')
g_\Delta(\textbf{x}'-\textbf{y}')
+\bar{\sigma}
\int_{\mathcal{D}}d^2\textbf{x}d^2\textbf{y}
\int_{\mathcal{R}^2}d^2\textbf{x}'
g_\eta(\textbf{x}-\textbf{x}')g_\eta(\textbf{y}-\textbf{x}')
\ee

\be
\langle \Gamma_R^2\rangle=\frac{2 \pi \bar{\sigma}^2 \Delta^2\eta^2}{\Delta^2+\eta^2}
\int_{\mathcal{D}}d^2\textbf{x}d^2\textbf{y}
\int_{\mathcal{R}^2}d^2\textbf{x}'
g_\eta(\textbf{x}-\textbf{x}')g_{\sqrt{\Delta^2+\eta^2}}(\textbf{x}'-\textbf{y})
+\pi \bar{\sigma}\eta^2
\int_{\mathcal{D}}d^2\textbf{x}d^2\textbf{y}
g_{\sqrt{2}\eta}(\textbf{x}-\textbf{y})
\ee

\be
\langle \Gamma_R^2\rangle=\frac{4 \pi^2 \bar{\sigma}^2 \Delta^2\eta^4}{\Delta^2+2\eta^2}
\int_{\mathcal{D}}d^2\textbf{x}d^2\textbf{y}
g_{\sqrt{\Delta^2+2\eta^2}}(\textbf{x}-\textbf{y})
+\pi \bar{\sigma}\eta^2
\int_{\mathcal{D}}d^2\textbf{x}d^2\textbf{y}
g_{\sqrt{2}\eta}(\textbf{x}-\textbf{y})
\ee

We can integrate in domain $\mathcal{D}$ up to the firs non trivial order in $R$, in this case is $R^4$.

\be
\langle \Gamma_R^2\rangle=\frac{4 \pi^4 \bar{\sigma}^2 \Delta^2\eta^4}{\Delta^2+2\eta^2}R^4
+\pi^3 \bar{\sigma}\eta^2 R^4
\ee

Recovering the dimensional units of circulation and using $R=\eta_K\bar{R}$.

\be
\langle \Gamma_R^2\rangle=\frac{\epsilon_0}{3\nu}\xi^2_0
\pi^3 \bar{\sigma} \eta^6_K a^2 \bar{R}^4 f_0(\alpha,a,\bar{\sigma},\eta_K)+\mathcal{O}(\bar{R}^6)
\ee

Where.

\be
f_0(\alpha,a,\bar{\sigma},\eta_K)=
1+4\pi\bar{\sigma}\eta^2_K\left(
\frac{a^2\alpha^2}{\alpha^2+2a^2}
\right)
\ee

In the last equation were used that $\eta=a\eta_K$. Now we will try the same approach to the fourth statistical moment of circulation, keeping second order in $\bar{\sigma}$, now we will keep the second non trivial order in $\bar{R}$ to evaluate the first non trivial approximation to the kurtosis as function of $\bar{R}$.

\be
\langle \Gamma_R^4\rangle=
\frac{\epsilon_0^2}{9\nu^2}\langle \xi^4_R\rangle
\int_{\mathcal{D}}d^2\textbf{x}_1d^2\textbf{x}_2
d^2\textbf{x}_3d^2\textbf{x}_4
\int_{\mathcal{R}^2}d^2\textbf{x}'_1d^2\textbf{x}'_2
d^2\textbf{x}'_3d^2\textbf{x}'_4
g_\eta^{1-1'}g_\eta^{2-2'}g_\eta^{3-3'}g_\eta^{4-4'}\times
\ee
\be\times
\langle
\sigma(\textbf{x}'_1)\sigma(\textbf{x}'_2)
\sigma(\textbf{x}'_3)\sigma(\textbf{x}'_4)
\rangle
\langle
\tilde{\omega}(\textbf{x}'_1)\tilde{\omega}(\textbf{x}'_2)
\tilde{\omega}(\textbf{x}'_3)\tilde{\omega}(\textbf{x}'_4)
\rangle
\ee

Density fourth order correlation up to second order in the mean density give us.

\be
\langle
\sigma(\textbf{x}'_1)\sigma(\textbf{x}'_2)
\sigma(\textbf{x}'_3)\sigma(\textbf{x}'_4)
\rangle=
\bar{\sigma}\delta_{12}'\delta_{13}'\delta_{14}'+
4\bar{\sigma}^2\delta_{12}'\delta_{13}'+
3\bar{\sigma}^2\delta_{12}'\delta_{34}'
\ee

As $\tilde{\omega}$ is Gaussian, we can use Wick's theorem to evaluate 4-point statistics in terms of 2-point statistics. We will show now the identities which will be usefull.

\be
\langle
\tilde{\omega}^4(\textbf{x}'_1)
\rangle=3\langle
\tilde{\omega}^2(\textbf{x}'_1)
\rangle=3
\ee

\be
\langle
\tilde{\omega}^3(\textbf{x}'_1)
\tilde{\omega}(\textbf{x}'_2)
\rangle=3\langle
\tilde{\omega}^2(\textbf{x}'_1)
\rangle
\langle
\tilde{\omega}(\textbf{x}'_1)
\tilde{\omega}(\textbf{x}'_4)
\rangle\approx 3g_{\Delta}(\textbf{r}'_1-\textbf{r}'_2)
\ee

\be
\langle
\tilde{\omega}^2(\textbf{x}'_1)
\tilde{\omega}^2(\textbf{x}'_2)
\rangle=
\langle
\tilde{\omega}^2(\textbf{x}'_1)
\rangle
\langle
\tilde{\omega}^2(\textbf{x}'_2)
\rangle
+
2\langle
\tilde{\omega}(\textbf{x}'_1)
\tilde{\omega}(\textbf{x}'_2)
\rangle^2
\approx 1+2g_{\Delta/\sqrt{2}}(\textbf{r}'_1-\textbf{r}'_2)
\ee

We can shaft the computation into 4 integrals, the whole flatness will be given by the following relation.

\be\label{19}
\langle \Gamma_R^4\rangle=
\frac{\epsilon_0^2}{9\nu^2}\langle \xi^4_R\rangle\left(
3\bar{\sigma} I_1+
3\bar{\sigma}^2 I_2+
6\bar{\sigma}^2 I_3+
12\bar{\sigma}^2 I_4
\right)
\ee

Where.

\be
I_1=\int_{\mathcal{D}}d^2\textbf{x}_1d^2\textbf{x}_2
d^2\textbf{x}_3d^2\textbf{x}_4
\int_{\mathcal{R}^2}d^2\textbf{x}'_1
g_\eta^{1-1'}g_\eta^{2-1'}g_\eta^{3-1'}g_\eta^{4-1'}
\ee

\be
I_2=\int_{\mathcal{D}}d^2\textbf{x}_1d^2\textbf{x}_2
d^2\textbf{x}_3 d^2\textbf{x}_4
\int_{\mathcal{R}^2}d^2\textbf{x}'_1d^2\textbf{x}'_2
g_\eta^{1-1'}g_\eta^{2-1'}g_\eta^{3-2'}g_\eta^{4-2'}
\ee

\be
I_3=\int_{\mathcal{D}}d^2\textbf{x}_1d^2\textbf{x}_2
d^2\textbf{x}_3 d^2\textbf{x}_4
\int_{\mathcal{R}^2}d^2\textbf{x}'_1d^2\textbf{x}'_2
g_\eta^{1-1'}g_\eta^{2-1'}g_\eta^{3-2'}g_\eta^{4-2'}g_{\Delta/\sqrt{2}}^{1'-2'}
\ee

\be
I_4=\int_{\mathcal{D}}d^2\textbf{x}_1d^2\textbf{x}_2
d^2\textbf{x}_3 d^2\textbf{x}_4
\int_{\mathcal{R}^2}d^2\textbf{x}'_1d^2\textbf{x}'_2
g_\eta^{1-1'}g_\eta^{2-1'}g_\eta^{3-1'}g_\eta^{4-2'}g_{\Delta}^{1'-2'}
\ee

Starting whit $I_1$ we get that.

\be
I_1=\frac{\pi \eta^2}{2}\int_{\mathcal{D}}d^2\textbf{x}_1d^2\textbf{x}_2
d^2\textbf{x}_3d^2\textbf{x}_4
e^{-\frac{3\sum_{i=1}^4 \textbf{r}^2_i+\mathcal{O}(\textbf{r}_i\cdot\textbf{r}_j)}{8\eta^2}}
\ee

Now we can expand the exponential up to $\mathcal{O}(\textbf{r}^2)$, one can argue that the scalar products do not contribute at least at this order, all those integral will be of the type $rr'\cos{(\theta-\theta')}$ which will vanish in integration around $2\pi$.

\be
I_1=\frac{\pi \eta^2}{2}\int_{\mathcal{D}}d^2\textbf{x}_1d^2\textbf{x}_2
d^2\textbf{x}_3d^2\textbf{x}_4
\left(1-\frac{3}{8\eta^2}\sum_{i=1}^4 \textbf{r}^2_i
\right)
\ee

\be
I_1=\frac{\pi \eta^2}{2}
\left(
(\pi R^2)^4-\frac{3}{2\eta^2}(\pi R^2)^3 \frac{\pi R^4}{2}
\right)=
\frac{\pi^5 \eta^2 R^8}{2}
\left(
1-\frac{3}{4}\frac{ R^2}{\eta^2}
\right)
\ee

The second integral can be but in the form.

\be
\sqrt{I_2}=\int_{\mathcal{D}}d^2\textbf{x}d^2\textbf{y}
\int_{\mathcal{R}^2}d^2\textbf{r}
g_\eta(\textbf{x}-\textbf{r})g_\eta(\textbf{y}-\textbf{r})
\ee

\be
\sqrt{I_2}=\pi \eta^2 \int_{\mathcal{D}}d^2\textbf{x}d^2\textbf{y}
g_{\sqrt{2}\eta}(\textbf{x}-\textbf{y})
\ee

\be
\sqrt{I_2}=\pi \eta^2 \left(
(\pi R^2)^2 -\frac{2}{4\eta^2}\pi R^2 \frac{\pi R^4}{2}
\right)
\ee

\be
I_2=\pi^6 \eta^4 R^8\left(
1-\frac{R^2}{2\eta^2}
\right)
\ee

For the third integral the thing is more complicated than the first one and the second one, first we will call $\Delta'=\Delta/\sqrt{2}$ to make the calculation simpler, integrating in $\textbf{r}'_2$ we get that.

\be
I_3=\frac{2\pi \eta^2 \Delta'^2}{\eta^2+2\Delta'^2}\int_{\mathcal{D}}d^2\textbf{x}_1d^2\textbf{x}_2
d^2\textbf{x}_3 d^2\textbf{x}_4
\int_{\mathcal{R}^2}d^2\textbf{x}'_1
g_\eta^{1-1'}g_\eta^{2-1'}
e^{-\frac{\Delta'^2(\textbf{r}_3-\textbf{r}_4)^2+
\eta^2(2\textbf{r}'^2_1-2\textbf{r}'_1\cdot\textbf{r}_3-2\textbf{r}'_1\cdot\textbf{r}_4+\textbf{r}^2_3+\textbf{r}^2_4)}
{2\eta^2(2\eta^2+\Delta'^2)}}
\ee

\be
I_3=\frac{\pi^2 \eta^4 \Delta'^2}{\eta^2+\Delta'^2}\int_{\mathcal{D}}d^2\textbf{x}_1d^2\textbf{x}_2
d^2\textbf{x}_3 d^2\textbf{x}_4
e^{-\frac{3\eta^2+2\Delta'^2}
{8\eta^2(\eta^2+\Delta'^2)}
\sum_{i=1}^4 \textbf{r}^2_i+\mathcal{O}(\textbf{r}_i\cdot\textbf{r}_j)}
\ee

\be
I_3=\frac{\pi^2 \eta^4 \Delta'^2}{\eta^2+\Delta'^2}\left(
(\pi R^2)^4
-\frac{3\eta^2+2\Delta'^2}
{8\eta^2(\eta^2+\Delta'^2)}
4(\pi R^2)^3 \frac{\pi R^4}{2}
\right)
\ee

\be
I_3=\frac{\eta^4 \Delta^2}{2\eta^2+\Delta^2}\pi^6 R^8\left(
1
-\left(\frac{3\eta^2+\Delta^2}
{2\eta^2+\Delta^2}\right)
\frac{R^2}{2\eta^2}
\right)
\ee

Now the last integral is then, $I_4$, integrating over $\textbf{r}'_2$ we get.

\be
I_4=\frac{2\pi \eta^2 \Delta^2}{\eta^2+\Delta^2}
\int_{\mathcal{D}}d^2\textbf{x}_1d^2\textbf{x}_2
d^2\textbf{x}_3 d^2\textbf{x}_4
\int_{\mathcal{R}^2}d^2\textbf{x'}_1
g_\eta^{1-1'}g_\eta^{2-1'}g_\eta^{3-1'}g_g_{\sqrt{\Delta^2+\eta^2}}^{4-1'}
\ee

\be
I_4=\frac{4\pi^2 \eta^4 \Delta^2}{4\eta^2+3\Delta^2}
\int_{\mathcal{D}}d^2\textbf{x}_1d^2\textbf{x}_2
d^2\textbf{x}_3 d^2\textbf{x}_4
e^{-\frac{3\eta^2+2\Delta'^2}
{2\eta^2(4\eta^2+3\Delta'^2)}
\sum_{i=1}^3 \textbf{r}^2_i}
e^{-\frac{3\eta^2}
{2\eta^2(4\eta^2+3\Delta'^2)}\textbf{r}^2_4
+\mathcal{O}(\textbf{r}_i\cdot\textbf{r}_j)}
\ee

\be
I_4=\frac{4\pi^2 \eta^4 \Delta^2}{4\eta^2+3\Delta^2}\left(
(\pi R^2)^4-\frac{3}{2}
\left(\frac{3\eta^2+2\Delta^2}
{4\eta^2+3\Delta^2}\right)
(\pi R^2)^3\frac{\pi R^4}{2\eta^2}
-\frac{3}{2}
\left(\frac{\eta^2}
{4\eta^2+3\Delta^2}\right)
(\pi R^2)^3\frac{\pi R^4}{2\eta^2}
\right)
\ee

\be
I_4=\frac{\eta^4 \Delta^2}{4\eta^2+3\Delta^2}4 \pi^6 R^8
\left(
1
-\frac{3}{2}\left(\frac{2\eta^2+\Delta^2}
{4\eta^2+3\Delta^2}\right)\frac{R^2}{\eta^2}
\right)
\ee

Using (\ref{19}), the relation between $\Delta$ and $\eta_K$ and $\eta$ and $\eta_K$ we get that.

\be
\langle \Gamma_R^4\rangle=
\frac{\epsilon_0^2}{9\nu^2}\langle \xi^4_R\rangle\pi^5 a^2 \eta^{10}_K \bar{R}^8
\bar{\sigma}
\left(
f_1(\alpha,a,\bar{\sigma},\eta_K)-f_2(\alpha,a,\bar{\sigma},\eta_K)\bar{R}^2
\right)+\mathcal{O}(\bar{R}^{12})
\ee

Where

\be
f_1(\alpha,a,\bar{\sigma},\eta_K)=\frac{3}{2}+\pi\bar{\sigma}\eta^2_K a^2\left(
3+\frac{48\alpha^2}{4a^2+3\alpha^2}+\frac{6\alpha^2}{2a^2+\alpha^2}
\right)
\ee

\be
f_2(\alpha,a,\bar{\sigma},\eta_K)=\frac{9}{8 a^2}+\pi\bar{\sigma}\eta^2_K \left(
\frac{3}{2}+
\frac{72\alpha^2(\alpha^2+2a^2)}{4a^2+3\alpha^2}+
\frac{3\alpha^2(\alpha^2+3a^2)}{2a^2+\alpha^2}
\right)
\ee

Kurtosis is given from the following equation.

\be
\frac{\langle \Gamma_R^4\rangle}{\langle \Gamma_R^2\rangle^2}=
\frac{\langle \xi_R^4\rangle}{\langle \xi_R^2\rangle^2}
\frac{f_0^{-2}}{\pi\bar{\sigma}\eta^2_K a^2}\left(
f_1-f_2\bar{R}^2
\right)
\ee

\be
\frac{\langle \Gamma_R^4\rangle}{\langle \Gamma_R^2\rangle^2}=
\frac{R_{\lambda}^{3\mu/2}}{15^{3\mu/4}}
\frac{f_0^{-2}}{\pi\bar{\sigma}\eta^2_K a^2}\left(
f_1-f_2\bar{R}^2
\right)
\ee

Constant $\mu \approx 0.17$ comes from the log normal model for $\xi$. For $\bar{R}\rightarrow 0$ circulation's variance must be proportional to vorticity's variance.

\be
\langle \Gamma_R^2\rangle=\frac{\epsilon_0}{3\nu}\xi^2_0
\pi^3 \bar{\sigma} \eta^6_K a^2 \bar{R}^4 f_0\approx \langle \omega^2\rangle(\pi R^2)^2
\ee

But we know that $\langle \omega^2\rangle=\epsilon_0/3\nu$, then we get.

\be
\xi^2_0
\pi \bar{\sigma} \eta^2_K a^2 f_0\approx1
\ee

\be
\xi^2_0\approx \frac{1}{\pi \bar{\sigma} \eta^2_K}
\frac{1}{a^2 f_0(\alpha,a,\bar{\sigma},\eta_K)}
\ee

Let's estimate how good is the diluted gas approximation, in the limit that $\bar{R}$ goes to zero, circulation's kurtosis must approach a value of $C_4 R_{\lambda}^{\alpha_4}$, $C_4\approx 1.16$ and $\alpha_4\approx 0.41$.

\be
\frac{\langle \Gamma_R^4\rangle}{\langle \Gamma_R^2\rangle^2}=
\frac{R_{\lambda}^{3\mu/2}}{15^{3\mu/4}}
\frac{f_0^{-2}f_1}{\pi\bar{\sigma}\eta^2_K a^2}\approx C_4 R_{\lambda}^{\alpha_4}
\ee

\be
\frac{f_0^{-2}f_1}{\pi\bar{\sigma}\eta^2_K a^2}=15^{-3\mu/4} C_4 R_{\lambda}^{\alpha_4-3\mu/2}
\ee

How does the number of particles in circle of radius $\eta_K$ change with the Reynolds number. One may recall that $f_0$ and $f_1$ are functions of the argument $x=\pi\bar{\sigma}\eta^2_K a^2$, we can try to invert the above relation.

\be\label{50}
\frac{f_0^{-2}f_1}{\pi\bar{\sigma}\eta^2_K a^2}=\frac{1}{x}\frac{1}{(1+c_1x)^2}
\left(\frac{3}{2}+c_2x\right)=F=0.82131 R_{\lambda}^{0.155}
\ee

Where

\be\label{51}
c_1=\frac{4\alpha^2 a^2}{\alpha^2+2a^2}
\ee

\be\label{52}
c_2=3+\frac{48\alpha^2}{4a^2+3\alpha^2}+\frac{6\alpha^2}{2a^2+\alpha^2}
\ee

From (\ref{50}) we get.

\be\label{53}
c_1^2 F x^3+2Fc_1 x^2+(F-c_2)x-\frac{3}{2}=0
\ee

Equation (\ref{53}) can be solved analytically if we keep the second order on density, therefore, we get two solutions from the Bhaskara equation, but there is a solution that gives negative density, which forces us to select the positive one.

\be\label{54}
\pi\bar{\sigma}\eta^2=\frac{1}{4c_1F}\left(
c_2-F+\sqrt{(c_2-F)^2+12c_1F}
\right)
\ee

\begin{figure}[H]
\hspace{0.0cm}
\center
\includegraphics[width=0.70\textwidth]{VortexGas-draft/surf.jpeg}
\caption{
$\pi \sigma \eta^2$ as functions of $a$ ($y$-axis) and $R_\lambda$ ($x$-axis). Green surface is the solution of (\ref{53}) with $c_1=c_2=0$, yellow surface is given by (\ref{54}) and the blue surface is the numerical solution of (\ref{53}) (this is the unique solution which lies in $0<x<1$).
}
\end{figure}

We note that the "real" solution (Blue) differs from the approximated one (Yellow), only for low $a$ ($a<5$). From (\ref{54}) we can calculate the minimal $R_\lambda$ that the number of vortex in a cell of radius $\eta$ is less than $1$.

\be
4c_1F+F-c_2=\sqrt{(c_2-F)^2+12c_1F}
\ee

\be
8c_1 F \left(
2c_1 F +F-c_2 +\frac{3}{2}
\right)=0
\ee

For $F \neq 0$ it means that $R_\lambda \neq 0$ we get that.

\be
F=\frac{2c_2-3}{4c_1+2}
\ee

Using (\ref{50}), (\ref{51}) and (\ref{52}) in the limit of $a\rightarrow \infty$ we get the "critical" Reynolds.

\be
R^c_\lambda\approx 58324
\ee

When $R_\lambda \geq R^c_\lambda$, for any positive value of $a$, the mean number of vortices in a circle of radius $\eta$ is less or equal to $1$.

\section{Effects of $I(\textbf{r})-g_\Delta(\textbf{r})$ on the evaluation of Circulation's statistical moments}

The substitution of the $\tilde{\omega}(\textbf{r})$'s 2-point correlation from hypergeometric functions to gaussian ones, allow us to calculate analytically the circulation's statistical moments in an easy way. Now we will try to see how those moments changes in higher orders approximations, let's work with the hypergeometric function.

\be
I(\textbf{r})=2F1(4/6,7/6,1,-r^2/\eta_K^2)=\sum_{n=0}^{\infty}(-1)^n\frac{ \Gamma(2/3+n) \Gamma(7/3+n)}
{\Gamma(2/3)\Gamma(7/3)\Gamma(1+n)^2} 
\left(\frac{r}{\eta_K}\right)^{2n}
\ee

As we saw in previous calculations, the following integrals comes naturally in the arguments.

\be\label{60}
A=
\int_{\mathcal{D}}d^2\textbf{x}d^2\textbf{y}
\int_{\mathcal{R}^2}d^2\textbf{x}'g_\eta(\textbf{x}-\textbf{x}')
\int_{\mathcal{R}^2}d^2\textbf{y}'
g_\eta(\textbf{y}-\textbf{y}')
I(\textbf{x}'-\textbf{y}')
\ee

Let's define.

\be
\delta A_n=\frac{1}{\eta_K^{2n}}
\int_{\mathcal{D}}d^2\textbf{x}d^2\textbf{y}
\int_{\mathcal{R}^2}d^2\textbf{x}'g_\eta(\textbf{x}-\textbf{x}')
\int_{\mathcal{R}^2}d^2\textbf{y}'
g_\eta(\textbf{y}-\textbf{y}')
|\textbf{x}'-\textbf{y}'|^{2n}
\ee

Rescaling all the lenghts of the problem by $\eta$, we have $g_\eta(\textbf{r}) \rightarrow g_1(\textbf{r}) := g(\textbf{r})$, and we get.

\be
\delta A_n=\frac{\eta^{2n+8}}{\eta_K^{2n}}
\int_{\mathcal{D}'}d^2\textbf{x}d^2\textbf{y}
\int_{\mathcal{R}^2}d^2\textbf{x}'
g(\textbf{x}-\textbf{x}')
\int_{\mathcal{R}^2}d^2\textbf{y}'
g(\textbf{y}-\textbf{y}')
|\textbf{x}'-\textbf{y}'|^{2n}
\ee

Shifting $\textbf{r}=\textbf{x}'-\textbf{y}'$ we do not generate any jacobian and defining $\textbf{r}'=\textbf{x}'-\textbf{y}$ we get the following equation.

\be
\delta A_n=\eta_K^{8}a^{2n+8}
\int_{\mathcal{D}'}d^2\textbf{x}d^2\textbf{y}
\int_{\mathcal{R}^2}d^2\textbf{x}'
g(\textbf{x}-\textbf{x}')
\int_{\mathcal{R}^2}d^2\textbf{r}\ 
g(\textbf{r}'-\textbf{r})
r^{2n}
\ee

We can integrate in $\textbf{r}$ to get.

\be\label{64}
\delta A_n=2 \pi \eta_K^{8}a^{2n+8} 2^n \Gamma(n+1)
\int_{\mathcal{D}'}d^2\textbf{x}d^2\textbf{y}
\int_{\mathcal{R}^2}d^2\textbf{x}'
g(\textbf{x}-\textbf{x}') \mathcal{L}^n\left(-\frac{|\textbf{x}'-\textbf{y}|^2}{2}\right)
\ee

Laguerre's function $\mathcal{L}^n(x)$ comes from the integration in \textbf{r}. Integral over $\textbf{x}'$ must be done carefully, let's call.

\be
F_n(\textbf{x},\textbf{y})=
\int_{\mathcal{R}^2}d^2\textbf{x}'
g(\textbf{x}-\textbf{x}') \mathcal{L}^n\left(-\frac{|\textbf{x}'-\textbf{y}|^2}{2}\right)
\ee

Shifting $\textbf{r}=\textbf{x}'-\textbf{y}$ we do not generate any jacobian and defining $\textbf{r}'=\textbf{x}-\textbf{y}$ we get the following equation.

\be
F_n(\textbf{x},\textbf{y})=
\int_{\mathcal{R}^2}d^2\textbf{r}
g(\textbf{r}'-\textbf{r}) \mathcal{L}^n\left(-\frac{r^2}{2}\right)=\int_0^{2\pi} d\theta \int_0^{\infty} dr r e^{-r^2/2}e^{-r'^2/2}
e^{r r' \cos{(\theta)}}
\mathcal{L}^n\left(-\frac{r^2}{2}\right)
\ee

\be\label{67}
F_n(\textbf{x},\textbf{y})=
2 \pi\int_0^{\infty} dr r e^{-r^2/2}e^{-r'^2/2}
I_0(rr')
\mathcal{L}^n\left(-\frac{r^2}{2}\right)
\ee

We can't go further with the equation (\ref{67}) but using that $R \ll \eta_K \approx \eta$ we can expand $e^{-r'^2/2}I_0(rr')$ in a power series in (\ref{64}). 

\be
\delta A_n=4 \pi^2 \eta_K^{8}a^{2n+8} 2^n \Gamma(n+1)
\int_{\mathcal{D}'}d^2\textbf{x}d^2\textbf{y}
\int_0^{\infty} dr r e^{-r^2/2}e^{-r'^2/2}
I_0(rr')
\mathcal{L}^n\left(-\frac{r^2}{2}\right)
\ee

Changing variables to normalized position center and normalized relative position, I meant:

\be
\textbf{R}_+=\frac{\textbf{x}+\textbf{y}}{\sqrt{2}}
\ee

\be
\textbf{R}_-=\frac{\textbf{x}-\textbf{y}}{\sqrt{2}}
\ee

The Jacobian of this transformation is unit (one might think that this transformation is a simple rotation), we identify $\textbf{R}_-=\textbf{r}'$ and one can note that the integral does not depend on $\textbf{R}_+$ which generates the area of the region $\mathcal{D}'$, $A(\mathcal{D}')=\pi R^2/\eta^2=\pi \bar{R}/a^2$ and we get that.

\be
\delta A_n=4 \pi^3 \eta_K^{8}a^{2n+6} 2^n \Gamma(n+1) \bar{R}^2
\int_0^{2\pi} d\theta \int_0^{\bar{R}/a} dr' r'
\int_0^{\infty} dr r e^{-r^2/2}e^{-r'^2/2}
I_0(rr')
\mathcal{L}^n\left(-\frac{r^2}{2}\right)
\ee

\be
\delta A_n=8 \pi^4 \eta_K^{8}a^{2n+6} 2^n \Gamma(n+1) \bar{R}^2
\int_0^{\infty} dr \ 
r e^{-r^2/2}
\mathcal{L}^n\left(-\frac{r^2}{2}\right)
\int_0^{\bar{R}/a} dr'
r' e^{-r'^2/2}  I_0(rr')
\ee

As said before, the latter integral lies in a really small interval where $\bar{R}/a \ll 1$, then we can expand the function inside the integral up to second non trivial order, which gives us.

\be
r' e^{-r'^2/2}  I_0(rr') = r'+\frac{(r^2-2)}{4}r'^3+\mathcal{O}(r'^5)
\ee

\be
\delta A_n=8 \pi^4 \eta_K^{8}a^{2n+6} 2^n \Gamma(n+1) \bar{R}^2
\int_0^{\infty} dr \ 
r e^{-r^2/2}
\mathcal{L}^n\left(-\frac{r^2}{2}\right)
\left(\frac{\bar{R}^2}{2a^2}+
\frac{(r^2-2)}{16}\frac{\bar{R}^4}{a^4}+
\mathcal{O}(\bar{R}^6)
\right)
\ee

Now, we get exactly the normalization factor of the Laguerre's function.

\be
\int_0^{\infty} dr \ 
r e^{-r^2/2}
\mathcal{L}^n\left(-\frac{r^2}{2}\right)=2^n
\ee

\be
\int_0^{\infty} dr \ 
r (r^2-2) e^{-r^2/2}
\mathcal{L}^n\left(-\frac{r^2}{2}\right)=n 2^n
\ee

Finally our full solution for the deviance of the theory is.

\be
\delta A_n=\delta A_{n,1}+\delta A_{n,2}+\mathcal{O}(\bar{R}^6)
\ee

\be
\delta A_{n,1}=4 \pi^4 \eta_K^{8}a^{2n+4} 2^{2n} \Gamma(n+1) \bar{R}^4
\ee

\be
\delta A_{n,2}=\frac{\pi^4}{2} \eta_K^{8}a^{2n+2} 2^{2n} n\Gamma(n+1) \bar{R}^6
\ee

The latter result shows that we are missing a lot of information when we approximate the hypergeometric functions to gaussian ones $I(\textbf{r})-g_{\Delta}(\textbf{r})=\mathcal{O}(|\textbf{r}|^4)$, in fact we are missing information even in the lowest order in $\bar{R}$ and $\bar{\sigma}$ for all orders in $|\textbf{r}|^n$ to $n\geq 4$. We can try to restore the exact solution for the equation (\ref{60}) function calculating the sum of $\delta A_n's$.

\be
A=\sum_{n=0}^{\infty}(-1)^n
\frac{ \Gamma(2/3+n) \Gamma(7/3+n)}
{\Gamma(2/3)\Gamma(7/3)\Gamma(1+n)^2} 
\delta A_n(\bar{R})
\ee

Truncating $A_n(\bar{R})$ up to $\mathcal{O}(\bar{R}^4)$ we get that.

\be
A=4 \pi^4 \eta_K^{8} a^4 \bar{R}^4
\sum_{n=0}^{\infty}(-1)^n
\frac{ \Gamma(2/3+n) \Gamma(7/3+n)}
{\Gamma(2/3)\Gamma(7/3)\Gamma(n+1)} 
(2a)^{2n}
\ee

The factor of $\Gamma(n+1)$ in $\delta A$ destroys the reconstruction of the Hypergeometric function. =/

\end{appendix}

\end{widetext}


\title{Supplemental Material: Circulation statistics and the mutually excluding behavior of vortical turbulent structures}

\author{L. Moriconi$^{1}$, R.M. Pereira$^{2}$\footnote{Corresponding author: e-mail}, and V.J. Valadão$^{1}$}
\affiliation{$^{1}$Instituto de F\'\i sica, Universidade Federal do Rio de Janeiro, \\
C.P. 68528, CEP: 21945-970, Rio de Janeiro, RJ, Brazil}
\affiliation{$^{2}$Instituto de Física, Universidade Federal Fluminense, 24210-346 Niterói, RJ, Brazil}

\maketitle

\clearpage

\section{Density corrections for the point vortex model}
\renewcommand{\theequation}{S\arabic{equation}}

In order to motivate the introduction of interaction among vortices, consider the point vortex model put forward in \cite{PRE20} where the velocity circulation is carried out in a circular domain of radius $R$ due to analytical simplicity,
\be
\Gamma (\mathcal{D}) =\xi_\mathcal{D} \int_{\mathcal{D}} d^2 \textbf{r}\int_\gamma d^2 \textbf{r}' g_\eta(\textbf{r}-\textbf{r}')\tilde{\omega}(\textbf{r}')\sigma(\textbf{r}') \label{gamma_3} \ . \
\ee
where $g_\eta(\bfr)$, $\tilde \omega(\bfr)$, $\xi_\mathcal{D}$, $\gamma$ and $\eta$ have exactly the same meaning as in the main text. Interested in the asymptotic region $R/\eta_K\ll1$ we assume the first few terms of the gaussian field correlation to be the most relevant to the calculations,
\be\label{corr}
\langle\tilde{\omega}(\textbf{x})\tilde{\omega}(\textbf{y}) \rangle\approx 1 -\frac{|\textbf{x}-\textbf{y}|^2}{2\Delta^2}+\mathcal{O}(|\textbf{x}-\textbf{y}|^4)\approx e^{-\frac{|\textbf{x}-\textbf{y}|^2}{2\Delta^2}}=g_\Delta(|\textbf{x}-\textbf{y}|),
\ee
where $\Delta^2=\alpha^2\eta_K^2$ with $\alpha^2=9/14$. Employing the dilute gas approximation where $\sigma (\bfr) = \bar \sigma + \phi(\bfr) $, the statistic of density fluctuation can be built as a Poisson process up to fourth-order,
\bea
&&\langle \phi(\bfr_1 ) \phi(\bfr_2) \rangle = \bar \sigma \delta_{12}, \\
&&\langle \phi(\bfr_1 ) \phi(\bfr_2) \phi(\bfr_3) \rangle = \bar \sigma \delta_{12} \delta_{13}, \\
&&\langle \phi(\bfr_1 ) \phi(\bfr_2) \phi(\bfr_3) \phi(\bfr_4) \rangle = \bar \sigma \delta_{12}\delta_{13}\delta_{14} + \bar \sigma^2 ( \delta_{12} \delta_ {34} + \delta_{13} \delta_ {24} + \delta_{14} \delta_ {23} ),
\eea
where $\delta_{ij}\equiv\delta^3(\textbf{x}_i-\textbf{x}_j)$, contributions to circulation's variance, up to second order in $\bar\sigma$, are then
\be\label{varapp}
\langle \Gamma^2(\mathcal{D})\rangle=\xi^2_0
\ \lr{\bar{\sigma} \pi\eta^2_K a^2}
\ \lr{\pi R^2}^2
\ \lr{f_0-f_1\frac{R^2}{\eta^2_K}}
+\mathcal{O}\lr{\frac{R}{\eta_K}}^{8},
\ee
where we defined $\xi_0^2\equiv \langle\xi_\mathcal{D}^2\rangle$, and
\be
f_0=
1+4N_\eta\left(
\frac{\alpha^2}{\alpha^2+2a^2}
\right)
\ee
and
\be
f_1=\frac{1}{4 a^2}+2N_\eta
\left(\frac{\alpha^2}{(\alpha^2+2a^2)^2}\right)
,
\ee
where $N_\eta=\bar{\sigma} \pi\eta^2_K a^2$ is the mean number of vortices in a disk of radius $\eta=a\eta_K$, in this sense, the above expression can be regarded as a virial corrections of the statistical moments of circulations, which are supposed to be small if the gas is sufficiently dilute. After a long string of calculations using Wick's theorem to write higher order correlations of the gaussian field $\tilde \omega(\bfr)$ as combinations of (\ref{corr}), the fourth order circulation's moment,
\be
\langle \Gamma^4(\mathcal{D})\rangle=
\langle \xi^4_\mathcal{D}\rangle N_\eta \lr{\pi R^2}^4
\left(
f_2-f_3\frac{R^2}{\eta_K^2}
\right)+\mathcal{O}\lr{\frac{R}{\eta_K}}^{12},
\ee
where,
\be
f_2=\frac{3}{2}+N_\eta\left(
3+\frac{48\alpha^2}{4a^2+3\alpha^2}+\frac{6\alpha^2}{2a^2+\alpha^2}
\right),
\ee
and
\be
f_3=\frac{9}{8 a^2}+\frac{N_\eta}{a^2}\left(
\frac{3}{2}+72\alpha^2
\frac{\alpha^2+2a^2}{(4a^2+3\alpha^2)^2}+
6\alpha^2
\frac{\alpha^2+3a^2}{(2a^2+\alpha^2)^2}
\right),
\ee
Firstly, we fix $\xi_0$ by taking the limit of the circulation's variance, $\lim_{R\rightarrow0}\langle\Gamma^2_C\rangle=\langle \omega^2\rangle A^2=\pi^2R^4$ in Eq. (\ref{varapp}), $\xi^{-2}_0=N_\eta f_0$. Secondly, we examine circulation's flatness in order to fix the mean density as functions of the Reynolds number,
\be\label{kurtapp}
\frac{\langle \Gamma^4(\mathcal{D})\rangle}{\langle \Gamma^2(\mathcal{D})\rangle^2}\approx
\frac{\langle \xi_\mathcal{D}^4\rangle}{\langle \xi_\mathcal{D}^2\rangle^2}
\frac{f_0^{-2}f_2}{N_\eta}\left[
1-\left(\frac{f_3}{f_2}-2\frac{f_1}{f_0}\right)\left(\frac{R}{\eta_K}\right)^2
\right]+\mathcal{O}\left(\frac{R}{\eta_K}\right)^{4}
\ee
where the flatness of the $\xi_\mathcal{D}$ field is modeled through the OK62 framework, leading to a Reynolds number dependency $\langle \xi_\mathcal{D}^4\rangle/\langle \xi_\mathcal{D}^2\rangle^2=(R_\lambda/\sqrt{15})^{3\mu/2}$, where $R_\lambda$ is the Taylor based Reynolds number and the intermittency coefficient $\mu= 0.17\pm0.01$ \cite{tang_etal}. An empirical expression for the asymptotic limit of circulation's flatness is reported in \cite{Iyer_etal}, with a power law behavior $\lim_{R\ll\eta_K}F(R)\approx C_4 R_{\lambda}^{\alpha_4}$ with $C_4\approx 1.16$ and $\alpha_4\approx 0.41$, then, using Eq. (\ref{kurtapp})
\be
\frac{f_2}{f_0^{2}N_\eta}=15^{3\mu/4} C_4 R_{\lambda}^{\alpha_4-3\mu/2}\equiv F_0,
\ee
which can be stated as the roots of a cubic polynomial function of $N_\eta$, since $f_0$ and $f_2$ were linearly approximated,
\be\label{poly}
F_0c_1^2N_\eta^3+2F_0c_1N_\eta^2+(F_0-c_2)N_\eta-\frac{3}{2}=0
\ee
where
\be\label{c1}
c_1=\frac{4\alpha^2}{\alpha^2+2a^2},
\ee
and,
\be\label{c2}
c_2=3+\frac{48\alpha^2}{4a^2+3\alpha^2}+\frac{6\alpha^2}{2a^2+\alpha^2},
\ee
results of \cite{PRE20} are recovered by setting $c_1=c_2=0$ and solving the linear equation $N_\eta F_0=3/2$ to find the relation among $\bar{\sigma}$, $a$ and $R_\lambda$. 

Lastly, the expression $f_3/f_2-2f_1/f_0$ controls the curvature of the parabolic approximation for the flatness, which is indeed highly sensitive to the value of the parameter $a$. Simple numerical computation of the unique positive real solution of Eq. (\ref{poly}) shows a huge increase in $\bar{\sigma}$ of about $5.5$ times higher than first order solution, which is indeed unreal. An estimation of the mean intervortex distance is $\bar{\sigma}^{-1/2}\approx1.57\eta$ for the lowest Reynolds number $(R_\lambda=240)$, overlaping structures start to play a role and the dilute gas approximation must be taken carefully. Moreover, a decreased pack width parameter $a$ is observed, going from $a=3.3$ to $a\approx2.6$. This is a very sensitive point since the corrections are supposed to increase the value of $a$ accordingly to the vortex radii distribution showed in the main text having a mean value around $a=3.85$ and a maximum close to $a=3.1$. 

The latter remarks motivates the investigation of different mechanisms for preventing overrated intermittency to the point vortex model at small scales. The natural approach to the problem is indeed to address interactions among vortices since any finite range energy is negligible in the dilute approximation, i.e. high Reynolds number, the point vortex model can be seen as an asymptotic Reynolds approximation.

